\documentclass[preprint,11pt]{elsarticle}

\usepackage{lineno}
\modulolinenumbers[5]

\usepackage{charter}
\usepackage{wrapfig}
\usepackage{arydshln}
\renewcommand{\arraystretch}{1.3}

\usepackage[colorlinks,citecolor=blue,linktoc=all,linkcolor=cyan]{hyperref}
\usepackage{graphicx}
\usepackage[usenames]{color}

\usepackage[T1]{fontenc}
\usepackage{dsfont}               
\usepackage{mathrsfs}             
\usepackage{slashed}              
\usepackage[utf8]{inputenc}
\usepackage{amsmath}
\usepackage{amssymb}
\usepackage{amsbsy}
\usepackage{amsfonts}
\usepackage{booktabs}

\usepackage{tabularx}
\newcolumntype{R}{>{\raggedleft\arraybackslash}X} 
\newcolumntype{L}{>{\raggedright\arraybackslash}X}
\newcolumntype{C}{>{\centering\arraybackslash}X}  

\usepackage{nicefrac}
\usepackage{upgreek}
\usepackage{esint}
\usepackage{multirow}

\usepackage{import}

\numberwithin{equation}{section}
\numberwithin{table}{section}
\numberwithin{figure}{section}

\biboptions{numbers,sort&compress}

    \usepackage{sectsty}
    \usepackage{calc}

\usepackage[usenames]{color}
\usepackage{colortbl}

\definecolor{violet}{RGB}{111,0,255}
\definecolor{webgreen}{rgb}{0,0.75,0}
\definecolor{webred}{rgb}{0.75,0,0}
\definecolor{webblue}{rgb}{0,0,0.75}
\definecolor{darkblue}{rgb}{0,0,0.6}
\definecolor{darkgreen}{rgb}{0,0.5,0.5}
\definecolor{darkpurple}{rgb}{0.5,0,0.5}
\definecolor{darkorange}{rgb}{1,0.5,0}
\definecolor{darkgrey}{rgb}{0.4,0.4,0.4}
\definecolor{lgray}{rgb}{0.95,0.95,0.95}
\definecolor{lgreen}{rgb}{0.95,1.00,0.90}
\definecolor{lred}{rgb}{1.00,0.90,0.80}
\definecolor{lblue}{rgb}{0.2,0.35,1.00}
\definecolor{shadecolor}{rgb}{1.00,0.92,0.82}

\def\longlongrightarrow{\relbar\joinrel\relbar\joinrel\relbar\joinrel\relbar\joinrel\rightarrow}
\def\longlonglongrightarrow{\relbar\joinrel\relbar\joinrel\relbar\joinrel\relbar\joinrel\relbar\joinrel\relbar\joinrel\rightarrow}
\def\longlonglonglongrightarrow{\relbar\joinrel\relbar\joinrel\relbar\joinrel\relbar\joinrel\relbar\joinrel\relbar\joinrel\relbar\joinrel\relbar\joinrel\relbar\joinrel\relbar\joinrel\relbar\joinrel\rightarrow}

\usepackage[vcentermath]{youngtab}

\usepackage{enumitem}
\setitemize{noitemsep,topsep=0ex,parsep=0.25ex,partopsep=0pt,leftmargin=20pt}
\setenumerate{topsep=0pt,parsep=0pt,partopsep=0pt} 

\usepackage{bm}
\newcommand{\vect}[1]{\bm{#1}}

\newcommand{\mc}[1]{\mathcal{#1}}
\newcommand{\conjg}[1]{\ensuremath{\hspace{1pt}\overline{\hspace{-1pt}#1\hspace{-1pt}}}\hspace{1pt}}
\def\p{\partial}

\def\Slash#1{\setbox0=\hbox{$#1$} 
\dimen0=\wd0 
\setbox1=\hbox{/} \dimen1=\wd1 
\ifdim\dimen0>\dimen1 
\rlap{\hbox to \dimen0{\hfil/\hfil}} 
#1 
\else 
\rlap{\hbox to \dimen1{\hfil$#1$\hfil}} 
/ 
\fi}

\usepackage{titlesec}
\usepackage{sectsty}
\titleformat{\section}{\normalfont\Large\bfseries}{\thesection}{1em}{}
\titleformat{\subsection}{\normalfont\large\bfseries}{\thesubsection}{1em}{}
\titleformat{\subsubsection}{\normalfont\normalsize\bfseries}{\thesubsubsection}{1em}{}

\newcommand{\beq}{\begin{equation}}
\newcommand{\eeq}{\end{equation}}
\newcommand{\beqa}{\begin{align}}
\newcommand{\eeqa}{\end{align}}

\def\Slash#1{\setbox0=\hbox{$#1$} 
\dimen0=\wd0 
\setbox1=\hbox{/} \dimen1=\wd1 
\ifdim\dimen0>\dimen1 
\rlap{\hbox to \dimen0{\hfil/\hfil}} 
#1 
\else 
\rlap{\hbox to \dimen1{\hfil$#1$\hfil}} 
/ 
\fi}

\newcommand{\kslash}{\Slash{k}}

\begin{document}

\journal{Progress in Particle and Nuclear Physics}

\begin{frontmatter}

\title{Baryons as relativistic three-quark bound states}

\author[L2]{Gernot~Eichmann~\fnref{myfootnote2}}
\author[L1]{H\`elios~Sanchis-Alepuz~\fnref{myfootnote5}}
\author[L2]{Richard~Williams~\fnref{myfootnote4}}
\author[L1]{Reinhard Alkofer~\fnref{myfootnote1}}
\author[L2,L3]{Christian~S.~Fischer~\fnref{myfootnote3}}

\address[L2]{Institut f\"ur Theoretische Physik, Justus-Liebig--Universit\"at Giessen, 35392 Giessen, Germany}
\address[L1]{Institute of Physics, NAWI Graz, University of Graz,
Universit\"atsplatz 5, 8010 Graz, Austria}
\address[L3]{HIC for FAIR Giessen, 35392 Giessen, Germany}

\fntext[myfootnote2]{gernot.eichmann@physik.uni-giessen.de}
\fntext[myfootnote4]{richard.williams@physik.uni-giessen.de.}
\fntext[myfootnote5]{helios.sanchis-alepuz@uni-graz.at}
\fntext[myfootnote1]{reinhard.alkofer@uni-graz.at}
\fntext[myfootnote3]{christian.fischer@physik.uni-giessen.de}

\begin{abstract}
We review the spectrum and electromagnetic properties of baryons described as
relativistic three-quark bound states within QCD.
The composite nature of baryons results in a rich excitation spectrum,
whilst leading to highly non-trivial structural properties explored by the coupling to external (electromagnetic and other) currents.
Both present many unsolved problems despite decades of experimental and theoretical research.
We discuss the progress in these fields from a theoretical perspective, focusing on nonperturbative QCD as encoded
in the functional approach via Dyson-Schwinger and Bethe-Salpeter equations.
We give a systematic overview as to how results are obtained in this framework and explain technical connections to lattice QCD.
We also discuss the mutual relations to the quark model, which still serves as a reference to distinguish `expected' from `unexpected' physics.
We confront recent results on the spectrum of non-strange and strange baryons, their form factors and the issues of two-photon processes and
Compton scattering determined in the Dyson-Schwinger framework with those of lattice QCD and the available experimental data. The general
aim is to identify the underlying physical mechanisms behind the plethora of observable phenomena in terms of the underlying quark and gluon
degrees of freedom.
\end{abstract}


\end{frontmatter}

\newpage

\thispagestyle{empty}
\tableofcontents


%
%
\newpage

\section{Introduction}\label{intro}
Baryons make up most of the visible mass of the universe. They are highly nontrivial objects governed by the
strong interaction. Their complicated internal structure is far from understood and even supposedly trivial
properties like the charge radius of the proton pose disturbing puzzles~\cite{Carlson:2015jba}.
Historically, baryons have been the focus of both experimental and theoretical interest long before the
dawn of Quantum Chromodynamics (QCD) and will likely continue to be so for many decades to come.

Within the naive but indisputably successful quark model, baryons consist of three massive constituent quarks
that interact with each other by a mean-field type potential. In such a phenomenological picture, QCD's gluons
have been `integrated out' and dissolved within the parameters of the model for the strong~forces between
the quarks. The constituent quarks are much heavier than the current quarks appearing in the renormalized
Lagrangian of QCD\@, and the reason for this discrepancy -- dynamical chiral symmetry breaking -- cannot be described
within the quark model but is subject to the underlying QCD\@. Thus, while the simple quark model proved
surprisingly successful in the past, it is generally accepted that it is far from the full story.

The reasons for this are well known. For one, dynamics is an issue. Light quarks are genuinely relativistic
objects and their dynamics in principle cannot be fully captured by non-relativistic approximations and leading-order
corrections. This is already important for the spectrum but becomes even more apparent when form factors
or scattering processes probe the internal baryon dynamics at medium to large momentum transfer. Then there are
structural reasons. Some of the experimentally observed states with baryon number $B=\pm1$ simply may not
resemble three-quark states. Instead, they may be hybrids, i.e.\ states with quantum numbers generated by three
valence quarks plus an additional gluonic excitation, or they may be pentaquark states, i.e.\ states with three
valence quarks and an additional valence quark-antiquark pair. Especially the latter possibility has generated
a tremendous amount of interest over the years with many renewed activities after the recent report of an
experimental signature of a potential pentaquark at LHCb~\cite{Aaij:2015tga}. In addition, well-established states
such as the $\Lambda(1405)$ have been discussed to contain a strong `penta' $\bar{K}N$-component that may be responsible
for its unexpectedly low mass.

The latter case also shows the importance of the quark model as a standard by which one may define and distinguish the
`expected' from the `unexpected'. Despite its shortcomings, the quark model has dominated the theoretical
discussion of baryons for a long time and generated a list of `standard phenomenological problems' which
have been extensively discussed in the literature, see e.g.~\cite{Capstick:2000qj,Klempt:2009pi} for reviews.
Amongst these are: (i) The problem of missing resonances, which may be defined as states that are predicted by (symmetric) quark
models but have not yet been identified in experiments.
(ii) The question of three-quark vs.\ quark-diquark states, which is somewhat related and often discussed
within the framework of point-like diquarks. The corresponding quark-diquark states then show a clearly different
(and less overpopulated) spectrum than its three-quark counterpart. However, even such a reduced spectrum has
not been fully seen in experiments so far.
(iii) The role of meson cloud effects and corresponding meson-exchange forces between quarks on the structure
and the dynamical properties of baryons, which are visible at small momentum transfer and for small quark masses.

From a theoretical perspective it is highly desirable to bridge the gap between QCD and baryon phenomenology. In
the quark model language this entails understanding in detail how masses of constituent
quarks are generated by dynamical chiral symmetry breaking. To some extent the matter has been clarified already. In
fact, as we will see in Sec.~\ref{spec}, the answer to the question relates the small and large momentum
region and leads to a natural and close connection between the notions of `constituent' and `current' quarks.
Furthermore, the nature of `effective' forces between the quarks encoded in the quark potential models needs
to be inferred from the underlying QCD\@. In the heavy quark sector, this is a task partly solved by heavy quark
effective theory, see e.g.~\cite{Brambilla:2010cs}. In the light quark sector no such mapping exists due to
the problems with relativity as discussed above. Thus the question has to be reformulated in a relativistic context
and remains an interesting, important and open problem which we will discuss in more detail in the main body
of this review.

Another point of interest is the role that confinement plays in the structure and spectrum of baryons.
Confinement, together with unitarity, leads as one of its consequences to quark-hadron duality, for a review see e.g.\ Ref.~\cite{Melnitchouk:2005zr}.
Beyond the absence of coloured states in the physical spectrum this duality is the clearest experimental signature for confinement,
with the perfect orthogonality of the quark-glue versus the hadronic states providing an attractive way to express the subject.
As beautiful as this result might be it leaves us with a perplexing consequence:  confinement is so efficient that the determination of hadronic observables alone does not allow definite conclusions on the physical mechanism behind it.
If a purely hadronic ``language'' is sufficient to describe all of hadronic physics, then there is no \emph{a priori} way to
access the QCD degrees of freedom on purely observational grounds without any input from theory.
An archetype to picture confinement is the linear rising potential found in pure Yang-Mills theory and
its relation to the spinning stick~\cite{Greensite:2003bk,Alkofer:2006fu}.
In QCD, however, string breaking causes any interquark potentials to flatten out at large distances,
thus leaving only remnants of the linear behaviour in the intermediary distance region.
Furthermore, although this property may be relevant for heavy quarks, in the light quark sector the
relativistic dynamics of fast moving quarks renders a description in terms of a potential questionable at the least.

In order to generate a more fundamental understanding of the static and dynamic properties of hadrons,
sophisticated theoretical tools are necessary. These have to be genuinely nonperturbative in nature in
order to be able to deal directly with QCD\@. Lattice QCD is a prime candidate in this respect and has seen
several decades of continuous advances. These have improved upon the framework of quenched QCD, with its suppressed
quark loop effects, towards contemporary fully dynamical simulations of a range of observables at or close to
physical pion masses. With a potentially bright future ahead, these simulations are today mostly restricted to
ground states and static quantities, whereas the investigation of excited states and form factors at the physical point
remains a challenge.
On the other hand, there is certainly a need for well-founded continuum methods that are capable to shed light on
the qualitative physical mechanisms that drive baryon phenomenology. Functional methods like the Dyson-Schwinger- (DSE),
Bethe-Salpeter- (BSE) and Faddeev-equation approach have been developed in the past years beyond the quark-diquark
approximation. They have been used to determine static as well as dynamical information on baryons in terms of quark
and gluon $n$-point functions. On a fundamental level, these calculations are restricted by truncation assumptions which
can, however, be evaluated and checked in a systematic manner.
Thus, in principle, both lattice QCD and the functional framework are capable of delivering answers to many of
the questions posed above. To what extent this promise is already realised in the available literature is the subject of the
present review.
We will try to elucidate upon the inner workings of these frameworks without being too technical, so
that the non-expert reader may appreciate the individual strengths and the complementarity of these approaches.
To this
end we will also highlight the interrelations of these frameworks with each other and discuss their agreement with
experimental results.

Of course, we also have our personal views on the subject; we tried to earmark these clearly when they occur in the
text in order to distinguish them from generally accepted positions in the community. Furthermore, a review of this size
cannot be complete, and the choice of material reflects our personal interests. Many equally interesting topics cannot be
properly done justice within the given amount of space and time. In particular we did not touch upon important subjects
such as parton distribution functions and the transverse momentum structure of baryons, the proton spin puzzle or intrinsic heavy quark components.

The review is organized as follows. In Sec.~\ref{exp} we give a short overview on the status of the experimental
identification of the baryon spectrum and discuss open problems, explain methods and techniques used in the analysis
of experimental data and give a very brief survey of existing and planned experiments.
In Sec.~\ref{spec} we focus on the spectrum of ground and excited state baryons.
Therein we give a general and method-independent discussion of QCD and its correlators.
We explain the techniques used in functional methods to extract baryon masses and contrast them
with the corresponding ones in lattice QCD.
We carefully discuss the approximations involved in both approaches with particular attention on the physics aspects behind
the truncation schemes applied in the functional approach.
We compare results from both approaches with the experimental spectrum and discuss successes and open questions.
In Sec.~\ref{form} we then proceed to baryon form factors. Again, we first discuss method-independent properties
of the corresponding correlators of QCD and then explain the techniques used
to calculate form factors. We discuss the state of the art of quark model,
lattice QCD and functional method calculations
for form factors and relate the individual strengths and potential drawbacks of the different methods with each other.
We then discuss the electromagnetic and axial structure of a selection of different baryons in turn. Sec.~\ref{compton}
focuses on the general framework that is necessary to extract results for two-photon processes and other scattering amplitudes in the functional approach.
We review the model-independent relation of the hadronic and the quark-level description of these processes and discuss
the current progress towards a description of Compton scattering with functional methods. We conclude the review with a brief
outlook in Sec.~\ref{sec:summary}.


\newpage

\section{Experimental overview}\label{exp}

\subsection{The nucleon and its resonances}\label{sec:exp-nucleon-res}
The proton is the only truly stable hadron and as such it is
an ubiquitous ingredient to hadron structure experiments: from elastic and deep inelastic $ep$ scattering
to $pp$ and $p\bar{p}$ reactions, $N\pi$ scattering, pion photo- and electroproduction, nucleon Compton scattering and more;
even searches for physics beyond the Standard Model are typically performed on protons and nuclei.
To say that we have understood the structure of the nucleon, 55 years after R. Hofstadter won the Nobel prize for discovering its non-pointlike nature,
would be a gross overstatement in light of, for example, the recent proton
radius puzzle.
The nucleon is neither round nor simple but rather a complicated conglomerate of quarks and gluons,
and it is the complexity of their interaction that encodes yet unresolved phenomena such as confinement and spontaneous chiral symmetry breaking.
In addition, experiments typically create resonances too -- either meson resonances in $p\bar{p}$ annihilation or baryon resonances in most of the other reactions listed above,
and if they are not produced directly they will at least contribute as virtual particles to the background of such processes.
Although most of what we know about the quarks and gluons inside a hadron comes from our knowledge of the nucleon,
the same underlying features of QCD produce the remaining meson and baryon spectrum as well.
A combined understanding of the nucleon and its resonances is therefore a major goal in studying the strong interaction.

\paragraph{Nucleon resonances.}
A snapshot of our current knowledge about the baryon spectrum is presented in Table~\ref{tab:nucleondeltaresonances}, which lists the two-, three- and four-star resonances below $2$ GeV
quoted by the Particle Data Group (PDG)~\cite{Agashe:2014kda}.
There are currently 13 four-star nucleon and $\Delta$ resonances below 2 GeV;
however, many more have been predicted by the quark model and only a fraction of those have been observed to date~\cite{Capstick:2000qj}.
The search for resonances that are `missing' with respect to the quark model has been a major driving force in
designing and carrying out hadron physics experiments in the past decades.

Traditionally, the existence and basic properties of most of the known nucleon and
$\Delta$ resonances have been extracted from partial-wave analyses of $\pi N$ scattering.
However, this reaction makes no use of the rich information contained in electromagnetic transition amplitudes:
even if resonances couple weakly to the $\pi N$ channel their electromagnetic couplings to $\gamma N$ can still be large.
In recent years, a large amount of data on photoproduction on several final states ($\gamma N \to \pi N, \,\eta N, \,\pi\pi N$, etc.)
has been accumulated at ELSA/Bonn, GRAAL/Grenoble, MAMI/Mainz, Jefferson Lab and Spring-8 in Osaka; see~\cite{Klempt:2009pi,Crede:2013sze} for reviews.
The data set obtained by measurements of high-precision cross sections and polarisation observables in pion photoproduction is coming close towards
a complete experiment, where one is able to extract the four complex amplitudes involved in the process unambiguously, and
combining precision data with the development of multichannel partial-wave analyses has led to the addition
of several new states to the PDG.

In addition, electroproduction of mesons through the absorption of a virtual photon by a nucleon provides information on the internal structure of resonances.
Their electromagnetic couplings at spacelike momentum transfer $Q^2$ are described by transition form factors or, alternatively, the helicity amplitudes $A_{1/2}$, $A_{3/2}$ and $S_{1/2}$.
A big step forward has been made at Jefferson Lab in the last decade where
precise data over a large~$Q^2$ range have been collected in pion electroproduction experiments.
The interpretation of the electroproduction data and helicity amplitudes using different analyses is reviewed in~\cite{Aznauryan:2011qj,Tiator:2011pw}.
Currently, both photo- and electroproduction of
mesons off nucleons constitute the main experimental approaches for the
discovery of new nucleon resonances and our
understanding of electromagnetic baryon structure in the spacelike momentum region.
In the following we will give a very brief survey of the most prominent resonances and their basic properties;
more details can be found in the dedicated reviews~\cite{Pascalutsa:2006up,Klempt:2009pi,Aznauryan:2011qj,Tiator:2011pw,Aznauryan:2012ba,Crede:2013sze}.

\begin{table}[t]
\footnotesize
\begin{center}
\begin{tabular}{cr @{\qquad} cccccc}
\toprule
        $I$                        & $S$                  & $J^P=\tfrac{1}{2}^+$             & $\tfrac{3}{2}^+ $            & $\tfrac{5}{2}^+ $         & $\tfrac{1}{2}^-$          & $\tfrac{3}{2}^-$         & $\tfrac{5}{2}^-$   \\
\midrule
\multirow{4}{*}{$\tfrac{1}{2}$}    &\multirow{4}{*}{$0$}  & $\mathbf{N(940)}$            & $\mathbf{N(1720)}$           & $\mathbf{N(1680)}$        & $\mathbf{N(1535)}$        & $\mathbf{N(1520)}$       & $\mathbf{N(1675)}$       \\
                                   &                      & $\mathbf{N(1440)}$           & $N(1900)$                    & $\color{darkgrey}N(1860)$ & $\mathbf{N(1650)}$        & $N(1700)$                &          \\
                                   &                      & $N(1710)$                    &                              &                           & $\color{darkgrey}N(1895)$ & $N(1875)$                &           \\
                                   &                      & $\color{darkgrey}N(1880)$    &                              &                           &                           &                          &              \\
\midrule
\multirow{3}{*}{$\tfrac{3}{2}$}    &\multirow{3}{*}{$0$}  & $\mathbf{\Delta(1910)}$      & $\mathbf{\Delta(1232)}$      & $\mathbf{\Delta(1905)}$   & $\mathbf{\Delta(1620)}$   & $\mathbf{\Delta(1700)}$  & $\Delta(1930)$           \\
                                   &                      &                              & $\Delta(1600)$               &                           & $\color{darkgrey}\Delta(1900)$ & $\color{darkgrey}\Delta(1940)$  &          \\
                                   &                      &                              & $\Delta(1920)$               &                           &                           &                          &                    \\
\midrule
\multirow{3}{*}{$0$}               &\multirow{3}{*}{$-1$} & $\mathbf{\Lambda(1115)}$     & $\mathbf{\Lambda(1890)}$     & $\mathbf{\Lambda(1820)}$  & $\mathbf{\Lambda(1405)}$  & $\mathbf{\Lambda(1520)}$ & $\mathbf{\Lambda(1830)}$ \\
                                   &                      & $\Lambda(1600)$              &                              &                           & $\mathbf{\Lambda(1670)}$  & $\mathbf{\Lambda(1690)}$ &          \\
                                   &                      & $\Lambda(1810)$              &                              &                           & $\Lambda(1800)$           &                          &                    \\
\midrule
\multirow{2}{*}{$1$}               &\multirow{2}{*}{$-1$} & $\mathbf{\Sigma(1190)}$      & $\mathbf{\Sigma(1385)}$      & $\mathbf{\Sigma(1915)}$   & $\Sigma(1750)$            & $\mathbf{\Sigma(1670)}$  & $\mathbf{\Sigma(1775)}$  \\
                                   &                      & $\Sigma(1660)$               &                              &                           &                           & $\Sigma(1940)$           &          \\
                                   &                      & $\color{darkgrey}\Sigma(1880)$&                              &                           &                           &                          &          \\
\midrule
\multirow{1}{*}{$\tfrac{1}{2}$}    &\multirow{1}{*}{$-2$} & $\mathbf{\Xi(1320)}$         & $\mathbf{\Xi(1530)}$         &                           &                           & $\Xi(1820)$              &             \\
\midrule
\multirow{1}{*}{$0$}               &\multirow{1}{*}{$-3$} &                              & $\mathbf{\Omega(1672)}$      &                           &                           &                          &          \\
\bottomrule
\end{tabular}
\end{center}
 \caption{Two- to four-star baryon resonances below 2 GeV and up to $J^P = \tfrac{5}{2}^\pm$ from the PDG~\cite{Agashe:2014kda}, labeled by their quantum numbers isospin $I$, strangeness $S$, spin $J$ and parity $P$.
          The four-star resonances are shown in bold font and the two-star resonances in gray.
          Historically the $N$ and $\Delta$ resonances are labelled by the incoming partial wave $L_{2I,2J}$ in elastic $\pi N$ scattering,
          with $L=P,P,F,S,D,D$ for $J^P=\tfrac{1}{2}^+ \dots \tfrac{5}{2}^-$ from left to right.}
\label{tab:nucleondeltaresonances}
\end{table}

\paragraph{$\Delta(1232)$ resonance.}
The $\Delta(1232)$ with $J^P=\nicefrac{3}{2}^+$ is undoubtedly the best studied nucleon resonance.
It is the lightest baryon resonance, about 300 MeV heavier than the nucleon, and despite its width of about 120 MeV it is well separated from other resonances.
It almost exclusively decays into $N\pi$ and thus provides a prominent peak in $N\pi$ scattering,
whereas its electromagnetic decay channel $\Delta\to N\gamma$ contributes less than $1\%$ to the total decay width.
Although the electromagnetic $\gamma N\to\Delta$ transition is now well measured over a large $Q^2$ range, several open questions remain.
The process is described by three transition form factors: the magnetic dipole transition $G_M(Q^2)$, which is dominated by the spin flip of a quark in the nucleon
to produce the $\Delta$, and the  electric and Coulomb quadrupole ratios $R_{EM}$ and $R_{SM}$.
The prediction of the $\gamma N\to \Delta$ transition magnetic moment was among the first successes of the constituent-quark model,
which relates it to the magnetic moment of the proton via $\mu(\gamma p\to\Delta) = 2\sqrt{2}\,\mu_p/3$~\cite{Beg:1964nm}.
However, the quark-model prediction also underestimates the experimental value by about $30\%$ and entails $R_{EM}(Q^2) = R_{SM}(Q^2)=0$~\cite{Harari:1965zzb,Becchi:1965zz}.
Dynamical models assign most of the strength in the quadrupole transitions to the meson cloud that `dresses' the bare $\Delta$.
We will return to this issue in Sec.~\ref{form-res-trans} and also present a different viewpoint on the matter.

\paragraph{Roper resonance.}
    The lowest nucleon-like state is the Roper resonance $N(1440)$ or $P_{11}$ with $J^P=\nicefrac{1}{2}^+$, which has traditionally been a puzzle for quark models.
    The Roper is unusually broad and not well described within the non-relativistic constituent-quark model (see~\cite{Capstick:1986bm} and references therein),
    which predicts the wrong mass ordering between the Roper and the nucleon's parity partner $N(1535)$
    and the wrong sign of the $\gamma p\to N(1440)$ transition amplitude.
    Although some of these deficiencies were later remedied by relativistic quark models~\cite{Capstick:1986bm,Richard:1992uk,Capstick:1993kb,Capstick:1994ne,Glozman:1995fu},
    they have led to longstanding speculations about the true nature of this state being the first radial excitation of the nucleon
    or perhaps something more exotic.

\begin{figure*}[t]
            \centerline{%
            \includegraphics[width=0.99\textwidth]{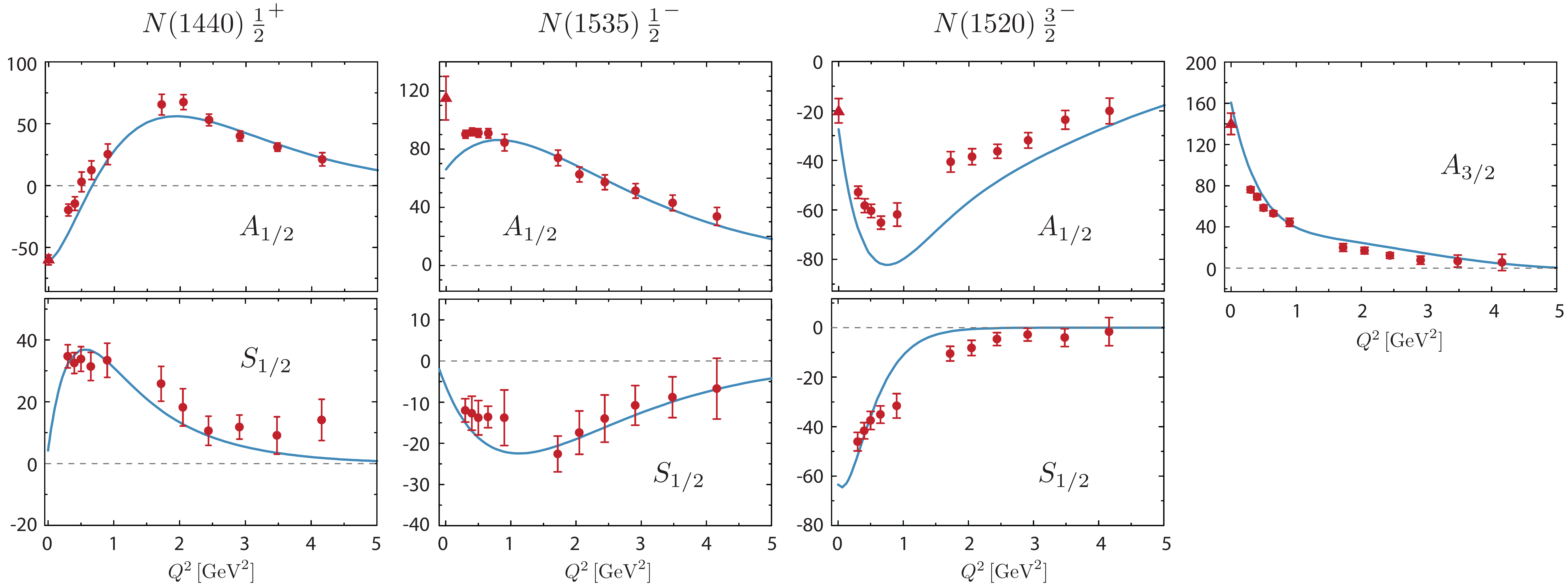}}
            \caption{$\gamma^\ast p\to N^\ast$ helicity amplitudes for the Roper, $N(1535)$ and $N(1520)$ resonances.
                     The data points (circles) correspond to the Jefferson Lab analysis of single-pion electroproduction with CLAS~\cite{Aznauryan:2009mx} and the curves are the MAID parametrizations~\cite{Drechsel:2007if}.
                     The triangles at the real photon point are the PDG values~\cite{Agashe:2014kda}.
                     The helicity amplitudes carry units of $10^{-3}$~GeV$^{-1/2}$. }
            \label{fig:helicity-amps}
            \end{figure*}

    The Jefferson Lab/CLAS measurements of single and double-pion electroproduction allowed for the determination of the electroexcitation
    amplitudes of the Roper resonance in a wide range of $Q^2$.
    The helicity amplitudes obtained from the Jefferson Lab and MAID analyses are shown in Fig.~\ref{fig:helicity-amps}.
    They exhibit a strong $Q^2$ dependence of the transverse helicity amplitude $A_{1/2}$ including a zero crossing, which also translates
    into a zero of the corresponding Pauli form factor $F_2(Q^2)$.
      Such a behavior is typically expected for radial excitations and it has been recovered by a number of approaches,
      from constituent~\cite{Obukhovsky:2011sc} and light-front constituent-quark models~\cite{Aznauryan:2007ja} to
      Dyson-Schwinger calculations~\cite{Segovia:2015hra},
      effective field theory~\cite{Bauer:2014cqa}, lattice QCD~\cite{Lin:2008qv} and AdS/QCD~\cite{Gutsche:2012wb}.
      Although none of them has yet achieved pointwise agreement with the data they all predict the correct signs and orders of magnitude of the amplitude.
      Taken together, consensus in favor of the Roper resonance
     as predominantly the first radial excitation of the three-quark ground state is accumulating
     and we will return to this point in Sec.~\ref{spec:results}.

\paragraph{Other nucleon resonances.}
The mass range around $1.5$ GeV is the so-called second resonance region and features
a cluster of three nucleon resonances:
the Roper, the nucleon's putative parity partner $N(1535) \,S_{11}$ with $J^P=\nicefrac{1}{2}^-$, and
the $N(1520) \, D_{13}$ resonance with $J^P = \nicefrac{3}{2}^-$. The $N(1535)$ has large branching ratios to both $\pi N$ and $\eta N$ channels
and was extensively studied in $\pi$ and $\eta$ electroproduction off protons; until recent years
it had been the only other state apart from the $\Delta(1232)$ for which transition form factors were measured over a similarly large $Q^2$ range.
Its helicity amplitude $S_{1/2}$ shows an unusually slow falloff with $Q^2$, which translates into a cancellation of the corresponding Pauli form factor $F_2(Q^2)$ that is consistent with zero above $Q^2\sim 2$ GeV$^2$~\cite{Ramalho:2011fa}
and rapidly rises below that value, cf. Figs.~\ref{fig:helicity-amps} and~\ref{fig:ff-vs-helicity-0}.
By comparison, the $\gamma^\ast p\to N(1520)$ helicity amplitudes look rather ordinary
and suggest a dominant three-quark nature;
they are well described by quark models although quantitative agreement is only achieved when meson cloud effects are included.
Results for several higher-lying resonances are also available, and the extension to the mass range up to $2.5$~GeV as well as $Q^2$  up to $12$~GeV$^2$
is part of the experimental $N^\ast$ program with CLAS12 at Jefferson Lab~\cite{Burkert:2016dxc,Mokeev:2016hqv}.

\paragraph{Strangeness.}

The situation in the strange sector is less developed than in the non-strange sector in the sense that even more states are missing
compared to naive quark model expectations. Matters are additionally complicated by the fact that singlet and octet states are present
and the assignments of experimentally extracted states to the different multiplets are certainly ambiguous. The lowest-lying state in the
negative parity sector, the $\Lambda(1405)$, is highly debated since its mass is surprisingly low; in fact, despite its strange quark content
it is even lower than the ground state in the corresponding non-strange channel. Its spin and parity quantum numbers have only
been identified unambiguously from photoproduction data at Jefferson Lab~\cite{Moriya:2014kpv}. Quark models assign a dominant flavour singlet nature to this state,
which seems confirmed by exploratory lattice calculations~\cite{Menadue:2011pd,Engel:2012qp}.
On the other hand, the $\Lambda(1405)$ has  long since been viewed as
a prime candidate for a state that is generated dynamically via coupled channel effects, see~\cite{Hyodo:2011ur} for a review.
In the coupled channel chiral unitary approach there is even evidence for two states sitting close together, mostly appearing as a single resonance
in experiment~\cite{Jido:2003cb}. Other states in the negative parity sector, the $\Lambda(1670)$ and the $\Lambda(1800)$, may be predominantly
flavour octets and agree well with quark model classifications. In the positive parity sector, there is the Roper-like $\Lambda(1600)$ and then
one encounters the three-star $\Lambda(1810)$, which may either be one of the octet states or the parity partner of the $\Lambda(1405)$.

Although the ground-state hyperon masses are well known, their interactions and internal structure remain a largely unknown territory.
While there is abundant experimental information on nucleon electromagnetic structure, our experimental knowledge of
hyperons is limited to their static properties~\cite{GoughEschrich:2001ji,Taylor:2005zw,Keller:2011aw,Keller:2011nt}. First measurements
of hyperon form factors at large timelike photon momenta have been recently presented by the CLEO collaboration~\cite{Dobbs:2014ifa}. The study
of hyperon structure is also one of the main goals of the CLAS collaboration.

\begin{figure*}[t]
        \begin{center}
        \includegraphics[width=0.99\textwidth]{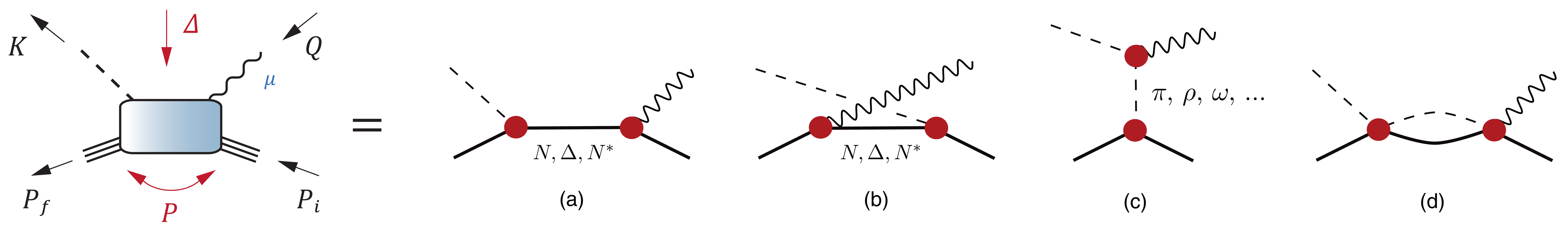}
        \caption{Pion electroproduction amplitude. Diagrams (a) and (b) constitute the $s$- and $u$-channel nucleon and nucleon resonance
                 contributions to photo- and electroproduction. The non-resonant $t$-channel meson exchanges (c) contribute to the background of the process.
                 Figure (d) is an example for intermediate meson-baryon channels which provide the necessary cut structure.}\label{fig:one-photon-electroproduction}
        \end{center}
\end{figure*}

\subsection{Photo- and electroproduction processes}\label{sec:electroproduction}

In this section we summarise the necessary elements to describe single-meson photo- and
electroproduction experiments, such as the kinematical phase space, choice of amplitudes and
observables. We will also briefly describe the models used to extract
information on the resonance spectrum and their properties from the scattering
amplitudes.

        Consider the process $e^- +N \to e^- +N+\pi$. In the one-photon exchange approximation the amplitude factorizes into a leptonic and a hadronic part
        and it is sufficient to consider the reaction $\gamma^\ast + N \to N + \pi$, which is the pion electroproduction amplitude depicted in Fig.~\ref{fig:one-photon-electroproduction}.
        It depends on three independent momenta:
        the average nucleon momentum $P=(P_f+P_i)/2$, the virtual photon momentum $Q$, and the pion momentum $K$.
        In addition we denote the $t$-channel momentum transfer by $\Delta=Q-K=P_f-P_i$.
        Both nucleons and the pion are onshell ($P_f^2=P_i^2=-m^2$, $K^2=-m_\pi^2$) and therefore the process is described by three Lorentz-invariant kinematic
        variables. For later convenience and also in view of the analogous situation in Compton scattering discussed in Sec.~\ref{compton}, we choose them as\footnote{We use
        Euclidean conventions throughout this review, but since Lorentz-invariant scalar products differ from their Minkowski counterparts only by minus signs
        these variables are the same in Minkowski space if one defines them as
        $\tau = -q^2/(4m^2)$, $\eta=-k\cdot q/m^2$, etc.,
cf.\ App.~\ref{app:conventions} for more details.}
        \begin{equation}\label{ep-variables}
           \tau=\frac{Q^2}{4m^2}\,, \qquad
           \eta = \frac{K\cdot Q}{m^2}\,, \qquad
           \lambda = -\frac{P\cdot Q}{m^2} = -\frac{P\cdot K}{m^2}\,,
        \end{equation}
        where $\lambda$ is the crossing variable and $m$ the nucleon mass.
        Naturally the description through any other combination of three independent
        Lorentz invariants is equivalent; for example in terms of the three Mandelstam variables~$\{s,u, \tilde t\}$:
        \begin{equation}\label{ep-su}  \renewcommand{\arraystretch}{1.0}
           \left\{ \begin{array}{c} s \\ u \end{array}\right\} = -\left(P \pm \frac{Q+K}{2}\right)^2 = m^2 \,( 1- \eta \pm 2\lambda),  \qquad
           \tilde t= - \Delta^2 =   - Q^2 + 2m^2 \eta + m_\pi^2 \,.
        \end{equation}
        These Mandelstam variables satisfy the usual relation
$s+\tilde t + u = 2m^2 + m_\pi^2 - Q^2$ where the  minus sign reflects the Euclidean convention for the virtual photon momentum.
The fact that $\tilde t$ is negative in the experimental region and that it usually appears in combination with a factor $4m^2$ motivates to slightly redefine the Mandelstam variable in this channel as
        \begin{equation} \label{ep-tDef}
         t = \frac {\Delta^2}  {4m^2}  = \tau - \frac{\eta}{2} - \mu\,,
        \end{equation}
where we used the abbreviation $\mu = m_\pi^2/(4m^2)$.

       At the hadronic level, the electroproduction amplitude is expressed by the sum of Born terms for the nucleon and its resonances (the $\Delta$ resonance, Roper, etc.),
       illustrated by the diagrams (a, b) in Fig.~\ref{fig:one-photon-electroproduction}, which are augmented by $t$-channel meson exchanges in diagram (c) as well as hadronic loops (d).
       If one has sufficiently good control over the `QCD background' stemming from the latter two topologies, this is ultimately
       how information on nucleon resonance masses and their transition form factors can be extracted from experiments.
       The relevant information is encoded in the six electroproduction amplitudes $A_i(\tau,\eta,\lambda)$ which enter in the covariant decomposition of the full amplitude:
       \begin{equation}\label{ep-amp}
          \mc{M}^{\mu}(P,K,Q) = \conjg{u}(P_f) \,\Bigg( \sum_{i=1}^6 A_i(\tau,\eta,\lambda)\,M^\mu_i(P,K,Q) \Bigg)\, u(P_i)\,.
       \end{equation}
       Electromagnetic gauge invariance entails that the amplitude is transverse with respect to the photon momentum, $Q^\mu\mc{M}^\mu(P,K,Q)=0$, which leaves six independent amplitudes
       in the general case and four amplitudes in photoproduction where the photon is real ($\tau=0$).

        \begin{figure}
        \centering
          \includegraphics[width=0.93\textwidth]{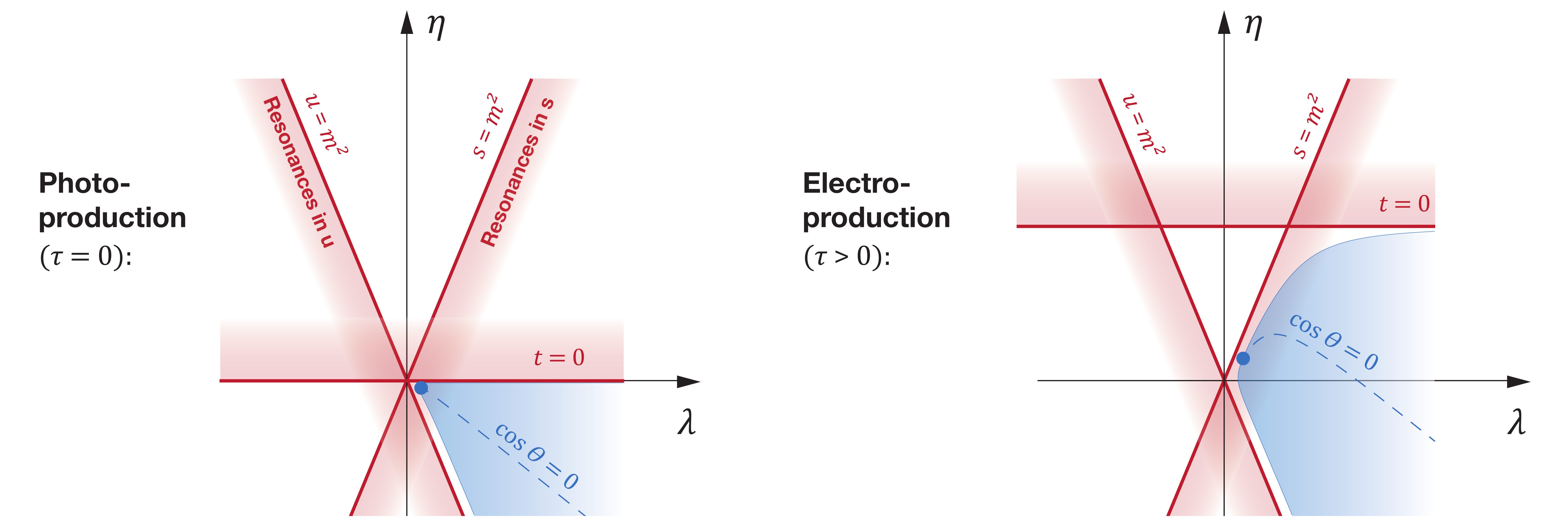}
        \caption{Phase space of the pion electroproduction amplitude in the variables $\eta$ and $\lambda$, with $\tau$ held fixed.
                 The left panel shows the case of photoproduction ($\tau=0$) and the right panel the analogous situation in electroproduction ($\tau>0$).
                 The $s$- and $u$-channel nucleon Born poles are shown by the thick (red) lines together with
                 the nucleon resonance regions. The horizontal lines and bands indicate the $t$-channel region $t<0$.
                 The shaded (blue) area in the bottom right is the physical region; the dot marks the threshold and the dashed line is the limit $\cos\theta=0$.}
        \label{fig:electroprod-phasespace}
        \end{figure}

        The description in terms of the variables~\eqref{ep-variables} has the advantage that the nucleon resonance positions at fixed $s=s_R$ and $u=u_R$
        do not change with $\tau$, so they only depend on two Lorentz-invariant kinematic variables as can be seen from~\eqref{ep-su} and Fig.~\ref{fig:electroprod-phasespace}.
        The two diagrams illustrate the phase space for photoproduction ($\tau=0$, left panel) and for electroproduction ($\tau>0$, right panel)  in the variables $\lambda$ and $\eta$.\footnote{
Note that the phase space for $\pi N$ scattering is identical
       if $Q^2$ is held fixed at $Q^2=-m_\pi^2$, or equivalently $\tau = -\mu$.}
        The Born poles appear at $s=m^2$ and $u=m^2$ corresponding to $\lambda = \pm \eta/2$. The resonance regions are indicated by the shaded (red) areas in the plot, where
       at larger $s$ and $u$ the resonances are eventually washed out because their hadronic decay widths
       shift their poles into the complex plane.
       The horizontal lines mark the onset of the timelike $t$-channel regions for $t<0$,
where one has in addition the pion pole
       stemming from diagram (c) in Fig.~\ref{fig:one-photon-electroproduction} as well as other meson poles.
       In addition, at fixed $\eta$ one has branch cuts from multiparticle $N\pi$, $N\pi\pi$, $\dots$ production:
       the right-hand cut starts at the threshold $s=(m+m_\pi)^2$ and extends to infinity and the left-hand cut begins at $u=(m+m_\pi)^2$.
       In the $t$-channel region there are additional cuts from multipion production starting at $t=-4\mu$.

       The shaded (blue) areas in the bottom right show the physical regions that are accessible in pion electroproduction experiments,
       defined by $s>(m+m_\pi)^2$, $\tau>0$ and $-1<\cos\theta<1$, where $\theta$ is  the CM scattering angle from~\eqref{ep-cm-angle} below.
       They start at the thresholds
       \begin{equation}
          \lambda_\text{thr} = \frac{2\sqrt{\mu}}{1+2\sqrt{\mu}}\left(\tau+(1+\sqrt{\mu})^2\right), \qquad
          \eta_\text{thr} = \frac{4\sqrt{\mu}}{1+2\sqrt{\mu}}\left(\tau-\mu-\sqrt{\mu}\right)
       \end{equation}
       with $\mu = m_\pi^2/(4m^2)$.
       Note that both thresholds vanish in the chiral limit $m_\pi=0$.
       In the physical region the amplitudes $A_i(\tau,\eta,\lambda)$ are necessarily complex functions due to the cut structure.
        In practice one performs multipole expansions for their angular dependence in $\cos\theta$
       around the central value $\cos\theta=0$, which we will discuss further below,
       so that the remaining multipole amplitudes only depend on $s$ and $\tau$.
       In principle one can then extract the various $N\gamma^\ast\to N^\ast$ transition form factors, which are functions of $\tau$ only, from the resonance locations $s=s_R$.

       Ideally one would like to work with electroproduction amplitudes $A_i(\tau,\eta,\lambda)$ that only have physical poles and cuts and are otherwise
       free of kinematic singularities or constraints.
       In principle this can be achieved by choosing an appropriate tensor basis constructed along
       the lines of Lorentz covariance, gauge invariance, analyticity and charge-conjugation invariance. The simplest such basis is given by
       \begin{equation}\label{ep-basis}
          M_{1\dots 6}^\mu(P,K,Q) =
          i\gamma_5  \left\{ \;
          [\gamma^\mu, \slashed{Q}]\,, \quad
          t^{\mu\nu}_{QK}\,P^\nu, \quad
          t^{\mu\nu}_{QQ}\,P^\nu, \quad
          t^{\mu\nu}_{QP}\,i\gamma^\nu, \quad
          \lambda\, t^{\mu\nu}_{QK}\,i\gamma^\nu, \quad
          \lambda\, t^{\mu\nu}_{QQ}\,i\gamma^\nu  \; \right\},
       \end{equation}
       where we abbreviated
       \begin{equation}\label{tAB}
          t^{\mu\nu}_{AB} = A\cdot B\,\delta^{\mu\nu} - B^\mu A^\nu\,.
       \end{equation}
       Because $A^\mu \,t^{\mu\nu}_{AB}=0$, one immediately verifies that all tensors are transverse to the photon momentum.
       They are free of kinematic singularities \textit{and} feature the lowest possible powers in $Q^\mu$, i.e., the basis is `minimal' with respect to the photon momentum.
       Furthermore, the factors $\lambda$ ensure that each basis element is invariant under charge conjugation: $M_i^\mu(P,K,Q) = -\mc{C}\,M_i^\mu(-P,K,Q)^T\,\mc{C}^T$, where
       $\mc{C}=\gamma^4 \gamma^2$ is the charge-conjugation matrix
(cf.\ App.~\ref{app:conventions}), because the same invariance must hold for the full amplitude as well.
       As a consequence, the amplitudes $A_i(\tau,\eta,\lambda)$ are symmetric in $\lambda$
       and thus they only depend on $\lambda^2$,
       so we really only need to discuss the right half of the phase space ($\lambda>0$) in Fig.~\ref{fig:electroprod-phasespace}.
       In the case of real photons, $M^\mu_3$ and $M^\mu_6$ decouple from the cross section and one is left with four independent amplitudes.

       As a side remark, we note that the covariant tensors defined in~\cite{Fubini:1958zz,Dennery:1961zz}, which are frequently used in
       theoretical descriptions of pion electroproduction,
       \begin{equation}\label{ep-dennery}
       \widetilde{M}_{1\dots 6}^\mu = \left\{ \;-\frac{M_1^\mu}{2} , \quad M_3^\mu-2M_2^\mu, \quad \frac{M_5^\mu}{\lambda}, \quad m M_1^\mu + 2M_4^\mu, \quad \frac{4\tau M_2^\mu-\eta M_3^\mu}{\lambda},\quad \frac{M_6^\mu}{\lambda} \;\right\},
       \end{equation}
       do not form a minimal basis due to the element $\widetilde{M}_5^\mu = i\gamma_5\,t^{\mu\nu}_{QQ}\,K^\nu$.
       The relation between~\eqref{ep-basis} and~\eqref{ep-dennery} shows
       that two of the corresponding `Dennery amplitudes' $\widetilde{A}_2$ and $\widetilde{A}_5$ have a kinematic singularity
       at the pion pole location $t=-\mu$,
which is outside of the physical region but still has to be subtracted in dispersion integrals~\cite{Pasquini:2007fw}.
       In any case, $\widetilde{M}_5^\mu$ and $\widetilde{M}_6^\mu$ drop out in photoproduction
       where the remaining amplitudes are kinematically safe.
       Note also that $\widetilde{A}_{3,5,6}$ are antisymmetric in $\lambda$ and therefore they vanish for $\lambda=0$.

\begin{figure}[t]
 \centerline{ \includegraphics[width=0.7\textwidth]{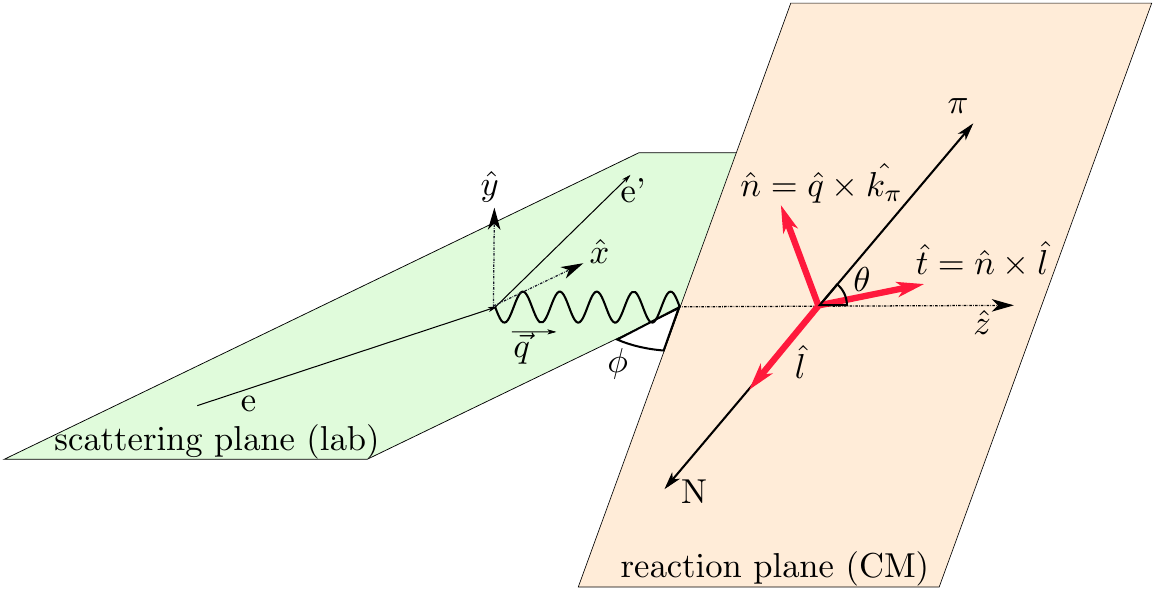} }
 \caption{Kinematics and reference axes of a meson-production
experiment.}\label{fig:scattering_plane-new}
\end{figure}

\paragraph{Pion electroproduction in the CM frame.}
In the one-photon approximation the reaction $e^- +N \to e^- +N+\pi$ can be split into a leptonic and a hadronic part.
It is common to evaluate the former in the laboratory frame
and the latter in the nucleon-pion CM frame, which is illustrated in Fig.~\ref{fig:scattering_plane-new}. The leptonic reaction takes place
in the scattering plane and the hadronic reaction in the reaction plane, where $\theta$ is the scattering angle between the virtual
photon and the pion in the CM frame.
Using a Euclidean notation, the virtual photon, pion and nucleon momenta in the CM frame are given by
\begin{equation} \renewcommand{\arraystretch}{1.0}
    Q = \left[ \begin{array}{c}  \vect{q} \\ iE_q \end{array}\right], \quad
    K = \left[ \begin{array}{c}  \vect{k} \\ iE_k \end{array}\right], \quad
    P_i = \left[ \begin{array}{c} -\vect{q} \\ iE_i \end{array}\right], \quad
    P_f = \left[ \begin{array}{c} -\vect{k} \\ iE_f \end{array}\right].
\end{equation}
If we introduce a variable $\delta$ by writing $s=m^2(1+4\delta)$, then
their relations to the Lorentz invariants $s$ and $\tau$
are given by
\begin{equation}\label{ep-cm-quantities}
\begin{array}{rl}
  E_q &= \displaystyle\frac{2m^2}{\sqrt{s}}\,(\delta-\tau)\,, \\[2mm]
  E_k &= \displaystyle\frac{2m^2}{\sqrt{s}}\,(\delta+\mu)\,,
\end{array}\qquad
\begin{array}{rl}
  E_i &= \displaystyle\frac{2m^2}{\sqrt{s}}\,\big(\tfrac{1}{2}+\delta+\tau\big)\,, \\[2mm]
  E_f &= \displaystyle\frac{2m^2}{\sqrt{s}}\,\big(\tfrac{1}{2}+\delta-\mu\big)\,,
\end{array}\qquad
\begin{array}{rl}
  |\vect{q}| &= \displaystyle \frac{2m^2}{\sqrt{s}}\,\sqrt{(\delta+\tau)^2+\tau}\,, \\[2mm]
  |\vect{k}| &= \displaystyle \frac{2m^2}{\sqrt{s}}\,\sqrt{(\delta-\mu)^2-\mu}\,,
\end{array}\qquad
\end{equation}
with $\mu=m_\pi^2/(4m^2)$.
The CM scattering angle defined by $\vect{q}\cdot\vect{k} = |\vect{q}||\vect{k}|\,\cos\theta$ additionally depends on $t$
and it is related to the Lorentz invariants via
\begin{equation}\label{ep-cm-angle}
 t = \frac{E_i\,E_f - \vect{q}\cdot\vect{k} - m^2}{2m^2} \quad \Rightarrow \quad
 \cos\theta = \frac{(\delta+\tau)(\delta-\mu) + \tfrac{1}{2}(\tau-\mu) - (\delta+\tfrac{1}{4})\,2  t}{\sqrt{\left((\delta+\tau)^2+\tau\right)\left((\delta-\mu)^2-\mu\right)}}\,.
\end{equation}
The unphysical point $|\vect{q}|=0$ is called the pseudo-threshold or Siegert limit \cite{Siegert:1937yt}.

The differential cross section for pion virtual photoproduction in the CM frame is given by
\begin{equation}\label{ep-cross-section}
   \frac{d\sigma}{d\Omega} = \sigma_T\,+ \epsilon\,\sigma_L + \epsilon\,\sigma_{TT}\,\cos\,2\phi + \sqrt{2\epsilon\,(1+\epsilon)}\,\sigma_{LT}\,\cos\phi + h\sqrt{2\epsilon(1-\epsilon)}\,\sigma_{LT'}\,\sin\phi\,.
\end{equation}
It carries traces from the leptonic part of the process: the angle $\phi$ between the scattering and reaction plane,
the helicity $h$ of the incident electron and the transverse polarisation $\epsilon$ of the virtual photon.
The cross section is characterized by five `structure functions' $\sigma_T$, $\sigma_L$, $\sigma_{TT}$, $\sigma_{LT}$ and $\sigma_{LT'}$
of the process $\gamma^\ast N\to N\pi$ which can be expressed by the amplitudes $A_i(\tau,\eta,\lambda)$ and depend on the same three variables.
We stated the formula for an unpolarised target and a longitudinally polarised beam without recoil polarisation detection;
the general expression and a discussion of polarisation observables can be found in Refs.~\cite{Raskin:1988kc,Drechsel:1992pn,Knochlein:1995qz}.

While the decomposition of the electroproduction amplitude in terms of~\eqref{ep-amp} makes its Lorentz covariance and gauge invariance properties
explicit, in experiments one typically has a certain control over the polarisation of the initial or final states and therefore one employs
alternative decompositions, either in terms of helicity amplitudes or
Chew-Goldberger-Low-Nambu (CGLN) amplitudes.
The CGLN amplitudes~\cite{Chew:1957tf} are defined in the CM frame via
\begin{equation}
   \varepsilon^\mu \mathcal{M}^\mu(P,K,Q) = \frac{4\pi\sqrt{s}}{m}\,\chi_f^\dag\,\mathcal{F}\,\chi_i\,, \qquad
   \varepsilon^\mu = -\frac{e}{Q^2}\,\conjg{u}(K_f)\,i\gamma^\mu\,u(K_i)
\end{equation}
where $\chi_i$ and $\chi_f$ denote the initial and final Pauli spinors, $\varepsilon^\mu$ with $\varepsilon\cdot Q = 0$
is the transverse polarisation vector of the virtual photon stemming from the leptonic part of the process, and $K_i$, $K_f$ are the electron momenta with $Q = K_i-K_f$.
Defining $\vect{a} = \vect{\varepsilon} - (\vect{\varepsilon}\cdot\vect{\hat{q}})\,\vect{\hat{q}}$,
the amplitude $\mc{F}$ can be written as
\begin{equation}\label{eq:CGLN_amplitudes}
  \mc{F} = i(\vect{\sigma}\cdot\vect{a})\,\mc{F}_1
           +(\vect{\sigma}\cdot\vect{\hat{k}})\,\vect{\sigma}\cdot(\vect{\hat{q}}\times\vect{a})\,\mc{F}_2
           +i(\vect{\hat{k}}\cdot\vect{a})  \,\vect{\sigma}\cdot (\vect{\hat{q}}\,\mc{F}_3 + \vect{\hat{k}}\,\mc{F}_4)
           -i\,\frac{Q^2}{E_q^2}\,(\vect{\hat{q}}\cdot\vect{\varepsilon})\, \vect{\sigma}\cdot (\vect{\hat{q}}\,\mc{F}_5 +  \vect{\hat{k}}\,\mc{F}_6)
\end{equation}
with the unit three-vectors $\vect{\hat{q}}$ of the photon and $\vect{\hat{k}}$ of the pion.
For photoproduction only the amplitudes $\mathcal{F}_{1\dots 4}$ contribute.
The relations between the structure functions and CGLN amplitudes (as well as the helicity amplitudes) are stated in Ref.~\cite{Knochlein:1995qz};
note the kinematical factor between $\epsilon$ with $\epsilon_L$ therein which can be equally absorbed in the longitudinal structure functions.
The relations between the CGLN and Dennery amplitudes can be found in Refs.~\cite{Dennery:1961zz,Pasquini:2007fw}.

The next step in the analysis of experimental data is to project the CGLN amplitudes onto a partial-wave basis by separating the angular dependence in $z=\cos\theta$ (cf.~Fig.~\ref{fig:electroprod-phasespace}) through a polynomial expansion.
The respective coefficients are the multipole amplitudes which depend on the variables $s$ and $\tau$.
These are the transverse amplitudes $M_{\ell\pm}$ and $E_{\ell\pm}$ and longitudinal (or scalar) amplitudes $L_{\ell\pm} = E_q\,S_{\ell\pm}/|\vect{q}|$,
which are related to photons of the magnetic, electric and Coulomb type, respectively; $\ell$ is the angular momentum of the $\pi N$ final state.
For example, the decomposition for the CGLN amplitudes reads~\cite{Dennery:1961zz}
\begin{align}\label{eq:CGLN_multipoles}
\mc{F}_1=&\sum_{\ell\geq 0}\left[(\ell M_{\ell+}+E_{\ell+}) \,P'_{\ell+1} + ((\ell+1) \,M_{\ell-}+E_{\ell-}) \,P'_{\ell-1}\right], &
\mc{F}_2=&\sum_{\ell\geq 1}\left[(\ell+1) M_{\ell+}+\ell M_{\ell-}\right] P'_{\ell}\,,\nonumber\\
\mc{F}_3=&\sum_{\ell\geq 1}\left[(E_{\ell+}-M_{\ell+})\, P''_{\ell+1} + (E_{\ell-}+M_{\ell-})\, P''_{\ell-1} \right],&
\mc{F}_4=&\sum_{\ell\geq 2}(M_{\ell+}-E_{\ell+}-M_{\ell-}-E_{\ell-})\, P''_{\ell}\,,\nonumber\\
\mc{F}_5=&\sum_{\ell\geq 0}\left[(\ell+1)\,L_{\ell+}\,P'_{\ell+1}-\ell L_{\ell-}\,P'_{\ell-1}\right], &
\mc{F}_6=&\sum_{\ell\geq 1}\left[\ell L_{\ell-}-(\ell+1)\, L_{\ell+}\right] P'_{\ell}\,,
\end{align}
where $P_\ell(\cos\theta)$ are Legendre polynomials and primes denote their derivatives, and
the relations can be inverted using the orthogonality properties of the Legendre polynomials.
The multipole amplitudes are linear combinations of the transverse partial wave helicity amplitudes,
which at the resonance locations are related to the $N\gamma^\ast \to N^\ast$ helicity amplitudes
(see e.g.~\cite{Pascalutsa:2006up,Drechsel:2007if,Aznauryan:2011qj} for the explicit formulas).

            \begin{figure*}[t]
            \centerline{%
            \includegraphics[width=0.8\textwidth]{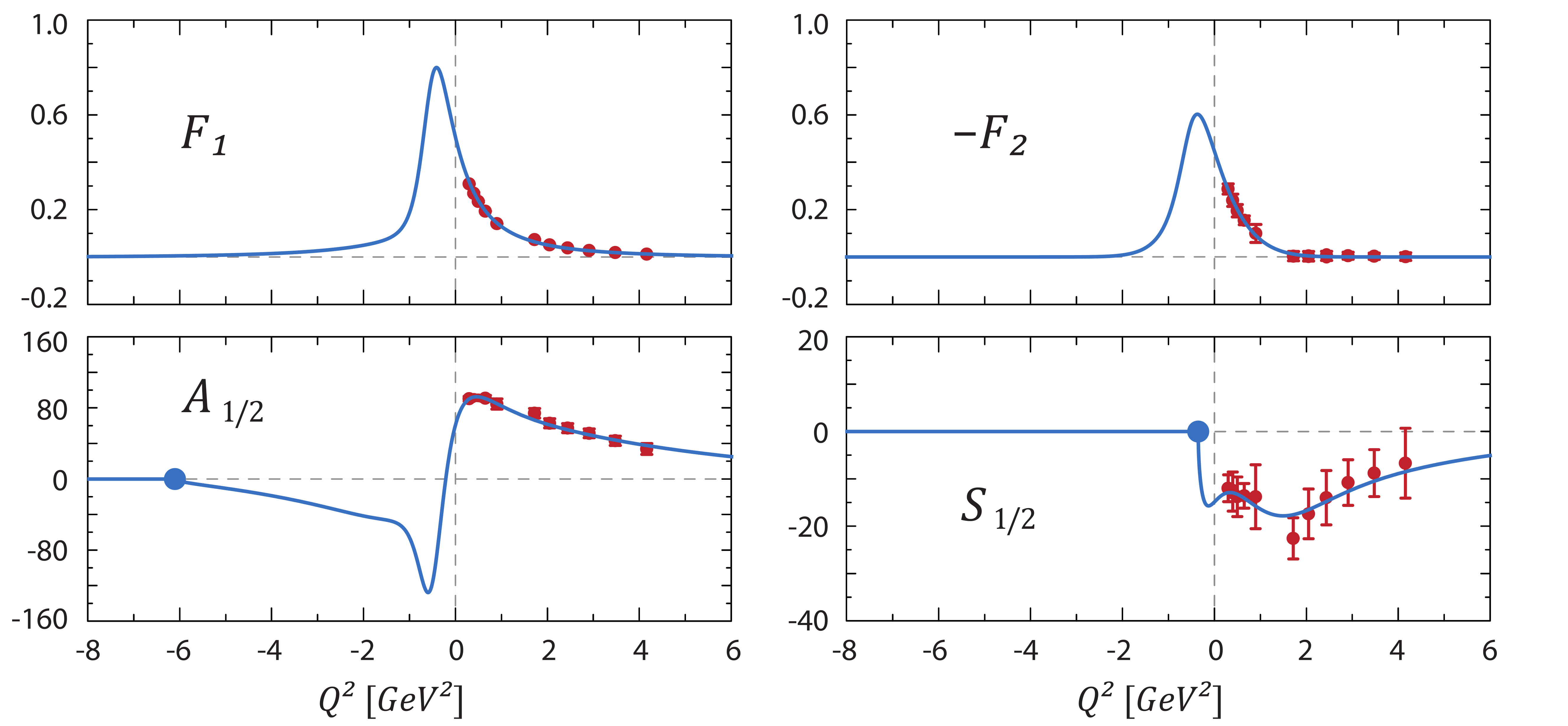}}
            \caption{CLAS data for the $p\gamma^\ast\to N(1535)$ transition form factors and helicity amplitudes~\cite{Aznauryan:2009mx}, together with a
simple parametrization
                     including a vector-meson bump (adapted from Ref.~\cite{Eichmann:2016jqx}). The helicity amplitudes carry units of $10^{-3}$ GeV$^{-1/2}$. }
            \label{fig:ff-vs-helicity-0}
            \end{figure*}

\paragraph{Helicity amplitudes vs. transition form factors.}
        The $N\gamma^\ast \to N^\ast$ electromagnetic transitions are conventionally expressed in terms of $Q^2$-dependent helicity amplitudes.
        Resonances with $J^P = \nicefrac{1}{2}^\pm$ depend on two helicity amplitudes ($A_{1/2}$ and $S_{1/2}$)
        and $J^P = \nicefrac{3}{2}^\pm$ and higher resonances on three ($A_{3/2}$, $A_{1/2}$ and $S_{1/2}$).
        From a theoretical point of view it is more convenient to work with the transition form factors that constitute the corresponding currents.
        To this end we write the $\gamma^\ast N \to N^\ast$ transition current for a $J^P=\nicefrac{1}{2}^\pm$ resonance as
        \begin{equation}\label{1/2-current} \renewcommand{\arraystretch}{1.0}
           \mc{J}^\mu = i\conjg{u}(P')\left[\begin{array}{c} \mathds{1} \\ \gamma_5 \end{array}\right]\left( \frac{F_1(Q^2)}{m^2}\,t^{\mu\nu}_{QQ}\gamma^\nu + \frac{iF_2(Q^2)}{4m}\,[\gamma^\mu,\slashed{Q}]\right) u(P_i)\,,
        \end{equation}
        where $P_i$ is the incoming momentum and $P'$ the momentum of the resonance; $t^{\mu\nu}_{AB}$ was defined in~\eqref{tAB}.
        Our definition slightly deviates from the conventions in the literature: the Dirac-like form factor $F_1(Q^2)$ differs by a factor $Q^2/m^2$ from the usual one, thus removing the kinematic zero at $Q^2=0$;
        and the factor $m$ in the second term is often written as $m\to (m+m_R)/2$, where $m_R$ is the resonance mass.
        To arrive at compact relations between the form factors and helicity amplitudes, let us first abbreviate
        \begin{equation}
           \delta_\pm = \frac{m_R \pm m}{2m}\,, \quad
           \delta = \delta_+ \delta_- = \frac{m_R^2-m^2}{4m^2}\,, \quad
           \lambda_\pm = \delta_\pm^2 + \tau\,, \quad
           R_\pm = e\sqrt{\frac{\lambda_\pm}{2m\delta}}\,, \quad
           \gamma = \frac{m}{m_R}\sqrt{\frac{\lambda_+\lambda_-}{2}}\;,
        \end{equation}
        where $e$ is the electric charge and $\lambda_+\lambda_- = (\delta+\tau)^2+\tau$.
        The relations between the helicity amplitudes $A_{1/2}$ and $S_{1/2}$
        and the form factors in~\eqref{1/2-current} are then given by~\cite{Devenish:1975jd}
        \begin{equation}
           F_1 = \frac{1}{4R_\mp\,\lambda_\pm}\left( A_{1/2} \pm \frac{\delta_\pm}{\gamma}\,S_{1/2}\right), \qquad
           F_2 = \pm\frac{1}{R_\mp\,\lambda_\pm}\left(\delta_\pm\,A_{1/2} \mp \frac{\tau}{\gamma}\,S_{1/2}\right), \qquad J^P = \tfrac{1}{2}^\pm\,.
        \end{equation}
        Being Lorentz invariant, they are again identical in Euclidean and Minkowski conventions.
        As illustrated in Fig.~\ref{fig:ff-vs-helicity-0} for the $N(1535)$ transition,
        if the form factors are free of kinematic constraints the helicity amplitudes must have kinematic zeros:
        a naive parametrization of the experimental form factors $F_1$ and $F_2$ by a vector-meson bump produces
        kinematic zeros for $A_{1/2}$ and $S_{1/2}$ at $\lambda_\pm=0 \Leftrightarrow \tau = -\delta_\pm^2$ and beyond those points they become imaginary.
        The analogous relations for the $J^P=\nicefrac{3}{2}^\pm$ transition currents defined later in~\eqref{onshell-3/2-current-0}, expressed in terms of the Jones-Scadron form factors
        $G_M(Q^2)$, $G_E(Q^2)$ and $G_C(Q^2)$, read~\cite{Jones:1972ky,Devenish:1975jd}
        \begin{equation} \renewcommand{\arraystretch}{1.0}
            \left[\begin{array}{c} G_M \\ G_E \end{array}\right] = -\frac{A_{1/2}+\sqrt{3}\,A_{3/2}}{2\delta_\pm R_\mp}\,, \qquad
            \left[\begin{array}{c} G_E \\ G_M \end{array}\right] =  \frac{A_{1/2}-\frac{1}{\sqrt{3}} A_{3/2}}{2\delta_\pm R_\mp}\,, \qquad
            G_C = \frac{m_R}{\gamma m}\,\frac{S_{1/2}}{2\delta_\pm R_\mp}\,.
        \end{equation}

\begin{figure*}[t]
        \begin{center}
        \includegraphics[width=0.99\textwidth]{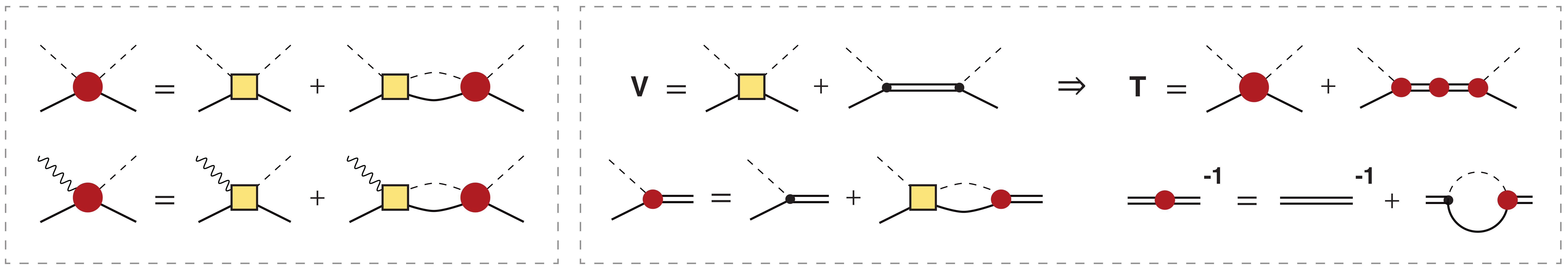}
        \caption{\textit{Left:} Scattering equations for $N\pi$ scattering and pion electroproduction amplitudes. The filled circles denote the T-matrices and the squares are the potentials.
                 \textit{Right:} Decomposition of the potentials (here for the $N\pi$ case) into non-resonant and resonant parts, which leads to the same separation for the T-matrix.
                        The $N\to N^\ast$ transition vertices and dressed propagators are determined from the equations at the bottom. Note that the loop diagram in the vertex equation
                        can be equally written in terms of the background $N\pi$ scattering matrix and a bare vertex (instead of the background $N\pi$ potential and a dressed vertex).}\label{fig:reaction-models}
        \end{center}
\end{figure*}

\paragraph{Analysis of experimental results.}
While the bump landscape in the experimentally measured structure functions in~\eqref{ep-cross-section} provides a basic indication
of the underlying baryon spectrum, the detailed extraction of baryon properties requires a more sophisticated toolbox.
Several analysis tools have been developed and are
still under development to achieve this task. They can be roughly categorised as
reaction models, which assume a certain reaction mechanism and determine
resonance observables by fitting a large set of parameters to the experimental multipole amplitudes,
and dynamical coupled-channel models which
aim at a self-consistent description of the reaction dynamics.
In the following we will sketch the basic ideas behind these approaches and refer to Refs.~\cite{Burkert:2004sk,Pascalutsa:2006up,Klempt:2009pi,Aznauryan:2011qj,Beck:2016hcy}
for details and a comprehensive list of references.

The common goal is to calculate the T-matrix or, equivalently,
its multipole expansion in terms of interaction potentials $V_{ij}$, which are split
into a non-resonant background and resonant contributions.
The background potentials are typically derived from the tree-level diagrams of chiral effective Lagrangians
 and contain the nucleon Born terms together with the
 $u$-channel resonances and $t$-channel  meson exchanges in Fig.~\ref{fig:one-photon-electroproduction};
 the resonant $s$-channel diagrams encode the $N^\ast$ exchanges together with their couplings to the photons and mesons.
Upon selecting the channel space ($N\gamma$, $N\pi$, $N\eta$, $\Delta\pi$, $N\rho$, $N\sigma$ etc.),
one can establish a system of coupled-channel equations for the T-matrix.
For example, keeping only the $N\gamma$ and $N\pi$ channels in the low-energy region leads to the scattering equation
\begin{equation}\label{coupled-channel-eqs} \renewcommand{\arraystretch}{1.0}
  \mathbf{T} = \mathbf{V} + \mathbf{V}\,\mathbf{G}\,\mathbf{T}\,, \qquad
  \mathbf{T} = \left(\begin{array}{cc} T_{\pi\pi} & T_{\pi\gamma} \\ T_{\gamma\pi} & T_{\gamma\gamma} \end{array}\right), \quad
  \mathbf{V} = \left(\begin{array}{cc} V_{\pi\pi} & V_{\pi\gamma} \\ V_{\gamma\pi} & V_{\gamma\gamma} \end{array}\right), \quad
  \mathbf{G} = \left(\begin{array}{cc} G_{\pi} & 0 \\ 0 & G_{\gamma} \end{array}\right),
\end{equation}
where $G_\pi$ and $G_\gamma$ are the two-body nucleon-pion and nucleon-photon propagators and
the scattering matrices correspond to $N\pi$ scattering ($\pi\pi$), pion electroabsorption/electroproduction ($\pi\gamma$, $\gamma\pi$)
and nucleon Compton scattering ($\gamma\gamma$).
Neglecting also electromagnetic effects leaves two equations for $T_{\pi\pi}$ and $T_{\gamma\pi}$ which are shown in the left of Fig.~\ref{fig:reaction-models}:
here only the integral equation for the $N\pi$ scattering amplitude has to be solved and everything else is in principle determined by a one-loop calculation.

There are two standard ways to rewrite~\eqref{coupled-channel-eqs}. One is to split the propagator into two parts, which leads to
the distinction between `T-matrix' and `K-matrix':
\begin{equation}\label{k-matrix}
   \mathbf{T} = \mathbf{V} + \mathbf{V}\,(\mathbf{G}_1+\mathbf{G}_2)\,\mathbf{T}\,, \qquad
   \mathbf{K} = \mathbf{V} + \mathbf{V}\,\mathbf{G}_1\,\mathbf{K} \qquad\Rightarrow\qquad
   \mathbf{T} = \mathbf{K} + \mathbf{K}\,\mathbf{G}_2\,\mathbf{T}\,.
\end{equation}
It allows one to separate the two-body propagator into a principal-value integral and an onshell pole contribution,
where the former goes into the equation for $\mathbf{K}$ and the latter into $\mathbf{T}$, thus eliminating the need for an integration.
The other modification is to explicitly pull out the $s$-channel $N^\ast$ resonance contributions from the potential
as illustrated in the right panel of Fig.~\ref{fig:reaction-models}, so that the remainder only contains non-resonant diagrams.
The consequence is that
also the T-matrix is now the sum of non-resonant and resonant parts, where the former satisfy the same relations as before (left panel in Fig.~\ref{fig:reaction-models})
but the dressing effects for the resonance vertex and propagator are now calculated separately.
The advantage is that one can study the effects of meson-baryon interactions explicitly:
if the full T-matrix is fitted to the electroproduction data, switching off the dressing effects for the masses and transition form factors provides an estimate for the `quark core' of the resonance.

In practice the simplest course would be to set $\mathbf{T}\approx\mathbf{V}$, so that final-state interactions are neglected completely
and the experimental data are fitted to the tree-level expressions. However, this does not preserve unitarity
and the scattering amplitude does not have the correct cut from the elastic threshold $s=(m+m_\pi)^2$ to infinity.
An approximative way to include rescattering effects
is the K-matrix formulation of~\eqref{k-matrix}, which has found widespread applications and reduces the integral equations to a set of algebraic relations.
For example, the unitary isobar models employed by MAID~\cite{Drechsel:1998hk,Drechsel:2007if,Tiator:2011pw} and the JLab group~\cite{Aznauryan:2002gd,Aznauryan:2011qj} separate the electroproduction amplitude
into a background and a resonant part,
\begin{equation}\label{uim}
   T_{\gamma\pi} = T_{\gamma\pi}^B + T_{\gamma\pi}^R = K_{\gamma\pi}^B\,(1+iT_{\pi\pi}^B) + T_{\gamma\pi}^R \approx V_{\gamma\pi}^B\,(1+iT_{\pi\pi}^B)  + \sum_{R} c_R\,\frac{m_R\,\Gamma}{s-m_R^2+im_R\,\Gamma}\,,
\end{equation}
where the former is expressed through the K-matrix which is subsequently approximated by the background potential $V_{\gamma\pi}^B$.
The resonant contributions are parametrized by a Breit-Wigner form including the resonance masses $m_R$ and the total decay width $\Gamma$, where the coefficients $c_R$ include the transition form factors and their fit parameters.
Knowledge of the pion-nucleon amplitude $T_{\pi\pi}$ in terms of the $\pi N$ scattering phase shifts and inelasticities then completes the formula
and allows one to extract helicity amplitudes (such as those in Fig.~\ref{fig:helicity-amps}) by fitting to the electroproduction data.
Also based on different K-matrix approximations but including a larger number of channels are for example the SAID parametrization~\cite{Arndt:2002xv,Workman:2012jf}, the KSU model~\cite{Manley:1992yb,Shrestha:2012ep},
the Giessen coupled-channel approach~\cite{Penner:2002ma,Penner:2002md,Shklyar:2006xw,Shklyar:2012js} and the Bonn-Gatchina model~\cite{Anisovich:2005tf,Anisovich:2009zy,Anisovich:2011fc}.

The original set of equations in Fig.~\ref{fig:reaction-models} is solved in dynamical coupled-channel models which also take into account the dispersive parts from intermediate channels.
Among those are the Sato-Lee model~\cite{Sato:1996gk}, which has been extended by the EBAC~\cite{Matsuyama:2006rp,JuliaDiaz:2007kz,Suzuki:2009nj} and ANL/Osaka collaborations~\cite{Kamano:2013iva}
and extensively applied to analyze pion photo- and electroproduction data.
The Dubna–Mainz–Taiwan (DMT)~\cite{Kamalov:2000en,Chen:2007cy}, Valencia~\cite{Sarkar:2004jh}, J\"ulich-Bonn ~\cite{Doring:2010ap,Ronchen:2014cna} and GSI models~\cite{Lutz:2001yb,Gasparyan:2010xz,Lutz:2015lca}
have been developed along similar lines.
Theoretical constraints like unitarity and analyticity of the S-matrix are manifest in these approaches.
The potentials are derived from phenomenological Lagrangians and the parameters of the models are then fitted to the scattering data.
The resonances can be determined through their poles in the complex energy plane together with their residues and electro\-couplings.
Such determinations do not need to assume a specific resonance line shape such as the Breit-Wigner parametrization,
and they produce quite spectacular figures for the resonance pole structures in the complex plane, see e.g. Ref.~\cite{Tiator:2011pw},
and the movement and conjunction of poles if rescattering effects are switched off~\cite{Suzuki:2009nj}.
An open issue in coupled-channel approaches is how many `bare' states one has to include. In general one can distinguish two cases: a hadron resonance
is generated from coupled channel corrections to a bare state that may be accounted for as a quark core, or the corresponding analytic structure
is just generated by the coupled channel dynamics itself without such a core. This leaves even room for the extreme case that all resonances
are generated dynamically. This hadrogenesis scenario has been explored in e.g.~\cite{Kolomeitsev:2003kt,Heo:2014cja}
and references therein.

What our discussion makes clear is that
the extraction of masses and transition form factors of excited baryons requires model input.
Common to all models is the approximation of the driving potentials as tree-level terms of effective Lagrangians.
The dimensionality of the equations is usually reduced from four to three,
and since the dynamical models represent hadronic integral equations for the $N\pi$ and electroproduction amplitudes they necessarily depend on offshell hadronic effects.
Another issue is electromagnetic gauge invariance, which is difficult to maintain for loop diagrams because in principle the photon has to be attached to every line that carries charge.
In any case, these approaches provide a unified theoretical framework
that allows for a combined and consistent analysis of $N\pi$ scattering and meson electroproduction.

\subsection{Experimental facilities}
In the following we briefly describe the current and future experimental facilities focused on the study of
baryon properties. An enormous amount of data on electromagnetically induced reactions have been accumulated
over the years, mainly from experiments at Jefferson Lab, ELSA, MAMI, GRAAL, LEGS, MIT-Bates and Spring-8.
We summarise here some of the currently active and planned experiments that have as a main goal the
study of baryon structure. This list is not meant to be exclusive; other experiments such as BES-III,
COMPASS, J-PARC, LHCb or BELLE II (will) also contribute important data on specific aspects of baryon physics.

\paragraph{Jefferson Lab.} With the $12$~GeV upgrade of the CEBAF electron accelerator at the
Thomas Jefferson Laboratory (JLab) all experiments in Halls A, B, and C have been upgraded and the new Hall D established.
Concerning baryon physics, Hall A experiments will focus on the electromagnetic structure of the proton including
their form factors at large $Q^2$ and deeply virtual Compton scattering. In Hall B, experiments performed with the $12$~GeV upgrade CLAS12 focus
on the $N^*$ program (that is, the search and study of nucleon resonances), with particular emphasis on the
$N\rightarrow N^*$ electromagnetic transition form factors and the problem of missing resonances.
Approved experiments on light-meson electroproduction will determine the $Q^2$-evolution of electrocoupling parameters
for $N^*$ states with masses below $2$~GeV up to momentum transfers $Q^2\sim 10$~GeV$^2$, via the study of the major
meson-production channels ($eN\rightarrow eN\pi$, $eN\eta$, $eKY$). At the CLAS experiment all polarisation observables
(beam, target and recoil) are in principle accessible. In addition, it is planned to measure two-pion production processes
$ep\rightarrow ep\pi^+\pi^-$. In particular, the invariant mass distributions and angular distributions for the three
different choices of final hadron pairs will be measured. Hall C experiments will study the pion form factor at large
momentum transfer and, amongst other activities, the EMC effect.

\paragraph{ELSA.} At the 3.5 GeV ELSA electron accelerator facility in Bonn a (polarised) electron beam is used
to generate highly energetic photons via bremsstrahlung. These can be polarised and are subsequently used for
meson photoproduction experiments on different targets. The workhorse of the Crystal-Barrel/TAPS collaboration, the
Crystal Barrel, is a nearly $4\pi$ photon spectrometer. The experimental setup was upgraded with the addition of the
TAPS detector in the forward region to improve resolution and solid angle coverage, and with the possibility of using
polarised photon beams. In the last upgrade it was further supplemented with charged-particle detectors in the forward
direction and the possibility of using polarised targets. This allowed for the (very successful) measurement of single-
and double-polarisation observables. Baryon resonances up to masses of $2.5$~GeV have been investigated and generated several
additions to the PDG in the past years. The new BGO-OD experiment was commissioned recently~\cite{Bella:2014rba} and data
taking has started. The BGO-OD experiment consists of the BGO calorimeter and a magnetic spectrometer at forward angles.
The physics covered in future experiments includes especially processes with mixed charged final states.

\paragraph{MAMI.} Two experiments at the Mainz Microtron (MAMI), which provides a continuous wave, high intensity, polarised
electron beam with an energy up to $1.6$~GeV, deal with the structure of baryons. Experiment A1 is an electron scattering
experiment, equipped with three high-resolution spectrometers and large acceptance. It aims at studying spacelike electromagnetic
form factors of nucleons at high precision and nucleon polarisabilities via virtual Compton scattering. The experiment A2 is a
photoproduction experiment, with photons of energy up to $1.5$~GeV. Both experiments have access to beam, target and recoil polarisation
observables.

\paragraph{PANDA.}
The FAIR accelerator facility at the location of GSI in Darmstadt will provide a high intensity and high energy antiproton beam.
In the $\bar{\mbox{P}}\mbox{ANDA}$ detector the beam will be scattered on fixed proton and nucleon targets in the momentum range
of $1.5\text{--}15$~GeV. $\bar{\mbox{P}}\mbox{ANDA}$ will be a unique anti-proton facility and therefore a highly welcome complementary addition
to existing experiments. The detector consists of a target and a forward spectrometer each with several layers of specialized
subcomponents~\cite{Lutz:2009ff}. The $\bar{\mbox{P}}\mbox{ANDA}$ physics program in the baryon sector includes in particular
the spectrum and properties of (multiple-) strange and charmed baryons, the dynamics of $p\bar{p}$ reactions, and structure physics
such as timelike form factors in the range between $5$ and $28$~GeV$^2$ as well as transverse parton distributions~\cite{Wiedner:2011mf}.


\newpage

\section{Baryon spectrum: theory overview}\label{spec}
The hadron spectrum serves as an important reference in understanding the strong interaction, since its
mass hierarchy and decay patterns express the underlying symmetries and dynamics of the theory. At
the low energies relevant to the bound-state properties of hadrons, their constituent particles -- the quarks
(or valence quarks) -- were postulated as the end result of such a study, leading to the so-called quark model.
At the same time, the parton model was introduced to understand reactions of hadrons at high energies such as
deep inelastic scattering. Although these reactions probe the short-distance properties of the strong
interaction, it became apparent that partons and quarks are the same objects. Subsequent to the introduction
of `colour', the quantum theory of the strong force was established in what we now know as QCD.

These two perspectives lead to two very `different quarks': one is light, of the order of a few MeV and essentially
point-like; the other is an extended object confined within colourless bound states with an effective mass of
several hundred MeV. These are the \emph{current} and \emph{constituent} quarks, respectively, with the high-energy
interactions of the former being described well by perturbation theory owing to the property of asymptotic freedom,
whilst the latter requires non-perturbative techniques or modelling. It is obvious to ask at this stage how the
dynamics of QCD account for this dichotomy and we will address the question in due course.

\subsection{The quark model}\label{spec:quarkmodel}
To fully appreciate the masterstroke that is the quark model
(see e.g.\ Refs.~\cite{Hey:1982aj,Capstick:2000qj,Klempt:2009pi,Richard:2012xw} for reviews)
in the development of hadron physics, one has to keep in mind that it was
constructed without knowledge of the underlying dynamics of QCD\@.
The very concept of quarks themselves was not apparent and instead, until the 1950s at
least, the number of known hadrons was still few enough that they could be considered
elementary. They were the proton and neutron (plus their antiparticles) together
with the three pions, classified according to their Fermi
statistics into baryons and mesons. Whereas it was the inter-nucleon force necessitating
the need for further meson exchanges (vectors and scalars), in addition to pion exchange~\cite{Yukawa:1935xg},
also the new resonances subsequently discovered in
$\pi N$ scattering could hardly be considered  elementary.
An attempt to understand the experimental data was provided by the bootstrap method, a
coupled-channel or molecular model in which the hadrons are built up out of one
another, e.g.\ $\Delta=\pi N +\cdots$, $N = \pi \Delta+\cdots$ etc. Nuclear
democracy demanded that mesons were similarly composed of a baryon and an
antibaryon, which proved difficult to accommodate in practice and despite some successes
it remained unsatisfactory. The situation was
only made worse by the discovery of the $\Lambda$ and $\Sigma$ baryons and the
kaons, requiring the introduction of a new quantum number --
strangeness -- conserved by the strong interaction~\cite{Nakano:1953zz,Gell-Mann:1956iqa}.

\begin{figure}[t]
    \centering\includegraphics[scale=0.8]{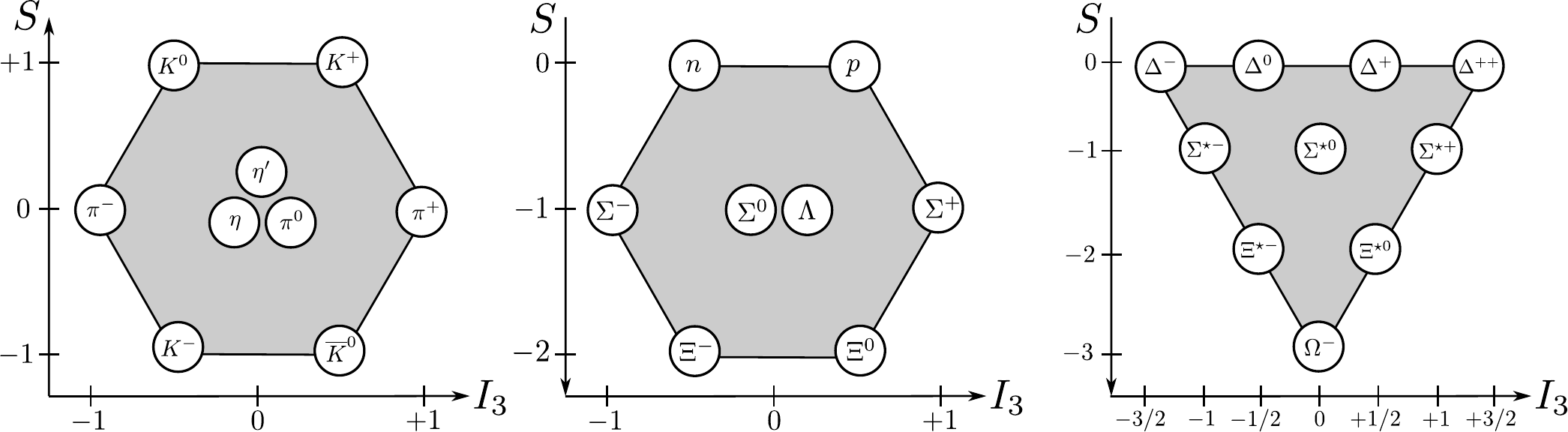}
    \caption{Pseudoscalar meson nonet, baryon octet and baryon decuplet for three light quarks.\label{fig:nonetoctetdecuplet}}
\end{figure}

\paragraph{Eight-fold way.}
Including strangeness, a classification scheme for hadrons was established based upon the
notion of (approximate) symmetries of the strong interaction and thus manifest
in the appearance of near-degenerate multiplets in the hadronic spectrum. It
generalised the already successful $SU{(2)}_f$ isospin symmetry to
flavour $SU{(3)}_f$, characterized now by hypercharge $Y$ in addition to the
$I$ of isospin. The $J=\nicefrac{1}{2}$ baryons were grouped
into an octet while the $J=\nicefrac{3}{2}$ baryons belonged in a decuplet;
the mesons could be grouped into an octet and a singlet, see
Fig.~\ref{fig:nonetoctetdecuplet}.
This is the famous eight-fold way of
Gell-Mann~\cite{GellMann:1961ky,Neeman:1961cd,GellMann:1962xb}, which ultimately
led to the proposal that hadrons be composed of three different flavours of
quarks~\cite{GellMann:1964nj,Zweig:1981pd,Zweig:1964jf}, named \emph{up},
\emph{down} and \emph{strange}, which are spin-$\nicefrac{1}{2}$ fermions with
fractional electric charge.

The baryons are made up of three such valence quarks, whilst the mesons are comprised of
a quark and an antiquark. With quarks (antiquarks) in the fundamental
representation $\mathbf{3}$ ($\mathbf{\overline{3}}$), these are the direct products
\begin{equation}
\mathbf{3} \otimes \mathbf{\overline{3}} = \mathbf{1}\oplus\mathbf{8}\;, \qquad
\mathbf{3} \otimes \mathbf{3} \otimes \mathbf{3} = \mathbf{10}_S\oplus\mathbf{8}_{M_S}\oplus\mathbf{8}_{M_A}\oplus\mathbf{1}_A\;,
\end{equation}
where $S$ and $A$ denote symmetric and antisymmetric and $M_S$,
$M_A$ refer to mixed symmetry (symmetric or antisymmetric with respect to two quarks).
The decuplet corresponds to the ten possibilities to construct a symmetric state of
three quarks: three for equal quarks ($uuu$, $ddd$, $sss$); six where one quark
is different ($uud$, $ddu$, $uus$, $dds$, $ssu$, $ssd$); and one for all quarks
different ($uds$). The states with mixed symmetry can be combined into eight permutation-group doublets:
six for two different quarks and two for all quarks different.
Finally, there is one antisymmetric combination $uds$ where all quarks are different.

\paragraph{Quark model classification scheme.}
To distinguish between the spin-$\nicefrac{1}{2}$ and spin-$\nicefrac{3}{2}$
baryons we combine $SU{(3)}_f$ with the $SU{(2)}$ of quark spin,
\begin{equation}
    \mathbf{2}\otimes\mathbf{2} = \mathbf{3}_{S} \oplus \mathbf{1}_{A}\;, \qquad
    \mathbf{2}\otimes\mathbf{2}\otimes\mathbf{2} = \mathbf{4}_{S} \oplus \mathbf{2}_{M_S}\oplus \mathbf{2}_{M_A} \;.
\end{equation}
Assuming that $\mathbf{\Psi}_{\mathrm{spin}}\otimes\mathbf{\Psi}_{\mathrm{flavour}}$
is symmetric, the possible spin-flavour combinations produce a spin-$\nicefrac{1}{2}$ octet $(\mathbf{2},\mathbf{8})$
and a spin-$\nicefrac{3}{2}$ decuplet $(\mathbf{4},\mathbf{10})$
which correspond to the observed ground states in Fig.~\ref{fig:nonetoctetdecuplet}.
To satisfy the Pauli principle the overall wave function
$\mathbf{\Psi}= \mathbf{\Psi}_{\mathrm{space}}\otimes\mathbf{\Psi}_{\mathrm{spin}}\otimes\mathbf{\Psi}_{\mathrm{flavour}}$
should be antisymmetric, which would imply that its spatial part is antisymmetric and thus not $s$-wave, contrary to our
expectations for ground states.
One could construct antisymmetric spin-flavour combinations, although this would yield $(\mathbf{2},\mathbf{8})$ and $(\mathbf{4},\mathbf{1})$
and thus no decuplet baryons in orbital ground states (the $\Delta^{++}$ is a particularly apposite example). The situation can be remedied by introducing a new
quantum number, namely the $SU(3)$ of colour~\cite{Greenberg:1964pe,Han:1965pf,GellMann:1981ph},
and hence an antisymmetric component $\mathbf{\Psi}_{\mathrm{colour}}$ to the baryon wave function:
\begin{align}\label{eqn:baryonwavefunctiondecomposition}
\mathbf{\Psi}= \mathbf{\Psi}_{\mathrm{space}}\otimes\mathbf{\Psi}_{\mathrm{spin}}\otimes\mathbf{\Psi}_{\mathrm{flavour}}\otimes\mathbf{\Psi}_{\mathrm{colour}}\;.
\end{align}

To treat excited states we need to include quark orbital angular momentum by folding in $O(3)$,
the orthogonal rotation group in coordinate space,
thus requiring the extension of the above construction to all possible spin-flavour symmetries.
The baryons then belong to multiplets
\begin{align}
\mathbf{6}\otimes\mathbf{6}\otimes\mathbf{6} = \mathbf{56}_S \oplus \mathbf{70}_{M_S} \oplus \mathbf{70}_{M_A} \oplus \mathbf{20}_A\;,
\end{align}
where the $56$-plet corresponds to the orbital ground states $(\mathbf{2},\mathbf{8})$ and $(\mathbf{4},\mathbf{10})$ discussed above, whereas the remainder
requires spatial excitations to ensure that the full wave function remains antisymmetric.
Unfortunately, the number of states predicted by the $SU(6)\otimes O(3)$ model
far exceeds the number of observed states in experiment. This is the so-called
problem of missing resonances~\cite{Isgur:1978xj,Isgur:1978wd,Isgur:1979be}, to
which two solutions are typically given. The first is that such states are
difficult to measure because they couple weakly in the scattering cross sections
typically explored; this suggests investigating different channels. The second
is that within the quark model we are making an explicit assumption about the
dominant effective degrees of freedom, namely the constituent quarks. This may be
an oversimplification that does not hold for low-lying excited states where, e.g.,
two quarks could be strongly coupled and form an effective quark-quark correlation,
thus reducing the number of degrees of freedom and hence the number
of excited states.

\paragraph{Dynamical models.}
There are many excellent reviews on the development of quark
potential models~\cite{Hey:1982aj,Capstick:2000qj,Klempt:2009pi,Richard:2012xw}
to which we refer the interested reader for details; here we only summarise a few key points.
At the heart of the quark model is the assumption that the effective degrees of
freedom are the constituent quarks, whose origin is not explicated and
whose masses instead enter as parameters of the theory. Then, the Hamiltonian is
comprised of a kinetic part describing these constituents together with a potential
for the inter-quark forces. Non-relativistically this has the
form
\begin{align}
H = \sum_i \left(m_i + \frac{\vect{p}_i^2}{2m_i} \right) + V\left(\vect{r}_1,\vect{r}_2,\vect{r}_3\right)\,,
\end{align}
where $m_i$ are the constituent quark masses as parameters of the model,
$\vect{p}_i$ the momenta and $\vect{r}_i$ the individual quark coordinates.
The simplest prototypical example of a confining potential is
that of a harmonic oscillator where the quark-quark potential is of the form
\begin{align}
    V\left(\vect{r}_1,\vect{r}_2,\vect{r}_3\right) = \sum_{i<j}V_{\mathrm{conf}}\left(\vect{r}_{ij}\right) \;,\qquad V_{\mathrm{conf}}\left(\vect{r}_{ij}\right) = \frac{1}{2} K r_{ij}^2\;,
\end{align}
i.e., a sum of two-body confining potentials where
$\vect{r}_{ij} = \vect{r}_i - \vect{r}_j$, $r_{ij} = \left| \vect{r}_{ij}\right|$ and $K$ is some constant.
Another possibility is a linearly rising potential, first achieved by the introduction
of anharmonic perturbations to the harmonic oscillator potential~\cite{Gromes:1976cr}.
Such a potential was motivated by early lattice calculations~\cite{Kogut:1974ag,Dosch:1975gf}
and Bethe-Salpeter studies~\cite{Schnitzer:1975qm}. At first sight, both potentials lead to a
very different spectrum with a dependence on energy $E$ vs. $E^2$. However, as clarified recently,
this depends on the context: within the front form Hamiltonian dynamics a quadratic potential leads
to the same qualitative spectrum than a linear potential in the instant form dynamics, see
Ref.~\cite{Trawinski:2014msa} for details.

Additional terms in the potential can be motivated by one-gluon exchange and relativistic
corrections, see e.g.~\cite{DeRujula:1975qlm,Chodos:1974je,Lichtenberg:1975bi,Isgur:1978wd,Isgur:1979be,Bhaduri:1981pn,Hey:1982aj,Theussl:2000sj} and
references therein. The progress of ref.~\cite{Gromes:1976cr},
enabling the first few
energy levels of non-strange baryons for a linearly rising potential to be
determined, was helped by the assumed link between the quark-antiquark
potential in mesons and the quark-quark potential appearing in baryons.
Early potential models paved the way to
more sophisticated analyses by the introduction of converged variational
methods~\cite{Kukulin:1977ux,Varga:1995dm,Papp:1998yt} and the use of Faddeev
equations~\cite{Faddeev:1961cu,SilvestreBrac:1985ic,Richard:1992uk,Papp:1998yt}.
Connections to the heavy quark sector of QCD have been established on the one hand via
heavy quark effective theory and nonrelativistic QCD, see e.g.~\cite{Neubert:1993mb,Brambilla:2004jw}
for reviews, and on the other hand via lattice QCD, see e.g.~\cite{Koma:2006si,Koma:2006fw}.
For baryons the question of $Y$ vs. $\Delta$-shaped potentials has been discussed e.g.
in~\cite{Alexandrou:2002sn,deForcrand:2005vv,Suganuma:2015ycu,Andreev:2015riv}.
Relativistic corrections to inter-quark potentials
have been provided in Refs.~\cite{Basdevant:1984rk,Godfrey:1985xj,Capstick:1986bm}, whilst attempts
at formulating a covariant picture have been made for mesons~\cite{Koll:2000ke,Ricken:2000kf}
and baryons~\cite{Loring:2001kv,Loring:2001kx,Loring:2001ky}.

In these dynamical models the problem of too many predicted states (`missing resonances') remains. In addition, the level ordering hardly agrees with experiment. A notorious problem is
posed by the Roper resonance: the lowest nucleon resonance possesses also positive parity. For hyperons this is not the case, and thus one has to conclude that, if the picture of
mutually interacting constituents quarks is a good description of reality, these interactions are
flavour dependent. In particular, such interactions can be motivated as instanton-induced
forces~\cite{Blask:1990ez,Loring:2001kx,Loring:2001ky}, or spin-spin forces via one-boson
exchange~\cite{Glozman:1995fu,Glozman:1997ag,Glantschnig:2004mu}.

Another question concerns the role of so-called diquark degrees of freedom. As already
mentioned above, in such a picture the number of low-lying states is significantly reduced
and flavour dependencies become more transparent; see e.g.~\cite{Anselmino:1992vg} for
an early review. But can such a picture be derived from QCD, and if so, how?

\paragraph{Connecting to QCD.}
The $SU(3)$ colour symmetry can be identified with the extra symmetry of non-Abelian
gauge theories~\cite{Fritzsch:1972jv,Fritzsch:1973pi,Gross:1973ju,Weinberg:1973un}.
This yields QCD, an asymptotically free theory of quarks and self-interacting gluon
fields. It requires a non-perturbative treatment owing to the strength of the coupling
constant at the low-energy scales relevant for bound states. The lack of isolated
quarks and gluons is then resolved by introducing the concept
of \emph{confinement}. On the one hand, confinement is of utter importance
to understand hadron physics. On the other hand, it is an elusive phenomenon and even
a precise definition within quantum field theory has proven to be highly non-trivial.
It is probably fair to say that there is even no general agreement (yet) on several
aspects of how to define confinement. We come back to this point in Sec.~\ref{spec:green}.

One may then ask: what are the actual differences between quark models and a dynamical description of baryons within QCD?
What becomes of the Schr\"odinger equation and its potential, the quark masses and the wave function in quantum field theory?
The dynamical theory  contains several elements that can indeed be traced back to their quark model analogues and allow one to make such connections.
In short, these are:

(i) The \textit{quark propagator}.
One typically employs the notion that, at hadronic scales relevant to bound states, the
current quarks  strongly couple to the QCD vacuum and essentially surround
themselves with a cloud of quark-antiquark pairs and
gluons. This leads to a massive quasiparticle or constituent quark, which is
the effective degree of freedom in quark models.
As we will discuss in Sec.~\ref{spec:green}, this connection is provided by the dressed quark propagator in QCD, as it
encodes the change from a current quark to a `constituent quark' through its momentum-dependent
mass function and thus establishes a link between the quarks of QCD and the constituent quarks in the quark model.

(ii) The \textit{wave function}. The fully covariant Bethe-Salpeter wave function, discussed
in Sec.~\ref{spec:BS}, has a similar structure to the wave function~\eqref{eqn:baryonwavefunctiondecomposition}
in the quark model. In fact, the colour and flavour parts are identical,
whereas the spin and spatial parts are no longer separate in a covariant framework but intertwined.
Thus their structure is much more complicated than in the non-relativistic case, leading to
the natural presence of higher orbital angular momentum components including $s$-, $p$- and $d$-wave contributions
for the nucleon and even $f$~waves for the $\Delta$ baryon. In turn, this means that the non-relativistic
quark model description is not complete. Another manifestation of this fact is the sector
of `exotic' quantum numbers. Whereas in the non-relativistic classification certain quantum numbers
cannot be built from two constituent quarks and therefore count as  exotic, this is no longer
true in the relativistic framework. Thus, unlike in the quark model, exotic states do not
\textit{necessarily} contain valence contributions beyond the quark-antiquark picture.
We come back to this point in Sec.~\ref{spec:BS}.

(iii) The \textit{equation of motion}. In full QCD the Bethe-Salpeter equation is an exact equation for the
bound state wave function. In practice it has to be approximated to be useful in actual calculations.
These truncations can take place on different levels of sophistication; we discuss corresponding details
in Sec.~\ref{spec:approx}. However, in the non-relativistic limit of (very) heavy quarks, the Bethe-Salpeter
equation simplifies~\cite{Watson:2012ht,Swanson:2012zz} and establishes the connection to the
Dirac and Schr\"odinger equations used in the quark model.

(iv) The \textit{quark-(anti-)quark interaction}. In full QCD the interaction between the valence quarks is encoded in the $q\bar{q}$ and $qqq$
interaction kernels, see Sec.~\ref{spec:BS} for a detailed discussion. These kernels are fully
momentum dependent and in practice it turns out that momenta of the order of $0.5$~--~$2$~GeV are the most important
ones in bound state equations. By contrast, in the quark model this covariant object is replaced by the interquark
potential, which is again only appropriate in the non-relativistic limit of small quark velocities,
i.e.\ at momenta much smaller than the ones typically probed in bound state equations.

These connections can be made more explicit and in part we will do so in the course of this section.
They also imply that some of the central pressing questions of the quark model can be
reformulated and answered in the covariant QCD framework. For example,
what is the precise connection between the current quarks introduced by the Lagrangian of
QCD and the quasi-particle constituent quarks that surface as the relevant degrees of freedom in quark model calculations?
And, is there a dynamical arrangement of the three valence quarks in a baryon such that an
effective quark-diquark state dominates the baryon bound-state amplitude?

\subsection{Correlators and non-perturbative methods in QCD}\label{spec:green}
To discuss the methods required to answer the questions posed at the end of the previous section, we need to introduce
the basics of QCD. We will be very brief here because the details can be found in numerous textbooks.
Recall that we use Euclidean conventions throughout the text; corresponding conventions can be found in App.~\ref{app:conventions}.

The physics of the strong interaction is expressed by QCD's classical action
\begin{align}\label{qcd-action}
     S_\text{QCD} = \int \! d^4x \, \conjg{\psi}\,\mathbf{D}\,\psi + S_\text{YM} \,, \qquad
     \mathbf{D} = \slashed{\partial} + ig\slashed{A} + \mathsf{m}\,, \qquad
     S_\mathrm{YM} = \int \! d^4x \, \frac{1}{4} F^a_{\mu \nu} \,F^a_{\mu \nu} \,,
\end{align}
which depends on the quark and antiquark fields $\psi$, $\conjg{\psi}$ and the vector
gauge fields $A^\mu_a$ representing the gluons.
$\mathbf{D}$~is the Dirac operator that enters in the matter part of the action and contains the quark-gluon coupling term with coupling constant $g$,
and $\mathsf{m} = \text{diag}\,(m_u,m_d,m_s,\dots)$ is the quark mass matrix for $N_f$ flavours.
The Yang-Mills part $S_\mathrm{YM}$ sums over the colour index $a$ of the gluon field strength tensor
$F_{\mu \nu}^a = \partial_\mu A_\nu^a - \partial_\nu A_\mu^a - g f^{abc} A_\mu^b A_\nu^c$
with the structure constants $f^{abc}$ of the gauge group $SU(3)_c$. Its explicit form is
\begin{align}
   S_\text{YM} = \int \! d^4x \left[ -\frac{1}{2}\,A^\mu_a\,( \Box\,\delta^{\mu\nu} - \partial^\mu \partial^\nu)\,A^\nu_a
                                 - \frac{g}{2}\,f_{abc}\,(\partial^\mu A^\nu_a - \partial^\nu A^\mu_a)\,A^\mu_b\,A^\nu_c
                                 + \frac{g^2}{4}\,f_{abe}\,f_{cde}\,A^\mu_a\,A^\nu_b\,A^\mu_c\,A^\nu_d \right].
\end{align}
We have stated the action in terms of bare fields, masses and couplings. Their renormalized counterparts are related
through renormalization constants for the quark and gluon fields, the mass and the coupling:
\begin{equation}\label{renormalization-constants}
   \psi = Z_2^{1/2}\,\psi_\text{R}\,, \qquad
   A = Z_3^{1/2} A_\text{R}\,, \qquad
   \mathsf{m} = Z_m\,\mathsf{m}_\text{R}\,, \qquad
   g = Z_g\,g_\text{R}\,.
\end{equation}
In the following we will drop the index `R' again and, unless stated otherwise, work with renormalized quantities instead.
We will often also use a symbolic notation where we simply write $S_\text{QCD} = \conjg{\psi}\,\mathbf{D}\,\psi + S_\mathrm{YM}$.

The naive physical content of the theory is plain: quarks interact with gluons
which also interact amongst themselves. In terms of tree-level
Feynman diagrams, this is fully described by the propagation of quarks
and gluons and their interactions through the quark-gluon, three-gluon and four-gluon vertices.
Alas, as we all know, the situation is far from being that simple.

\paragraph{Correlation functions.}
The dynamical content of QCD as a quantum field theory is encoded in its partition function, defined by the path integral $\mathcal{Z}$,
which is also the common starting point for nonperturbative methods such as lattice QCD and Dyson-Schwinger equations.
The physical properties of the theory are then extracted from the partition function in terms of Green functions or
correlation functions, which are the expectation values of the corresponding operators $\mathcal{O}$:
\begin{align}\label{correlators}
\mathcal{Z} = \int \mathcal{D} A\,\mathcal{D}\psi \,\mathcal{D} \conjg{\psi} \, e^{-(\conjg{\psi}\,\mathbf{D}\,\psi + S_\mathrm{YM})}
\qquad \rightarrow \qquad
\langle \mathcal{O} \rangle = \frac{1}{\mathcal{Z}} \int \mathcal{D} A\, \mathcal{D}\psi\, \mathcal{D} \conjg{\psi} \, e^{-(\conjg{\psi}\,\mathbf{D}\,\psi + S_\mathrm{YM})}\, \mathcal{O}\,.
\end{align}
The operators $\mathcal{O}$ can contain combinations of fundamental fields or also gauge-invariant composite fields.
Green functions are the central elements of any quantum field theory; they are at the heart of any calculation of scattering and decay
processes and carry all the physical, gauge-invariant information of the theory.
Thus, once we know all correlation functions we have solved the theory.

What is necessary to extract such information? Suppose we work in a weakly
coupled theory such as QED\@. The gauge invariant, physical states of QED are
transverse photons and electrons (or muons and taus) supplemented with photon
clouds, but also bound states such as positronium. Since the
coupling is small, perturbation theory is sufficient to calculate the
correlation functions that are relevant for scattering processes and decay rates up to
the desired precision. The machinery is complemented by suitable methods to
calculate the properties of bound states using Schr\"odinger or Dirac equations
or, in a quantum-field theoretical framework, Bethe-Salpeter equations. The success of
this toolbox is apparent in the spectacular agreement between theory and experiment
for the anomalous magnetic moment of the
electron at the level of ten significant digits. In addition, the correlation functions encode
interesting information on the offshell behaviour of the physical particles of QED such
as the momentum-dependent mass function of the electron or the running of the
QED coupling `constant'.

\paragraph{Confinement.}
In QCD the situation is far more complicated and a prime culprit for this is confinement. Although there is no
doubt that QCD is the underlying theory of hadron physics and that it exhibits confinement, an understanding of this
phenomenon is missing. This contrasts sharply with the comprehension of QCD in the high-energy regime.
QCD's property of asymptotic freedom is the reason for Bjorken scaling \cite{Bjorken:1968dy}, and the logarithmic running of the
coupling leads to scaling violations which have been verified over a huge range of kinematical variables.
Referring to the picture in which at large momentum scales the observed approximate scale independence is
accounted for by the presence of quasi-free and point-like constituents of hadrons, quarks and gluons provide
the most effective language to describe hadronic processes. These QCD degrees of freedom are
complete in the sense that they allow for a description of every hadronic observable.
Changing the perspective to an attempt to understand the strong interaction coming from the low-energy regime,
a possible way of phrasing the existence of confinement in QCD is the statement that all
possible hadronic degrees of freedom will form a complete set of physical states. Therefore, the equivalence
of descriptions of observables in either quark and glue or hadronic degrees of freedom is a direct
consequence of confinement and unitarity. This line of argument can be sophisticated and applied to many
different phenomena involving hadrons. It is known under the name ``quark-hadron duality'', and its
consequences have been verified on the qualitative as well as the semi-quantitative level, for a review see,
e.g., Ref.~\cite{Melnitchouk:2005zr}. The verification of this duality is (beyond the trivial fact of the absence
of  coloured states) the clearest experimental signature for confinement. To appreciate such
a scenario it is important to note that the perfect orthogonality of the quark-glue on the one hand and hadronic states on
the  other hand (and thus the perfect absence of ``double-counting'' in any of the two ``languages'') is
nothing else but another way to express confinement.

One further important aspect of confinement is the occurrence of the linear rising potential in the limit of
infinitely heavy quarks and the property of N-ality of its asymptotics at large distances
found in pure Yang-Mills theory~\cite{Greensite:2003bk,Alkofer:2006fu}. With dynamical quarks, however,
string breaking takes place at large distances, thus leaving only remnants of the linear behaviour in
the intermediary distance region.
Another argument invoked in favour of `confinement by a linearly rising potential' is the existence of Regge
trajectories with an universal slope in the hadron spectrum. However, to this end one should note that
also other types of interactions may lead to Regge trajectories, see, e.g., Ref.~\cite{Fischer:2014xha} for
a discussion.

As they have no immediate consequences for the following sections we will refrain here
to discuss aspects of the manifestations of confinement such as the analytic properties of
the Green's functions of the theory or the structure of its asymptotic state space but refer to~\cite{Alkofer:2000wg}.
An approach that strengthens the evidence for confinement being related to the behaviour of Green's
function in the extreme infrared only is based on the study of the gluon's
quantum equation of motion and how its saturation in the infrared by physical or unphysical degrees of freedom
distinguishes between the Coulomb, the Higgs and the confining phase of a gauge theory~\cite{Schaden:2013ffa}.
This investigation  also further elucidates the role of the BRST symmetry of the gauge-fixed theory for several
types of confinement scenarios, see also the reviews~\cite{Maas:2011se,Vandersickel:2012tz} and references therein.
Other related central questions are the one about the nature of the confining field configurations, the relation
of confinement to dynamical chiral symmetry breaking and the axial anomaly, etc., see
Refs.~\cite{Greensite:2003bk,Alkofer:2006fu} and references therein.

\paragraph{Chiral symmetry.}
Now let us briefly discuss the flavour symmetries of QCD's action.
Noether's theorem states that for each global symmetry of the action there is a conserved current and a conserved charge.
In the massless limit $\mathsf{m}=0$, the classical action of QCD exhibits a chiral symmetry
\begin{equation}
   U(1)_V \otimes SU(N_f)_V \otimes SU(N_f)_A \otimes U(1)_A
\end{equation}
which leads to conserved vector and axialvector currents.
Expressed in terms of the renormalized quark fields, they are given by
\begin{equation}\label{qqbar-currents}
    j^\mu_a(z) = Z_2\,\conjg{\psi}(z)\,i\gamma^\mu\,\mathsf{t}_a\,\psi(z)\,, \qquad
    j^\mu_{5,a}(z) = Z_2\,\conjg{\psi}(z)\, \gamma_5 \gamma^\mu\,\mathsf{t}_a\,\psi(z) \,,
\end{equation}
where $\mathsf{t}_a$ are the $SU(N_f)$ flavour matrices:
the Pauli matrices ($\mathsf{t}_a = \tau_a/2$, $a=1,2,3$) in the two-flavour case, the Gell-Mann matrices ($\mathsf{t}_a = \lambda_a/2$, $a=1 \dots 8$) for three flavours,
and the unit matrix for the flavour-singlet currents ($\mathsf{t}_0=\mathds{1}$).
In the general case $\mathsf{m}\neq 0$ the divergences of the Noether currents become
\begin{equation}\label{vcc-pcac-0}
   \p_\mu \,j^\mu_a = Z_4\,i\conjg{\psi} \left[ \mathsf{m},\mathsf{t}_a\right]\psi \;\stackrel{\mathsf{m} \,=\, m_q}{\longlongrightarrow} \;0\,, \qquad
   -i\p_\mu \,j^\mu_{5,a} = Z_4\,i\conjg{\psi} \left\{ \mathsf{m},\mathsf{t}_a\right\} \gamma_5\,\psi \;\stackrel{\mathsf{m} \,=\, m_q}{\longlongrightarrow} \; 2m_q\,j_{5,a}\,,
\end{equation}
where $Z_4=Z_2\,Z_m$ and $j_{5,a}(z) = Z_4\,\conjg{\psi}(z)\,i\gamma_5\,\mathsf{t}_a\,\psi(z)$ is the pseudoscalar density.
The first relation is the statement of vector current conservation:
the $U(1)_V$ symmetry corresponds to baryon number conservation and holds in general,
whereas $SU(N_f)_V$ is QCD's flavour symmetry discussed in Sec.~\ref{spec:quarkmodel} which is still preserved for equal quark masses ($\mathsf{m}=m_q$).
Even in the case of flavour breaking due to unequal quark masses, the diagonal currents corresponding to $\mathsf{t}_0$, $\mathsf{t}_3$ and $\mathsf{t}_8$
are still conserved and thus their corresponding charges (baryon number, third isospin component, hypercharge) are good quantum numbers.
The multiplet classification in Fig.~\ref{fig:nonetoctetdecuplet} is therefore general  and not tied to the quark model, except that when the symmetry is broken
the states with same $I_3$ and $S$ can mix between the multiplets.
The axial symmetries $SU(N_f)_A$ and $U(1)_A$, on the other hand, are classically only preserved in the chiral limit
because a mass term $\conjg{\psi}\,\mathsf{m}\,\psi$ is not invariant under axial rotations.
This is encoded in the PCAC (partially conserved axialvector current) relation, the second identity in~\eqref{vcc-pcac-0}, which relates
the axialvector current with the pseudoscalar density. There are multiple consequences of these relations which we will explore later.

Symmetries can also be broken in different ways and this leads to  two prominent features of QCD:
the spontaneous breaking of $SU(N_f)_A$ and the axial anomaly or anomalous breaking of $U(1)_A$.
As we will see below, the former is a dynamical and nonperturbative effect that is
generated by the strong interactions between quarks and gluons, hence often also termed dynamical chiral symmetry breaking (DCSB).
In practice it is visible in Green functions containing external quark and antiquark legs, as they do not share the symmetry
of the Lagrangian and nonperturbatively generate  tensor structures that would be forbidden if chiral symmetry were preserved.
DCSB has numerous practical consequences for QCD phenomenology: it is the mechanism that transforms
the current quark in the QCD action into a `constituent quark' and thereby contributes a large portion to a light hadron's mass;
it is responsible for the unnaturally light pseudoscalar mesons because they are QCD's Goldstone bosons in the chiral limit;
and it generates spin-dependent forces which would otherwise be absent.
We will explore the effects of DCSB in various places throughout this review:
its consequences for the quark propagator are discussed in the present section
and those for the pion in Sec.~\ref{spec:BS}, and in Sec.~\ref{sec:ff-vertices}
we will give a quick proof of the Goldstone theorem \cite{Goldstone:1961eq} in QCD.
Note that spontaneous chiral symmetry breaking does not affect the PCAC relation in~\eqref{vcc-pcac-0} which is still valid in that case,
in contrast to the axial anomaly which breaks the $U(1)_A$ symmetry as a necessity of regularisation and leads to an additional term in the flavour-singlet PCAC relation.
For a discussion of the axial anomaly in the context of functional approaches see e.g. \cite{Bhagwat:2007ha,Alkofer:2008et,Alkofer:2013bk}.

In the following we concentrate on a selection of methods that aim at the determination of baryon properties from the underlying theory
of quarks and gluons.

\paragraph{Lattice QCD.}
The basic idea of lattice QCD \cite{Wilson:1974sk} is to extract QCD's
correlation functions by formulating the theory on a
discrete lattice of spacetime points. Here the
formulation~\eqref{correlators} of the partition function on a Euclidean
spacetime manifold is necessary for the fields to form a statistical ensemble
that can be treated with Monte-Carlo methods. The statistical nature of the
approach entails that lattice results always carry a statistical error;
sources for additional systematic errors are the finite lattice spacing $a$
and the finite lattice volume.
In the course of this review we summarise
aspects of lattice QCD that are useful for the later discussion of the
baryon spectra and form factors. Naturally, in a non-expert review we skip over
most technical details and refer the interested reader to specialised lattice
reviews such as~\cite{Gattringer:2010zz,Colangelo:2010et,Fodor:2012gf,Aoki:2013ldr}.

Most of the calculations performed within lattice QCD utilise the path integral in
its gauge-invariant version~\eqref{correlators}, which guarantees that the results are
gauge-invariant from the start. However, it is also possible to
fix the gauge on the lattice and there are a number of good reasons why this is
of interest. On the one hand, there is a long tradition of studying fundamental
problems like confinement on a level that gives access to individual gauge
field configurations, see e.g.~\cite{Greensite:2003bk,Alkofer:2006fu} for
reviews. On the other hand, working in a gauge-fixed formulation allows one to study
non-perturbative correlation functions such as propagators and vertices,
which are the basic building blocks for functional methods. As discussed below,
functional methods almost always involve truncations which need to be
controlled, and one important cross check which has proven extremely
fruitful in the past is a direct comparison with
correlation functions obtained on the lattice.
We discuss this point further below and come back to gauge-invariant lattice QCD in
Sec.~\ref{spec:extracting}, where we deal with the extraction of hadron
properties from lattice simulations.

\paragraph{Light-front QCD and holographic QCD.}
In the traditional canonical quantization of quantum field theories the concept of the Fock space as the Hilbert space of (onshell) multiparticle states
is employed only for the asymptotic in and out states in order to circumvent mathematical subtleties; see e.g.~\cite{Haag:1992hx}.
The properties of the interacting theory are then encoded in its matrix elements:
the (elementary or composite) Green functions of the theory and its physical scattering amplitudes.
In that way the `Fock states' refer to external, physical particles on their mass shells whereas the intermediary `virtual particles'
carrying the diagrammatic content of Feynman diagrams are represented by the Green functions of the theory.
These are either the tree-level propagators and vertices in perturbation theory or, nonperturbatively,
their dressed counterparts together with the hadronic Bethe-Salpeter wave functions discussed below.

Light-front Hamiltonian methods (see e.g.~\cite{Brodsky:1997de,Burkardt:1995ct,Heinzl:2000ht,Miller:2000kv,Cruz-Santiago:2015dla,Hiller:2016itl} for reviews) take a different route in this respect.
Motivated by the triviality of the vacuum in light-cone quantization,
the goal is to calculate the eigenstates of the gauge-fixed QCD Hamiltonian directly by solving large-scale eigenvalue problems.
In that way the light-cone wave functions move into the center of interest: once calculated, in principle they can be used to compute
hadronic matrix elements of currents as overlaps of light-cone wave functions.
Although corresponding methods have been successfully employed to a variety of quantum field theories including QED,
the systematic application of this machinery to QCD is quite nontrivial and awaits to be further explored~\cite{Hiller:2016itl}.

On the other hand, light-front based ideas have found widespread applications
in the context of hard exclusive processes which are dominated by their behavior close to the light cone.
They provide, for example, a connection of the parton model with QCD, but also a systematic treatment of two-photon processes such as
deeply virtual Compton scattering in terms of generalized parton distributions~\cite{Brodsky:1981kj,Brodsky:1989pv,Jaffe:1996zw,Belitsky:2005qn,Guidal:2013rya,Mueller:2014hsa,Kumericki:2016ehc}.

Instead of diagonalizing the QCD Hamiltonian directly, light-front techniques can also be used
to construct quark models, particularly in combination with the AdS/QCD approach~\cite{Brodsky:2014yha}.
The latter is based on the conjecture that a quantum gauge field theory in four dimensions
corresponds to a classical gravitational theory in five dimensions (the holographic dual)~\cite{Maldacena:1997re}.
In principle this allows one to compute physical observables in a strongly coupled gauge theory in terms of a weakly coupled classical gravity theory,
which encodes information of the boundary theory.
In recent years progress has been achieved in the description of the baryon spectrum from holographic QCD,
see e.g.~\cite{deTeramond:2011qp,Gutsche:2011vb,deTeramond:2014asa,Brodsky:2014yha,Deur:2014qfa,Dosch:2015nwa,Berenstein:2002ke,Hong:2006ta,Forkel:2007tz} and references therein.

\paragraph{Dyson-Schwinger equations (DSEs).}
Dyson-Schwinger equations~\cite{Dyson:1949ha,Schwinger:1951ex,Schwinger:1951hq} are the quantum equations of motion of a quantum field theory;
see~\cite{Roberts:1994dr,Alkofer:2000wg,Maris:2003vk,Fischer:2006ub,Binosi:2009qm,Swanson:2010pw,Bashir:2012fs} for reviews of their applications to QCD.
The starting point for the derivation of DSEs is the same as in lattice QCD, namely the generating functional or partition function~\eqref{correlators}.
The basic idea is that instead of extracting Green functions directly from the path integral
one derives relations among them from the partition function. These are the quantum equations of motion which follow from
the expectation values of the classical equations of motion in analogy to~\eqref{correlators}, and they form
an infinite tower of integral equations that couple Green's functions to
one another in a hierarchical fashion. By construction they are nonperturbative
since they resum an infinite number of diagrams without recourse to a weak
coupling expansion.

Functional methods like the Dyson-Schwinger
approach or the functional renormalisation group~\cite{Berges:2000ew,Pawlowski:2005xe,Gies:2006wv}
mainly work with gauge-fixed correlation functions, which requires a
gauge-fixing procedure such as the Faddeev-Popov approach. Its main
effect is the addition of gauge fixing terms $S_\mathrm{gf}$ to the action, which
introduce a pair of additional fields (the Faddeev-Popov ghosts)
plus an extra bilinear term into the gluonic part of the action.
Gauge fixing beyond perturbation theory is a very delicate (but highly
interesting) issue which is reviewed e.g.\ in~\cite{Sobreiro:2005ec,Maas:2011se,Vandersickel:2012tz,vonSmekal:2013cla}.
We note here only  that,
fortunately, it turns out that the associated `Gribov copies' \cite{Gribov:1977wm}
are only relevant for momenta below $\sim 100$~MeV \cite{Sternbeck:2005tk,Bogolubsky:2009dc},
so their impact upon the calculation of observable quantities is in practice very small compared
to other sources of errors, see e.g. \cite{Blank:2010pa}.

In practice one is mainly interested in the proper one-particle irreducible (1PI)
Green functions, i.e., the connected and amputated vertex functions.
They are the fully dressed inverse propagators and $n$-point vertices of the theory --
the inverse quark and gluon propagators, the
three-point functions such as the quark-gluon and three-gluon
vertex, the four-point functions and so on. They contain all possible loop corrections
and thus they are the exact $n$-point functions in
contrast to the ones calculated in perturbation theory.
The 1PI vertex functions are encoded in the 1PI effective action $\Gamma_{1\mathrm{PI}}[\tilde{\Phi}]$,
which is the Legendre transform of the partition function and follows from introducing external sources $J$ for the fields $\Phi \in \left\{A,\psi,\ldots\right\}$ in the Lagrangian
\begin{align}\label{eqn:legendretransformationonepi}
\Gamma_{1\mathrm{PI}}[\tilde{\Phi}] = -\ln \mathcal{Z}[J] - J_i\,\tilde{\Phi}_i\;,
\end{align}
where the \emph{averaged field} $\tilde{\Phi}$ is the
expectation value of $\Phi$ in the presence of sources $J$.
The effective action can be expanded in the
1PI vertex functions $V^{(N)}_{i_1,\ldots,i_N}$ and, vice versa, they can be derived from it via functional derivatives
\begin{align}\label{action}
\Gamma_{1\mathrm{PI}}[\tilde{\Phi}] = \sum_N \frac{1}{N!} \left(\tilde{\Phi}_{i_1}\ldots \tilde{\Phi}_{i_N}\right) V^{(N)}_{i_1,\ldots,i_N}\;,\qquad\qquad
V^{(N)}_{i_1,\ldots,i_N} =\frac{\delta^N \Gamma_{1\mathrm{PI}}[\tilde{\Phi}]}{\delta \tilde{\Phi}_{i_1}\ldots\delta \tilde{\Phi}_{i_N}}\bigg|_{\tilde{\Phi}=0}\;.
\end{align}
In addition to the
$n$-point functions that appear in QCD's Lagrangian, the effective
action also contains vertex functions without a tree-level counterpart such as
e.g.\ the (1PI) four-quark vertex. If one were to expand such quantities in
perturbation theory one would not find tree-level contributions but loop
`corrections' only.
From the effective action one can derive the Dyson-Schwinger equations by
taking further functional derivatives, see e.g.~\cite{Roberts:1994dr,Alkofer:2000wg} for details, and the resulting equations
relate the various 1PI Green functions to each other.

\begin{figure*}[t]
        \begin{center}
        \includegraphics[width=0.95\textwidth]{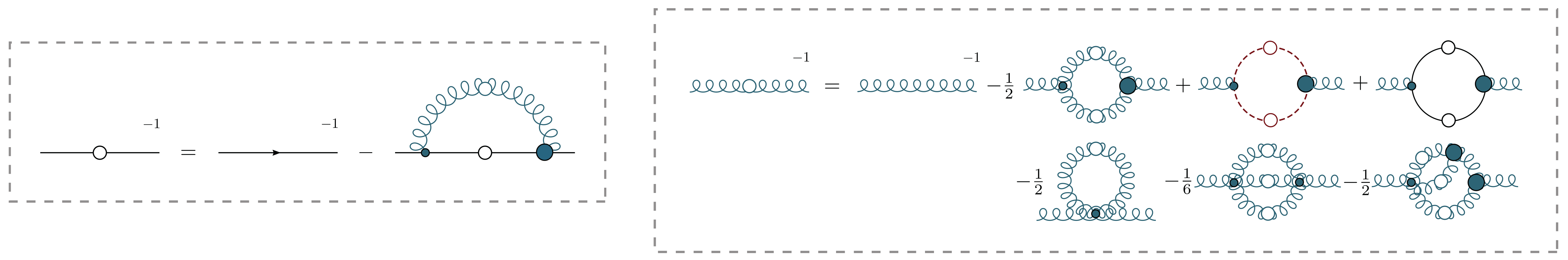}
        \caption{DSEs for the quark propagator (left) and the gluon propagator (right). Solid, curly and dashed lines represent quarks, gluons and ghosts, respectively.}\label{fig:DSEs}
        \end{center}
\end{figure*}

As an example, consider the quark DSE in the left panel of Fig.~\ref{fig:DSEs}.
It is an exact equation that determines the dressed propagator (the line with the circle)
moving in spacetime from one point to another.
The nonperturbative loop diagram contains the sum of all possible processes where the quark emits
and reabsorbs a gluon. The gluon is also fully dressed and includes all possible splittings and
reunifications into two gluons, three gluons, a $q\bar{q}$ pair
and so on, and the dressed quark-gluon vertex subsumes all possible vertex
corrections. Finally, the appearance of a dressed quark propagator in the
loop indicates that the equation has to be solved self-consistently.
To do so, one needs to know the dressed gluon propagator and the dressed quark-gluon
vertex. These satisfy their own Dyson-Schwinger equations (such as the gluon DSE in
the right panel of Fig.~\ref{fig:DSEs}) which contain other
three- and four-point functions with their own DSEs.
Hence we arrive at an infinite tower of self-consistent equations. Apart from special
limits~\cite{Alkofer:2004it,Fischer:2006vf,Fischer:2009tn} it cannot be
solved without approximations, but these can be made systematic by consistently
neglecting or approximating Green functions higher than a given order. A simplistic
example would be to set the three-point vertices bare and neglect all higher Green functions.
From the experience with such a truncation the obvious next step is to take into account
the full three-point functions but approximate the four-point functions, and
so on. To systemetise this procedure until apparent convergence is reached
is at the heart of modern treatments of the tower of DSEs, see
e.g. \cite{Blum:2014gna,Eichmann:2014xya,Aguilar:2014lha,Cyrol:2014kca,Williams:2015cvx,Huber:2016tvc}
and references therein.
Other strategies are possible and are discussed below and in the course of this
review.

\paragraph{Functional Renormalisation Group (FRG).}
A similar but nevertheless completely independent approach
is the functional renormalisation group, see
e.g.~\cite{Berges:2000ew,Pawlowski:2005xe,Gies:2006wv} for reviews. Here the idea
is to introduce an infrared momentum cutoff into the effective action
$\Gamma_{1\mathrm{PI}}$ and study its variation with the cutoff scale~$k$. This
leads to a differential equation with a characteristic one-loop structure which is exact
\cite{Wetterich:1992yh,Morris:1993qb}. In practice it
needs to be approximated, e.g., by keeping a finite number of
terms in the expansion~\eqref{action} of the effective action and/or neglecting
momentum dependencies of the $n$-point functions in question. Again,
by taking further derivatives with respect to the fields one derives coupled
integro-differential equations which relate different Green functions
to each other. Compared to the tower of DSEs, FRGEs only feature (nonperturbative) one-loop diagrams and the extra
dependence on the cutoff scale makes the equations more local than the DSEs.
Due to these differences, combinations of DSEs and FRGEs can be used
to study, control and thus minimize artefacts from the necessary truncations.
This strategy has been successfully employed in the past and FRGEs and DSEs
together have been essential in the determination of QCD's Green functions, see
\cite{Braun:2014ata,Fischer:2006vf,Fischer:2009tn,Fischer:2013eca,Mitter:2014wpa,Cyrol:2016tym} and references therein.
With respect to the main topic of this review,
however, although FRG studies have begun to make progress in the meson sector they
have not yet contributed to our understanding of the spectrum and properties of baryons.

\paragraph{Quark propagator and dynamical chiral symmetry breaking.}
Green functions are vital in the Dyson-Schwinger approach as building blocks
for the calculation of gauge-invariant bound states, form
factors and production processes and therefore we need to cover the essentials
before discussing methods to extract the hadron spectrum in Sec.~\ref{spec:extracting}.
In the following we briefly summarise the status of results for the
basic Green functions of QCD: the quark propagator, gluon propagator, quark-gluon vertex and three-gluon
vertex.

The quark is a spin-$\nicefrac{1}{2}$ particle and its dressed propagator is given by
\begin{align}\label{quark-propagator}
S(p) = \frac{1}{A(p^2)}\frac{-i\slashed{p}+ M(p^2)}{p^2 + M^2(p^2)} = -i\slashed{p}\,\sigma_v(p^2) + \sigma_s(p^2)\;.
\end{align}
The quark mass function $M(p^2)$ and inverse quark `wave function' $A(p^2)$ are momentum dependent,
which reflects the scale dependence of the quantum corrections encoded within.
They connect the current quarks probed in
deep inelastic scattering processes at large momentum transfer with the constituent-quark
picture at small momenta, and
thus the quark propagator serves to elucidate the connection of the quark model with QCD.
Non-zero values of $M(p^2)$ in the chiral limit $m_q=0$ are an indication for spontaneous chiral
symmetry breaking: for massless quarks the Lagrangian exhibits a chiral symmetry but the Green function at the quantum level, the quark propagator, does not.
This can be seen in the renormalization point dependent chiral condensate, obtained from the trace of the quark propagator in the chiral limit
\begin{align}
-\left<\bar{q}q\right> =  Z_2Z_m N_c \int \!\!\frac{d^4p}{\left(2\pi\right)^4}\,\mathrm{Tr} \,S(p)\;.
\end{align}
A non-zero value signals the presence of dynamical chiral symmetry breaking.
In fact, we will argue below that dynamical chiral symmetry breaking leaves its traces in the
tensor structures of every Green function with external quark and antiquark legs.

The explicit form of the quark DSE in Fig.~\ref{fig:DSEs} reads
\begin{align}\label{baryontheory:eqn:quarkgapequation}
S^{-1}(p) = Z_2 \,S^{-1}_0(p) - Z_{1f} \,g^2 \,C_F  \int \!\! \frac{d^4q}{(2\pi)^4}\, i\gamma^\mu S(q) \,\Gamma_\mathrm{qg}^\nu(q,p) \,D^{\mu\nu}(k)\;,
\end{align}
where $Z_2$ and $Z_{1f}=Z_g \,Z_2 \,Z_3^{1/2}$ are the renormalization constants for the quark propagator and quark-gluon vertex, see~\eqref{renormalization-constants},
and $C_F=4/3$ is the colour Casimir for $N_c=3$. The inverse tree-level propagator is given by
$S^{-1}_0(p) = i\slashed{p}+Z_m \,m_q$, where $m_q$ is the renormalized quark mass from the QCD action. The
unknown quantities in this expression are the dressed gluon propagator
$D^{\mu\nu}(k)$ with gluon momentum $k=q-p$ and the dressed quark-gluon vertex $\Gamma_\mathrm{qg}^\mu(q,p)$. Since
the nonperturbative gluon propagator is a rather well-known
object by now, it is the choice of the quark-gluon vertex (equivalently the form of the four-quark interaction kernel) that ultimately
determines the quality of the truncation, as we will explain later in Sec.~\ref{spec:approx}.

\begin{figure*}[t]
        \begin{center}
        \includegraphics[width=0.99\textwidth]{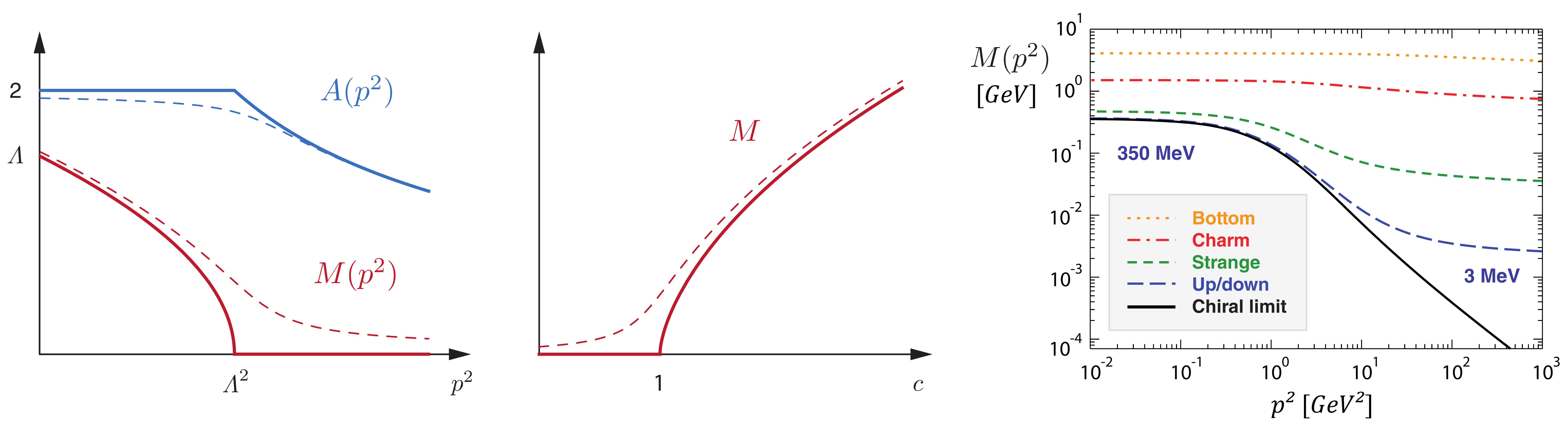}
        \caption{Quark propagator in the Munczek-Nemirovsky model (\emph{left}) and NJL model (\emph{center}).
         The solid lines are solutions in the chiral limit $m_q=0$ and the dashed lines show the qualitative behaviour at $m_q\neq 0$.
         The right diagram contains typical DSE solutions for the quark mass function from realistic truncations.}\label{fig:quark-mn}
        \end{center}
\end{figure*}

For the sake of a pedagogical introduction, we follow Ref.~\cite{Roberts:2007jh} and elucidate upon
the properties of the quark propagator with two toy models before moving on to
more realistic calculations. Both of these neglect dressing effects in the
quark-gluon vertex, assuming $\Gamma_\mathrm{qg}^\mu(q,p)=i \gamma^\mu$, and use a very simple
ansatz for the gluon propagator which allows one to explore the qualitative consequences of
dynamical chiral symmetry breaking. Without loss of generality one can drop all renormalization constants.\\[-3mm]
\begin{itemize}
\item \textbf{Munczek-Nemirovsky model}~\cite{Munczek:1980cj,Munczek:1983dx}: Here
the Feynman-gauge gluon propagator is replaced by a $\delta$-function peaked at
the origin with an associated mass scale $\Lambda$:
\begin{align}\label{baryontheory:eqn:munczeknemirovsky}
\frac{g^2\, C_F}{(2\pi)^4}\,D^{\mu\nu}(k) = \delta^{\mu\nu}\,\Lambda^2\, \delta^4(k)\;.
\end{align}
Consequently, the integration can be performed analytically and
for $m_q=0$ one obtains two solutions,
\begin{equation}
\left\{
\begin{array}{rl}
   M(p^2)\!\! &= \sqrt{\Lambda^2-p^2}  \\
   A(p^2)\!\! &= 2
\end{array}\right\},\qquad
\left\{
\begin{array}{rl}
   M(p^2)\!\! &= 0  \\
   A(p^2)\!\! &= \frac{1}{2}\left(1+\sqrt{1+8\Lambda^2/p^2}\right)
\end{array}\right\},
\end{equation}
which are connected at $p^2 = \Lambda^2$ and lead to the non-trivial quark mass function shown in
the left panel of Fig.~\ref{fig:quark-mn}. For $m_q\neq 0$ the transition
between the infrared and ultraviolet is smoothed out. Although the model
does not feature a critical coupling for the onset of spontaneous symmetry breaking, it
already captures an essential part of realistic DSE calculations: the momentum
dependence of the quark propagator reflects dynamical mass
generation, which kicks in around some typical hadronic scale $\Lambda$ and
turns the `current quark' into a dressed constituent quark in the infrared. \\[-3mm]
\item \textbf{NJL model:} Instead of localizing the gluon
propagator in momentum space, one can take the opposite extreme and localize it
in coordinate space. This results in an effective four-fermi interaction between
two quarks, which is the Nambu--Jona-Lasinio (NJL)
model~\cite{Nambu:1961fr,Nambu:1961tp} with
\begin{align}\label{baryontheory:eqn:contactinteraction}
\frac{g^2\, C_F}{(2\pi)^2}\,D^{\mu\nu}(k) = \delta^{\mu\nu} \frac{c}{\Lambda^2}\;.
\end{align}
The resulting momentum integral in the quark DSE is now divergent
and requires the introduction of a cutoff~$\Lambda$ above which the gluon vanishes.
As a consequence, both dressing functions remain
momentum independent and the equation is solved by $A=1$ and $M=m_q+cM\left(1-a\,\ln(1+1/a)\right)$, with
$a=M^2/\Lambda^2$.
However, now we see the desired critical behaviour with respect to the coupling~$c$:
in the chiral limit the quark mass $M$ remains zero for $c<1$ but above that value chiral symmetry is
spontaneously broken, as can be seen in the middle diagram of
Fig.~\ref{fig:quark-mn}. \\[-3mm]
\end{itemize}

The gluon propagator in QCD is neither a $\delta-$function nor a constant and the quark-gluon vertex has a complicated tensor
structure and momentum dependence. Thus, the resulting quark mass function is
richer in structure than the ones generated by these models. Using
ans\"atze for the three- and four-point vertices of the theory, first
self-consistent solutions for the gluon and quark propagator have been obtained
in~\cite{Fischer:2003rp}. Modern truncations also include numerical solutions
for the three- and four-point functions~\cite{Mitter:2014wpa,Williams:2015cvx,Cyrol:2016tym} whose results
compare well with those from lattice QCD~\cite{Skullerud:2003qu,Sternbeck:2005tk,Bowman:2005vx,Ilgenfritz:2006he,Cucchieri:2008qm,Cucchieri:2011ig}.
The typical behaviour of the quark
mass function for different quark flavours is shown in the right panel of
Fig.~\ref{fig:quark-mn}. The quark mass generated in the infrared connects
smoothly with the logarithmic running in the ultraviolet as dictated by
perturbation theory, whereas in the chiral limit it roughly falls like $1/p^2$, see
e.g.~\cite{Fischer:2003rp} for details. The associated scale
is not put in by hand (as in the models) but dynamically generated in
the Yang-Mills sector of the theory. Clearly, the light quarks are dominated by
the dynamical effects whereas for heavy quarks explicit chiral symmetry
breaking starts to take over.

\begin{figure*}[t]
        \begin{center}
        \includegraphics[width=0.99\textwidth]{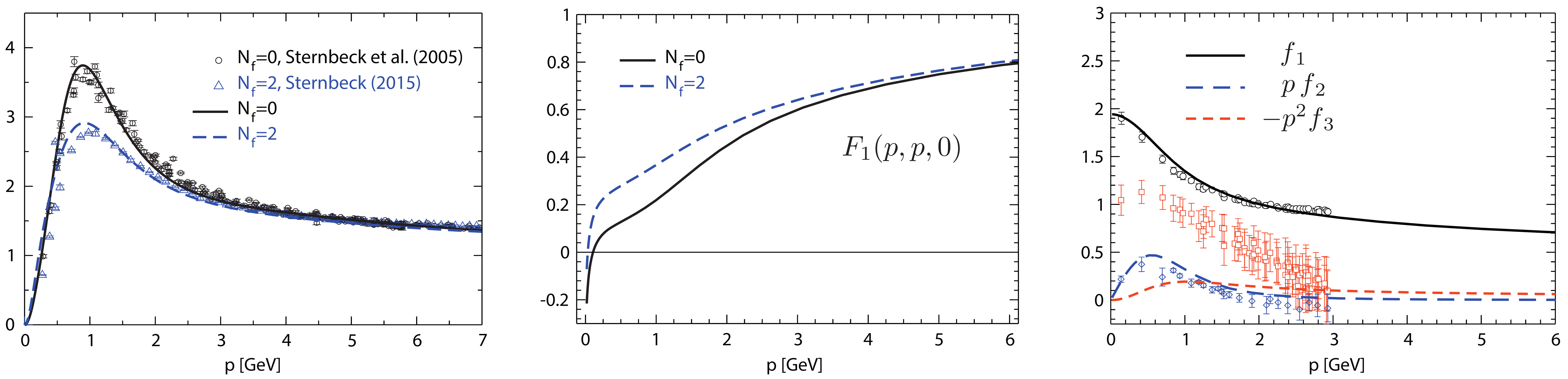}
        \caption{\textit{Left:} Gluon dressing function $Z(p^2)$ for $N_f=0$ and $N_f=2$ calculated from DSEs~\cite{Williams:2015cvx} and compared to lattice
 calculations~\cite{Sternbeck:2005tk,Sternbeck:2016}.
 \textit{Center:} Leading dressing function of the three-gluon vertex evaluated on a symmetric momentum configuration~\cite{Eichmann:2014xya}.
 \textit{Right:} Three dressing functions of the quark-gluon vertex evaluated in the soft gluon limit $k^\mu=0$. The (quenched)
 DSE results of~\cite{Williams:2015cvx} are compared with lattice results of Ref.~\cite{Skullerud:2003qu}.}\label{fig:greens}
        \end{center}
\end{figure*}

\paragraph{Other $n$-point functions.}
Naturally, nonperturbative effects are not exclusive to the quark
propagator but appear in all other Green functions as well.
For example, the gluon propagator in Landau gauge is given by
\begin{align}\label{gluon}
D^{\mu \nu}(k) = \left(\delta^{\mu \nu} - \frac{k^\mu k^\nu}{k^2}\right) \frac{Z(k^2)}{k^2}\,,
\end{align}
where the gluon dressing function $Z(k^2)$ carries all the nontrivial information.
At large momenta $Z(k^2)$ shows the usual logarithmic
behaviour generated by the interactions of the asymptotically free theory.
Below $2$~GeV this changes drastically and nonperturbative effects take over. This can be seen in the left
diagram of Fig.~\ref{fig:greens}, where we compare recent results from
DSEs~\cite{Williams:2015cvx} and gauge-fixed lattice
calculations~\cite{Sternbeck:2005tk,Sternbeck:2016}. The gluon propagator has
been studied intensely in the past both on the lattice, with DSEs and the
FRG due to its supposed connection with
confinement of the non-Abelian gauge theory, see e.g.~\cite{vonSmekal:1997ern,Aguilar:2008xm,Fischer:2008uz,Dudal:2008sp,Cucchieri:2008fc,Boucaud:2008ky,Bogolubsky:2009dc,Cucchieri:2011ig,Strauss:2012dg,Aguilar:2015bud,Cyrol:2016tym} and references therein.

Due to its much more complicated structure, the nonperturbative three-gluon
vertex has only been begun to be explored in numerical simulations in recent
years, see e.g.~\cite{Maas:2007uv,Cucchieri:2008qm,Binosi:2013rba,Blum:2014gna,Eichmann:2014xya,Mitter:2014wpa,Williams:2015cvx,Cyrol:2016tym}
and references therein. In the numerical calculations it turned out that the
tensor structure of the bare vertex,
\begin{align}
\Gamma^{\alpha\beta\gamma}_{\mathrm{3g}(0)}(p_1,p_2,p_3) &=
 \delta^{\alpha\beta}\left(p_1 - p_2\right)^\gamma
+\delta^{\beta\gamma}\left(p_2 - p_3\right)^\alpha
+\delta^{\gamma\alpha}\left(p_3 - p_1\right)^\beta\,,
\end{align}
also dominates the nonperturbative dressed vertex by far~\cite{Eichmann:2014xya}.
It may have the interesting feature of a zero crossing at finite momenta, cf. Fig.~\ref{fig:greens}.
Phenomenologically, the three-gluon vertex could be an important component of (coloured) three-body
forces, although it turns out that the vertex is suppressed inside baryons due to the colour algebra~\cite{Sanchis-Alepuz:threebodyinprep}.

As will be discussed further in Sec.~\ref{spec:approx},
the quark-gluon vertex is probably the most relevant Green function regarding the quality of
truncations used in the DSE framework. It features twelve tensor
structures,
       \begin{equation}\label{qgv}
           \Gamma^\mu_\mathrm{qg}(p,k) = f_1\,i\gamma^\mu  - f_2\,p^\mu + f_3\,ip^\mu\slashed{p} -  f_4\,p\cdot k\,\tfrac{1}{2}[\gamma^\mu,\slashed{p}] + \Gamma^\mu_\perp(p,k)\,,
       \end{equation}
with a transverse part that is structurally identical to that of the quark-photon vertex in~\eqref{qpv-transverse-basis}.
Here we have expressed the vertex in terms of the total (gluon) momentum $k$ and a relative or average quark momentum $p$ (in~\eqref{baryontheory:eqn:quarkgapequation} the relative momentum would be $(q+p)/2$),
and for the sake of consistency we will continue to do so from now on.

In contrast to the Abelian theory, the momentum dependence of the vertex is potentially very different due to strong
non-Abelian contributions that enter in the DSE, and the Slavnov-Taylor identities \cite{Taylor:1971ff,Slavnov:1972fg} are of
limited help since they cannot be solved exactly.
Exploratory lattice results are available in the quenched
approximation~\cite{Skullerud:2003qu}, with unquenched updates under way. In
the DSE framework the quark-gluon vertex has been tackled with three different
strategies: (i) nonperturbative ans\"atze for selected tensor structures,
constructed along continuations of resummed perturbation theory into the infrared and
supplemented with information from the Slavnov-Taylor
identities~\cite{Fischer:2003rp,Bhagwat:2004kj,Fischer:2005en,Fischer:2006ub,Chang:2009zb,Chang:2010hb,Chang:2011ei,Fischer:2012vc};
(ii) explicit solutions of approximated Slavnov-Taylor
identities~\cite{Aguilar:2014lha}; and (iii) numerical solutions of truncations
of the quark-gluon vertex DSE and FRGE~\cite{Bender:2002as,Bhagwat:2004hn,Matevosyan:2006bk,Alkofer:2008tt,Fischer:2008wy,Fischer:2009jm,Hopfer:2014szm,Windisch:2014lce,Williams:2014iea,Mitter:2014wpa,Braun:2014ata,Williams:2015cvx}.
At least on a quantitative level the results from these approaches are not in very good agreement
with each other. On the other hand, recent truncations of
DSEs and FRGEs for the propagators \textit{and} vertices are very
encouraging and begin to show quantitative agreement with each other and also with lattice results.
In Fig.~\ref{fig:greens} we compare results from the calculation of Ref.~\cite{Williams:2015cvx} with lattice
data~\cite{Skullerud:2003qu}.
The agreement is particularly interesting for $f_2$ which can only be
present once chiral symmetry is broken and thus it is on the same
footing as the quark mass function, clearly indicating that dynamical chiral
symmetry breaking is more than dynamical mass generation. There is also a marked disagreement
for $f_3$ which is hard to extract on the lattice as is visible from the large error bars in the plot.
A reevaluation of the quark-gluon vertex on the lattice is currently in progress and
will be an important benchmark for the DSE/FRG calculations.

\subsection{Extracting the hadron spectrum from QCD}\label{spec:extracting}
In this section we deal with the gauge-invariant content of correlation
functions and in particular with the extraction of the hadron spectrum from
these objects. We start with general considerations about which
correlators are suitable for the extraction of baryon masses and thereafter
discuss lattice QCD\@. The corresponding procedure for functional methods is discussed in
Sec.~\ref{spec:BS}.

\begin{figure*}[t]
    \begin{center}
    \includegraphics[width=0.7\textwidth]{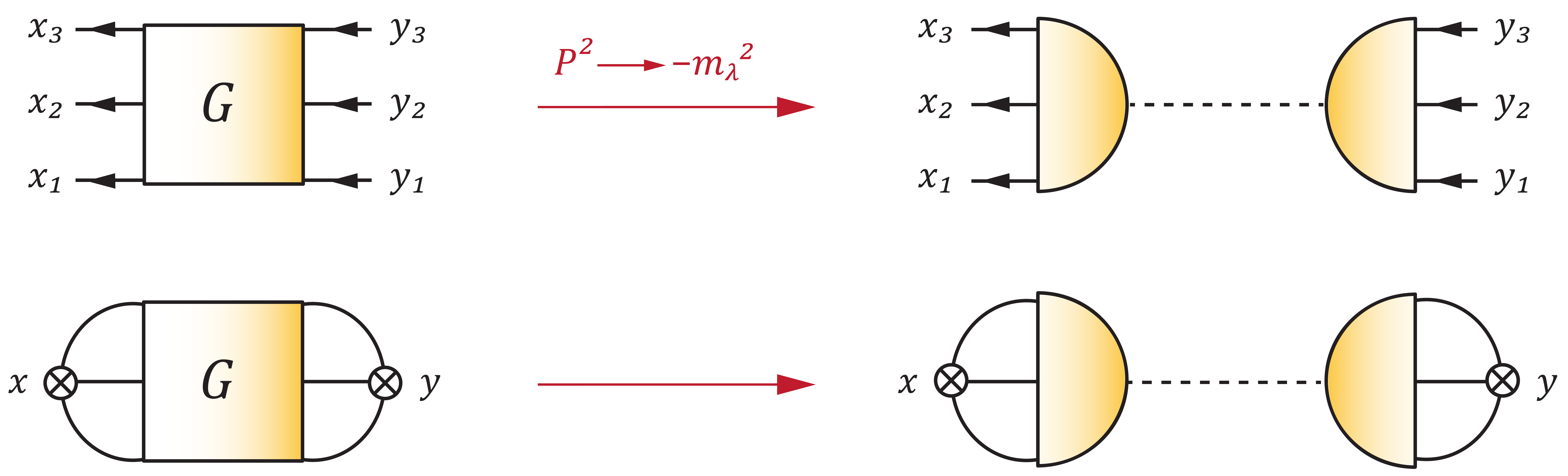}
    \caption{\textit{Top:} Six-point green function for the interaction between three quarks and its pole behaviour.
             The half-circles are the baryons' Bethe-Salpeter wavefunctions and the dashed lines are Feynman propagators.
             \textit{Bottom:} Two-point current correlator that is evaluated in lattice calculations of the baryon spectrum. }\label{fig:current-correlator-baryons}
    \end{center}
\end{figure*}

\paragraph{Current correlators.}
Hadron properties are encoded in QCD's Green functions and hence in scattering
amplitudes and cross-sections. Bound states and resonances are colour singlets
and they can appear as poles in $n-$point functions through their spectral
representation. In practice, the same information is contained in many Green
functions, however the effort to extract it can vary greatly. Take for example
the quark six-point function made of three incoming and three outgoing quarks,
which is illustrated in Fig.~\ref{fig:current-correlator-baryons}
\begin{align}\label{baryon-G6}
\mathbf G_{\alpha\beta\gamma,\alpha'\beta'\gamma'}(x_1,x_2,x_3;y_1,y_2,y_3) :=
\langle 0 | \mathsf{T}\,\psi_\alpha(x_1)\,\psi_\beta(x_2)\,\psi_\gamma(x_3)\,\conjg{\psi}_{\alpha'}(y_1)\,\conjg\psi_{\beta'}(y_2)\,\conjg\psi_{\gamma'}(y_3)\, | 0 \rangle\,.
\end{align}
Because a composite operator $\psi_\alpha \,\psi_\beta \,\psi_\gamma$ can
produce colour-singlet quantum numbers
($\textbf{3}\otimes \textbf{3} \otimes \textbf{3} =  \mathbf{1} \oplus \textbf{8} \oplus \textbf{8} \oplus \textbf{10}$),
inserting a complete set of eigenstates of QCD's Hamiltonian produces
bound-state poles. The resulting spectral decomposition in momentum space is
given by
\begin{align}\label{baryon-G6-spectral}
    \mathbf G_{\alpha\beta\gamma,\alpha'\beta'\gamma'}(p_f,q_f,P;p_i,q_i,P) \simeq \sum_{\lambda}
        \frac{ \mathbf\Psi^\lambda_{\alpha\beta\gamma}(p_f,q_f,P)\,\conjg{\mathbf\Psi}^{\lambda}_{\alpha'\beta'\gamma'}(p_i,q_i,P)}{P^2+m_\lambda^2}
\end{align}
plus further crossed-channel topologies that are non-resonant at
$P^2=-m_\lambda^2$, where $p_{f,i}$, $q_{f,i}$ are relative momenta and $P$ is
the total momentum. The sum over $\lambda$ is formal because it contains not
only single-particle states but also multi-particle continua which involve
integrations over relative momenta. The residue defines the baryon's
Bethe-Salpeter wave function $\mathbf\Psi^\lambda_{\alpha\beta\gamma}(p,q,P)$
whose definition in coordinate space is
\begin{align}\label{bs-wf-0}
     \mathbf\Psi^\lambda_{\alpha\beta\gamma}(x_1,x_2,x_3,P) = \langle 0 | \mathsf{T}\,\psi_\alpha(x_1)\,\psi_\beta(x_2)\,\psi_\gamma(x_3)\,| \lambda \rangle\,.
\end{align}
It is the time-ordered matrix element of the product of three quark fields
between the vacuum and a baryon state $|\lambda\rangle$ with onshell momentum
$P$. Thus, in principle one could determine the gauge-invariant masses of
colour-singlet baryons directly from the coloured and gauge-dependent quark
six-point function.

However, there is a much simpler way. We can define interpolating fields for
baryons as gauge-invariant spinors of the form
\begin{align}\label{baryon-current-1}
    J_\sigma(x) = \Gamma_{\alpha\beta\gamma\sigma}\,\psi_\alpha(x)\,\psi_\beta(x)\,\psi_\gamma(x)\,,
\end{align}
where $\Gamma_{\alpha\beta\gamma\sigma}$ carries a chosen Dirac structure
together with the appropriate flavour and colour wave functions to provide the
quantum numbers of the baryon under consideration. In Sec.~\ref{spec:BS} we will
see that the form of $\Gamma_{\alpha\beta\gamma\sigma}$ corresponds to the
possible tensor structures in the baryon's Bethe-Salpeter wave function.
Then, setting all $x_i=x$ and $y_i=y$ in~\eqref{baryon-G6}, a contraction
with two such operators yields current correlators of the form
\begin{align}\label{2-current-correlator}
      \mathbf G_{\sigma\sigma'}(x,y) = \langle 0 | \mathsf{T}\,J_\sigma(x)\,\bar J_{\sigma'}(y)\, | 0 \rangle \,.
\end{align}
These are two-point functions and can be viewed as effective baryon propagators,
since they contain the composite fields $J_\sigma(x)$ which describe baryons
propagating from the source $y$ to the sink $x$. Two-point functions are the
simplest correlators to deal with in any approach and therefore they are very
convenient. Poles in these functions will only emerge from those states that
coincide with the quantum numbers of the currents in question and therefore they
can be used to select the channels of interest. Another advantage is that the
Green functions in~\eqref{2-current-correlator} are gauge-invariant (in contrast
to the quark six-point function) since they contain gauge-invariant, local
products of quark fields. As will be detailed below, these two-point correlators
are frequently used in lattice calculations since the properties of such
correlators at the pole (in coordinate space: the large Euclidean time
behaviour) can be calculated directly from the QCD partition function.

The spectral decomposition~\eqref{baryon-G6-spectral} becomes particularly
simple for the two-point correlator. Setting all $x_i=x$ in coordinate space is
equivalent to an integration over the relative momenta in momentum space, and
due to translational invariance the dependence on $x$ can only enter through a
phase
\begin{align}\label{rlambda-def}
    \Gamma_{\alpha\beta\gamma\sigma}\,\mathbf\Psi^\lambda_{\alpha\beta\gamma}(x,x,x,P) = \langle 0 | \, J_\sigma(x)\,| \lambda \rangle
    = \langle 0 | \, J_\sigma(0)\,| \lambda \rangle \,e^{-i x\cdot P} = r_\lambda\,u_{\sigma}^{(\lambda)}(\vect{P})\,e^{-i x\cdot P}\,.
\end{align}
The overlap factor $r_\lambda$ is the integrated Bethe-Salpeter wave function
which measures the overlap of the operator $\Gamma_{\alpha\beta\gamma\sigma}$
with the wave function, cf.~Fig.~\ref{fig:current-general}, and
$u_\sigma^{(\lambda)}(\vect{P})$ is the respective onshell spinor that carries
the quantum numbers of the baryon. The resulting spectral decomposition becomes
\begin{align}\label{2-current-correlator-3}
           \mathbf G_{\sigma\sigma'}(z) =
            \int \!\! \frac{d^4P}{{(2\pi)}^4}\,e^{ iP \cdot  z} \,\mathbf G_{\sigma\sigma'}(P) \,, \qquad
            \mathbf G_{\sigma\sigma'}(P) \simeq \sum_{\lambda}  \frac{|r_\lambda|^2}{P^2+m_\lambda^2}\,u_{\sigma}^{(\lambda)}(\vect{P})\,\conjg{u}_{\sigma'}^{(\lambda)}(\vect{P})\,,
\end{align}
where the phases conspire with the time orderings to produce the Feynman propagator pole. Note also that due to translational
invariance the two-point correlator only depends on the separation $z=x-y$.
Hence, these overlap factors determine the importance of the respective operator
and therefore the strength of the signal.

In practice the location of the ground-state pole can be extracted from the
current correlator from its behaviour at large spacelike separations of the
baryon fields, which means going to large `Euclidean times' $\tau=x_4-y_4$.
Taking the Fourier transform of~\eqref{2-current-correlator-3} with respect to
$P_4$ yields
\begin{align}\label{2-current-correlator-2}
            \mathbf G_{\sigma\sigma'}(\vect{P},\tau) \simeq \sum_\lambda \int \! \frac{dP_4}{2\pi} \,  \frac{e^{ iP_4 \tau}}{P_4^2+\vect{P}^2 + m_\lambda^2} \, |r_\lambda|^2 \,u_\sigma^{(\lambda)}(\vect{P})\,\conjg u_{\sigma'}^{(\lambda)}(\vect{P})
            = \sum_{\lambda} \frac{e^{-E_\lambda |\tau| }}{2E_\lambda}\,|r_\lambda|^2\, u_\sigma^{(\lambda)}(\vect{P})\,\conjg u_{\sigma'}^{(\lambda)}(\vect{P})
\end{align}
with $E_\lambda^2 = \vect{P}^2 + m_\lambda^2$. A timelike pole in momentum space
corresponds to an exponential Euclidean time decay, and therefore at
$\tau\rightarrow\infty$ the ground state dominates the current correlator.

The application to mesons is analogous; in that case~\eqref{baryon-current-1}
must be replaced by appropriate fermion bilinears and the contraction is applied
to the quark-antiquark four-point function. The vector-meson correlator, for
example, is just the hadronic vacuum polarisation discussed later
in~\eqref{hvp}. The correlator method has the advantage that one is not
restricted to $q\bar{q}$ or $qqq$ operators: one could test combinations with
four- or five-quark operators, which would originate from corresponding higher
$n$-point functions, and by inserting covariant derivatives one can explore the
gluonic content of hadrons. In any case, let us emphasise again that these
considerations are independent of the chosen nonperturbative method to extract
the hadron masses from the correlators. The same gauge-invariant information of
a given state can be extracted from suitable gauge-invariant or gauge-dependent
correlators. Below we will detail the corresponding procedure within lattice QCD
which deals with the former type, whereas in Sec.~\ref{spec:BS} we discuss the
approach via functional methods which treats the gauge-dependent version.

\paragraph{Lattice QCD and the extraction of hadron properties.}
Correlation functions on the lattice are determined by evaluating expectation
values of corresponding operators, see~\eqref{correlators}. Due to conservation
laws, fermions always appear in powers of bilinears of the fermion fields
$\conjg{\psi},\psi$ in any combination $\mc{O}$ and consequently the fermions
can be integrated out into expressions that contain the fermion determinant
$\det \mathbf{D}$. For instance, for a single fermion bilinear
$\psi \conjg{\psi}$ one finds (omitting any indices or arguments for brevity)
\begin{align}
\int\mathcal{D}\psi\, \mathcal{D} \conjg{\psi} \,\,\psi \conjg{\psi}\,\, e^{- \conjg{\psi}\,\mathbf{D}\,\psi } = (\det \mathbf{D}) \,\mathbf{D}^{-1}\,,
\end{align}
which needs to be evaluated in the background of the gauge field ensemble. The
diagrammatic content of the fermion determinant is that is produces closed
fermion loops which are quantitatively important for most observables. The
computational time needed to evaluate $\det \mathbf{D}$ scales very badly with
the quark masses present in the QCD Lagrangian, so that its computation is
prohibitively expensive even for moderate lattice sizes. Thus, early lattice
studies used the so-called `quenched approximation' where this determinant is
simply neglected. Contemporary studies are carried out by taking into account
closed loops of \emph{up} and \emph{down} quarks ($N_f=2$), or additionally \emph{strange}
($N_f=2+1$) and even \emph{charm} quarks ($N_f=2+1+1$). We will use the same notation
below and in the results part of this section. Note that similar approximations
on a diagrammatic level can also be made in the framework of Dyson-Schwinger and
Bethe-Salpeter equations, see Sec.~\ref{spec:approx}.

Lattice QCD has been developed continually over the past decades and decisive
progress has been made due to both breakthroughs in algorithmic methods as well
as a tremendous increase in the available computer power. Still, a serious and
well-known problem is the dramatic increase in computational cost that is needed
to perform simulations with realistic bare quark masses, so that hadronic
observables can be evaluated at the physical point and directly compared with
experiment. A measure of this effect is the pion mass, which is related to the
quark mass via the Gell-Mann-Oakes-Renner relation discussed in
Sec.~\ref{sec:ff-vertices}. Simulations at or very close to the physical point,
i.e.\ with $m_\pi = 138$ MeV, have been performed in the past years for some
selected observables, see e.g.~\cite{Colangelo:2010et,Fodor:2012gf,Aoki:2013ldr,Green:2012ud}
for reviews and results. Many other quantities, however, are just too
cost-intensive and will be discussed later also for a range of heavier pion
masses, some even as large as $m_\pi = 600$ MeV. In many of these cases chiral
extrapolations from heavy pion masses to the physical point can be performed
using well-known results from chiral perturbation theory.

Although this is not a lattice review and our focus is not on technicalities,
we would like to give a short (non-expert) summary of the methods that are used
on the lattice to extract the spectrum of ground and excited baryon states. The
basic method to determine the hadron spectrum is to evaluate matrices of
correlation functions as in~\eqref{2-current-correlator}, i.e.\ between
time-ordered hadronic creation and annihilation operators $J_i$ with $i=1\dots N$
carrying specific quantum numbers. In practice the number of operators is always
finite and limited, but it has to be chosen sufficiently large so as to be able
to disentangle different states with the same quantum numbers. On the lattice
one averages over the spatial dependencies $\vect{x}$ and $\vect{y}$, which is
equivalent to the expression~\eqref{2-current-correlator-2} for the case
$\vect{P}=0$. Using energy as well as parity projection operators (see
e.g.~\cite{Mahbub:2009nr} for details), the resulting correlation matrix has the
form
\begin{align}\label{l1}
  G_{ij}(\tau) = \sum_\lambda r_\lambda^i\,\conjg{r}_\lambda^j\,e^{-m_\lambda \tau}\,,
\end{align}
where $r_\lambda^i$ describes the coupling of the baryon created by $J_i$ to the
eigenstate $|\lambda\rangle$ of the Hamiltonian. Thus the correlator
$G_{ij}(\tau)$ is completely described by a linear combination of couplings to
hadrons with masses $m_\lambda$. Many eigenstates may contribute to a given
operator. For sufficiently large times, the sum in~\eqref{l1} is always
dominated by the ground state with lowest mass in a given channel and its mass
can be extracted as long as the statistics is good enough. However, access to
the excited state masses requires additional efforts.

The basic idea of the well-established variational method \cite{Michael:1982gb,Luscher:1990ck}
is to generate linear
combinations of the interpolating operators $J_i$ which are close to the
physical states of the theory such that the corresponding correlation matrix
is effectively diagonalised. This is always possible given that the number $N$
of interpolating operators is sufficiently complete. The correlator then
satisfies a generalised eigenvalue equation
\begin{align}
G_{ij}(\tau) \,v_j^\lambda(\tau) = e^{-m_\lambda \tau} G_{ij}(0) \,v_j^\lambda(\tau)
\end{align}
from where the energies of the ground and excited states up to a maximum of $N$
states can be extracted. This procedure is limited by the statistical noise of
the Monte-Carlo calculation. Unfortunately, the noise increases both with $\tau$
and the number of operators used; in practice this enforces a guided selection
of operators with maximal overlap with the ground and lowest excited states in a
given channel. There are a number of technical tools
(`quark source smearing'~\cite{Gusken:1989ad,Gusken:1989qx},
`distillation'~\cite{Peardon:2009gh}, AMIAS~\cite{Alexandrou:2014mka}, etc.)
to increase the signal-to-noise ratio that have been suggested, refined and
successfully applied over the years.

\begin{figure*}[t]
        \begin{center}
        \includegraphics[width=0.75\textwidth]{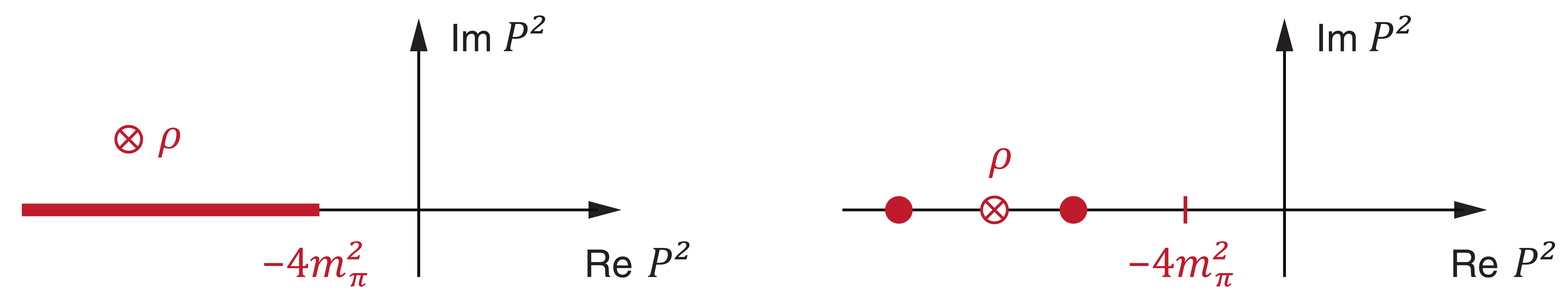}
        \caption{\textit{Left:} Sketch of the analytic structure of a vector-meson correlator, with a two-pion branch cut and the $\rho-$meson pole in the second Riemann sheet.
                 \textit{Right:} Analogous situation in a finite volume, where the bound state is accompanied by scattering states. }\label{fig:correlator-analytic}
        \end{center}
\end{figure*}

Here we would like to address another important issue. In scattering amplitudes
and current correlators one encounters apart from the discrete energy spectrum
of bound states and resonances also a continuum of scattering states once
corresponding thresholds are crossed, cf.~Fig.~\ref{fig:correlator-analytic}.
Due to the finite volume on the lattice this continuum of states turns into a
finite number of discrete scattering states, which need to be carefully
considered in the process of identifying the masses of single particle states.
In general it depends on the details of the lattice simulation (such as the pion masses employed,
or the lattice volume) as to whether these states have to
be taken into account or not~\cite{Engel:2010my}.

Consider for example the $\rho$ vector meson, which acquires its width from the
coupling to $\pi\pi$ in a relative p-wave. In a lattice simulation with heavy pion masses
($m_\rho < 2m_\pi$) such a coupling is kinematically forbidden, and consequently
the $\rho$-meson ground state can be extracted from the lowest energy level in
the $1^{--}$ channel without much disturbance. The higher energy levels,
however, may be contaminated with scattering states. For realistic pion masses the
situation is different, cf.~Fig.~\ref{fig:correlator-analytic}: the lowest two-pion
scattering state has an energy close to $2 m_\pi$ plus one unit of lattice momentum
$2\pi/L$, which depending on the lattice volume $V=L^3$ may now be smaller than that
of the $\rho$ meson. The next higher scattering state comes with one more unit of
lattice momentum and consequently more such states may be found in the
region of the $\rho$ mass. Whether they are actually seen in a simulation also
depends on the choice of the operator basis $J_i$; if the basis contains only
operators with minimal valence quark content, like fermion bilinears for the
case of mesons, it may have poor overlap with scattering states.

A way to study this is to vary the lattice volume: whereas the scattering states
are highly dependent on $L$, bound states may be less so and therefore they can be
distinguished. However, there is a good reason why one would like to
study scattering states in detail: they are responsible for and can be used to
determine the width of an unstable state such as the $\rho$-meson. It turns out
that the energies of these states can be used to extract phase shifts in the
infinite volume, which allow for a reconstruction of the S-matrix as a function
of energy and consequently a determination of the resonance pole positions. In
the literature this  is known as the L\"uscher method~\cite{Luscher:1990ck,Luscher:1991cf};
originally formulated for the case of a particle at rest, the framework has been
generalised to moving frames~\cite{Rummukainen:1995vs,Christ:2005gi,Kim:2005gf}
and to multi-channel problems, see e.g.
\cite{He:2005ey,Hansen:2012tf,Briceno:2012yi,Briceno:2014tqa,Lu:2015riz,Hansen:2015azg}
and references therein.

Whereas the method has been successfully applied in the meson sector of QCD (see
e.g.~\cite{Dudek:2014qha,Wilson:2014cna,Briceno:2015dca} and references therein),
the baryon sector still waits to be explored. One reason is the worse signal to noise
ratio as compared to the baryon sector. Another reason are the
large operator bases involved in such extractions and the need for taking into
account multiquark operators with the same quark content as the scattering states
to provide maximal overlap. For baryons such bases have only been begun to be
studied~\cite{Lang:2012db,Kiratidis:2015vpa}. The influence of scattering states,
however, may be important already in the positive- and negative-parity nucleon
channel. For example, the first radial excitation of the nucleon, the Roper,
will be affected by the presence of (p-wave) $\pi N$ scattering states and
therefore meson-nucleon interpolators should be included in the operator
basis~\cite{Edwards:2011jj}. Whether these have a material impact is not clear
from an outsider's perspective since potentially conflicting results have been
found by different lattice groups~\cite{Lang:2012db,Kiratidis:2015vpa}. We will
return to the issue in the results part of this section.

In addition to providing maximal overlap with scattering states, multi-particle
operators may also be important when it comes to the determination of the
structure of states that go beyond the conventional quark-model picture in terms
of the number of valence quarks. Such exotic states are heavily studied in the
meson sector, where a number of candidates for $qq\conjg{q}\conjg{q}$ states have
been detected in recent experiments such as Belle, BABAR, BES and LHCb, see
e.g.~\cite{Esposito:2014rxa,Chen:2016qju} for recent reviews. In the baryon sector, five-quark
states are of considerable interest especially since the experimental signature
of a potential pentaquark has been reported at LHCb~\cite{Aaij:2015tga}. However,
besides this new state (and its potential cousins detected in the future) there
are also well-established states such as the $\Lambda(1405)$ that are heavily
debated since their location in the baryon spectrum does not agree with
(potentially naive) expectations from the quark model. One  possible explanation
is therefore a strong meson-baryon (i.e.\ five-quark) component in its
wave function. Recent studies of the $\Lambda(1405)$ on the lattice are discussed
below in the results section.

Finally, we need to address how the scale is set in lattice QCD\@. In order to match to
experiment one tunes the bare parameters of lattice QCD -- the coupling $\beta=6/g^2$
and the bare quark masses. Most calculations are done in the isospin symmetric
limit $m_u=m_d$, and the coupling as well as the light-quark mass are then fixed
by the pion mass and another quantity that sets the scale. For the latter any
observable is convenient that depends less on the light quark mass than
$m_\pi^2 \sim m_{u,d}$. Frequent choices in hadron spectroscopy are the masses of
stable baryons such as $M_N$, $M_\Xi$, $M_\Omega$, average masses of baryon
multiplets, the pseudoscalar decay constant, or distance measures in the heavy-quark
potential that are determined from $\Upsilon$-spectroscopy~\cite{Fodor:2012gf}.

When going beyond the aforementioned isospin limit one should, in addition to
the mass difference between the up and down quarks, also take into account
electromagnetic corrections to the hadron masses due to the electric charge of
the valence quarks. On the lattice this is done by including the QED action into
the functional integral. Due to the involved computational costs this is a very
recent endeavour, with results being presented in~\cite{Borsanyi:2013lga,Borsanyi:2014jba}.
We will not discuss further the topic here due to constraints of space.

\subsection{Bound-state equations}\label{spec:BS}
In the previous section we introduced current correlators as a means
by which hadronic poles could be extracted from QCD's Green functions
and we highlighted their role in lattice QCD calculations.
In principle one could employ the same methods in functional approaches, although this would necessitate the solution of
at least a four-point function for mesons ($q\bar{q}\to q\bar{q}$) or six-point function for baryons ($qqq \to qqq$)
from a set of functional relations, such as Dyson-Schwinger equations, to be able to isolate the pole behaviour therein.
In the course of the discussion around Eq.~\eqref{baryon-G6-spectral}, we briefly touched upon the Bethe-Salpeter wave function $\mathbf{\Psi}^\lambda_{\alpha\beta\gamma}(p,q,P)$
as the residue of hadronic poles contained within QCD's Green functions.
In practice it is then much simpler to derive self-consistent relations for these objects since they contain the full information about the hadron on its pole.
The resulting integral equations are known as homogeneous Bethe-Salpeter equations~\cite{Salpeter:1951sz,LlewellynSmith:1969az,Nakanishi:1969ph},
or alternatively Faddeev equations in the case of baryons~\cite{Faddeev:1960su,Taylor:1966zza}, or simply hadronic bound-state equations.
In the following we discuss their properties and demonstrate how Bethe-Salpeter equations are obtained in the functional framework.
The result will be used to discuss baryons as bound states
of three quarks as well as explicating the quark-diquark approximation.

While some of the derivations and representations of the structure of the baryon's wave function are somewhat technical, we took care to provide
short \underline{\it summaries} of the most relevant points from time to time. Thus the reader not interested in the technical details may very well
gloss over the equations and merely pick up the summaries.

\begin{figure*}[t]
    \begin{center}
    \includegraphics[width=0.9\textwidth]{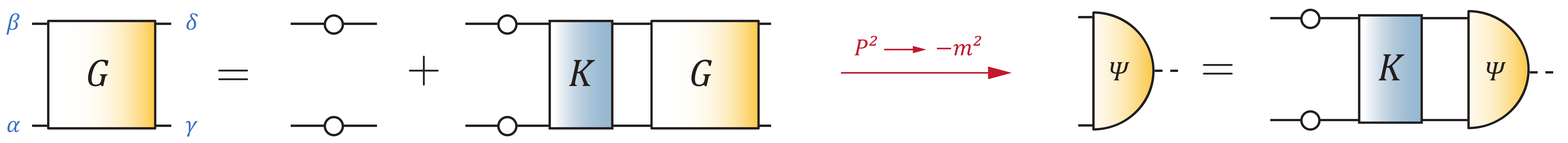}
    \caption{Dyson equation~\eqref{dyson-eq} and onshell Bethe-Salpeter equation~\eqref{bse-general} in graphical form. }\label{fig:bse}
    \end{center}
\end{figure*}

\paragraph{Bethe-Salpeter equations.}
The derivation of the Bethe-Salpeter equation (BSE) relies upon the classification of graphs
according to their (ir)reducibility together with the Dyson equation, see e.g. Ref.~\cite{Loring:2001kv}
for a detailed and pedagogical discussion. Let us start with mesons and introduce the quark four-point function
$\mathbf{G}_{\alpha\beta,\delta\gamma}$, the analogue of the six-point function of~\eqref{baryon-G6}, as illustrated in Fig.~\ref{fig:bse}.
Dropping indices and momentum integrations in the notation,\footnote{We will frequently use such a compact notation
but emphasise that it is only for notational convenience. In practice all four-dimensional loop momentum integrations are
performed explicitly and in principle the quantities $\mathbf{T}$, $\mathbf{K}$, etc. carry their full tensor structure.
With regard to the similar relations in hadronic coupled-channel approaches, $\mathbf{K}$ plays the role of the potential $\mathbf{V}$ in~\eqref{coupled-channel-eqs}.}
we denote the disconnected product of a dressed quark and antiquark propagator by $\mathbf{G}_0$ and construct the scattering matrix $\mathbf{T}$ --
the amputated and connected part of the four-point function -- by
$\mathbf{G} = \mathbf{G}_0 + \mathbf{G}_0\, \mathbf{T}\,\mathbf{G}_0$.
Both of these four-point functions satisfy Dyson equations
\begin{align}\label{dyson-eq}
    \mathbf{G} = \mathbf{G}_0 + \mathbf{G}_0\,\mathbf{K}\,  \mathbf{G}  \qquad  \Leftrightarrow \qquad
    \mathbf{T} = \mathbf{K}   + \mathbf{K}\,  \mathbf{G}_0\,\mathbf{T}\;,
\end{align}
which introduce the four-quark scattering kernel $\mathbf{K}$ as two-particle irreducible with respect to the quark propagators.
That is, the kernel contains all possible diagrams except those that fall apart by cutting one horizontal quark and one antiquark line, because those are generated by the iteration.
We can easily see that the result of the Dyson equation is to enact a resummation of the scattering kernel:
\begin{align}
    \mathbf{T} &= \mathbf{K}   + \mathbf{K}\, \mathbf{G}_0\, \mathbf{K}
                               + \mathbf{K}\, \mathbf{G}_0\, \mathbf{K}\, \mathbf{G}_0\, \mathbf{K} +\cdots
                               = \mathbf{K} \left( \mathds{1} + \mathbf{G}_0\, \mathbf{T}\right).
\end{align}
Notice, however, that $\mathbf{K}$ is not required to be small: the equations~\eqref{dyson-eq} are nonperturbative and they can be derived from an effective action,
much like a geometric series where $f(x)=1+x  f(x)$ defines the original equation with solution $f(x) = 1/(1-x)$, whereas its perturbative expansion
$f(x) = 1 + x + x^2 + \dots$ is only valid for small $|x|<1$. The difference is that $f(x) = 1 + x f(x) = 1 + x + x^2 f(x) = \dots$ always contains a nonperturbative term
that restores the exact result in the series.

Of course, the equation \emph{per se} does not provide us with any new information because we have merely shifted the unknown content from $\mathbf{T}$ to $\mathbf{K}$,
i.e., we have essentially defined the kernel according to our needs.
Diagrammatically  it still contains infinitely many terms, but as we will see below there are ways to derive those terms systematically
or find general constraints for the functional form of the kernel based on symmetry relations.
To proceed, we treat $\mathbf{K}$ as an unknown black box and in that sense all following relations
(and also those for current matrix elements in Sec.~\ref{sec:ff-currents} and scattering amplitudes in Sec.~\ref{sec:cs-hadronic-vs-quark}) are exact.
For later reference we collect the (symbolic) equations for the inverse quantities,
\begin{equation}\label{Ginv-Tinv}
    \mathbf{G}^{-1} = \mathbf{G}_0^{-1} - \mathbf{K} \qquad \Leftrightarrow \qquad
    \mathbf{T}^{-1} = \mathbf{K}^{-1} - \mathbf{G}_0\,,
\end{equation}
and note that a generic derivative induces $(\mathbf{G}\,\mathbf{G}^{-1})' = 0$ $\Rightarrow$ $(\mathbf{G}^{-1})' = -\mathbf{G}^{-1} \mathbf{G}' \,\mathbf{G}^{-1}$.

The connection to the Bethe-Salpeter equation for the amplitude is straightforward. Following~\eqref{baryon-G6-spectral},
the four-point functions contain all possible meson poles and at the respective pole locations they become
\begin{align}\label{pole-in-G-or-T}
    \mathbf{G} \rightarrow \frac{\mathbf{\Psi}\,\overline{\mathbf{\Psi}}}{P^2+M^2} \qquad \Leftrightarrow \qquad \mathbf{T} \rightarrow \frac{\mathbf\Gamma\,\overline{\mathbf\Gamma}}{P^2+M^2}\,, \qquad \mathbf{\Psi} = \mathbf{G}_0\,\mathbf\Gamma
\end{align}
where the Bethe-Salpeter amplitude $\mathbf\Gamma$ is the amputated wave function $\mathbf\Psi$ and appears together with its conjugate as the residue of the onshell particle pole at $P^2 = -M^2$.
In principle this also includes the case of resonances, where the pole is located in the complex plane and the `mass' $M$ is thus complex.
Inserting this into the Dyson equations~\eqref{dyson-eq} above, we see that the meson pole can only appear in $\mathbf{G}$ and $\mathbf{T}$ but not in $\mathbf{K}$.
Comparing the residues on both sides of the equation yields the homogeneous Bethe-Salpeter equation at the pole, either formulated in terms of the wave function or the amplitude (see Fig.~\ref{fig:bse}):
\begin{align}\label{bse-general}
    \mathbf{\Psi} = \mathbf{G}_0\,\mathbf{K}\,\mathbf{\Psi} \qquad \Leftrightarrow \qquad   \mathbf\Gamma = \mathbf{K}\,\mathbf{G}_0\,\mathbf\Gamma \,.
\end{align}

Being homogeneous, the equation is equipped with an auxiliary normalisation
condition. It follows from taking the derivative of $\mathbf{G}$ at $P^2=-M^2$ and
comparing the most singular terms:
\begin{equation}\label{eqn:normalisation}
    \frac{d\mathbf{G}}{dP^2} = - \mathbf{G}\,\frac{d \mathbf{G}^{-1}}{dP^2}\,\mathbf{G}  \quad \Rightarrow \quad
    \conjg{\mathbf{\Psi}}\,\frac{d \mathbf{G}^{-1}}{dP^2}\,\mathbf{\Psi} \,\bigg|_{P^2=-M^2} = 1 \quad \Leftrightarrow \quad
    \conjg{\mathbf\Gamma} \left( \frac{d\mathbf{G}_0}{dP^2} + \mathbf{G}_0\,\frac{d\mathbf{K}}{dP^2}\,\mathbf{G}_0\right)\mathbf\Gamma\,\bigg|_{P^2=-M^2} = -1\,.
\end{equation}
To arrive at the last form we used~\eqref{Ginv-Tinv} and the derivative property for $\mathbf{G}_0^{-1}$.
The normalisation condition is an important component of the Bethe-Salpeter approach since it is ultimately also what ensures the correct charge of a hadron.
This can be intuitively understood by comparison with~\eqref{emcurrent-gauging}: the electromagnetic current at vanishing momentum transfer \textit{becomes} the normalization condition,
and therefore charge normalization is not enforced by hand but follows automatically.

  \begin{figure}[t]
  \centering
    \includegraphics[width=0.99\textwidth]{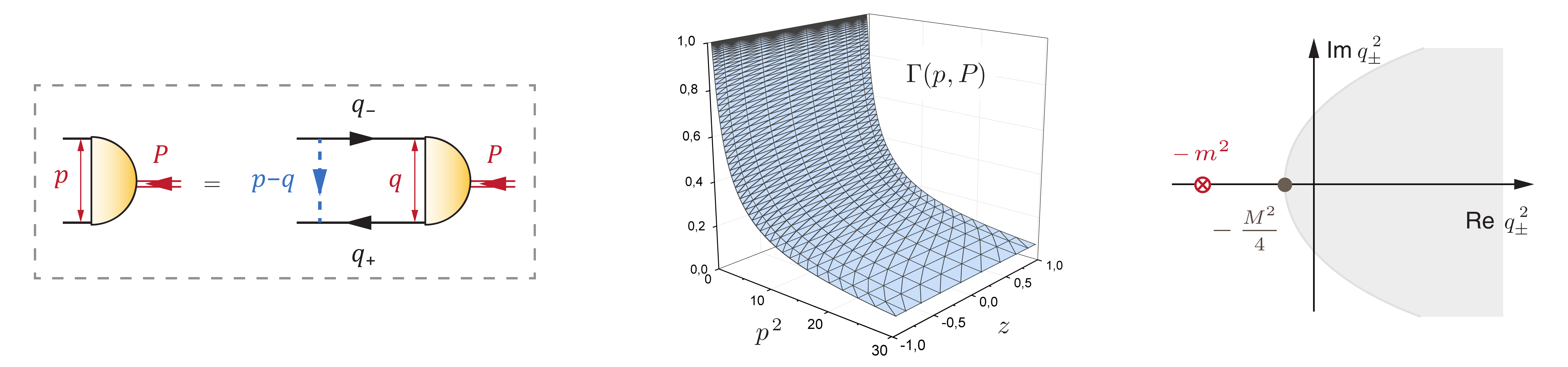}
  \caption{Scalar Bethe-Salpeter equation (\textit{left}) and corresponding solution for the amplitude (\textit{center}). $p^2$ carries arbitrary units and $z$ is dimensionless.
          The right panel shows the parabolic domain of the propagator that is integrated over in the equation.}
  \label{fig:scalar-bse-1}
  \end{figure}

 \paragraph{Basic properties and solution techniques.}
    Before applying these relations to QCD,
    let us take a step back and illustrate the basic solution techniques for Euclidean Bethe-Salpeter equations,
    since this may be helpful for the non-expert reader and the standard methods are quite different from analogous treatments in Minkowski space~\cite{Carbonell:2014dwa,Carbonell:2016ekx,Gutierrez:2016ixt}.
    We consider a simple toy model: a scalar bound state (mass $M$) of two scalar constituents (with equal masses $m$ for simplicity),
    which are bound by a scalar exchange particle (mass~$\mu$); all propagators are at tree level.
    For $\mu=0$ one recovers the well-known Wick-Cutkosky model which admits analytic solutions~\cite{Wick:1954eu,Cutkosky:1954ru}.
    The corresponding BSE is illustrated in Fig.~\ref{fig:scalar-bse-1} and reads
    \begin{equation}\label{scalar-bse}
       \mathbf\Gamma(p,P) = g^2 \!\int \!\! \frac{d^4q}{(2\pi)^4}\,D(k^2,\mu^2)\,D(q_+^2,m^2)\,D(q_-^2,m^2)\,\mathbf\Gamma(q,P)\,, \qquad D(k^2,m^2) = \frac{1}{k^2+m^2}\,.
    \end{equation}
    The scalar Bethe-Salpeter amplitude depends on two momenta: the relative momentum $p$ between the constituents and the total onshell momentum $P$ with $P^2=-M^2$.
    The internal momenta are $q_\pm = q \pm P/2$ for the constituents and $k=p-q$ for the exchange particle, and one can define a dimensionless coupling $c=g^2/(4\pi m)^2$.

    Note that the equation is fully Lorentz invariant, so in principle we never need to specify a frame.
    Nevertheless it is instructive to work in the rest frame:
    \begin{equation}\label{rest-frame} \renewcommand{\arraystretch}{1.0}
        P=\left(\begin{array}{c}  0  \\ 0 \\ 0 \\ iM \end{array}\right), \quad
        p = \sqrt{p^2} \left( \begin{array}{c} 0 \\ 0 \\ \sqrt{1-z^2} \\ z \end{array}\right), \quad
        q = \sqrt{q^2} \left( \begin{array}{c} 0 \\ \sqrt{1-{z'}^2}\,\sqrt{1-y^2} \\ \sqrt{1-{z'}^2}\,y \\ z' \end{array}\right).
    \end{equation}
    We used hyperspherical coordinates where the volume integral becomes
    \begin{equation}
       \int d^4q = \frac{1}{2}\int_0^\infty dq^2 q^2 \int_{-1}^1 dz' \sqrt{1-{z'}^2} \int_{-1}^1 dy \int_0^{2\pi} d\psi\,.
    \end{equation}
    The Bethe-Salpeter amplitude depends on the Lorentz invariants $p^2$, $p\cdot P$ and $P^2=-M^2$ and thus we could express the equation entirely in terms of the invariants
    \begin{equation}
        p^2, \qquad z = \hat{p}\cdot \hat{P}, \qquad q^2, \qquad z' = \hat{q}\cdot\hat{P}, \qquad y = \widehat{p_\perp}\cdot \widehat{q_\perp}\,,
    \end{equation}
    where a hat denotes a normalized momentum and $p_\perp^\mu = p^\mu - (p\cdot \hat{P})\,\hat{P}^\mu$ a transverse projection with respect to the total momentum.
    The Euclidean domain of the equation is the two-dimensional phase space defined by $p^2 > 0$ and $-1<z<1$. For given parameters $c$ and $\mu$, the numerical solution
    is straightforward and the resulting amplitude is plotted in Fig.~\ref{fig:scalar-bse-1}. Observe that the dependence on the angular variable $z$ is extremely weak;
    since the integral measure in $z$ is the weight for Chebyshev polynomials of the second kind, a Chebyshev expansion would converge rapidly.

    Although the propagators in~\eqref{scalar-bse} have an `onshell' form, they are practically always sampled at offshell momenta during the integration.
    In fact, their poles in the integrand will pose restrictions on the kinematic domain. Since $p$ and $q$ are real four-vectors, the denominator of the exchange propagator
    is always real and positive and thus its pole at $k^2 = (p-q)^2=-\mu^2$ does not enter in the integration domain.
    We do need to worry, however, about the constituent poles. They depend on the imaginary momentum $P$, and therefore the constituent propagators are sampled within
    complex parabolas, see Fig.~\ref{fig:scalar-bse-1}:
    \begin{equation}\label{scalar-poles}
        q_\pm^2 = q^2 -\frac{M^2}{4} \pm iM\sqrt{q^2}\,z\,.
    \end{equation}
    The onshell poles appear at $q_\pm^2 = -m^2$ and therefore we recover the condition $-m^2 < -M^2/4 \Rightarrow M < 2m$.
    In general the momentum partitioning is arbitrary and we could also distribute the momenta differently,
    for example by writing $q_+ = q + \eta P$ and $q_-= q-(1-\eta) P$
    which has the effect that one parabola shrinks and the other grows.\footnote{Note that
    this also changes the condition $M<2m$ to $M<\text{min} \left( m/\eta, m/(1-\eta)\right)$,
    which leads back to the ideal choice $\eta=\nicefrac{1}{2}$ that maximizes the domain in $M$.
    This was overlooked in Ref.~\cite{Carbonell:2010tz}.}
    Since the bound state mass is independent of $\eta$, this option is especially useful if one deals with constituents of different masses.

 \begin{figure}
 \centering
   \includegraphics[width=0.9\textwidth]{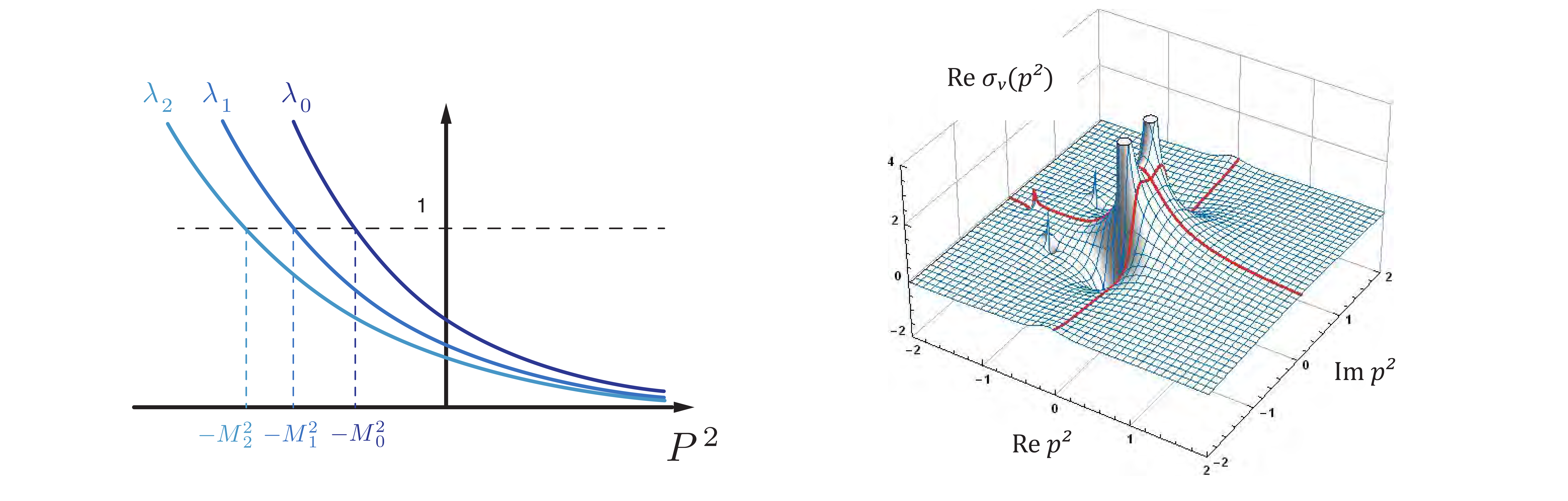}
 \caption{\textit{Left:} Sketch of the eigenvalue spectrum of a Bethe-Salpeter equation. \textit{Right:} Singularity structure of the quark propagator in a rainbow-ladder truncation.
                         The dimensions are  GeV$^2$ for $p^2$ and GeV$^{-2}$ for the vector dressing function $\sigma_v(p^2)$ defined in~\eqref{quark-propagator}. }
 \label{fig:scalar-bse-2}
 \end{figure}

    In general one is not only interested in the ground-state solution but also in the radially excited states.
    If we introduce an artificial eigenvalue $\lambda(P^2)$ and write the equation as
    \begin{equation}
       \mathbf{K\,G_0}\,\mathbf\Gamma = \lambda\,\mathbf\Gamma\,,
    \end{equation}
    we see that it has the structure of an eigenvalue equation.
    The vector space is spanned by the continuous variables $p^2$ and $z$ (or the discrete indices of their moments using adequate polynomial expansions),
    whereas $P^2$ is an external parameter.
    By determining the eigenvalue spectrum $\lambda_i(P^2)$ of the kernel $\mathbf{K\,G_0}$
    we can read off the masses $M_i$ of the ground and excited states from the intersections $\lambda_i(P^2=-M_i^2) = 1$,
    cf.~Fig.~\ref{fig:scalar-bse-2},
    and calculate their Bethe-Salpeter amplitudes $\mathbf\Gamma_i(p^2,z,-M_i^2)$ as the respective eigenvectors.

\underline{\it Summary:} Bethe-Salpeter equations are homogeneous eigenvalue equations $\mathbf{K\,G_0}\,\mathbf\Gamma = \lambda\,\mathbf\Gamma$ for the bound-state amplitudes.
The covariant equations for the amplitudes can be rewritten as Lorentz-invariant equations for their dressing functions.
The resulting eigenvalue spectrum determines the masses of the ground and excited states.
The auxiliary normalisation condition for the wave function ensures also
charge normalisation.

 \paragraph{Bethe-Salpeter amplitudes in QCD.}
    The features illustrated above are also present in the equations for bound states made of quarks and antiquarks in QCD.
    Instead of dealing with the poles of QCD's Green functions directly,
    one instead solves an integral equation for the Bethe-Salpeter amplitude which has solutions at discrete
    values of $P^2 = -M_i^2$.
    If such solutions exist, they correspond to poles in the T-matrix
    and therefore to bound states of mass $M_i$ with $M_1$ the ground state.
    If we drop for now the flavour and colour components, the meson Bethe-Salpeter equation in Fig.~\ref{fig:bse} reads explicitly:
    \begin{equation}\label{bse-meson}
    \left[\mathbf\Gamma^{\mu_1\ldots\mu_J}(p,P)\right]_{\alpha\beta}
    = \int\!\!\frac{d^4q}{\left(2\pi\right)^4} \left[\mathbf{K}(p,q,P)\right]_{\alpha\gamma;\delta\beta}\left[ S(q_+) \,\mathbf\Gamma^{\mu_1\ldots\mu_J}(q,P)\,S(q_-)\right]_{\gamma\delta}\,,
    \end{equation}
    where $p$ and $P$ are again the relative and total momenta, $\mathbf{K}$ is the quark-antiquark kernel, and $S(q_\pm)$ is the dressed quark propagator of~\eqref{quark-propagator}.
    Quarks are fermions and therefore their Bethe-Salpeter amplitudes necessarily carry Dirac indices.
          For mesons with spin-parity $J^P$ they are $4\times 4$ matrices in Dirac space
          and for $J>0$ they also carry Lorentz indices $\mu_1 \ldots \mu_J$. They can be expanded in corresponding tensor bases with scalar coefficients $f_i$,
          \begin{align}\label{theory:bsetensorstructures}
          \left[\mathbf\Gamma(p,P)\right]^{\mu_1\ldots\mu_J}_{\alpha\beta} = \sum_i f_i\left(p^2,p\cdot P; P^2=-M^2\right)\left[\tau_i(p,P)\right]^{\mu_1\ldots\mu_J}_{\alpha\beta}\,,
          \end{align}
          and appropriate projections allow one to cast the Bethe-Salpeter equations
          into coupled integral equations for the Lorentz-invariant dressing functions $f_i$.
          Similarly to the scalar system in Fig.~\ref{fig:scalar-bse-1}, it turns out that for systems with equally massive valence quarks the angular dependencies are usually extremely weak
          and polynomial expansions converge rapidly,  so that the resulting functions $f_i$ effectively only depend on the variable $p^2$.

    In contrast to the scalar example, however, the quark-antiquark kernel $\mathbf{K}$ and quark propagator cannot be chosen independently of each other
    because they are both intrinsically related.
    The propagator is the fully dressed
    two-point correlation function inclusive of all quantum corrections with the typical shape of Fig.~\ref{fig:quark-mn}.
    It carries a certain
    singularity structure (complex conjugate poles, branch cuts, etc.; see Fig.~\ref{fig:scalar-bse-2} for an exemplary result) which
    still leads to similar singularity restrictions as earlier.
    The nearest singularities define the `parabola mass' $m_P$, which imposes the restriction $M<2m_P$ for mesons and $M<3m_P$ for baryons, where typically $m_P \sim 0.5$ GeV for light quarks.
    Due to confinement these are not actual particle thresholds but they still impose practical difficulties.
    For larger bound state masses one would have to include their residues which would be analogous to solving the system in Minkowski space.
    However, such a task is very difficult to accomplish in actual calculations because the singularity structure of the quark is the result of a dynamical equation (the quark DSE)
    and usually not even known in the entire complex plane except in specific cases.

\paragraph{Spin structure.}
Before moving on to physical applications, let us discuss the spin structure of covariant Bethe-Salpeter amplitudes.
For a bound state of two quarks, e.g.\ a meson with spin-parity $J^P$, the
object we need to construct in~\eqref{theory:bsetensorstructures} is a mixed tensor composed of a Dirac index for each valence quark and, for states of non-zero integer spin, $J$ Lorentz indices.
It is a function of two independent momenta, the relative quark momentum $p$ and the total momentum $P$.
To determine the basis functions, we first construct a basis for $J=0$  that spans the Dirac components alone.
The set of totally antisymmetric tensors is sufficient~\cite{Joos:1962qq,Weinberg:1964cn,LlewellynSmith:1969az},
\begin{align}\label{eqn:diracmatrices}
\mathds{1}, \quad
\gamma^\mu, \quad
[\gamma^\mu, \gamma^\nu ], \quad
[\gamma^\mu, \gamma^\nu, \gamma^\rho], \quad
[\gamma^\mu, \gamma^\nu, \gamma^\rho, \gamma^\sigma]\,,
\end{align}
after saturating the indices with the momenta $p$ and $P$ and using $\gamma_5$ to distinguish the elements by parity.
The brackets denote a complete antisymmetrization of the encompassed objects (cf.~App.~\ref{app:conventions}).
For example, for a pion amplitude $\mathbf\Gamma(p,P)$ this yields
\begin{equation}\label{pion-basis}
   \tau_i(p,P) =  \left\{ \,\mathds{1}, \; i\slashed{P}, \; (p\cdot P)\,i\slashed{p}, \; [\slashed{p}, \slashed{P}]\,\right\} \gamma_5,
\end{equation}
where we inserted a factor $p\cdot P$ in the third element to ensure that the charge-conjugation invariance property $\conjg{\mathbf\Gamma}(p,P) = \mathbf\Gamma(p,-P)$
holds for each basis element separately  (see again App.~\ref{app:conventions}).
As a consequence, all dressing functions $f_i$ (often denoted by $E$, $F$, $G$, $H$ in the literature~\cite{Maris:1997tm}) are symmetric in $z=\hat{p}\cdot\hat{P}$
and their $z$ dependence is practically negligible, much like in the example of Fig.~\ref{fig:scalar-bse-1}.

In practice it has turned out extremely efficient to employ orthonormal tensor bases in the numerical solution of Bethe-Salpeter equations.
For a meson amplitude we have two momenta $p$ and $P$ which we can orthonormalize to each other
\begin{equation}\label{meson-orth-momenta}
   P^\mu \;\; \to \;\; \hat{P}^\mu, \qquad\quad
   p^\mu \quad \to \quad p_\perp^\mu = p^\mu - (p\cdot \hat{P})\,\hat{P}^\mu \quad \to \quad n^\mu = \widehat{p_\perp}^\mu = \frac{\hat{p}^\mu - z\,\hat{P}^\mu}{\sqrt{1-z^2}}\,.
\end{equation}
These are Lorentz-covariant definitions; in the rest frame defined by~\eqref{rest-frame}, $\hat{P}^\mu$ is the unit vector in the four-direction and $n^\mu$ in the three-direction, and
therefore $n^2=\hat{P}^2=1$ and $n\cdot \hat{P}=0$.
We also define the transverse projector $T^{\mu\nu}_P = \delta^{\mu\nu} - \hat{P}^\mu \hat{P}^\nu$
and transversalize the $\gamma-$matrices with respect to both $n$ and $\hat{P}$
\begin{equation}
   \gamma^\mu \quad \to \quad \gamma^\mu_\perp = T^{\mu\nu}_P\,\gamma^\nu = \gamma^\mu - \hat{P}^\mu \hat{\slashed{P}} \quad \to \quad \gamma^\mu_{\perp\perp} = \gamma^\mu - n^\mu \slashed{n} - \hat{P}^\mu \hat{\slashed{P}}\,,
\end{equation}
so that they anticommute with both $\slashed{n}$ and $\hat{\slashed{P}}$.
Using these definitions, the tensor bases for $J=0,1,2$ mesons can be cast in a very compact form:
\begin{equation}\label{meson-basis-1}
\begin{split}
  J=0: &\quad  \{ \, \mathds{1}, \; \hat{\slashed{P}} \,\} \, \{ \, \mathds{1}, \;\, \slashed{n} \, \}\,, \\
  J=1: &\quad  \{ \, \mathds{1}, \; \hat{\slashed{P}} \,\} \, \{ \, \mathds{1}, \;\, \slashed{n} \, \} \, \{ \, n^\mu, \;\; \gamma^\mu_{\perp\perp}   \} \,, \\
  J=2: &\quad  \{ \, \mathds{1}, \; \hat{\slashed{P}} \,\} \, \{ \, \mathds{1}, \;\, \slashed{n} \, \} \, \{ \, n^\mu n^\nu - \tfrac{1}{3}\,T_P^{\mu\nu}, \;\; \gamma^\mu_{\perp\perp} n^\nu + n^\mu \gamma^\nu_{\perp\perp} \,\} \\
\end{split}
\end{equation}
and similarly for higher spin, with appropriate attachments of $\gamma_5$ to distinguish between positive and negative parity.
It turns out that also for higher $J$ there are just two angular momentum tensors $T_1^{\mu_1\ldots\mu_J}$ and $T_2^{\mu_1\ldots\mu_J}$,
which are symmetric and traceless in the Lorentz indices and transverse to the momentum $P$~\cite{Zemach:1968zz},
\begin{equation}
    J \geq 2: \quad   \{ \, \mathds{1}, \; \hat{\slashed{P}} \,\} \, \{ \, \mathds{1}, \;\, \slashed{n} \, \} \, \{ \, T_1^{\mu_1\ldots\mu_J}, \;\; T_2^{\mu_1\ldots\mu_J} \,\}\,,
\end{equation}
so they constitute  eight components as well.
Details of their construction can be found in Refs.~\cite{LlewellynSmith:1969az,Krassnigg:2010mh,Fischer:2014xha}.

\paragraph{Partial-wave decomposition.}
The above notation is also useful for performing a `partial-wave decomposition', which in our context means to organize
the basis elements with respect to their quark spin and orbital angular momentum content in the hadron's rest frame.
The construction for baryons as quark-diquark or three-quark states is explained in detail in Refs.~\cite{Oettel:1998bk,Oettel:2000ig,Alkofer:2005jh,Eichmann:2009zx,Eichmann:2011vu}; here we illustrate
the analogous case for mesons.
Only the total spin $J$ is Poincar\'e-invariant
and it is described by the Pauli-Lubanski operator
\begin{equation}
   W^\mu = \frac{1}{2}\,\varepsilon^{\mu\alpha\beta\lambda}\,\hat{P}^\lambda\,J^{\alpha\beta} \quad \Rightarrow \quad
   W^2 = \frac{1}{2}\,T_P^{\mu\alpha}\,T_P^{\nu\beta} J^{\mu\nu} J^{\alpha\beta}\,,
\end{equation}
where the eigenvalues of $W^2 \to J(J+1)$ define the spin of the bound state.
$J^{\mu\nu}$ and $P^\mu$ are the generators of the Poincar\'e algebra which satisfy the usual commutator relations.
The interpretation in terms of quark spin and orbital angular momentum is frame- and gauge-dependent;
nevertheless, we can define covariantized spin and orbital angular momentum operators via
\begin{equation}\label{spin-oam}
   S^\mu = \frac{1}{4} \varepsilon^{\mu\alpha\beta\lambda}\,\hat{P}^\lambda\,\sigma^{\alpha\beta} = \frac{i}{2} \gamma^\mu_\perp \gamma_5\,\hat{\slashed{P}}\,, \qquad
   L^\mu = i\varepsilon^{\mu\alpha\beta\lambda}\,\hat{P}^\lambda \,p^\alpha \frac{\partial}{\partial p^\beta}\,,
\end{equation}
which reduce to their conventional forms in the rest frame: $\vect{S} = \vect{\Sigma}/2$ and $\vect{L} =  \vect{p}\times i\vect{\nabla}$.
Using the relations in App.~\ref{app:conventions},
taking their squares gives
\begin{equation}
   S^2 = \frac{3}{4}\,, \qquad
   L^2 = 2p_\perp\cdot \frac{\partial}{\partial p} + \left(p_\perp^\alpha \,p_\perp^\beta - p_\perp^2 \,T_P^{\alpha\beta}\right) \frac{\partial}{\partial p^\alpha} \frac{\partial}{\partial p^\beta}\,.
\end{equation}
For a quark-antiquark system with total momentum $P$ and relative momentum $p$,
the Pauli-Lubanski operator can be written as $W^\mu = \mathbf{S}^\mu + L^\mu$,
where $\mathbf{S}^\mu = S^\mu \otimes \mathds{1} - \mathds{1} \otimes S^\mu$, so that its square becomes
\begin{equation}\label{spin2}
\mathbf{S}^2 = \frac{3}{2}\,\mathds{1}\otimes\mathds{1} - \frac{1}{2}\,\gamma^\mu_\perp \gamma_5\,\hat{\slashed{P}} \otimes \hat{\slashed{P}}\,\gamma_5\,\gamma^\mu_\perp \,.
\end{equation}
The tensor product is understood as $(A\otimes B)_{\alpha\gamma,\delta\beta} = A_{\alpha\gamma}\,B_{\delta\beta}$ as in Fig.~\ref{fig:bse}.
It is straightforward to evaluate the eigenvalues of the spin $\mathbf{S}^2 \to s(s+1)$; for example, taking the leading tensor structure $\gamma_5$ of the pion:
\begin{equation}
   \mathbf{S}^2\, [\gamma_5] = \frac{3}{2}\,\gamma_5 - \frac{1}{2}\,\gamma^\mu_\perp \gamma_5\,\hat{\slashed{P}}\,\gamma_5\,\hat{\slashed{P}}\,\gamma_5\,\gamma^\mu_\perp = 0\,.
\end{equation}

The orbital angular momentum operator $L^2$, on the other hand, is blind to the Dirac structure and
also only acts on $n^\mu$ but not on $\hat{P}^\mu$. Observe that $L^2$ evaluated on a Lorentz invariant ($p^2$, $p\cdot P$, $P^2$)
gives zero, so when applying it to a Bethe-Salpeter amplitude it commutes through the dressing functions
and acts on the basis elements only.
Applied to the possible combinations $n^\alpha$, $n^\alpha\,n^\beta$, $n^\alpha\,n^\beta\,n^\gamma,\ldots$ that appear in~\eqref{meson-basis-1},
one can work out its eigenfunctions and eigenvalues $L^2 \to l(l+1)$:
\begin{equation}\label{oam-meson}
\begin{split}
   l=1: &\quad n^\alpha, \\
   l=2: &\quad n^\alpha\,n^\beta - \tfrac{1}{3}\,T_P^{\alpha\beta}, \\
   l=3: &\quad n^\alpha\,n^\beta\,n^\gamma - \tfrac{1}{5} \left( T_P^{\alpha\beta} n^\gamma + T_P^{\beta\gamma} n^\alpha + T_P^{\gamma\alpha} n^\beta\right),
\end{split}
\end{equation}
and higher combinations can be found using the relation
\begin{equation}
   L^2 (fgh) = f\,L^2(gh) + g\,L^2(fh) + h\,L^2(fg) - (gh)\,L^2 f - (fh)\,L^2 g - (fg)\,L^2 h\,.
\end{equation}
We see that these are the transverse, symmetric and traceless combinations of products of $n^\alpha$.
With these formulas it becomes very simple to arrange the tensor elements~\eqref{meson-basis-1} into eigenstates of $s$ and $l$:
\begin{equation} \renewcommand{\arraystretch}{1.0}
\begin{split}
   J=0: &\quad \{ \, \mathds{1}, \; \hat{\slashed{P}} \,\} \, \{ \, \mathds{1}, \;\; \slashed{n} \,\}\,,  \\
   J=1: &\quad \{ \, \mathds{1}, \; \hat{\slashed{P}} \,\} \, \{ \, \gamma^\mu_\perp, \;\; n^\mu, \;\; \gamma^\mu_\perp \,\slashed{n} - n^\mu, \;\; n^\mu \slashed{n} - \tfrac{1}{3} \gamma^\mu_\perp \,\}\,,
\end{split}
\end{equation}
etc. Only the elements in the brackets on the right are relevant for determining $s$ and $l$: $L^2$ does not act on the total momentum and $\hat{\slashed{P}}$ can be commuted to the left in~\eqref{spin2}
where it factorizes out. The same is true for the $\gamma_5$ matrix when attached to states with opposite parity.
Then, the two $J=0$ elements associated with the unit matrix carry $(s,l)=(0,0)$, i.e., they are $s$ waves, whereas those with $\slashed{n}$ have $(s,l)=(1,1)$ and are $p$~waves.
From left to right, the tensors for $J=1$ carry $(s,l)=(1,0)$, $(0,1)$, $(1,1)$ and $(1,2)$ and so
they represent two $s$~waves, four $p$ waves and two $d$ waves.

\underline{\it Summary:}
Poincar\'e covariance supplies hadrons with a rich and complicated tensor structure.
In particular, tensors that correspond to $p$ waves in the rest frame appear naturally in the Bethe-Salpeter amplitudes even for pseudoscalar and vector mesons.
This is in contrast to the non-relativistic quark-model classification, where the phenomenological relations $P=(-1)^{l+1}$ and $C=(-1)^{l+s}$
constrain pseudoscalar mesons to be $s$~waves, vector mesons to be made of $s$ and $d$ waves etc.,
whereas `exotic mesons' with $J^{PC}=0^{--}$, $0^{+-}$, $1^{-+}$, \dots \, are forbidden and need additional gluons in the form of hybrid configurations.
Relativistically all these options are allowed as $q\bar{q}$ states; for example, the analogue of~\eqref{pion-basis} for an exotic pseudoscalar with $J^{PC}=0^{--}$ reads
\begin{equation}\label{exotic-basis}
   \tau_i(p,P) =  \left\{ \,(p\cdot P)\,\mathds{1}, \; (p\cdot P)\,i\slashed{P}, \; i\slashed{p}, \; (p\cdot P)\,[\slashed{p}, \slashed{P}]\,\right\} \gamma_5.
\end{equation}
Notice the angular factors $p\cdot P$ which are a consequence of the opposite $C$ parity compared to the pion which has $J^{PC}=0^{-+}$.
Since all elements carry powers of the relative momentum, this induces suppression and effectively results in a much higher mass for the bound state.
There is however no principal reason that would forbid exotic $q\bar{q}$ states in a relativistic framework;
they merely do not survive the non-relativistic limit. Results for $q\bar{q}$ exotics
are routinely obtained in the Bethe-Salpeter framework, although they are much
more sensitive to truncation artefacts compared to pseudoscalar and vector mesons~\cite{Burden:2002ps,Krassnigg:2009zh,Qin:2011xq,Fischer:2014xha,Hilger:2016efh}.

\paragraph{Chiral symmetry and the pion.}
The pseudoscalar mesons play a special role in the strong interaction because they would be the massless Goldstone bosons of QCD
in the exact chiral limit where all current-quark masses vanish.
This distinguishes them from other hadrons and
the corresponding features are already encoded in their Bethe-Salpeter amplitudes.
Recalling the discussion in Sec.~\ref{spec:extracting}, the very same steps leading to the current correlator~\eqref{2-current-correlator-2}
can be applied to mesons as well, except that here one starts from the four-point function $\mathbf{G}_{\alpha\beta,\delta\gamma}$,
replaces the baryon's Bethe-Salpeter wave function by its meson counterpart, and the interpolating
fields~\eqref{baryon-current-1} by appropriate fermion bilinears such as in~\eqref{qqbar-currents}:
\begin{align}
\label{eqn:bs-wf-baryon}
     \mathbf\Psi^\lambda_{\alpha\beta}(x_1,x_2,P) = \langle 0 | \mathsf{T}\,\psi_\alpha(x_1)\,\conjg{\psi}_\beta(x_2)\,| \lambda \rangle\,, \qquad
     j^\Gamma(z) = \conjg{\psi}(z)\,\Gamma\,\psi(z)
\end{align}
where we dropped the flavour parts for simplicity.
As a consequence, the onshell residues $\langle 0 | \,j^\Gamma(0)\,| \lambda \rangle$ of the current correlators are the contracted meson Bethe-Salpeter wave functions,
which in momentum space entails an integration over the relative momentum:
\begin{equation}
    \langle 0 |\,j^\Gamma(x)\,| \lambda \rangle =
    \langle 0 | \,j^\Gamma(0)\,| \lambda \rangle \,e^{-ix\cdot P} =
    -\Gamma_{\beta\alpha}\,\mathbf\Psi^\lambda_{\alpha\beta}(x,x,P) = -\int \!\frac{d^4p}{(2\pi)^4}\,\text{Tr}\left\{\Gamma\,\mathbf\Psi^\lambda(p,P)\right\}\,e^{-ix\cdot P}\,.
\end{equation}
Take for example $\Gamma=\gamma_5\gamma^\mu$ and $i\gamma_5$ which produce the axialvector current $j^\mu_{5}$ and the pseudoscalar density $j_{5}$,
respectively.
          If we apply the tensor decomposition~\eqref{pion-basis} to a pseudoscalar meson's wave function and replace $\slashed{p}\to\slashed{p}_\perp$ therein,
          we see that the axialvector current projects out the scalar
          dressing function proportional to $i\slashed{P} \gamma_5$, integrated over $d^4p$, whereas the pseudoscalar density gives the one attached to $\gamma_5$.
          The resulting Lorentz covariants depend on $P^\mu$ with $P^2=-m_\lambda^2$ fixed:
          \begin{equation}\label{decay-constants}
              \langle 0 |\, j^\mu_5(x) \,| \lambda\rangle = - iP^\mu f_\lambda \,e^{-i x\cdot P}, \qquad
              \langle 0 |\, j_5(x) \,| \lambda\rangle = -ir_\lambda \,e^{-i x\cdot P}.
          \end{equation}
          The first quantity encodes the transition from a pseudoscalar meson to an axialvector current and thereby defines its electroweak decay constant $f_\lambda$.
          The pseudoscalar analogue $r_\lambda$ is not associated with a measurable quantity; still, we see that
          the gauge-\textit{invariant} quantities $f_\lambda$ and $r_\lambda$ contain the information carried by the integrated dressing functions of the gauge-\textit{dependent}
          Bethe-Salpeter wave function.

\begin{figure*}[t]
    \begin{center}
    \includegraphics[width=0.87\textwidth]{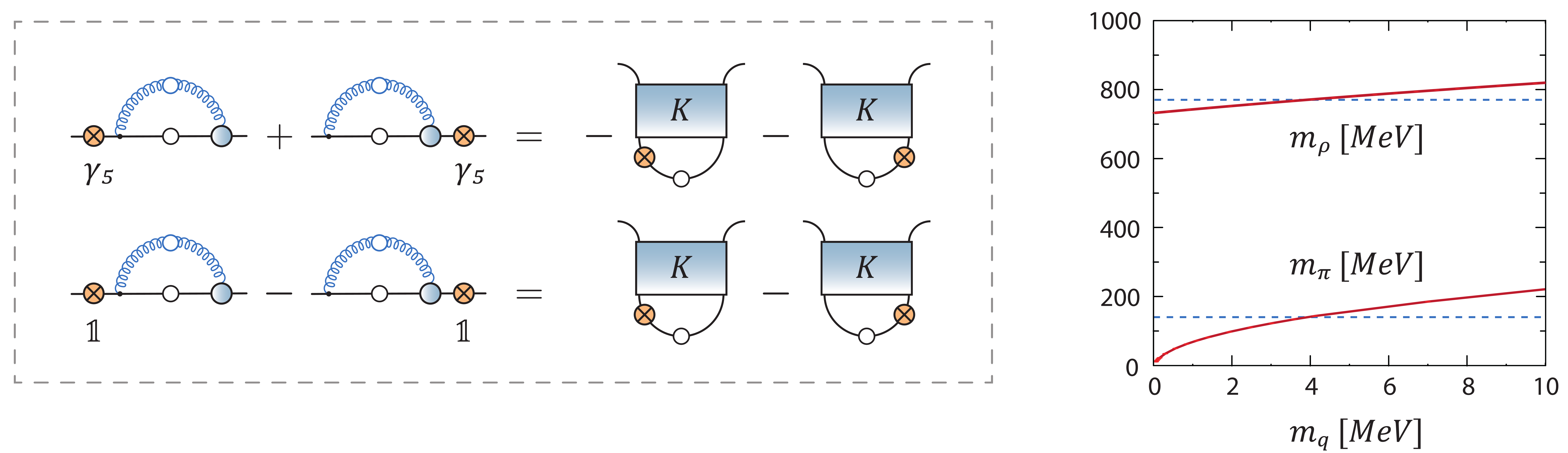}
    \caption{\textit{Left:} Consistency relations between the quark self energy and the Bethe-Salpeter kernel imposed by the axialvector (\textit{top}) and vector Ward-Takahashi identity (\textit{bottom}).
             The crossed boxes represent the injection of an external onshell momentum $P^\mu$ in combination with a Dirac matrix, either $\gamma_5$ (\textit{top}) or $\mathds{1}$ (\textit{bottom}).
             For $P^\mu \to 0$ one recovers the chiral limit.
             \textit{Right:} current-mass evolution of the pion and $\rho-$meson masses.}\label{fig:axwti}
    \end{center}
\end{figure*}

          An immediate consequence of the PCAC relation~\eqref{vcc-pcac-0} for equal quark masses is
          \begin{equation}\label{fm=2mr}
              f_\lambda \,m_\lambda^2 = 2m_q\,r_\lambda\,,
          \end{equation}
          which is valid for all flavour non-singlet pseudoscalar mesons (in the singlet case there would be an additional term from the axial anomaly).
          For unequal quark masses one has to evaluate the anticommutators in the original PCAC relation~\eqref{vcc-pcac-0}.
          For example, the equation relates the pion decay constant $f_\pi$ and pion mass $m_\pi$ with the pseudoscalar transition matrix element $r_\pi$.
          Although this already resembles  the Gell-Mann-Oakes-Renner (GMOR) relation, it has nothing to do with
          spontaneous chiral symmetry breaking but only tells us that in the chiral limit $m_q=0$ either the mass or the decay constant of a pseudoscalar meson must vanish.

          We postpone the proof of the Goldstone theorem to Sec.~\ref{sec:ff-vertices} because it is most easily derived using the properties of the quark-antiquark vertices which we have not yet discussed.
          Its essence is to show that the pion decay constant does \textit{not} vanish in the chiral limit as a consequence of dynamical chiral symmetry breaking.
          The precise relation is~\cite{Maris:1997hd}
          \begin{equation}\label{B-fpi-0}
             f_\pi\,\mathbf\Gamma_\pi(p,P=0) = A(p^2) \, M(p^2) \, \gamma_5\,,
          \end{equation}
          which relates the pion Bethe-Salpeter amplitude in the chiral limit with the quark mass function $M(p^2)$.
          If $A(p^2) \,M(p^2) \neq 0$, neither the pion amplitude nor its decay constant can vanish and therefore $m_\pi=0$.
          As another consequence, the properties of the pion are directly determined by the quark propagator
          and this is the crucial feature that distinguishes the pseudoscalar mesons from other hadrons.
          In turn, their radial excitations by definition have $m_\lambda \neq 0$, which entails that their decay constants must disappear for $m_q\to 0$~\cite{Holl:2004fr}.

          At the quantum level, the vector current conservation and PCAC relations from~\eqref{vcc-pcac-0}
          are realized through Ward-Takahashi identities \cite{Ward:1950xp,Takahashi:1957xn} to be discussed in Sec.~\ref{sec:ff-vertices}.
          They lead to consistency relations between the quark self energy and the Bethe-Salpeter kernel, which are illustrated in Fig.~\ref{fig:axwti}
          and must be respected in any symmetry-preserving truncation of the Bethe-Salpeter equation.
          This is what ultimately guarantees the above behaviour of the pion amplitude as well as the GMOR relation which follows as a corollary.
          For illustration, the right panel in Fig.~\ref{fig:axwti} shows  typical solutions for the resulting masses as a function of the current-quark mass $m_q$.
          The pion mass exhibits the square-root behaviour $m_\pi^2 \sim m_q$ from the GMOR relation whereas the $\rho-$meson mass
          goes to a nonzero constant in the chiral limit. Therefore, we can identify two necessary ingredients to reproduce such a behaviour model-independently:
          a truncation that preserves the consequences of chiral symmetry in the form of the PCAC relation, and a realization of dynamical chiral symmetry breaking
          that produces a nonvanishing quark mass function.

\paragraph{Baryons as three-quark bound-states.}
Let us now continue with the application of Bethe-Salpeter equations to baryons.
It is straightforward to generalize the formalism from~\eqref{dyson-eq}--\eqref{eqn:normalisation} to three-quark systems.
The six-quark Green function in Fig.~\ref{fig:current-correlator-baryons} contains all possible baryon poles
that are compatible with the valence-quark content.
For spin-$\nicefrac{1}{2}$ baryons the pole behaviour is
\begin{equation}\label{cj}
  \mathbf{G} \rightarrow \frac{\mathbf{\Psi}\,(-i\slashed{P}+M)\,\overline{\mathbf{\Psi}}}{P^2+M^2} = 2M\,\frac{\mathbf{\Psi}\,\Lambda_+(P)\,\overline{\mathbf{\Psi}}}{P^2+M^2} = 2M\,\frac{\mathbf{\Psi}\,\overline{\mathbf{\Psi}}}{P^2+M^2}\,,
\end{equation}
where $\Lambda_+(P)=(\mathds{1}+\hat{\slashed{P}})/2$ is the positive-energy projector.
For general spin $J=n+\nicefrac{1}{2}$ the generalized Rarita-Schwinger projector $\mathds{P}^{\mu_1\ldots\mu_n\nu_1\ldots\nu_n}(P)$ appears in the free baryon propagator,
but in all cases the projectors can be absorbed into the Bethe-Salpeter wave functions.

            \begin{figure*}[t]
            \centerline{%
            \includegraphics[width=0.75\textwidth]{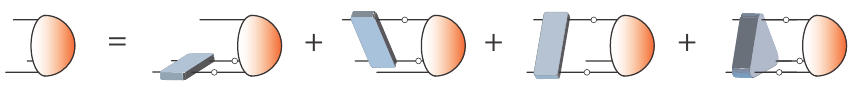}}
            \caption{Three-quark Faddeev equation.}
            \label{fig:faddeev}
            \end{figure*}

The resulting three-quark Bethe-Salpeter or Faddeev equation is depicted in Fig.~\ref{fig:faddeev}.
It is the sum of two-quark and three-quark irreducible contributions, formally written as
\begin{equation}
    \mathbf\Gamma = \mathbf{K}\,\mathbf{G}_0\,\mathbf\Gamma\,, \qquad
    \mathbf{K} = \mathbf{K}^{(3)} + \sum_i \left(\mathbf{K}_i^{(2)}\otimes S^{-1}_i\right), \qquad
    \mathbf{G}_0 = S \otimes S \otimes S\,,
\end{equation}
where $\mathbf{K}^{(2)}_i$ and $\mathbf{K}^{(3)}$ denote the two- and three-body irreducible kernels and $\mathbf{G}_0$
is the disconnected product of three quark propagators.
Omitting three-body forces, the equation is the sum of three permuted diagrams and reads explicitly:
\begin{align}
\big[\mathbf\Gamma(p_1,p_2,p_3)\big]_{\alpha\beta\gamma\sigma} = \int \!\! \frac{d^4k}{(2\pi)^4}  \Big\{
&\big[\mathbf{K}^{(2)}(p_1,p_2;k_1,\widetilde{k}_2)\big]_{\alpha\alpha^\prime\beta\beta^\prime} \big[S(k_1)\big]_{\alpha^\prime\alpha^{\prime\prime}} \big[S(\widetilde{k}_2)\big]_{\beta^\prime\beta^{\prime\prime}}
\big[\mathbf\Gamma(k_1,\widetilde{k}_2,p_3)\big]_{\alpha^{\prime\prime}\beta^{\prime\prime}\gamma\sigma} \nonumber \\
&\big[\mathbf{K}^{(2)}(p_2,p_3;k_2,\widetilde{k}_3)\big]_{\beta\beta^\prime\gamma\gamma^\prime} \big[S(k_2)\big]_{\beta^\prime\beta^{\prime\prime}} \big[S(\widetilde{k}_3)\big]_{\gamma^\prime\gamma^{\prime\prime}}
\big[\mathbf\Gamma(p_1,k_2,\widetilde{k}_3)\big]_{\alpha\beta^{\prime\prime}\gamma^{\prime\prime}\sigma} \\[1mm]
&\big[\mathbf{K}^{(2)}(p_3,p_1;k_3,\widetilde{k}_1)\big]_{\alpha\alpha^\prime\gamma\gamma^\prime}  \big[S(k_3)\big]_{\gamma^\prime\gamma^{\prime\prime}} \big[S(\widetilde{k}_1)\big]_{\alpha^\prime\alpha^{\prime\prime}}
\big[\mathbf\Gamma(\widetilde{k}_1,p_2,k_3)\big]_{\alpha^{\prime\prime}\beta\gamma^{\prime\prime}\sigma} \Big\}\;, \nonumber
\end{align}
with $p_i$ the external quark momenta and $k_i$, $\widetilde{k}_i$
the internal quark momenta (see Sec.~4.1 in~\cite{Eichmann:2009zx} for details).
We already see a strong connection between the two-body kernel appearing here
and in the corresponding Bethe-Salpeter equation~\eqref{bse-meson} for mesons.
Once a framework is provided that stipulates the form of the
two-body kernel, we have a consistent approach for tackling both the meson and
baryon spectrum which does not require any further input or parameters.
Such a course has already been exploited with some success in the calculations of Refs.~\cite{Eichmann:2009qa,SanchisAlepuz:2011jn,Sanchis-Alepuz:2014sca}.
The three-body irreducible kernel has been neglected in these works mainly for technical reasons;
however, its leading-order topology of a three-gluon vertex connecting all three quarks
vanishes to all orders in QCD as a result of the colour trace, and
explicit calculations for light baryons of the non-perturbative next-to-leading contributions
confirm the smallness of the term directly~\cite{Sanchis-Alepuz:threebodyinprep}.

Unsurprisingly, the three-body equation is much more complicated than its two-body analogue for mesons.
This is mainly due to the structure of the baryon amplitude
which depends on three independent momenta and many more tensor structures as we will see below.
Sophisticated methods based on permutation-group symmetries are necessary and have been developed to facilitate the problem,
see Refs.~\cite{Eichmann:2011vu,Eichmann:2015nra} for state-of-the-art solution techniques;
with their help it has recently also become possible to solve the analogous four-body equation for tetraquarks~\cite{Eichmann:2015cra}.
In Sec.~\ref{spec:approx} we review the truncations and approximations that have been made over the years in solving the equation,
from relativistic reductions to quark-diquark models and eventually returning to the full system.
In the remainder of this section we discuss the structure of the Bethe-Salpeter amplitude, its partial-wave decomposition,
and the quark-diquark approximation which -- with simplifying assumptions -- also relies upon the same common
two-body kernel as input.

\paragraph{Spin structure for baryons.}
Earlier we have explained the structure of meson Bethe-Salpeter amplitudes and their tensor decompositions.
Now, let us apply similar constructive principles to baryons of spin $j=n+\nicefrac{1}{2}$.
Their Bethe-Salpeter amplitudes are mixed tensors of rank $4+n$ whose general form is
\begin{align}
\left[\mathbf\Gamma(p,q,P)\right]^{\mu_1\ldots\mu_n}_{\alpha\beta\gamma\sigma} = \sum_i f_i(p^2, q^2, p\cdot q, p\cdot P, q\cdot P, P^2=-M^2) \left[\tau_i(p,q,P)\right]^{\mu_1\ldots\mu_n}_{\alpha\beta\gamma\sigma}\;.
\end{align}
Here we have expressed the amplitude in terms of two relative momenta $p$ and $q$ and the total momentum~$P$;
compared to mesons the $f_i$ are now more complicated scalar functions of five Lorentz invariants.
We have three valence quarks that contribute a Dirac index each ($\alpha$, $\beta$, $\gamma$),
plus a fourth spinor index~($\sigma$) corresponding to the spin-$\nicefrac{1}{2}$ part of the baryon,
combined with $n$ Lorentz indices for baryons with higher spin.

For spin-$\nicefrac{1}{2}$ baryons one can show
that a linearly independent basis is spanned by the 64 elements~\cite{Carimalo:1993aa,Eichmann:2009qa}
\begin{equation}
\begin{array}{rl}
    D_i\,\gamma_5 \,\mathcal{C} \; \otimes & \!\!\!\! D_j\,\Lambda_+(P)\,, \\
    \gamma_5\,D_i\,\gamma_5 \,\mathcal{C} \; \otimes & \!\!\!\!  \gamma_5\,D_j\,\Lambda_+(P)\,,
\end{array}\qquad
\begin{array}{rl}
   D_i &= \left\{  \mathds{1}, \,\, \slashed{p}, \,\, \slashed{q}, \,\, \slashed{P}, \,\, [\slashed{p},\slashed{P}], \,\,
        [\slashed{q},\slashed{P}], \,\, [\slashed{p},\slashed{q}], \,\, [\slashed{p},\slashed{q},\slashed{P}] \right\}, \\
   \Lambda_\pm(P) &= \tfrac{1}{2}\,(\mathds{1} \pm \hat{\slashed{P}})\,,
\end{array}
\end{equation}
where the tensor product is understood as
$\left(f\otimes g\right)_{\alpha\beta\gamma\sigma} = f_{\alpha\beta}\,g_{\gamma\sigma}$.
The factors $\gamma_5\,\mc{C}$, where $\mc{C}=\gamma^4 \gamma^2$ is the charge-conjugation matrix, follow as a consequence of dealing with two fermions (instead of a fermion and antifermion)
and even occurrences of $\gamma_5$ account for positive parity.
The positive-energy projector $\Lambda_+(P)$~enters through the Green function~\eqref{cj} and selects a positive-energy baryon;
one could equally contract the index~$\sigma$ with a nucleon spinor $u(P) = \Lambda_+(P)\,u(P)$.
Note that the projector eliminates half of the eight tensor structures because of $\slashed{P}\,\Lambda_+(P) = iM\,\Lambda_+(P)$,
and therefore we indeed end up  with 64 independent elements.

In analogy to the case of mesons that we discussed earlier, one can cast the basis into eigenstates of quark spin and orbital angular momentum in the baryon's rest frame.
Here we will just sketch the basic ideas and refer to App.~B.2 and B.3 of Ref.~\cite{Eichmann:2011vu} for the detailed construction.
The three-quark spin operator has the form $\mathbf{S}^\mu = S^\mu \otimes \mathds{1} \otimes \mathds{1} + \mathds{1} \otimes S^\mu \otimes \mathds{1} + \mathds{1} \otimes \mathds{1} \otimes S^\mu$
and the orbital angular momentum is the sum of the operators for both relative momenta: $L^\mu = L^\mu_{(p)} + L^\mu_{(q)}$, cf.~\eqref{spin-oam}.
The first step is to orthogonalize the momenta $p$, $q$ and $P$ that appear in the amplitude.
In principle this yields three mutually orthogonal unit vectors. However,
from three vectors one can construct another axialvector using $\varepsilon^{\mu\alpha\beta\gamma}$ that is orthogonal to all of them:
\begin{equation}
   P^\mu \;\; \to \;\; \hat{P}^\mu, \qquad\;
   p^\mu \;\; \to \;\; n_3^\mu = \widehat{p_\perp}^\mu\,, \qquad\;
   q^\mu \;\; \to \;\; n_2^\mu = \frac{\widehat{q_\perp}^\mu - y\,n_3^\mu}{\sqrt{1-y^2}}\,, \qquad\;
   n_1^\mu = \varepsilon^{\mu\alpha\beta\gamma} n_2^\alpha\,n_3^\beta\,\hat{P}^\gamma\,,
\end{equation}
where $y=\widehat{p_\perp}\cdot\widehat{q_\perp}$ and we employed the notation of~\eqref{meson-orth-momenta}.
In that way we have constructed four covariant and orthonormal momenta $n_1^\mu$, $n_2^\mu$, $n_3^\mu$ and $\hat{P}^\mu$ which,
in the rest frame of~\eqref{rest-frame}, simply become the four Euclidean unit vectors (and therefore $\slashed{n}_1=\gamma_1$, etc.). Such a strategy also
saves us from the trouble of putting explicit instances of $\delta^{\mu\nu}$ and $\gamma^\mu$ into the basis because they can be reconstructed from the unit vectors:
\begin{equation}\label{onb-baryon-0}
   T^{\mu\nu}_P = \delta^{\mu\nu} - \hat{P}^\mu \hat{P}^\nu = \sum_{i=1}^3 n_i^\mu\,n_i^\nu\,, \qquad
   \gamma^\mu_\perp = \gamma^\mu - \hat{P}^\mu\,\hat{\slashed{P}} = \sum_{i=1}^3 n_i^\mu\,\slashed{n}_i\,.
\end{equation}
This is clearly true in the rest frame but because these relations are covariant they hold in any frame.
Hence we can equally write the basis as
\begin{equation}\label{onb-baryon-1}
\begin{array}{rl}
    D_i'\,\Lambda_\pm(P)\,\gamma_5 \,\mathcal{C} \; \otimes & \!\!\!\! D_j'\,\Lambda_+(P)\,, \\
    \gamma_5\,D_i'\,\Lambda_\pm(P)\,\gamma_5 \,\mathcal{C} \; \otimes & \!\!\!\!  \gamma_5\,D_j'\,\Lambda_+(P)\,,
\end{array}\qquad
   D_i' = \left\{ \mathds{1}, \, \gamma_5\,\slashed{n}_1, \, \slashed{n}_2, \, \slashed{n}_3 \right\}, \qquad \gamma_5\,\slashed{n}_1 = \slashed{n}_2\,\slashed{n}_3\,\hat{\slashed{P}}\,.
\end{equation}

The next step is to find the eigenfunctions of $L^2$ when acting on products of the $n_i^\alpha$ which appear in the basis elements.
Similarly to~\eqref{oam-meson} one finds
\begin{equation}
   l=1: \;\; n_i^\alpha\,, \qquad\qquad
   l=2: \;\; n_i^\alpha\,n_i^\beta - \tfrac{1}{3}\,T_P^{\alpha\beta} \quad (i=2,3) \quad \text{and} \quad
             n_i^\alpha\,n_j^\beta + n_j^\alpha\,n_i^\beta \quad  (i \neq j)\,.
\end{equation}
For $l\geq 2$ these are again the transverse, symmetric and traceless combinations.
Take for example the nine combinations $n_i^\alpha\,n_j^\beta$:
they produce the five symmetric and traceless elements above which carry $l=2$;
three antisymmetric combinations
\begin{equation}
   n_i^\alpha\,n_j^\beta - n_j^\alpha\,n_i^\beta = \varepsilon^{\alpha\beta\gamma\delta}\,n_k^\gamma\,\hat{P}^\delta
\end{equation}
where $\{i,j,k\}$ is an even permutation of $\{1,2,3\}$,
which are equivalent to the three elements with $l=1$; and the sum in~\eqref{onb-baryon-0} gives the trace $T^{\alpha\beta}_P$ with $l=0$.
To cast the basis~\eqref{onb-baryon-1} into eigenstates of $L^2$, it is then sufficient to arrange the 16 combinations appearing in $D_i' \otimes D_j'$,
\begin{equation}
    \mathds{1}\otimes\mathds{1}\,, \qquad
    \mathds{1}\otimes \slashed{n}_j\,, \qquad
    \slashed{n}_j \otimes \mathds{1} \,, \qquad
    \slashed{n}_i \otimes \slashed{n}_j = (\gamma^\alpha_\perp \otimes \gamma^\beta_\perp) \,n_i^\alpha\,n_j^\beta\,,
\end{equation}
into eigenstates of $L^2$. We omitted the $\gamma_5$ attached to $\slashed{n}_1$ since it can always be put back in again in the final basis to give the correct parity.
The first element has $l=0$, the next six elements carry $l=1$, and the nine combinations $n_i^\alpha\,n_j^\beta$ produce
one $s$ wave, three $p$ waves and five $d$ waves as explained above. Combining this with the eigenfunctions of $\mathbf{S}^2$,
the 64 basis elements for a $J=\nicefrac{1}{2}$ Bethe-Salpeter amplitude are decomposed as
\begin{equation}\label{sl-nuc}
    (s,l)  \quad = \quad
    \big(\tfrac{1}{2}, 0\big) \, \times\, 8, \qquad
    \big(\tfrac{1}{2}, 1\big) \, \times\,24, \qquad
    \big(\tfrac{3}{2}, 1\big) \, \times\,12, \qquad
    \big(\tfrac{3}{2}, 2\big) \, \times\,20\,.
\end{equation}
That is, we arrive at eight $s$ waves, 36 $p$ waves and $20$ $d$ waves.

Once again we see that with respect to orbital angular momentum in the baryon's amplitude there are a number of $p$ waves, in contrast to the non-relativistic quark model which permits $s$ and $d$ waves only.
The $p$ waves are a consequence of Poincar\'e covariance and they would not be present in a non-relativistic description.
Of course, their true magnitude
is a dynamical question, but as we will see in Sec.~\ref{spec:results}
they can be as large as $30\%$ for the nucleon and an astonishing $60\%$ for the Roper resonance.
Hence, the dominant orbital angular momentum in the nucleon comes from $p$ waves and \textit{not} from $d$ waves.
This is perhaps not fully appreciated even in relativistic treatments of the nucleon.
With a relativistic spinor notation as in Ref.~\cite{Carimalo:1992ia},
the $p$ waves come disguised as three-spinors with negative parity;
however, the `wrong' parity is matched by the parity of the relative momenta which enter as unit vectors in the basis~\eqref{onb-baryon-1}
and do not produce visible momentum dependencies (see also the discussion around Eq.~(4.27) in Ref.~\cite{Eichmann:2009zx}).
In general only the $\gamma_5$ matrices in the basis can truly flip the parity of a baryon with quantum numbers $J^P$, i.e.,
by multiplying all elements with $\gamma_5$ either on the left or right side of the tensor product.

To accommodate baryons of higher spin $J=n+\nicefrac{1}{2}$, one has to replace in the basis~\eqref{onb-baryon-1}
\begin{equation}
   D_j'\,\Lambda_+(P) \; \to \; {D_j'}^{\mu_1 \dots \mu_n}\,\mathds{P}^{\mu_1\ldots\mu_n\nu_1\ldots\nu_n}(P)\,,
\end{equation}
where $\mathds{P}^{\mu_1\ldots\mu_n\nu_1\ldots\nu_n}(P)$ is the generalised Rarita-Schwinger projector of rank $2n$; for example
in the case $n=1$ for the $\Delta$ baryon:
\begin{align}\label{theory:normalraritaschwinger}
	\mathds{P}^{\mu\nu}(P) = \Lambda_+(P)\left(T^{\mu\nu}_P - \frac{1}{3}\gamma^{\mu}_\perp\gamma^{\nu}_\perp\right)\;.
\end{align}
In principle this implies that the ${D_j'}^{\mu_1 \dots \mu_n}$  must be constructed from a set of Dirac matrices.
However, our  analysis above entails that we can simply take over the $D_j'$ from before and multiply them with spinless objects that are expressed by the Lorentz vectors $n_i^\mu$ only.
The Rarita-Schwinger projector satisfies $P^\mu\, \mathds{P}^{\mu\nu} = 0$ and $\gamma^\mu \,\mathds{P}^{\mu\nu}=0$,
which from Eq.~\eqref{onb-baryon-0} implies that $n_1^\mu \,\mathds{P}^{\mu\nu}$ is linearly dependent, and therefore the vectors $n_2^\mu$ and $n_3^\mu$ are sufficient for the construction.
The recipe for obtaining the basis for $J=\nicefrac{3}{2}$ baryons then goes as follows: take the basis~\eqref{onb-baryon-1},
multiply it with $n_2^\mu$ and $n_3^\mu$,
work out the new eigenfunctions of the orbital angular momentum and put them back into the basis,
take appropriate linear combinations that are eigenfunctions of~$\mathbf{S}^2$,
contract the open Lorentz index with the Rarita-Schwinger projector,
and multiply with $\gamma_5$ to reinstate positive parity.
The construction is detailed in Ref.~\cite{SanchisAlepuz:2011jn} and yields 128 elements for the $\Delta$ baryon:
\begin{equation}\label{sl-delta}
    (s,l)  \;\; = \;\;
    \big(\tfrac{3}{2}, 0\big) \, \times\, 4, \quad
    \big(\tfrac{3}{2}, 1\big) \, \times\,12, \quad
    \big(\tfrac{3}{2}, 2\big) \, \times\,20, \quad
    \big(\tfrac{3}{2}, 3\big) \, \times\,28, \quad
    \big(\tfrac{1}{2}, 1\big) \, \times\,24, \quad
    \big(\tfrac{1}{2}, 2\big) \, \times\,40\,.
\end{equation}

\underline{\it Summary:} In a covariant three-body approach a baryon is a very complicated object. The relevant tensor structures carrying the
quantum numbers of the baryon are 64 for spin-$\nicefrac{1}{2}$ and 128 for
spin-$\nicefrac{3}{2}$ baryons. These can be grouped in $s$-, $p$- and $d$-wave
components
for spin-$\nicefrac{1}{2}$; in addition, even $f$ waves appear for
spin-$\nicefrac{3}{2}$ baryons. In principle, all of these have to be included
in a full calculation
and approaches that neglect part of this structure may deliver incomplete results.
\paragraph{Baryons as quark-diquark bound states.}\label{sec:quarkdiquarkapproximation}
           While the three-body Bethe-Salpeter equation Fig.~\ref{fig:faddeev} is conceptually simple and transparent, it also poses a numerical challenge
           due to the complicated structure of the amplitudes in terms of many Lorentz invariants and tensor structures.
           In this respect, quark-diquark models provide a welcome alternative to simplify the problem and offer additional insight into the underlying physics.
           A diquark clustering in baryons has been suggested long ago to resolve the problem of missing resonances predicted by the quark model;
           see~\cite{Anselmino:1992vg,Klempt:2009pi} for reviews.
           Simply speaking, if two quarks combine to a diquark they freeze out degrees of freedom and produce fewer possibilities to excite a baryon.
           Consequently, diquarks have been introduced under many different guises in the attempt to
           construct dynamical models of baryons, in particular with regards to spectroscopy,
           and quark-diquark models in the spirit of the quark model have enjoyed great popularity until today~\cite{Jaffe:2003sg,Forkel:2008un,Santopinto:2014opa}.
           On the other hand, some of those `missing' states have now started to appear with photoproduction experiments,
           and also recent mass spectra from lattice QCD find a multitude of states which are closer to those in the symmetric quark model~\cite{Edwards:2011jj,Dudek:2012ag,Pennington:2014qra},
           which has led to the occasional claim that diquarks cannot truly play an important role.

           Nevertheless, before drawing conclusions too quickly one should perhaps appreciate the different levels of sophistication encountered in quark-diquark approaches.
           They span the range from simple models to functional approaches derived from QCD's Green functions,
           where diquarks are introduced as quark-quark correlations that appear in the same T-matrices as those discussed in the beginning of this section.
           Those `diquarks' carry a rich dynamical structure and have more in common with the actual three-body system in Fig.~\ref{fig:faddeev}
           than the early quark-diquark models.
           In fact, in the following we will show that only a few assumptions are needed to derive the quantum field theoretical version of the quark-diquark model
           from the three-body equation.

            \begin{figure*}[t]
            \centerline{%
            \includegraphics[width=1\textwidth]{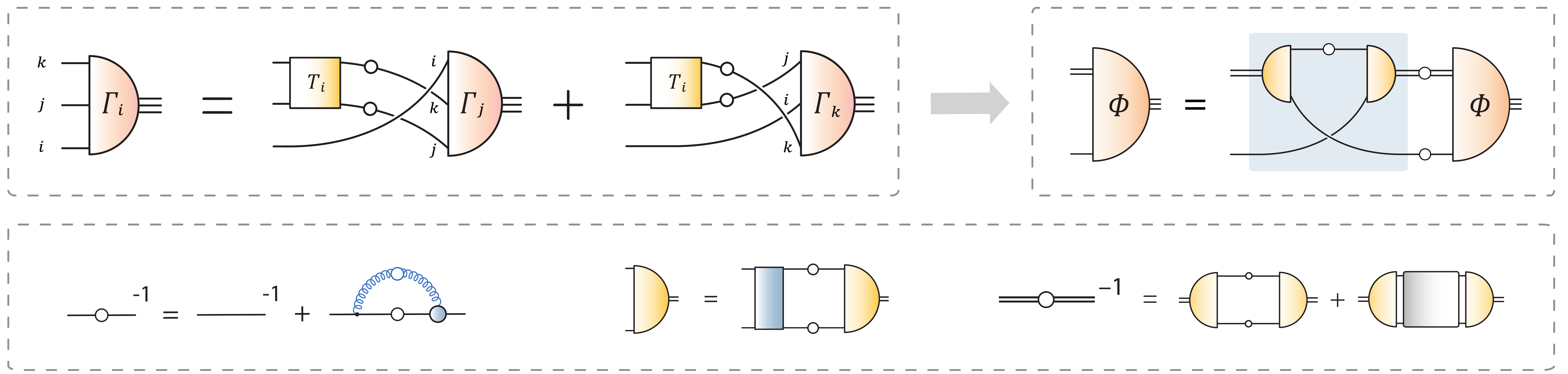}}
            \caption{Simplification of the Faddeev equation~\eqref{fe-rewritten} (\textit{top left}) to the quark-diquark Bethe-Salpeter equation~\eqref{qdq-bse} (\textit{top right}).
                     The bottom panel shows the ingredients that enter in the equation and in principle have to be calculated beforehand: the quark propagator, diquark Bethe-Salpeter amplitudes and diquark propagators.}
            \label{fig:quark-diquark}
            \end{figure*}

           The reduction of the description of baryons as bound states of three quarks to one that involves quark-diquark degrees of freedom is a two-step process.
           The first is simply a rewriting of the three-quark Bethe-Salpeter equation. Neglecting irreducible three-body forces, it takes the form
           \begin{equation}
             \mathbf{\Gamma} = \sum_{i=1}^3 \mathbf\Gamma_i = \sum_{i=1}^3 \mathbf{K}_i\,\mathbf{G}_0\,\mathbf{\Gamma}\,,
           \end{equation}
           where the Faddeev components $\mathbf\Gamma_i$ correspond to the three individual diagrams in Fig.~\ref{fig:faddeev}.
           The $\mathbf{K}_i$ are the two-body kernels and we temporarily refer to $\mathbf{G}_0$ as the product of two quark propagators.
           We can eliminate the two-body kernels in the equation in favor of the two-body T-matrices, which are related to each other via Dyson's equation~\eqref{dyson-eq}:
           \begin{equation}\label{dyson-eq-2}
             \mathbf{T}_i = (\mathds{1}+\mathbf{T}_i\,\mathbf{G}_0)\,\mathbf{K}_i\,.
           \end{equation}
           Applying this to the Faddeev amplitude gives
           \begin{equation}\label{fe-rewritten}
             \mathbf{T}_i\,\mathbf{G}_0\,\mathbf{\Gamma} = (\mathds{1}+\mathbf{T}_i\,\mathbf{G}_0)\,\mathbf{\Gamma}_i \quad \Rightarrow \quad
             \mathbf\Gamma_i = \mathbf{T}_i\,\mathbf{G}_0\,(\mathbf\Gamma-\mathbf\Gamma_i) = \mathbf{T}_i\,\mathbf{G}_0\,(\mathbf\Gamma_j+\mathbf\Gamma_k)
           \end{equation}
           with $\left\{i,j,k\right\}$ an even permutation of $\left\{1,2,3\right\}$.
           The equation is depicted in the upper left panel of Fig.~\ref{fig:quark-diquark} and this is what is usually referred to as `Faddeev equations'
           in the literature. Thus far we have made no further approximations beyond the omission of the
           irreducible three-body kernel; the Faddeev components $\mathbf{\Gamma}_i$ retain the same momentum
           dependence and Dirac tensor structure as the full baryon amplitude
           $\mathbf{\Gamma}$.

           However, the presence of the T-matrices enables one to
           consider a resonant expansion in terms of diquark correlations as a viable
           approximation. The \textit{quark-antiquark} scattering matrix contains all possible meson poles,
           which allows one to derive homogeneous Bethe-Salpeter equations at the poles.
           In the same spirit one can approximate the \textit{quark-quark} scattering matrix as a
            sum over separable (scalar, axialvector, \dots) diquark correlations:
           \begin{align}
           \big[\mathbf{T}(q,q',p_d)\big]_{\alpha\gamma;\beta\delta} &\simeq \sum  \big[\mathbf\Gamma^{\mu\ldots}_\text{D}(q,p_d)\big]_{\alpha\beta}\, D^{\mu\ldots\nu\ldots}(p_d^2)\,  \big[\conjg{\mathbf\Gamma}^{\nu\ldots}_\text{D}(q',p_d)\big]_{\delta\gamma}\;, \label{diquark-approximation} \\
           \big[\mathbf\Gamma_i(p,q,P)\big]_{\alpha\beta\gamma\sigma} &\simeq \sum \big[\mathbf\Gamma^{\mu\ldots}_\text{D}(q,p_d)\big]_{\alpha\beta} \,D^{\mu\ldots\nu\ldots}(p_d^2) \, \big[\Phi^{\nu\ldots}(p,P)\big]_{\gamma\sigma}\;.
           \end{align}
           Here, $\mathbf\Gamma^{\mu\ldots}_\text{D}$ is the diquark Bethe-Salpeter amplitude and
           $\overline{\mathbf\Gamma}^{\mu\ldots}_\text{D}$ its conjugate; the diquark
           propagator is $D^{\mu\ldots\nu\ldots}$; $p_d$ is the diquark momentum and $q$, $q'$ are the relative quark momenta in the diquark amplitudes.
           In the second line the same assumption
           was made for the Faddeev components, thus introducing the \textit{quark-diquark} Bethe-Salpeter amplitude $\Phi^{\nu\ldots}(p,P)$.
           What results is an approximation of the three-quark system to a two-body problem, namely
           to a coupled system of quark-diquark bound-state equations. They are illustrated in the upper right panel of Fig.~\ref{fig:quark-diquark}:
           \begin{equation}\label{qdq-bse}
              \big[\Phi^{\mu\ldots}(p,P)\big]_{\alpha\sigma} = \int \!\!\frac{d^4k}{(2\pi)^4}\,\big[\mathbf K_\text{Q-DQ}^{\mu\ldots \nu\ldots}\big]_{\alpha\beta} \big[S(k_q)\big]_{\beta\gamma}\,D^{\nu\ldots\rho\ldots}(k_d)\,\big[\Phi^{\rho\ldots}(k,P)]_{\gamma\sigma}\,,
           \end{equation}
           and the quark-diquark kernel is given by
           \begin{equation}
                \mathbf K_\text{Q-DQ}^{\mu\ldots \nu\ldots} =  \mathbf\Gamma_\text{D}^{\nu\ldots}(k_r,k_d)\,S^T(q)\,\conjg{\mathbf\Gamma}_\text{D}^{\mu\ldots}(p_r,p_d)\,.
           \end{equation}
           Here, $P$ is the baryon's total momentum, $p$ is the quark-diquark relative momentum and
           the remaining momenta can be inferred from the figure (see Sec.~5.2 in~\cite{Eichmann:2009zx} for details).
             Note that there are no additional gluon lines connecting the quarks and diquarks.
             The baryon is bound by quark exchange \cite{Cahill:1988dx,Cahill:1988zi,Reinhardt:1989rw};
             gluons no longer appear explicitly but they are rather absorbed into the building blocks: the quark propagator, diquark amplitudes and diquark propagators.
            The quark-diquark equation is apparently a great simplification, both in terms of kinematic variables and tensor structures,
            and the rather mild assumptions required to derive it suggest that it may still capture the essential dynamics of the three-body system.

             How good is then the approximation? Diquarks are of course not observable because they carry colour. Nevertheless,
             there are several observations that support a quark-diquark interpretation of baryons.
             As we already mentioned above, on a diagrammatic level the leading irreducible three-body force (the three-gluon vertex that connects three quarks)
             vanishes simply due the colour traces~\cite{Sanchis-Alepuz:threebodyinprep}, which suggests that two-quark correlations are more important.
             Second, the colour interaction between two quarks ($\mathbf{3}\otimes\mathbf{3} = \conjg{\mathbf{3}} \oplus \mathbf{6}$)
             is attractive in the colour-$\mathbf{\conjg 3}$ channel
             and supports the formation of diquarks as coloured `bound states' of quarks.
             The question is then whether the two-quark scattering matrix $\mathbf{T}$ indeed has such poles,
             because this is what allows one to  define a diquark Bethe-Salpeter amplitude from their residues and
             derive a diquark Bethe-Salpeter equation, which is analogous to~\eqref{bse-meson} for mesons
             and features the common exchange kernel $\mathbf{K}$:
            \begin{align}\label{bse-diquark}
            \left[\mathbf\Gamma_\text{D}^{\mu_1\ldots\mu_J}(p,P)\right]_{\alpha\beta}
            &= \int\!\!\frac{d^4q}{\left(2\pi\right)^4} \left[\mathbf{K}(p,q,P)\right]_{\alpha\gamma;\beta\delta}\left[ S(q_+)\,\mathbf\Gamma_\text{D}^{\mu_1\ldots\mu_J}(q,P)\,S^T(-q_-)\right]_{\gamma\delta}\;.
            \end{align}
            It turns out that the question is dynamical and depends on the form for the interacting kernel $\mathbf{K}$.
            The rainbow-ladder truncation discussed in Sec.~\ref{spec:approx} indeed generates diquark poles.
            The addition of crossed ladder exchange removes them from the spectrum, which is known to happen in the Munczek-Nemirovsky model~\cite{Bender:1996bb},
            although it was recently argued that an effective resummation of such diagrams can also bring them back again~\cite{Jinno:2015sea}.
            In any case, diquark correlations may well persist in one form or another simply due to the colour attraction:
            it is conceivable that the $qq$ scattering matrix has some quite
involved singularity structure, e.g., cuts, that allows one to identify diquark mass scales,
            and in that sense~\eqref{diquark-approximation} may well be a reasonable ansatz.

            The formal similarity of the meson and diquark Bethe-Salpeter equations
            allows one to compute diquark properties in close analogy to those of mesons from their BSEs.
            Also the structure of the diquark amplitudes is analogous to that in~\eqref{theory:bsetensorstructures}:
            scalar diquarks have the same tensor decomposition as pseudoscalar mesons except for an additional multiplication with the charge-conjugation matrix $\mc{C}$ (note that a two-fermion system has the opposite intrinsic parity as a fermion-antifermion system),
            axialvector diquarks are partnered with vector mesons, pseudoscalar diquarks with scalar mesons and so on.
            Scalar and axialvector diquarks are the lightest ones, hence they are most important for describing the positive-parity nucleon and $\Delta$ baryons.
            The typical mass scales obtained with rainbow-ladder calculations are about $800$ MeV for scalar diquarks, 1~GeV for axialvector diquarks,
            followed by pseudoscalar and vector diquarks~\cite{Maris:2002yu,Eichmann:2016jqx}.
            Thus, the mass pattern for mesons is repeated in the diquark spectrum,
            which also provides an underlying link between meson and baryon properties.
            This is clearly visible, for instance, in the NJL calculation of Ref.~\cite{Chen:2012qr}
            and we will return to this point in Sec.~\ref{spec:approx}.

          In contrast to mesons, the Pauli principle enforces that scalar diquarks carry isospin $I=0$ and axialvector diquarks $I=1$.
          Therefore, the nucleon and its resonances with $I=\nicefrac{1}{2}$ can feature both diquark isospins whereas $\Delta$ resonances ($I=\nicefrac{3}{2}$) can only consist of $I=1$ diquarks.
          An analysis of the nucleon and $\Delta$ excitation spectrum based on its diquark content, including also pseudoscalar and vector diquarks, can be found in Ref.~\cite{Eichmann:2016jqx}.
          Compared to their three-body analogues, the quark-diquark amplitudes $\Phi^{\mu\ldots}(p,P)$ for the nucleon and $\Delta$ depend on a manageable number of tensor basis elements, namely
          eight for the nucleon (two associated with scalar and six with axialvector diquarks) and also eight for the $\Delta$, coming from axialvector diquarks only~\cite{Oettel:1998bk}.
          Once again they can be classified with respect to their (quark-diquark) spin and orbital angular momentum in the rest frame~\cite{Oettel:1998bk,Oettel:2000ig,Alkofer:2005jh},
          leading to similar breakdowns as in~\eqref{sl-nuc} and~\eqref{sl-delta}. The explicit formulas are collected, for example, in Table I of Ref.~\cite{Eichmann:2011aa}.

            As we will discuss in Sec.~\ref{spec:approx}, the quark-diquark equation provides ample opportunities for further simplifications.
            One could employ pointlike diquarks and constituent propagators, which leads to the NJL version of the model,
            or momentum-dependent parametrizations for the propagators and amplitudes.
            However, a drawback of such strategies is that the approximation for $\mathbf{T}$ is designed to capture its timelike behaviour where the squared diquark momentum $p_d^2$ is negative,
            whereas the quark-diquark equation samples its structure predominantly at spacelike momenta.
            This is analogous to the parabola in Fig.~\ref{fig:scalar-bse-1} if it is interpreted as the complex plane in $p_d^2$.
            The onshell diquark poles never actually enter in the integration (and instead provide practical limits for the baryon masses),
            whereas on the spacelike side the scattering matrix must eventually become independent
            of $p_d$. The Dyson equation~\eqref{dyson-eq-2} implies $\mathbf{T}\to\mathbf{K}$ for $p_d^2 \to\infty$, because each intermediate quark propagator induces suppression,
            and $\mathbf{K}$ asymptotically becomes a one-gluon exchange which only depends on the gluon momentum but not on the total momentum of the two quarks.
            If the diquark amplitudes are provided by onshell solutions of their Bethe-Salpeter equations they are essentially fixed;
            however, a systematic way to implement the offshell behaviour is to put the
            diquark ansatz~\eqref{diquark-approximation} back into the scattering equation~\eqref{dyson-eq-2}, which gives equations for the diquark propagators:
            \begin{equation}
                D^{-1} = -\mc{N} + \frac{\mc{N}^2}{\mc{K}}\,, \qquad
                \mc{N} = \conjg{\mathbf{\Gamma}}_\text{D}\,\mathbf{G}_0\,\mathbf{\Gamma}_\text{D}\,, \qquad
                \mc{K} = \conjg{\mathbf{\Gamma}}_\text{D}\,\mathbf{G}_0\,\mathbf{K}\,\mathbf{G}_0\,\mathbf{\Gamma}_\text{D}\,.
            \end{equation}
            They are schematically illustrated in Fig.~\ref{fig:quark-diquark}. The resulting propagators still have onshell diquark poles but become constant at large spacelike momenta,
            which guarantees the correct \textit{integrated} behaviour of the scattering matrix that enters in the quark-diquark BSE. The precise relations can be found in App.~A.4 of Ref.~\cite{Eichmann:2009zx}.
            In that way the quark-diquark model can be put on solid footing within QCD and represents again an essentially parameter-free calculation
            for baryons whose ingredients are calculated consistently as shown in the bottom panel of Fig.~\ref{fig:quark-diquark}.
            Ultimately, this strategy is also responsible for the similarity between the quark-diquark results and those obtained with the three-body equation.

\underline{\it Summary:} The three-body Bethe-Salpeter equation for baryons may be approximated by a two-body quark-diquark Bethe-Salpeter equation. This can be done
in a systematic way, leading to diquarks with a rich internal structure. The quality of this approximation relies on the size of irreducible three-body forces
inside the baryon, which may very well be small, and the appearance of diquark correlations due to the attractive colour interaction between two quarks.

\newpage

\paragraph{Bethe-Salpeter equations vs. lattice QCD.}
Let us pause for a moment and review the information we have collected so far.
We have seen that both lattice QCD and functional methods are intrinsically nonperturbative. The great advantage of lattice QCD is that the nonperturbative
dynamics is fully contained within the path integral~\eqref{correlators}, which transforms
the classical action into the partition function and combinations of classical fields into their quantum expectation values. 
Given infinite precision, the calculation of observables on the lattice then amounts to their `extraction' from any suitable correlator.
Of course there are practical issues involved: volume and discretization artifacts,
the need for disentangling the effects from excited states and scattering states,
and the efforts that have been invested in constructing `good' lattice actions, reducing the statistical errors and reaching physical pion masses.

The sources for systematic errors in functional approaches are almost disjunct.
Statistical errors are absent and the numerical errors are generically below the $1\%$ level.
The dominant source for systematic errors is inherent in the approach itself:
the structure properties are generated dynamically from solving a set of self-consistent equations,
which in practice have  to be truncated to achieve numerical results.
Systematic truncation schemes are available but also numerically expensive to implement, and
most current applications concern the `art' of identifying the phenomenologically relevant features of the quark-gluon interaction.

Apart from systematizing the truncations, much effort in recent years has been invested in an efficient implementation of the model-\textit{independent} parts of the equations,
namely the spin and momentum structure of the Green functions and hadronic bound-state amplitudes.
In contrast to lattice QCD, where at least in principle any tensor structure that has enough overlap with the state is sufficient to extract a mass or some other observable,
the self-consistent nature of Bethe-Salpeter equations generates a baryon's spin and momentum structure dynamically.
It is therefore important not to `miss' anything,
which is the underlying motivation for constructing \textit{complete} tensor bases and implementing the full momentum dependencies in the equations.
This is quantitatively important because a reduction of the tensor basis to a subset can change the results, as we will see in the later sections.
The partial wave decompositions outlined above provide welcome organizing principles: $s$ waves are more important than $p$ waves, which are more important than $d$ waves, etc.
These efforts have been completed for the two- and three-body systems investigated so far.
For the full-fledged three-quark calculations all possible tensors carrying the baryons' quantum numbers in question are taken into account,
which means that the only remaining error source is the structure of the kernel itself.

In our opinion, the different sources for errors provide the main reason for the complementary nature
of the two approaches and make systematic comparisons interesting and fruitful.
Where lattice QCD can provide quantitative precision, a constructive approach provides insight in the underlying dynamics and may answer questions such as:
What is the origin of the level ordering in the baryon excitation spectrum?
What is the importance of two-quark versus three-quark correlations?
How does the structure of the quark-gluon vertex affect the spectrum?
What has to happen for dynamical chiral symmetry breaking to occur and what are the signatures for confinement?
We will address more such questions when discussing the structure properties of baryons in
Sec.~\ref{form} and~\ref{compton}, where the microscopic features play an even more cohesive role
for the dynamics of form factors and scattering amplitudes.

\subsection{Approximations and truncations}\label{spec:approx}
Having specified the general properties of QCD's correlation
functions and their reformulation in terms of Bethe-Salpeter equations in the vicinity of hadronic
poles, we will now review methods
to approximate these equations. While in principle truncation schemes are
available that can be made systematic within~QCD, many of the approximations done in practice
do not have this property and are instead tailored to meet
specific needs. The reason for this is the sheer analytical and numerical
complexity of the systematic schemes which have prohibited their applicability to
many interesting problems in the past and to some extent also in the present. In turn,
simple approximations that capture essential properties of QCD, such as
the pattern of dynamical chiral symmetry breaking, are still capable of
delivering qualitative explanations for many phenomena that we encounter within
hadron physics and therefore they have their merits despite potential
shortcomings. In the following survey we review the approximations made in connection to the Bethe-Salpeter equation in Fig.~\ref{fig:faddeev}
and its application to the baryon spectrum,
from its relativistic reduction to NJL and quark-diquark models to the full three-quark equation.

\paragraph{Relativistic reductions of the three-body system.}
One way to simplify the complicated Bethe-Salpeter equation for the baryon three-body
system is to sacrifice full covariance by a reduction to the so-called Salpeter
equation. This is performed by replacing the full interaction kernel between the
three quarks by an instantaneous potential with two-body and three-body parts.
Although retardation effects are neglected in this way, the resulting equations
can be made formally covariant such that they are frame-independent.
A study of the reliability of this type of approximation can be found
in~\cite{Oettel:2000uw}. In addition to the reduction, the fully dressed quark
propagators of the internal constituents are replaced by their tree-level counterparts
which, instead of the current quarks in the Lagrangian, depend on large constituent-quark masses as
model parameters~\cite{Koll:2000ke,Loring:2001kv,Loring:2001kx,Loring:2001ky,Ronniger:2011td}.
The purpose of such a scheme is twofold. On the one hand one tries to make
contact with non-relativistic quark models, while on the other hand one may
hope to capture essential QCD dynamics in the details of
the interaction potentials. To this end, the three-body part is chosen to
represent a potential linearly rising with distance and the two-body parts are
derived using an effective instanton induced interaction vertex between quarks,
see~\cite{Loring:2001kv} for details.

The resulting baryon spectrum has been calculated for a wide range of quantum
numbers and up to several radial excitations. From a general perspective, the
results share the successes but also the drawbacks of the quark models. In many
channels, the structure of the experimental spectrum is well reproduced and
linear Regge trajectories obey a universal slope for all flavours. In addition,
the instanton induced two-body interaction has been claimed to generate the low
mass of the Roper resonance. On the other hand, the level ordering between the
positive and negative parity nucleon states remains inverted as compared to
experiment, the $\Lambda(1405)$ with $J^P=\nicefrac{1}{2}^-$ remains a problem and several
states in the $\Delta$-spectrum are problematic~\cite{Loring:2001kv,Loring:2001kx,Loring:2001ky}.
A phenomenological extension to the potentials used
in~\cite{Loring:2001kv,Loring:2001kx,Loring:2001ky} with ten additional
parameters improved the situation but did not resolve all
issues~\cite{Ronniger:2011td}.

\paragraph{NJL-type models.}
The structure of NJL-type models has already been  discussed in Sec.~\ref{spec:green}
around~\eqref{baryontheory:eqn:contactinteraction}. The general idea is to
neglect the momentum dependence of the gluon propagator in the quark DSE and
thus make the equation algebraically solvable. The resulting quark propagator
has a mass function that is momentum independent and resembles a constituent
quark. The quarks interact with one another via a contact interaction, much like
in the Fermi theory of electroweak interactions. This comes at the expense of
asymptotic freedom and renormalisability and consequently the momentum
cutoff generates a scale that becomes one of the parameters of the model, in
addition to the coupling strength and the current-quark masses.

In turn, the NJL model is sufficiently simple so that many exploratory calculations are feasible
without great numerical effort. Since the essential features (although not the dynamical
details) of spontaneous chiral symmetry breaking are preserved, the model has
enjoyed great popularity until today. For reviews on the physics of the NJL model
see e.g.~\cite{Vogl:1991qt,Hatsuda:1994pi}; applications for ground states of
baryons are discussed in~\cite{Buck:1992wz,Kato:1993np,Ishii:1993rt,Cloet:2005pp,Carrillo-Serrano:2014zta}
and recent applications also to excited states and baryons with strangeness can be found in~\cite{Roberts:2011cf,Chen:2012qr}.
We discuss some of these results below in Sec.~\ref{spec:results}. In addition
to the NJL dynamics, all of these publications have used the quark-diquark
approximation of baryons as depicted in Fig.~\ref{fig:quark-diquark}. Studies on the impact of pion cloud effects in this
framework (also in the quark-diquark approximation) can be found
in~\cite{Ishii:1998tw,Cloet:2014rja}. Results for the genuine three-body system have not been
published so far.

In addition to the `conventional' description of baryons as three-quark states,
the NJL model has also been invoked to describe baryons as chiral solitons. We
will not touch upon this topic here since there are excellent specialised
reviews available, see e.g.~\cite{Alkofer:1994ph,Christov:1995vm,Weigel:2008zz}.

\paragraph{Momentum-dependent quark-diquark models.}
One of the generic problems of the NJL model is its lack of dynamics, leading
e.g.\ to drastically simplified wave functions for the baryon. An alternative
strategy that aims at the preservation of such dynamics has been pursued
in~\cite{Cahill:1988dx,Hellstern:1997pg,Oettel:1998bk} for ground-state
baryons and very recently also for the Roper~\cite{Segovia:2015hra}. Here one
works again with the quark-diquark Bethe-Salpeter equation from Fig.~\ref{fig:quark-diquark},
but where the quark and diquark propagators as well as the diquark amplitudes are modelled
by momentum-dependent ans\"atze. The merit of such
a treatment is that it enables one to go beyond pointlike diquark approximations
and explore the effects of momentum dependencies in the ingredients of the
equation on the properties and dynamics of baryon.
On the other hand, the drawback is that one never
solves a Dyson-Schwinger or Bethe-Salpeter equation for the baryon's constituents, thus losing the direct
connection to the underlying properties of QCD at the expense of introducing 
model parametrizations. Thus, similarly to the previous types of approximations, one
trades formal consistency for simplicity and practical feasibility of a wider
range of problems.
However, once the model parameters (diquark masses and couplings, etc.) are fixed to nucleon and $\Delta$ properties,
other states are genuine model predictions.
Beyond masses, this treatment also allows one to
calculate form factors with realistic momentum dependencies, see
Sec.~\ref{form} for details.
Again, we discuss some of the results obtained within this
framework in Sec.~\ref{spec:results} below.

\paragraph{Rainbow-ladder: from quark-diquark to three-body calculations.}
Moving to more and more complex approximation schemes, we now discuss the
rainbow-ladder truncation of QCD's Dyson-Schwinger equations.
This approximation is more sophisticated than those previously discussed since, rather than
constructing a phenomenological model of hadrons, the approach is based upon
the dynamics contained within the quark-gluon vertex of QCD\@.
While in principle the latter is a complicated object, see~\eqref{qgv},
the basic idea of rainbow-ladder is to reduce it to its leading tensor structure:
\begin{align}\label{baryontheory:eqn:rainbowladdervertex}
\Gamma_\mathrm{qg}^\mu(p,k) = i\gamma^\mu\, \Gamma(k^2)\,\mathsf{t}_a\;,
\end{align}
where $\mathsf{t}_a=\lambda_a/2$ are the colour generators for $SU(3)$.
The dressing function $\Gamma(k^2)$ is stripped of its full momentum
dependence to a function of the squared gluon momentum $k^2$ only, i.e.,
the dependence on the relative momentum $p$ is neglected. Upon insertion
into the quark DSE~\eqref{baryontheory:eqn:quarkgapequation},
it can be combined with the gluon dressing $Z(k^2)$ of~\eqref{gluon}
into a renormalisation-point independent effective running coupling:
\begin{equation}\label{alpha-eff}
 \alpha(k^2) = \frac{Z_{1f}}{Z_2^2}\,\frac{g^2}{4\pi}\,Z(k^2)\,\Gamma(k^2)\,.
\end{equation}
The momentum dependence of this function in the ultraviolet momentum region is
known from (resummed) perturbation theory, thus preserving the important property
of asymptotic freedom of QCD together with multiplicative renormalisability.

\begin{figure*}[t]
            \centerline{%
            \includegraphics[width=0.80\textwidth]{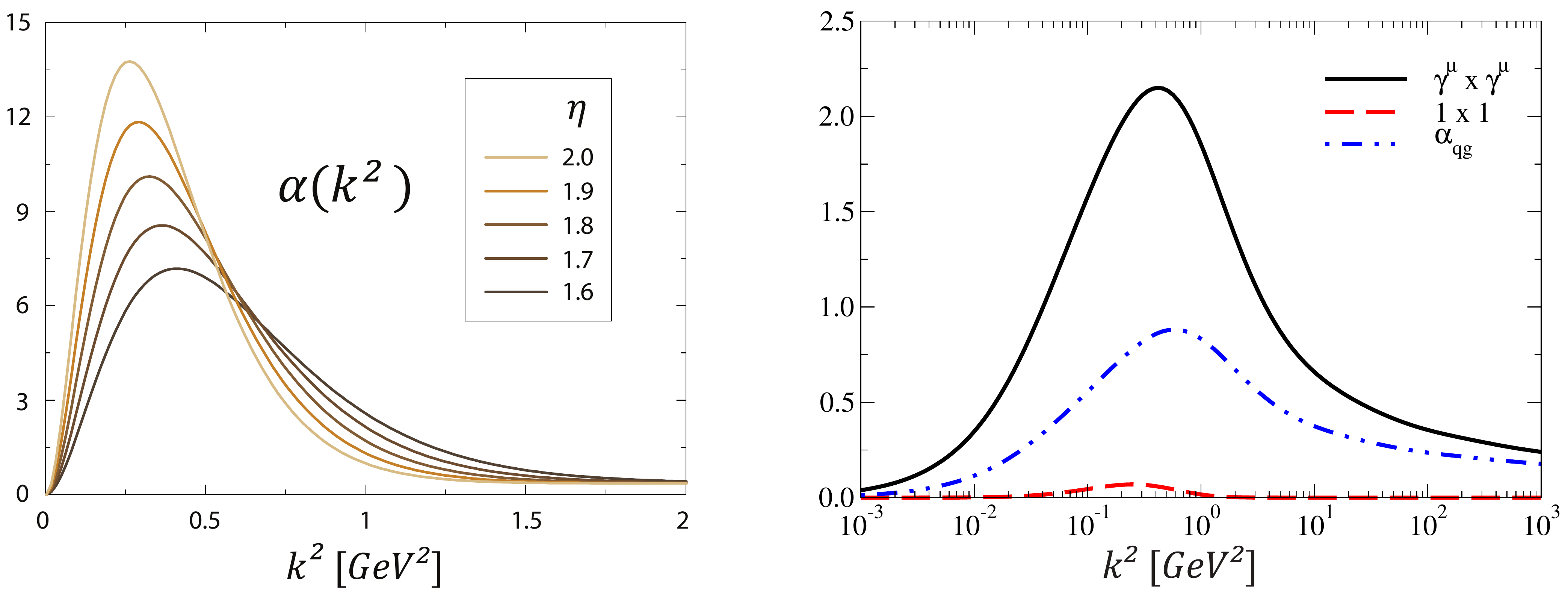}}
            \caption{\textit{Left:} Effective interaction of Eq.~\eqref{couplingMT} for a range of $\eta$ values. \textit{Right:} Analogous quantity $\alpha_{qg}$ of Eq.~\eqref{alpha-eff} extracted from realistic Dyson-Schwinger solutions for the gluon propagator and quark-gluon vertex, and the vector-vector vs. scalar-scalar projection of the Bethe-Salpeter kernel discussed in Ref.~\cite{Williams:2015cvx}.}
            \label{fig:coupling}
            \end{figure*}

In absence of self-consistent Dyson-Schwinger solutions for the quark-gluon vertex and gluon propagator
one subsequently employs a model parametrization for $\alpha(k^2)$.
One of the more frequently used effective interactions is that of Maris and
Tandy~\cite{Maris:1999nt} shown in Fig.~\ref{fig:coupling}, which is composed of an ultraviolet part
$\alpha_\text{UV}(k^2)$ matched to resummed perturbation theory and an infrared
component modelled by a Gaussian:
\begin{align}\label{couplingMT}
\alpha(k^2) = \pi \eta^7  x^2
e^{-\eta^2 x} + \alpha_\text{UV}(k^2) \,,\qquad
    \alpha_\text{UV}(k^2) = \frac{2\pi\gamma_m \big(1-e^{-k^2/\Lambda_t^2}\big)}{\ln \, \left[e^2-1+\big(1+k^2/\Lambda^2_{\mathrm{QCD}}\big)^2\right]}\,,
\end{align}
where the UV parameters are $\Lambda_t=1$~GeV, $\Lambda_\mathrm{QCD}=0.234\,{\rm GeV}$, and $\gamma_m=12/25$ for four active quark flavours.
The UV term ensures the correct perturbative running but is otherwise not essential;
one could neglect it without causing serious damage in the light-quark sector~\cite{Alkofer:2002bp}.
The nonperturbative physics is encoded in the first term with $x=k^2/\Lambda^2$, which is characterised by two parameters\footnote{The
relationship with the parameters $\{ \omega, D \}$ used in Ref~\cite{Maris:1999nt} is $\omega=\Lambda/\eta$ and $D=\eta \,\Lambda^2$. A typical value for $\Lambda$ would be $0.72$~GeV.}:
an infrared scale $\Lambda$ and a dimensionless parameter~$\eta$. Since the scale $\Lambda$ (together with the quark masses)
is fixed to experimental input, only one free parameter $\eta$ remains to which many observables
are insensitive within the range plotted in Fig.~\ref{fig:coupling}.

As discussed in connection with Fig.~\ref{fig:DSEwopions2} below, rainbow-ladder can be identified with the leading order term in a systematic truncation scheme.
The resulting Bethe-Salpeter kernel is a gluon exchange between quark and (anti-)\,quark, again with two tree-level vertices
and the same effective interaction $\alpha(k^2)$. That it is consistent with chiral symmetry can be easily verified from
the axialvector Ward-Takahashi identity in the top left of Fig.~\ref{fig:axwti}: a tree-level vertex anticommutes with $\gamma_5$, and because the quark-gluon vertex only
depends on the total (gluon) momentum $k^\mu$ and not on the relative momentum the injection of the external momentum passes through the vertex and thereby
ensures the equality. As a consequence, the Goldstone nature of the pion in~\eqref{B-fpi-0} is preserved.
Hence, with respect to meson and baryon Bethe-Salpeter equations,
the effect of rainbow-ladder is to compress the full structure of the quark-gluon
vertex into an effective dressing of the tree-level vertex and absorb
the full structure of the kernel into an effective gluon exchange.

Obviously, the reduction of the quark-gluon vertex to a single term leads to a
significant loss of the dynamics encoded therein, including the sizeable
quark-gluon couplings of subleading tensor structures as well as components
that are enhanced by dynamical chiral symmetry breaking.
This is illustrated in Fig.~\ref{fig:coupling}, where the model interaction is compared to the corresponding quantity in~\eqref{alpha-eff} extracted from a self-consistent calculation
of the gluon propagator and quark-gluon vertex~\cite{Williams:2015cvx}. For completeness
we also show the $\gamma^\mu\otimes\gamma^\mu$ and $\mathds{1}\otimes\mathds{1}$ projections of the Bethe-Salpeter kernel at a representative momentum slice.
Carefully noting the differences in scale, clearly the bump in the model interaction is not realistic but merely absorbs
the strength coming from the subleading vertex components, which is necessary to reproduce the phenomenologically required strength of
dynamical chiral symmetry breaking; see also the corresponding discussion in Ref.~\cite{Binosi:2014aea}.

We further note that the infrared $k^4$ behaviour in~\eqref{couplingMT} is only for technical convenience, as it cancels not only the $1/k^2$ factor in the gluon propagator~\eqref{gluon}
that enters in the quark Dyson-Schwinger equation but also the kinematic singularity stemming from the transverse projector in Landau gauge.
It is then straightforward to calculate the quark propagator in the entire complex plane, which leads for example to the plot in Fig.~\ref{fig:scalar-bse-2}.
Interactions with one less power in $k^2$ produce branch cuts in the integrand, which requires more sophisticated methods to evaluate the quark propagator for complex momenta (see App.~A.1 in~\cite{Eichmann:2009zx}
for a discussion and~\cite{Maris:1995ns,Fischer:2005en,Fischer:2008sp,Krassnigg:2009gd,Windisch:2012zd,Rojas:2014aka} for corresponding techniques).
In any case, many observables are insensitive to the deep-infrared behaviour of the interaction due to the freeze-out of the dynamical quark mass and for those such a change has almost no effect~\cite{Blank:2010pa}.
This is also true for the replacement $\pi\eta^7 x^2 \to 2\pi \eta^5 x$ which was suggested in Ref.~\cite{Qin:2011dd};
with a slight readjustment of parameters the results are essentially the same.

Certainly then, with respect to full QCD the rainbow-ladder truncation is still a severe approximation. However, it has some important properties
that are in marked contradistinction to phenomenological models:
it preserves asymptotic freedom and agrees with perturbation theory, and
it preserves chiral symmetry and the Goldstone nature of the pion.
While the physics of QCD's Yang-Mills sector and the quark-gluon vertex is encoded in one simple function,
it  allows for the subsequent calculation of the quark mass function or any kind of observable without further input.
Thus, rainbow-ladder may be viewed as the transition point between phenomenological
modeling and systematic truncations of QCD\@. Of course, the numerical effort to
determine observables such as the masses and wave functions of ground and excited
states, form factors and decays is highly increased compared for instance to the NJL
model or the momentum-dependent quark-diquark models discussed above. In turn,
however, rainbow-ladder allows for a calculation of those observables in close contact to QCD
and it serves as the starting point for a systematic improvement beyond rainbow-ladder, as detailed further below.

\vspace{1em}\par\noindent\underline{\emph{Quark-diquark approximation.}}
Rainbow-ladder also offers a highly welcome systematic method to connect the quark-diquark model with QCD.
As discussed in Sec.~\ref{sec:quarkdiquarkapproximation}, the rainbow-ladder interaction produces diquark
poles in the quark-quark scattering matrix, which in turn allows one to solve Bethe-Salpeter equations for the diquarks
and determine their masses and amplitudes in analogy with mesons~\cite{Maris:2002yu,Eichmann:2016jqx}.
The necessary steps are outlined in Fig.~\ref{fig:quark-diquark} and have been implemented in Refs.~\cite{Eichmann:2007nn,Eichmann:2008ef,Nicmorus:2008vb};
a detailed collection of formulae can be found in Ref.~\cite{Eichmann:2009zx}
and in Sec.~\ref{sec:ff-methods} we will also briefly discuss results for form factors.
It turns out that the resulting nucleon and $\Delta$ masses only differ by $5-10\%$ from the results obtained with the three-quark equation (cf.~Fig.~5 in~\cite{Eichmann:2011aa}).
Although this seems to corroborate the quark-diquark picture, we should note that the
consistent treatment of the ingredients plays a crucial role in this,
including the calculation of the diquark `propagators' which to some extent already absorb effects beyond the diquark approximation.

An interesting perspective on the baryon spectrum can be drawn from the inherent connection between meson and diquark properties~\cite{Roberts:2011cf}.
Their Bethe-Salpeter equations~\eqref{bse-meson} and~\eqref{bse-diquark} in rainbow-ladder are in fact so similar that after working out the Dirac, colour and flavour traces
they only differ by a factor~$\nicefrac{1}{2}$ -- diquarks are `less bound' than mesons.
In that way one has a correspondence between pseudoscalar, vector, scalar, axialvector, \dots mesons on the one hand
and scalar, axialvector, pseudoscalar, vector, \dots diquarks on the other hand.
Whereas pseudoscalar and vector mesons are well described in rainbow-ladder,
scalar and axialvector mesons come out too light~\cite{Maris:2002yu,Krassnigg:2009zh} --
which can be remedied with more sophisticated truncations~\cite{Chang:2009zb,Williams:2015cvx}.
Thus the diquarks will inherit this behaviour: the scalar and axialvector diquarks which are mainly responsible for positive-parity baryons
are reliable, whereas the pseudoscalar and vector diquarks will produce negative-parity baryons that are also too light and acquire repulsive shifts beyond rainbow-ladder.
These features are clearly visible in the contact interaction model of Refs.~\cite{Roberts:2011cf,Chen:2012qr}
and also persist in rainbow-ladder calculations as we will see in Sec.~\ref{spec:results}.
They explain why rainbow-ladder type approaches underestimate the negative-parity baryon spectrum
and therefore also lead to the wrong mass ordering between the Roper and the  parity partner $N^\ast(1535)$ of the nucleon.

On a more general note, we remark that the clustering of subsystems is a quite natural feature of Bethe-Salpeter equations.
A similar mechanism has been identified with four-body calculations for tetraquarks~\cite{Eichmann:2015cra}:
the rainbow-ladder four-body amplitude is dominated by the lowest-lying two-body poles (the pseudoscalar mesons),
whereas diquarks turn out to be much less important simply because of their higher mass scales.
This leads to a `meson-molecule' interpretation for the light scalar mesons,
which can be made explicit in a meson-meson/diquark-antidiquark approximation~\cite{Heupel:2012ua} --
the analogue of the quark-diquark model for baryons.
By contrast, the lightest subclusters in a baryon are the diquarks themselves; and although they are gauge-dependent and not observable,
the mere appearance of the quark-quark T-matrix in the Faddeev equation of Fig.~\ref{fig:quark-diquark} suggests
that diquark clustering, hidden in the analytic structure of the T-matrix,
will always play a certain role for the two-body interactions in baryons.

\vspace{1em}\par\noindent\underline{\emph{Genuine three-body approach.}}
Although the underlying diquark properties can provide insight into the structure of baryons,
there are also several good reasons to forgo the quark-diquark approximation.
Baryons with negative parity require more than scalar and axialvector diquark channels~\cite{Eichmann:2016jqx},
and the sheer effort to put the quark-diquark model on a solid basis in QCD
can become quite large as one can infer from Figs.~\ref{fig:quark-diquark} and~\ref{fig:ff-quark-diquark}.
In this respect, while the original three-body equation pictured in Fig.~\ref{fig:faddeev} is technically more demanding
it is also much cleaner on a conceptual level and therefore preferable from a microscopic point of view.
Here diquarks no longer appear explicitly and the equation instead depends on the original two- and three-body kernels made of quarks and gluons.
Clearly, the advantage is that
effects beyond rainbow-ladder but also irreducible three-body interactions can be implemented in a clean and systematic way.

The first solution of the three-body system has been achieved in Ref.~\cite{Eichmann:2009qa},
with an extension in~\cite{Eichmann:2011vu} where a remaining angular approximation in the Bethe-Salpeter amplitude was lifted to
provide the full rainbow-ladder result without further approximations.
Since then the methods have been applied to obtain the baryon octet and decuplet spectrum~\cite{SanchisAlepuz:2011jn,Sanchis-Alepuz:2014sca},
with further applications to form factors and Compton scattering (to be discussed in Sec.~\ref{form} and~\ref{compton}).
In such calculations only the flavour content of the valence quarks
is specified and no prejudice regarding how the baryon's internal structure may be
expressed in terms of diquarks is made. The only input is the effective coupling, for example~\eqref{couplingMT};
the dressed quark propagator is then determined from its Dyson-Schwinger equation and
the three-body Bethe-Salpeter equation solved explicitly, including the full set of covariant
tensor structures that constitute the amplitudes. Thus the framework is highly
self-consistent and can help judge where the quark-diquark approximation described
above is valid. Last but not least, this is also the setup that provides the majority of the results that we will discuss in the following sections.

\begin{figure*}[!t]
        \begin{center}
        \includegraphics[width=0.9\textwidth]{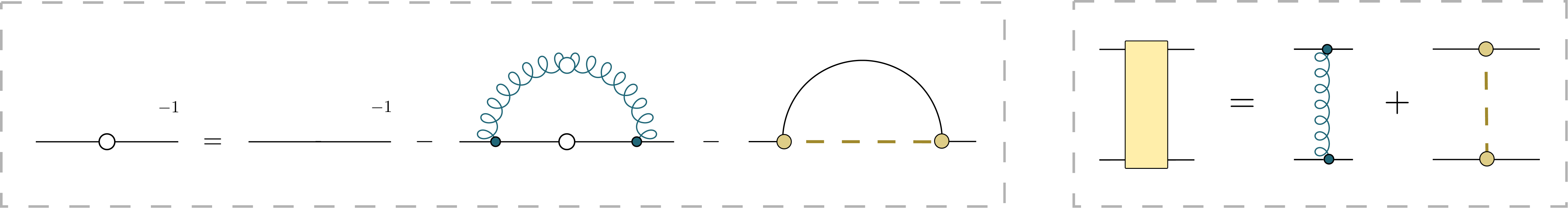}
        \caption{\textit{Left:} Pion self-energy correction to the quark propagator, stemming from hadronic resonance
                        effects in the dressed quark-gluon vertex.
                 \textit{Right:} Corresponding pion exchange in the Bethe-Salpeter kernel.}\label{fig:DSEwpions2}
        \end{center}
\end{figure*}

\paragraph{Pion cloud.}
A recent extension of rainbow-ladder type models has been discussed in
Refs.~\cite{Fischer:2007ze,Fischer:2008wy,Sanchis-Alepuz:2014wea}. Since the early
days of QCD, the general idea of a pion cloud surrounding the quark core of QCD bound
states has existed and been implemented in many phenomenological models, starting with
the cloudy bag model~\cite{Miller:1979kg,Theberge:1980ye,Thomas:1981vc}.
Within the framework of functional methods, the emission and absorption of pions from
quarks is encoded in the quark four-point function. It appears in the
Dyson-Schwinger equation for the quark-gluon vertex and therefore implicitly has an
impact on the quark propagator via vertex corrections. This has been worked out
in~\cite{Fischer:2007ze} together with an approximation scheme that enables these
contributions to be written in terms of a rainbow-ladder like pion-dressing of the
quark propagator, see Fig.~\ref{fig:DSEwpions2}. Constructing a Bethe-Salpeter kernel
in accord with the axial Ward-Takahashi identity leads to a pion exchange diagram analogous to the
rainbow-ladder gluon exchange. Therefore, although in principle the inclusion of
pion cloud effects goes beyond rainbow-ladder, in such an approximation they resemble
rainbow-ladder like corrections. Pion cloud effects on the masses
of mesons and baryons have been explored in this framework~\cite{Fischer:2008wy,Sanchis-Alepuz:2014wea}
and will be discussed below.
Note, however, that a complete inclusion of pion cloud effects has to take into account
not only contributions to the quark propagator and quark-quark two-body interaction but also resonant
contributions to the three-body interaction, e.g.\ in the form of $\pi N$ intermediate states. This is
work anticipated for the future.

\paragraph{Beyond Rainbow-Ladder towards full QCD.}\label{sec:beyondrainbowladder}
Thus far we have discussed the rainbow-ladder truncation and the inclusion of pion cloud effects,
both of which have been simultaneously applied to meson and baryons by employing the notion of
a universal two-body interaction kernel.
In these two cases, through specification of the quark-gluon vertex together with other simplying
assumptions, the form of the quark self-energy is fixed. Then, by inspection a Bethe-Salpeter kernel
can be constructed in accordance with the axial Ward-Takahashi identity, giving rise to the
gluon-ladder and pion exchange discussed above whilst preserving the pseudo-Goldstone character
of the pion. However, beyond the simplest self-energy corrections it quickly becomes irksome to
construct appropriate Bethe-Salpeter kernels.
Since our focus has been on Green functions and their Dyson-Schwinger equations derived
from the 1PI effective action~\eqref{eqn:legendretransformationonepi}, one might hope
that the Bethe-Salpeter kernel could similarly be obtained by deriving and solving the
four-quark DSE\@. Unfortunately, the Bethe-Salpeter kernel is two-particle irreducible (2PI)
with respect to the constituents, and is thus a very different object. Moreover, it is not clear
that blindly truncating at the level of vertices would respect e.g.\ current conservation, see~\eqref{qqbar-currents}. We therefore
briefly describe the transition from 1PI to 2PI (and subsequently to $n$PI) effective actions in the following.

Recall that the 1PI effective action $\Gamma_{1\mathrm{PI}}[\tilde{\phi}]$ is a functional of the one-point function $\tilde{\phi}$.
We can introduce the 2PI effective action $\Gamma_{2\mathrm{PI}}[\tilde{\phi},D]$, a functional of both $\tilde{\phi}$ and the
connected two-point function $D$ through an additional Legendre transformation with respect to some
bilocal source. This is the CJT effective action~\cite{Cornwall:1974vz}, see also Ref.~\cite{Berges:2000ew} for an overview, and can be further generalised
to $n$PI effective actions $\Gamma_{n\mathrm{PI}}[\tilde{\phi},D,U,\ldots]$ by introducing the appropriate source
terms for higher vertices $U$, $V$, and so on. In the absence of these additionally introduced sources the actions are equivalent
\begin{align}
    \Gamma_{1\mathrm{PI}}[\tilde{\phi}] = \Gamma_{2\mathrm{PI}}[\tilde{\phi},D=\overline{D}] = \cdots = \Gamma_{n\mathrm{PI}}[\tilde{\phi},D=\overline{D},U=\overline{U},\ldots]\;,
\end{align}
and lead to stationary conditions
\begin{align}
\frac{\delta\Gamma_{n\mathrm{PI}}}{\delta \tilde{\phi}}\bigg|_{\tilde{\phi}=0} =
\frac{\delta\Gamma_{n\mathrm{PI}}}{\delta D}\bigg|_{D=\overline{D}} =
\frac{\delta\Gamma_{n\mathrm{PI}}}{\delta U}\bigg|_{U=\overline{U}} = \cdots = 0\;,
\end{align}
from which the $n$-point functions are determined in the absence of sources. Note that this implies e.g.\ that we treat $D$, $U$ etc.\ as independent variables \emph{until} the source terms are set to zero.

\begin{figure*}[!t]
        \begin{center}
        \includegraphics[width=0.99\textwidth]{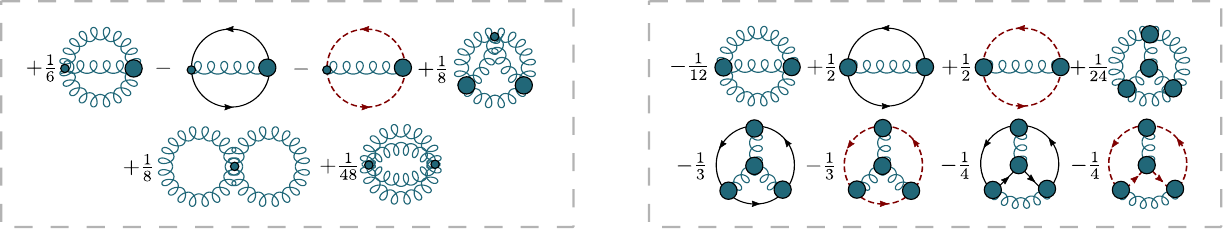}
        \caption{Non-interacting (\textit{left}) and interacting  part (\textit{right}) of the 3PI effective action to three-loop order.
        All propagators (dashed for ghost, straight for quark and springs for gluons) are dressed, with small circles denoting bare vertices and large circles dressed vertices.}\label{fig:effectiveaction3PI}
        \end{center}
\end{figure*}

In a compact notation, we write the $n$PI effective action as~\cite{Carrington:2010qq,York:2012ib,Carrington:2013koa}
\begin{align}
\Gamma_{n\mathrm{PI}}\left[\tilde{\phi},D,U,\ldots\right] = S_{\mathrm{cl}}[\tilde{\phi}]
                            +\frac{i}{2}\mathrm{Tr}\,\mathrm{Ln}D^{-1}
                            +\frac{i}{2}\mathrm{Tr}\left[D^{-1}_{(0)}D\right]
                            -i\Phi_{n\mathrm{PI}}[\tilde{\phi},D,U,\ldots]
                            +\mathrm{const}\;,
\end{align}
where $\Phi_{n\mathrm{PI}}[\tilde{\phi},D,U,\ldots]$ is usually separated into
$\Phi^0_{n\mathrm{PI}}[\tilde{\phi},D,\ldots]$ and $\Phi^{\mathrm{int}}_{n\mathrm{PI}}[\tilde{\phi},D,\ldots]$,
which are referred to as the non-interacting and interacting parts respectively.
Note, however, that these cannot be written exactly but are instead represented by e.g.\ a (nonperturbative) skeleton expansion in powers of $\hbar$. This introduces a very systematic and natural way for prescribing truncations of the action, see Fig.~\ref{fig:effectiveaction3PI} for
the non-interacting and interacting parts of the $3$PI effective action to three-loop order in QCD\@.

\begin{figure*}[!b]
        \begin{center}
        \includegraphics[width=0.85\textwidth]{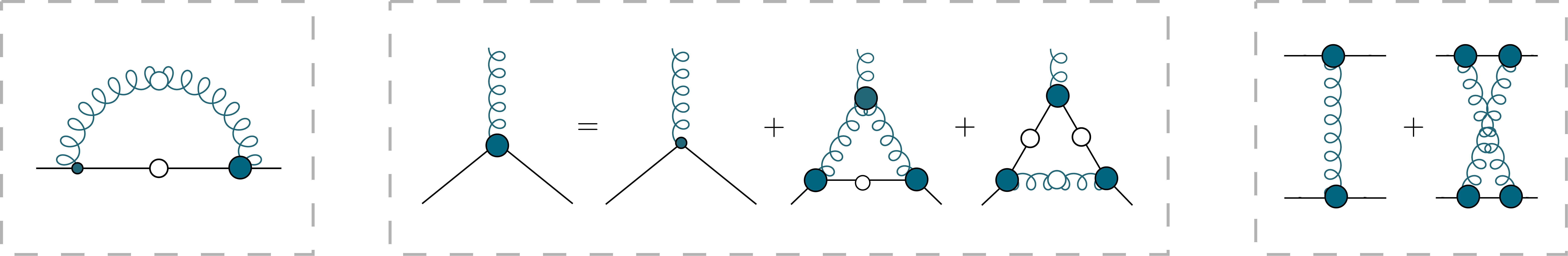}
        \caption{Quark self-energy (\textit{left}), quark-gluon vertex  (\textit{middle}) and Bethe-Salpeter kernel (\textit{right}) derived from the 3PI effective action to three-loop order, see Fig.~\ref{fig:effectiveaction3PI}.}\label{fig:kernel3PIexample}
        \end{center}
\end{figure*}

\begin{figure*}[!t]
        \begin{center}
        \includegraphics[width=0.99\textwidth]{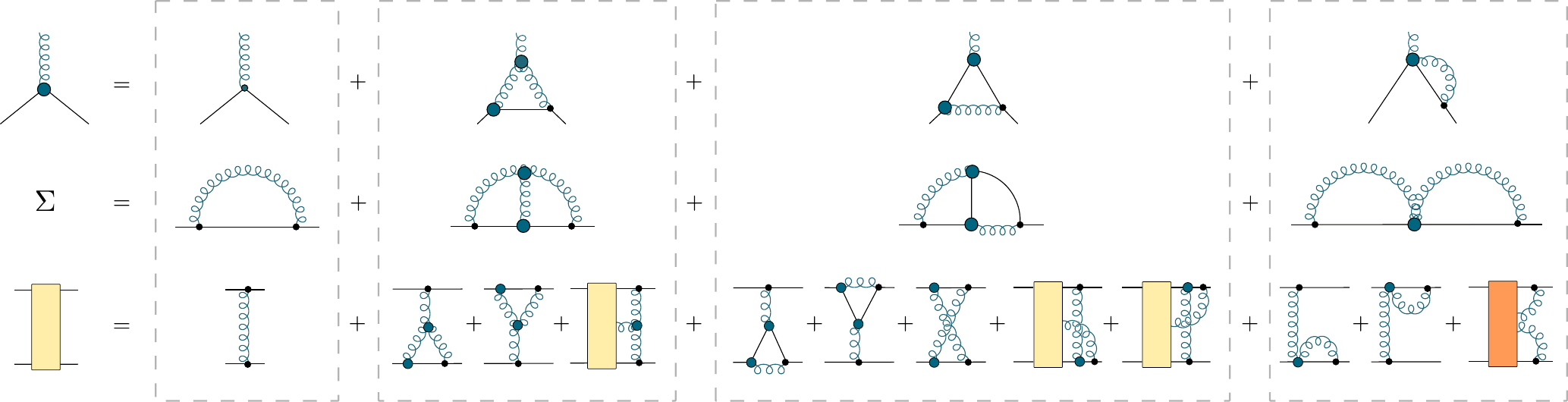}
        \caption{Contributions to the DSE for the quark-gluon vertex (\textit{top}); their corresponding self-energy correction to the quark DSE (\textit{middle});
        resultant term in the Bethe-Salpeter kernel following 2PI cutting~\cite{Munczek:1994zz} (\textit{bottom}) . The five-
        and six-point kernels arise from the implicit cutting of vertices and are determined by solving auxiliary
        Bethe-Salpeter equations.}\label{fig:DSEwopions2}
        \end{center}
\end{figure*}

\vspace{1em}\par\noindent\underline{\emph{nPI effective action.}}
Now, let us return to our original problem of the Bethe-Salpeter kernel $K$, and discuss it together with the quark-self energy
$\Sigma$, which is the loop integral component of~\eqref{quark-propagator}. In the language of the $n$PI effective
action, they are obtained by
\begin{align}
\Sigma[\overline{D},\overline{U},\ldots] &= \frac{\partial \Phi_{n\mathrm{PI}}^\mathrm{int}[D,U,\ldots]}{\partial D}\Bigg|_{D=\overline{D},\;U=\overline{U},\ldots}\;,\qquad
    K[\overline{D},\overline{U},\ldots] = \frac{\partial^2\Phi_{n\mathrm{PI}}^\mathrm{int}[D,U,\ldots]}{\partial D\partial D}\Bigg|_{D=\overline{D},\;U=\overline{U},\ldots}\;,
\end{align}
where after differentiation the stationary condition replaces $D$ by their solution $\overline{D}$, $U$ by $\overline{U}$ and so on. Similarly, DSEs for vertices would arise from a functional derivative of $\Phi_{n\mathrm{PI}}^\mathrm{int}[D,U,\ldots]$ with respect to $U$, followed by setting sources to zero.
Crucially, because the functional derivatives are strictly applied before imposition of the stationary condition, viz.\ $D,\;U,\;\ldots$ are independent variables, the DSEs and Bethe-Salpeter kernel do not contain higher $n$-point functions than those already present in the action. Thus they form a closed system of equations, a subset of which are shown in Fig.~\ref{fig:kernel3PIexample}; for an application of these methods to meson bound-states beyond
rainbow-ladder, see Ref.~\cite{Williams:2015cvx}. Note that to simplify the discussion somewhat, we restrict to the case of the two-body kernel featured in the flavour non-singlet channel.

\newpage

\vspace{1em}\par\noindent\underline{\emph{2PI effective action.}}
However, an interesting observation can be made in the case of the $2$PI effective action where the quark self-energy and kernel can be defined in relation to each other
\begin{align}
    \Sigma[\overline{D}] &= \frac{\partial \Phi_{2\mathrm{PI}}^\mathrm{int}[D]}{\partial D}\Bigg|_{D=\overline{D}}\;,\qquad
  K[\overline{D}] = \frac{\partial^2\Phi_{2\mathrm{PI}}^\mathrm{int}[D]}{\partial D\partial D}\Bigg|_{D=\overline{D}}
  = \frac{\partial}{\partial \overline{D}}\left[\frac{\partial\Phi_{2\mathrm{PI}}^\mathrm{int}[D]}{\partial D}\Bigg|_{D=\overline{D}}\right]
  = \frac{\partial\Sigma[\overline{D}]}{\partial \overline{D}}\;.
\end{align}
That is, and for this special case alone, we can impose the stationary condition before taking derivatives. Then, the Bethe-Salpeter kernel is equivalently defined
as the functional derivative of the quark self-energy with respect to the quark propagator. The benefit here is that we do not require an underlying effective action to be defined, although it is implied, and that regardless of the form it takes it is $2$PI in structure~\cite{Munczek:1994zz}.

The net result is that our quark self-energy is implied to be a loop expansion of dressed propagators but with bare perturbative vertices. Hence, functional
derivatives act explicitly upon this object and ``cut'' quark lines, thus promoting the two-point function to a four-point function that is identified with the
Bethe-Salpeter kernel. Now, for convenience we may \emph{define} a quark-gluon vertex as an auxiliary function that represents
the resummation of these higher order corrections~\cite{Bender:2002as,Bhagwat:2004hn,Watson:2004jq,Watson:2004kd,Matevosyan:2006bk,Fu:2015tdu,Gomez-Rocha:2015qga,Gomez-Rocha:2016cji}.
Then, being a functional of the quark propagator $\Gamma^\mu = \Gamma^\mu[S]$, it must be cut \emph{implicitly}, see~\cite{Heupel:2014ina,Binosi:2016rxz} for a discussion,
where an integral representation of the vertex is necessary so as to avoid an ambiguity in defining the momenta.  The downside of this is that
functionally differentiating a quark-gluon vertex leads to a five-point function that must satisfy an auxiliary Bethe-Salpeter equation, derivable from the
auxiliary equation for the vertex. If that vertex equation similarly depends upon higher $n$-point functions, as is the frequently the case in DSEs, then a second infinite hierarchy of coupled auxiliary Bethe-Salpeter equations is induced.

\begin{figure*}[!b]
        \begin{center}
        \includegraphics[width=0.95\textwidth]{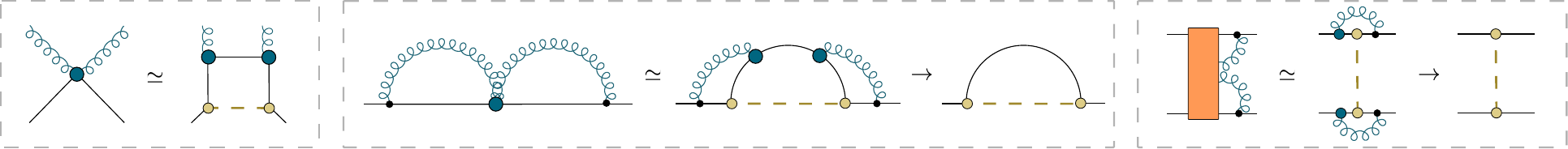}
        \caption{Resonant expansion of the glue-glue-quark-quark vertex (\textit{left}) ; its contribution to the quark-self energy after simplification (\textit{middle});
        resulting pion exchange in the Bethe-Salpeter kernel (\textit{right}).}\label{fig:vertexpionexchange}
        \end{center}
\end{figure*}

An example of this can be seen in the top row of Fig.~\ref{fig:DSEwopions2} where we give the exact DSE for the quark-gluon vertex, notably dependent upon the quark propagator, quark-gluon vertex and glue-glue-quark-quark vertex; the latter obviously comes equipped with its own DSE\@. In the second row,
we show the quark self-energy that this prescribes written in a convenient and symmetric form. Assuming
that the functional dependence is $2$PI in nature, the result of both explicit and implicit cuts
is given in the bottom row of Fig.~\ref{fig:DSEwopions2}. Note the appearance of the five- and six-point functions
that satisfy auxiliary Bethe-Salpeter equations (not shown).

Obviously, the structure is different to that of the kernel derived from the three-loop 3PI effective action, but this is a consequence of the action here being implicitly 2PI and, given the coupled resummation that the DSE for the quark-gluon vertex induces, is infinite rather than fixed order. As it stands, the system of equations is intractable until truncations are imposed. As an example, and in analogy with the pion cloud corrections
discussed in the previous section, we sketch in Fig.~\ref{fig:vertexpionexchange} how a resonant expansion of the glue-glue-quark-quark vertex gives rise to -- amongst several other diagrams not shown -- pion self-energy and pion exchange contributions in the quark DSE and Bethe-Salpeter kernel, respectively.

            \begin{figure*}[t]
                    \begin{center}
                    \includegraphics[width=0.9\textwidth]{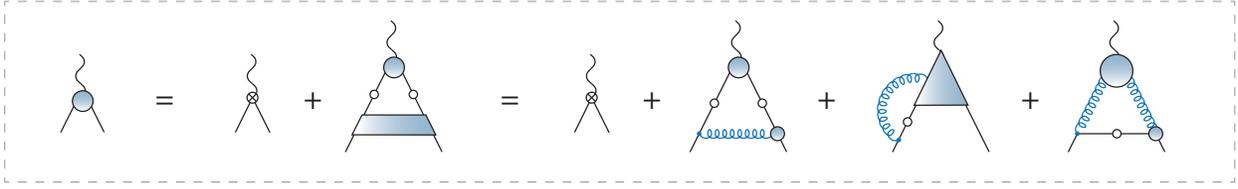}
                    \caption{Bethe-Salpeter equation for the dressed $q\bar{q}$ vertex. The first equality shows the usual equation depending on the kernel~$\mathbf{K}$,
                             whereas the second version is obtained by coupling the current to all ingredients of the quark DSE in Fig.~\ref{fig:DSEs}.
                             The last diagram only contributes to flavour-singlet quantum numbers.}\label{fig:bse-lambda}
                    \end{center}
            \end{figure*}

\vspace{1em}\par\noindent\underline{\emph{Other methods.}}
The kernel constructions discussed thus far rely upon functional derivatives of either
the effective action, or the quark self-energy in combination with an integral representation
of the quark-gluon vertex. There is, however, a different scheme~\cite{Chang:2009zb,Chang:2010hb} by which a Bethe-Salpeter kernel
can be constructed that is also in accord with the axial WTI\@. To explain the basic ideas we need to anticipate some of the discussion in Secs.~\ref{sec:ff-currents} and~\ref{sec:ff-vertices}.
There, Fig.~\ref{fig:vertex-def} shows how a Bethe-Salpeter equation for the quark-antiquark vertex follows from the observation that
the quantity $\langle 0 | \mathsf{T}\,\psi_\alpha\,\conjg{\psi}_\beta\,j^{[\mu]}\, | 0 \rangle$, defined in~\eqref{S-Smu},
is the contracted $q\bar{q}$ four-point function which satisfies a Dyson equation.
The vertex contains all possible meson poles with the quantum numbers prescribed by the index $[\mu]$, so one can extract their masses from it.
At the same time, however, this quantity is also the \textit{gauged} quark propagator, which allows one to derive the same equation by coupling
the external current to all ingredients of the quark Dyson-Schwinger equation, as illustrated in Fig.~\ref{fig:bse-lambda}:
       \begin{equation}\label{bse-with-lambda}
          \Gamma^{[\mu]} = \Gamma^{[\mu]}_0 + \mathbf{K\,G_0}\,\Gamma^{[\mu]} = \Gamma^{[\mu]}_0 + \mathbf{K'\,G_0}\,\Gamma^{[\mu]} + \Lambda^{[\mu]}\,.
       \end{equation}
Here, $\mathbf{K}$ is the usual Bethe-Salpeter kernel whereas $\mathbf{K}'$ denotes the combination of a dressed gluon propagator together with a bare and a dressed vertex.
For flavor-nonsinglet quantum numbers we can drop the last diagram in Fig.~\ref{fig:bse-lambda}, so that
the remaining quantity $\Lambda^{[\mu]}$ contains the four-point function $\Gamma_{\alpha\beta}^{\nu[\mu]}$,
the 1PI analogue of the quantity $\langle 0 | \mathsf{T}\,\psi_\alpha\,\conjg{\psi}_\beta\,A^\nu j^{[\mu]}\, | 0 \rangle$
which includes in addition the gluon field.

The key idea is then to make the equation self-consistent by reexpressing $\Lambda^{[\mu]}$ through the vertex $\Gamma^{[\mu]}$ itself.
To this end one exploits the PCAC relation~\eqref{vcc-pcac-0} to derive the axial WTI for $\Gamma^{\nu[\mu]}$ or, more precisely,
for the combination of the axialvector and pseudoscalar vertices analogous to Eq.~\eqref{axwti}. In momentum space it reads
\begin{align}
Q^\mu \,\Gamma^{\nu[\mu,5]} +2m\,\Gamma^{\nu[5]} &= \Gamma^\nu_{\mathrm{qg},+}\,i\gamma_5 + i\gamma_5\,\Gamma^\nu_{\mathrm{qg},-}\,,
\end{align}
where the quark-gluon vertex appears on the right-hand side and we dropped the momentum arguments for brevity.
Contracting also the gluon leg with Lorentz index $\nu$ with the gluon momentum then produces a sum of Slavnov-Taylor identites for the quark-gluon vertex.
With simplifying assumptions, among them that the quark-gluon vertex satisfies an Abelian-like Ward identity analogous to~\eqref{vwti}, one can
relate $\Gamma^{\nu[\mu]}$ and therefore also $\Lambda^{[\mu]}$ with the original $q\bar{q}$ vertex $\Gamma^{[\mu]}$ that is subject to the
equation~\eqref{bse-with-lambda} and thereby make it closed and solvable. As a result, the equation preserves chiral symmetry and relies on
the quark-gluon vertex and gluon propagator as model inputs.

While from the $n$PI point of view this approach is perhaps less systematic,
it exhibits greater flexibility since one can employ (Abelian-like) quark-gluon vertices with arbitrary transverse components and test their impact on the meson spectrum.
Notably, the importance of terms that are correlated with dynamical
chiral symmetry breaking and the anomalous chromomagnetic moment has been argued~\cite{Chang:2010hb,Bashir:2011dp,Chang:2011ei}.
Their coefficients are provided by ansatz, whose judicious choice enables the model
to improve upon rainbow-ladder in several respects; e.g.\  it reproduces the mass splitting between $a_1$, $b_1$ and the $\rho$ meson and generates
repulsive shifts for the scalar mesons.
So far, however, the framework has been limited to studies in the meson sector.

\subsection{Results for the baryon spectrum}\label{spec:results}

In the following we discuss results for the ground and excited
states of baryons in the non-strange and strange sectors of QCD. Due to
restrictions in space, our focus is on light quarks and we skip the
interesting topic of baryons with one or more charmed valence quarks.
Methodologically, we concentrate on results from functional methods and lattice
QCD (see e.g.~\cite{Lang:2015ljt} for a short review), with quark models acting as
a standard background for the distinction between ordinary and non-standard or
`exotic' states in the sense already discussed in the introduction. By comparing
these approaches, our general aim is to elucidate the underlying physics behind
the theoretical and experimental numbers.

Before we embark on our discussion we would first like to make clear what we
can expect from functional methods. In the previous section we discussed the
state of the art with respect to truncations. As of today, rainbow-ladder is the
most advanced approximation used in the baryon sector of QCD, in that it is
applied to \emph{both} spectroscopy and the study of form factors. However, a
programme for a systematic improvement upon it
beyond rainbow-ladder is currently  under way in both meson and baryon sectors.
A very significant step forward on the technical side was taken in
\cite{Sanchis-Alepuz:2015qra}, where for the first time a fully covariant
three-body calculation of the nucleon and $\Delta$ masses was performed using a
gluon exchange kernel with a full
quark-gluon vertex (i.e.\ with $12$ tensor structures). This work is the technical baseline for the implementation
of the systematic kernels discussed above.

\paragraph{Mesons.}

\begin{figure*}[t]
        \begin{center}
        \includegraphics[width=0.95\textwidth]{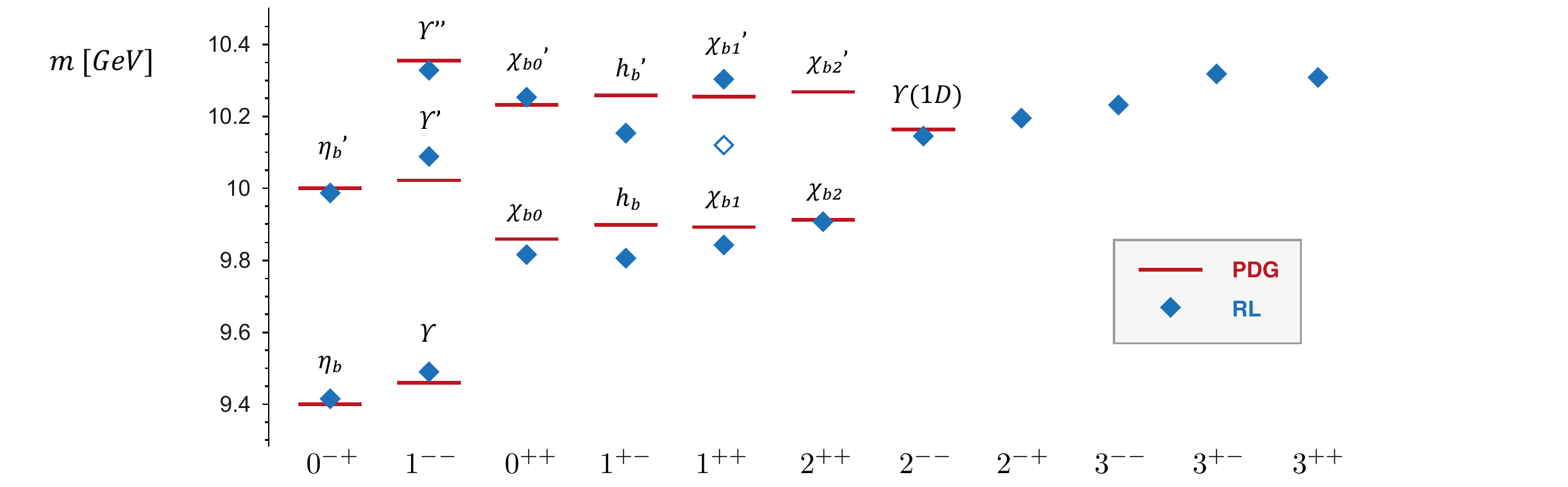}
        \caption{Bottomonium spectrum obtained from the $q\bar{q}$ Bethe-Salpeter equation with a rainbow-ladder truncation~\cite{Fischer:2014cfa}.}. \label{fig:bottom}
        \end{center}
\end{figure*}

We have already discussed the merits
and drawbacks of rainbow ladder from a systematic point of view in the previous section,
notably its reliance upon a phenomenologically constructed effective gluon exchange
that is restricted to vector-vector interactions. In order to appreciate the applicability
of such an approximation it is useful to take a glance at the complementary
meson spectrum.

For illustration, let us start with the bottomonium spectrum displayed in Fig.~\ref{fig:bottom}.
Bottomonia are sufficiently heavy so that a non-relativistic description in terms of quark potential models works very well.
There the potential is constructed as the sum of a linearly rising confining part
plus the non-relativistic limit of one-gluon exchange (the Breit-Fermi interaction) in analogy to positronium.
The latter consists of a Coulomb term and spin-dependent parts: the spin-spin contact interaction which produces
the hyperfine splitting between pseudoscalar and vector mesons, the spin-spin tensor force, and spin-orbit interactions.
Of those only the contact term contributes to $0^{-+}$, $1^{--}$ and
only the tensor force to $1^{+-}$, $2^{-+}$ states, whereas both tensor and spin-orbit interactions are important for the remaining channels.
On the other hand, the radial excitations within a given channel predominantly probe the linearly rising confining part of the
interaction.

In Fig.~\ref{fig:bottom} we compare the experimental spectrum with the rainbow-ladder Bethe-Salpeter calculation of Ref.~\cite{Fischer:2014cfa},
with similar results obtained in~\cite{Blank:2011ha,Hilger:2014nma}. We see an overall reasonable agreement in most channels,
especially for the pseudoscalar and vector mesons including their ground states and radial excitations.
In the quark model these are the `$s$ wave' states with vanishing orbital angular momentum.
Of the remaining states, the scalars and axialvectors show the largest discrepancies,
in particular the $1^{+-}$ channel which in the quark model is governed by the tensor force.
(The additional state in the $1^{++}$ channel turns out to be a truncation artifact that disappears beyond rainbow-ladder.)

These trends continue when lowering the current-quark mass;
the discrepancies are enhanced in the charmonium spectrum but they become especially striking in the light meson sector~\cite{Alkofer:2002bp,Krassnigg:2009zh,Fischer:2014xha}.
The light pseudoscalar (non-singlet) mesons are governed by spontaneous chiral symmetry
breaking and therefore automatically reproduced as long as the truncation respects
the axial-vector Ward-Takahashi identity~\cite{Maris:1997tm}.
The ground-state mass in the vector channel
is also in good agreement with experiment~\cite{Maris:1999nt}.
However, the deficiencies of rainbow-ladder are obvious for scalar and axialvector states  (the `$p$ waves' in the quark model).
In the scalar channel there is some evidence that the members of the lowest-lying nonet may be better described as
tetraquarks than quark-antiquark states, see e.g.~\cite{Jaffe:1976ig,Amsler:2004ps,Giacosa:2006tf,Ebert:2008id,Parganlija:2012fy,Eichmann:2015cra,Pelaez:2015qba}
and references therein. However,
also the axialvectors are underestimated by $20$--$40 \%$.
These shortcomings are well documented and have been frequently discussed in the literature.
Whereas in quark potential models it is the appropriate mixture of
spin-spin, tensor and spin-orbit forces that generates the splittings between different
quantum numbers, in the fully relativistic
framework these are encoded in the relative strengths of the different tensor
structures in the nonperturbative quark-gluon vertex and the quark-antiquark kernel~\cite{Qin:2011xq,Fischer:2014xha}.
Since rainbow-ladder retains only one of these structures, the $\gamma^\mu$ piece
of~\eqref{baryontheory:eqn:rainbowladdervertex}, it is certainly deficient in
this respect. Similar shortcomings are also observed in the heavy-light meson sector~\cite{Nguyen:2010yh,Gomez-Rocha:2014vsa,Rojas:2014aka}.

On the other hand, it is interesting to note that the
situation is still acceptable for the lowest-lying tensor mesons with $2^{++}$ and $3^{--}$~\cite{Krassnigg:2010mh,Fischer:2014cfa}; i.e.,
the states lying on the Regge trajectory $J^{PC}=\{1^{--},2^{++},3^{--}\}$ are well reproduced.
In the potential language, what distinguishes
these channels from the others is that the tensor force is either vanishing or particularly small.
Of course, the notion of a potential as such is highly questionable in
the light quark sector where matters are additionally complicated by the
consequences of spontaneous chiral symmetry breaking and the
sizeable widths of resonances.
Nevertheless one may conclude that the rainbow-ladder interaction
roughly reproduces the size of the spin-spin contact part and
the spin-orbit force, but materially overestimates the binding in the tensor
part of the spin-spin interaction.

The takeaway message of the discussion in view of the light baryon spectrum is the following:
rainbow-ladder works well for the pseudoscalar and vector ground states, together with some of their radial excitations.
While this does not speak for its broad applicability in the meson sector,
the same features also hold for their structure properties such as form factors, decays and more~\cite{Maris:2003vk,Maris:2005tt};
thus for processes involving pions, kaons, $\rho$ mesons etc. rainbow-ladder can be considered reliable.
It turns out that the situation is similar for baryons:
in the quark-model language also the ground-state octet and decuplet baryons are $s$~waves
and therefore we may expect rainbow-ladder to provide a reliable framework for the nucleon and $\Delta$ baryons.
These expectations are also corroborated in a quark-diquark interpretation, where the deficiencies for scalar and axialvector
mesons have direct consequences for negative-parity diquarks and therefore negative-parity baryons;
as we will see below, those states are indeed bound too strongly.

\begin{figure*}[t]
        \begin{center}
        \includegraphics[width=0.62\textwidth]{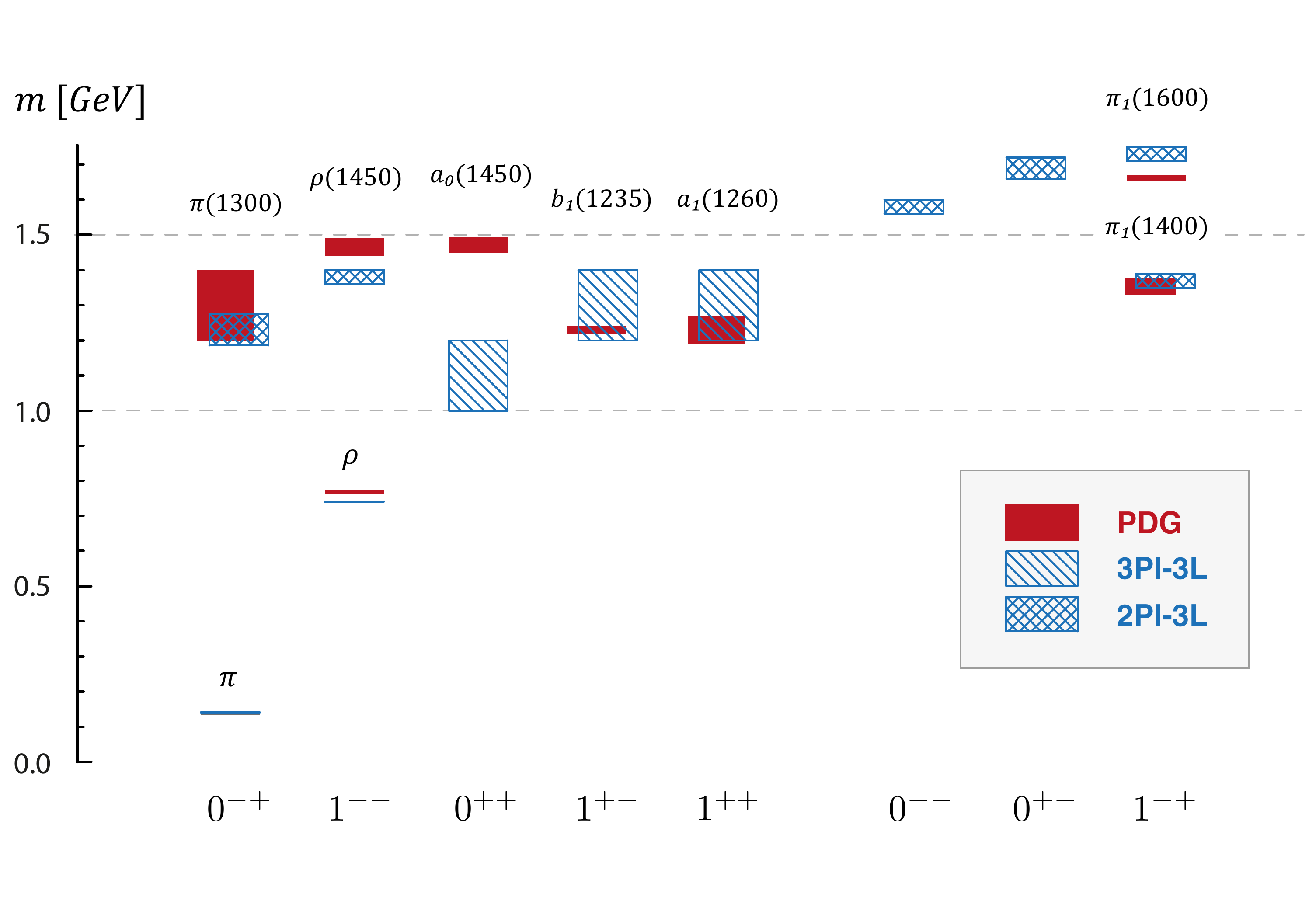}
        \includegraphics[width=0.35\textwidth]{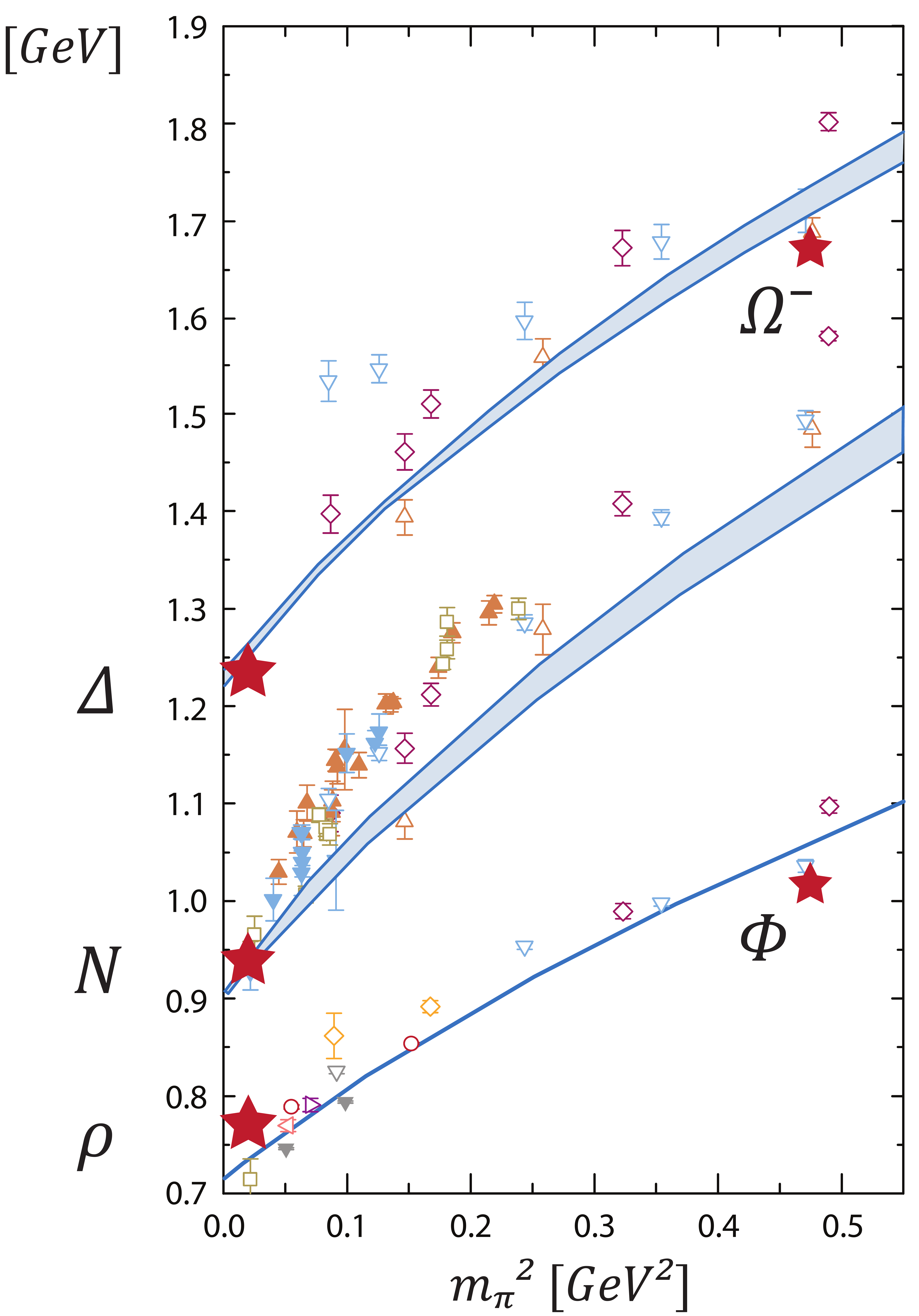}
        \caption{\textit{Left:} Isovector meson spectrum for $J=0$ and $J=1$ beyond rainbow-ladder,
                 obtained with 2PI and (where available) 3PI truncations~\cite{Williams:2015cvx}.
                 \textit{Right:} Masses of the $\rho$ meson, the nucleon and the $\Delta$ baryon as functions of the squared pion mass.
                 We compare results from the three-body Faddeev equation \cite{Eichmann:2009qa,Eichmann:2011vu,SanchisAlepuz:2011jn} with corresponding
                 lattice data. The lattice results for the $\rho$-meson mass below $m_\pi^2=0.2$~GeV$^2$ are extracted from resonance dynamics
                 and identical to those in Fig.~\ref{fig:delta-n-pi}. The remaining data are from LHPC (open~\cite{WalkerLoud:2008bp} and filled~\cite{Green:2014xba} triangles down),
                 ETMC (open~\cite{Alexandrou:2009hs} and filled~\cite{Alexandrou:2014sha} triangles up), PACS-CS (open diamonds~\cite{Aoki:2008sm}) and RQCD (open squares~\cite{Bali:2014nma}).} \label{fig:nucleon_delta}
        \end{center}
\end{figure*}

Finally, Fig.~\ref{fig:nucleon_delta} exemplifies the systematic improvements that have become possible when going beyond rainbow-ladder.
We display the isovector states with $J=0$ and $J=1$ including exotic quantum numbers, as well as the first radial excitations of the pion and $\rho$ meson.
In the 3PI scheme developed in~\cite{Williams:2015cvx} the quark-gluon vertex is explicitly
calculated from its Dyson-Schwinger equation which generates all independent
tensor structures dynamically. So far the approach has been applied to determine the ground-state masses
of light mesons, which in most channels are now in good agreement with the
experimental values and indicate a faithful representation of the relative magnitudes of the
vertex components. Similar results have been found in Ref.~\cite{Chang:2010hb} using a different framework.
Notice in particular the results for the three exotic channels obtained in the 2PI truncation: the $\pi_1(1400)$ and $\pi_1(1600)$ with $1^{-+}$,
which are nonrelativistically forbidden as $q\bar{q}$ states, are well reproduced beyond rainbow-ladder.
In any case, these truncations are very involved in terms of numerical complexity so
that -- at present -- no corresponding results for the baryon sector beyond the early $2$PI study
of~\cite{Sanchis-Alepuz:2015qra} are available.

\paragraph{Nucleon and Delta masses.}
Let us begin our discussion  of baryons with the mass evolution of the ground state nucleon
and $\Delta$ baryons shown in the right panel of Fig.~\ref{fig:nucleon_delta}. In lattice
QCD as well as in the functional framework, the current-quark masses appearing in
the Lagrangian of QCD are input parameters and may be tuned at will. As
discussed above, lattice calculations become prohibitively more expensive for
small quark masses and therefore current-mass evolutions have been frequently used in
the past to perform chiral extrapolations to the physical point~\cite{WalkerLoud:2008bp,Gattringer:2008vj,Alexandrou:2008tn,Alexandrou:2009qu,Engel:2010my}.
In recent years, simulations at or close to the physical point have become available
due to increasing computer power and improved algorithms~\cite{Durr:2008zz}.
Nevertheless, it is still interesting to study the
mass evolution of these states for systematic and physical reasons.
On the one hand, they allow one to extract the nucleon-sigma term from the Feynman-Hellman theorem:
$\sigma_{\pi X} = m_q \,\partial m_X/\partial m_q$, with baryon mass $m_X$ and quark
mass $m_q$. On the other hand, the curves may serve to study potential systematic
sources of errors when comparing different methods.

In this respect it is highly encouraging that the results from the three-body Bethe-Salpeter equation in
rainbow-ladder, shown in the right panel of Fig.~\ref{fig:nucleon_delta}, are comparable to the mass
evolution of the data from different lattice groups. The pion mass is calculated from the pion's
Bethe-Salpeter equation using the same interaction kernel and thereby mapped to the current-quark mass.
The bands in the figure show the variation with the parameter $\eta$ that appears in the
effective interaction~\eqref{couplingMT}, the same variation that is plotted in Fig.~\ref{fig:coupling},
thus demonstrating their insensitivity to this parameter as mentioned earlier.
The remaining scale $\Lambda$ is fixed to reproduce the pion decay constant,
and this is what yields the results $m_N=0.94$ GeV and $m_\Delta=1.22$ GeV at the physical point which are also quoted in Table~\ref{octet-decuplet}.
Similar results have also been obtained in the quark-diquark approximation~\cite{Eichmann:2008ef,Nicmorus:2008vb,Eichmann:2009zx}.

Observe that, as long as the kernel of the equation is flavour-independent,
one can also read off the mass of the $\Omega$ baryon from the interception
at the strange-quark mass. Although the flavour wave function of the $sss$ state differs from those of the $\Delta$ baryons,
the equation does not see this difference and the only change is the input quark mass.
There is no nucleon-like $sss$ state because the nucleon has mixed flavour symmetry,
but the analogous situation for mesons produces ideally mixed $s\bar{s}$ pseudoscalar and vector mesons.
The former does not exist in nature due to $\eta$-$\eta'$ mixing, but because the $\rho$, $\omega$ and $\phi$ are nearly ideally mixed
the latter can be identified with the $\phi$ meson.
The experimental masses of $\Omega$ and $\phi$ therefore serve as an additional check on the mass evolution.

The nucleon-sigma term can be extracted from the mass evolution via the Feynman-Hellman theorem
or from a direct calculation of the scalar form factor. Values around
$\sigma_{\pi N} \approx 37~\text{--}~44$~MeV~\cite{Bali:2011ks,Alexandrou:2014sha,Durr:2015dna,Yang:2015uis,Abdel-Rehim:2016won}
have been found with fairly small error bars. In particular, the most recent
evaluations have been performed at the physical point with realistic pion masses,
thus excluding systematic errors due to the chiral extrapolation. These values
are in tension with a phenomenological extraction based on analyticity,
unitarity and crossing symmetry, which results in
$\sigma_{\pi N} \approx 59$~MeV~\cite{Hoferichter:2015dsa} and leaves a
discrepancy on the level of $2.2~\text{--}~4.9$~$\sigma$ depending on which lattice
result is taken (see also \cite{Alarcon:2011zs} for an earlier, similar result).
The origin of this discrepancy is currently unclear and needs
to be resolved, see e.g.\ the detailed discussion in Ref.~\cite{Hoferichter:2016ocj}.
The rainbow-ladder BSE calculation typically provides values for the
nucleon and $\Delta$ sigma terms that are slightly too low~\cite{Sanchis-Alepuz:2014wea},
e.g. $\sigma_{\pi N}=30(3)$, as a result (amongst other effects) of the lack of
quark-mass dependence of the interaction.

Coming back to the mass evolution in Fig.~\ref{fig:nucleon_delta}, we wish to
discuss another important point.
In the coupled-channel approaches discussed in Sec.~\ref{sec:electroproduction}
the resonant nature of the $\Delta$, Roper etc. above the $N\pi$, $N\pi\pi$, \dots\, thresholds
is  dynamically generated by meson-baryon interactions.
This is inherent in the nature of these approaches as they start with bare hadronic states as input
and the dynamics is  produced entirely at the hadronic level.
Phenomenologically, this amounts to a picture of the baryon's `quark core' surrounded by a `meson cloud'.
From a fundamental point of view it is difficult to disentangle such effects because ultimately they must also originate from the quark-gluon level,
and their implementation would be a rather difficult task in a QCD's Green functions approach.
Rainbow-ladder is the first step towards a systematic description of baryons and as such
it does not reproduce the resonant nature of states above thresholds, at least not for quark-antiquark and three-quark systems.\footnote{By contrast,
in a four-body system rainbow-ladder automatically generates a threshold for the decay into two pions
and therefore the resulting tetraquarks are genuine resonances~\cite{Eichmann:2015cra}.}
Hence one may compare it with a quark core subsequently dressed by meson-cloud effects.
Such an interpretation will indeed arise in the discussion of form factors in Sec.~\ref{form}.

The transition from bound states to resonances depends on the quark mass.
For example, the $\Delta$ decays into $N\pi$; for small quark masses the
Goldstone nature of the pions makes them extremely light, whereas for heavy quark masses the `pion' becomes heavy
and the decay channel closes for $m_\pi >m_\Delta-m_N$, thus transforming the $\Delta$ resonance into a bound state.
Therefore, pion cloud effects will be most pronounced close to the chiral limit
where they change the internal structure of a hadron: the nucleon is surrounded by a pion cloud
and resonances may change their nature from being three-quark states to (partially) molecular states,
whereas for heavy quark masses such effects will decrease substantially.

How will such corrections be reflected in Fig.~\ref{fig:nucleon_delta}?
In the simplest manifestation meson-cloud effects are attractive and will reduce the masses of hadrons,
but moreover they also decrease the pion decay constant and the chiral condensate by a similar amount.
The pion decay constant sets the scale $\Lambda$ in a rainbow-ladder interaction such as~\eqref{couplingMT}.
Thus, instead of using the experimental value of $f_\pi$, 
the strategy adopted in Refs.~\cite{Eichmann:2008ef,Eichmann:2008ae} in view of modelling the quark core
was to inflate the scale to leave room for chiral corrections. 
Based on estimates for the $\rho$ meson this leads to consistently overestimated results for baryon masses, where
the inclusion of meson cloud effects would then reduce their masses as well as $f_\pi$.
On the other hand, such a strategy can become ambiguous when including beyond rainbow-ladder effects
because attractive and repulsive effects can compete nontrivially,
and thus the perhaps cleanest way (which we adopt here) is to use the experimental value for $f_\pi$ from the beginning.
As a consequence, those parts of the pion cloud and other beyond rainbow-ladder effects that reduce a hadron's mass and $f_\pi$ by the same amount
will drop out in plots like Fig.~\ref{fig:nucleon_delta} and only the net effects remain visible, such as for example chiral non-analyticities.

In any case, the size of meson-cloud effects is not clear \emph{a priori} but has to be determined by calculations.
Within the three-body framework, an explicit calculation of (a subset of) the diagrammatical contributions generated
by meson cloud effects has been presented in Ref.~\cite{Sanchis-Alepuz:2014wea}
following  similar studies for mesons~\cite{Fischer:2007ze,Fischer:2008wy}.
The net effect as described above was $\sim 80$~MeV attraction for the nucleon
together with a repulsive mass shift of  $\sim 70$~MeV for the $\Delta$ baryon.
This pioneering calculation is, however, only a first step towards a full inclusion
of meson could effects. It
must be further augmented by including also $K$ and $\eta$ effects as well
as resonant diagrams ($\pi N$, \ldots) in the three-quark
interaction and, ideally, the use of effective meson degrees freedom should
be lifted by including the
corresponding diagrams explicitly on the level of the quark-gluon vertex. Whether this has
a material quantitative impact on the results remains to be seen.

\paragraph{Octet and decuplet baryons.}
A selection of octet and decuplet masses (in the isospin limit) is given in Table~\ref{octet-decuplet},
including results from Bethe-Salpeter equations and
lattice QCD with $N_f=2+1+1$ quark flavours. The
lattice data are a typical example of such calculations~\cite{Alexandrou:2014sha};
many other groups have performed similar simulations with results that agree
very well with each other~\cite{Durr:2008zz,Aoki:2008sm,Bietenholz:2011qq,Alexandrou:2012xk,Edwards:2012fx,Mahbub:2013ala,Engel:2013ig},
see e.g.\ Fig.\ 26 of Ref.~\cite{Alexandrou:2014sha} for a graphical overview.
Within error bars, the lattice results agree well with the ones extracted from
experiment. Thus it is fair to say that ground state baryons are  under good control by
now in contemporary calculations.

\begin{table}[t]
\footnotesize
\begin{center}
\begin{tabular}{l @{\qquad\quad} llllllll}
\toprule

                                         & N 	& $\Lambda$ & $\Sigma$ 	& $\Xi$ 	& $\Delta$ 	& $\Sigma^*$& $\Xi^*$ & $\Omega$\\
\midrule
Quark-diquark model~\cite{Oettel:1998bk}
                                  &$0.94^{(\ast)}$& $1.13$     & $1.14$     & $1.32$   & $1.23^{(\ast)}$     & $1.38$     & $1.52$   & $1.67$ \\
Quark-diquark (RL)~\cite{Nicmorus:2008vb,Eichmann:2009zx}
                                  & $0.94$	&			&			&			& $1.28$		&			&			&		\\
Three-quark (RL)~\cite{Eichmann:2011vu,SanchisAlepuz:2011jn,Sanchis-Alepuz:2014sca}
                                  & $0.94$	& $1.07$		& $1.07$		& $1.24$		& $1.22$		& $1.33$		& $1.47$		& $1.65$	\\
Lattice~\cite{Alexandrou:2014sha}
   							 	  & $0.94^{(\ast)}$ 	& $1.12\,(2)$ 	& $1.17\,(3)$ 	& $1.32\,(2)$ 	& $1.30\,(3)$	& $1.46\,(2)$	& $1.56\,(2)$ 	& $1.67\,(2)$	\\
\midrule
Experiment (PDG)		        & $0.938$	& $1.116$		& $1.193$		& $1.318$		& $1.232$		& $1.384$			& $1.530$		& $1.672$	\\
\bottomrule
\end{tabular}
\end{center}
\caption{Octet and decuplet masses in different approaches. For the lattice simulation we quote the statistical error given in the original publication;
the nucleon mass is used to set the scale and therefore an input.
In the quark-diquark model the nucleon and $\Delta$ masses are used to fix input parameters.
All other numbers are predictions.}\label{octet-decuplet}
\end{table}

Let us now focus on the results from functional approaches. In
Table~\ref{octet-decuplet} we display results from a quark-diquark model,
where the corresponding quark-diquark BSE is solved using ans\"atze for the quark
propagators and diquark amplitudes~\cite{Oettel:1998bk}.\footnote{In the
NJL/contact model the corresponding spectrum has been determined
in~\cite{Chen:2012qr}, but with the above discussed philosophy to include
a guesstimated offset due to missing meson cloud effects.} This is contrasted
with a rainbow-ladder calculation in the quark-diquark approximation~\cite{Nicmorus:2008vb,Eichmann:2009zx}, where all
propagators and amplitudes are determined from an underlying effective
interaction on the quark-gluon level, cf.\ the discussion in Sec.~\ref{spec:approx}.
In the three-body Bethe-Salpeter calculations of Refs.~\cite{Eichmann:2009qa,Eichmann:2011vu,SanchisAlepuz:2011jn,Sanchis-Alepuz:2014sca}
the same underlying effective interaction has been used but the assumption of
diquark clustering inside baryons is absent. In the last two frameworks all
parameters (the scale $\Lambda$ and the current quark masses $m_{u/d}, m_s$) are
fixed in the meson sector (via $f_\pi$, $m_\pi$ and $m_K$) and $\eta=1.8$ is
inside the interval where results are independent of $\eta$ on the level below
5 MeV. Overall, all three approaches deliver very good results for the nucleon
and $\Delta$ and acceptable results for other states within a $5 \%$
range of experiment.

A very interesting topic is the potential flavour dependence
of the interactions inside baryons. It is well known that the
$\Lambda$-$\Sigma$ splitting is a dynamical effect which cannot be explained by
symmetry considerations alone. Thus, flavour-dependent forces need to generate
this splitting. In generic quark-models these are introduced by hand (see
e.g.\ the reviews~\cite{Glozman:1995fu,Capstick:2000qj,Klempt:2009pi} and refs.
therein), whereas in the functional approach the general idea is to explore the
origin of these forces from the underling quark-gluon interaction of QCD.
In the covariant three-body framework one clearly has to go
beyond the (flavour-blind) rainbow-ladder approximation,
which does not generate a $\Lambda$-$\Sigma$ splitting as one can see from Table~\ref{octet-decuplet}.
An explicit calculation of the non-perturbative quark-gluon
vertex shows that such a flavour dependence is indeed present~\cite{Williams:2014iea}
and first steps in such a direction were taken in~\cite{Sanchis-Alepuz:2014sca}.
In addition, pion exchange between the quarks has been taken into account,
introducing a flavour
dependence as discussed e.g.\ in~\cite{Glozman:1995fu};
it produces a $\Lambda$-$\Sigma$ splitting with the correct sign but its magnitude turns out to be much too small.
Since the pion exchange
contributions are also completely determined from the underlying interaction,
there is no adjustable parameter to remedy the situation. Nevertheless, the
setup of Ref.~\cite{Sanchis-Alepuz:2014sca} leaves room for improvements
(e.g., kaon exchange has not been included) and therefore one cannot
rule out Goldstone boson exchange as an underlying mechanism for the $\Lambda$-$\Sigma$ splitting yet.
In the quark-diquark approximation some of these forces are encoded in
the exchange of a quark between the quark and diquark constituents. Since
they may be of up/down or strange nature, some splitting is generated but the
magnitude is again much too small, cf.\ the first line of Table~\ref{octet-decuplet}.
Certainly also the available alternative, flavour-dependent three-body forces as for example in~\cite{Koll:2000ke,Loring:2001kv,Loring:2001kx,Loring:2001ky,Ronniger:2011td},
needs to be explored in the covariant framework.

\paragraph{Nucleon and $\Delta$ resonances.}

As discussed in Sec.~\ref{spec:extracting}, the extraction of baryon resonances on the lattice is a difficult endeavour that has seen great
progress in the past years but still needs to be refined in several aspects. This includes the use of a sufficiently large sets of operators,
the systematic identification of scattering states and the need to extend the simulations to the physical point. Results so far are therefore still on
the preliminary level, although very promising
\cite{Engel:2010my,Edwards:2011jj,Edwards:2012fx,Alexandrou:2012xk,Roberts:2013oea,Mahbub:2013ala,Engel:2013ig,Alexandrou:2014sha,Alexandrou:2014mka,Liu:2014jua}.
In the functional framework results for excited baryons are still scarce. In the NJL/contact model the first radial excitations of the
positive- and negative-parity $N$, $\Delta$ and $\Omega$ baryons have been studied in Ref.~\cite{Roberts:2011cf} and the properties of the Roper were
calculated within a momentum dependent diquark model~\cite{Segovia:2015hra}. Very recently, results for nucleon and $\Delta$ resonances have also become
available in the rainbow-ladder three-body framework \cite{roper:2016}. In the following we concentrate on the Roper $N(1440)$, whose nature has been debated since decades,
and the $N(1535)$.

In both the NJL/contact interaction based work~\cite{Roberts:2011cf} and the momentum-dependent quark-diquark model of Ref.~\cite{Segovia:2015hra} a similar
strategy to determine the properties of the Roper has been employed: anticipating attractive mass shifts from a meson cloud that is not explicitly contained in
the models, the model parameters (in particular the scale) were tuned to overestimate the nucleon mass, i.e. $m_N=1.14$
and $m_N=1.18$ GeV, respectively. As a consequence, also the mass of the first radial excitation comes out large:  $m_{N^*}=1.82$ and $m_{N^*}=1.73$ GeV,
respectively. The underlying idea is that these numbers represent the quark core contribution to the mass of the $N^*$ which needs to be
augmented by coupled-channel effects to generate the physical Roper. Indeed, such effects are predicted e.g. in the analysis of the EBAC
group~\cite{Suzuki:2009nj}, whereas in the J\"ulich framework no such bare state is needed \cite{Gasparyan:2003fp}. The result is interesting
because it shows that a mutually consistent description of quark-core states for the nucleon and its first radial excitation is  in principle possible
in such models. However, as already noted above, since there is no unambiguous reference value for the magnitude of pion cloud effects in the nucleon,
the size of detuning the model parameters is to some extent arbitrary and therefore the agreement between the models and EBAC results
may very well be accidental.

\begin{figure*}[t]
        \begin{center}
        \includegraphics[width=0.47\textwidth]{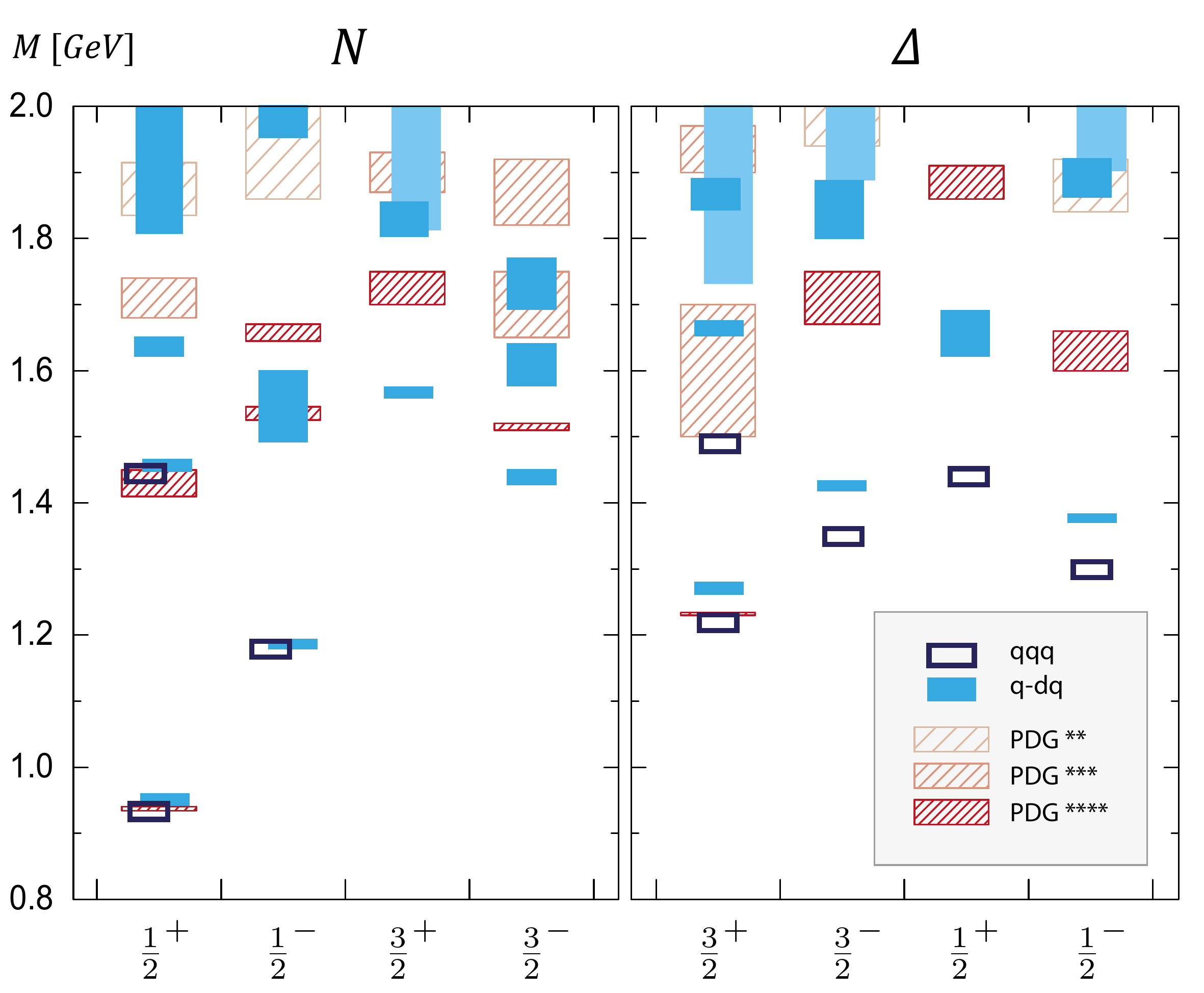}\qquad
        \includegraphics[width=0.47\textwidth]{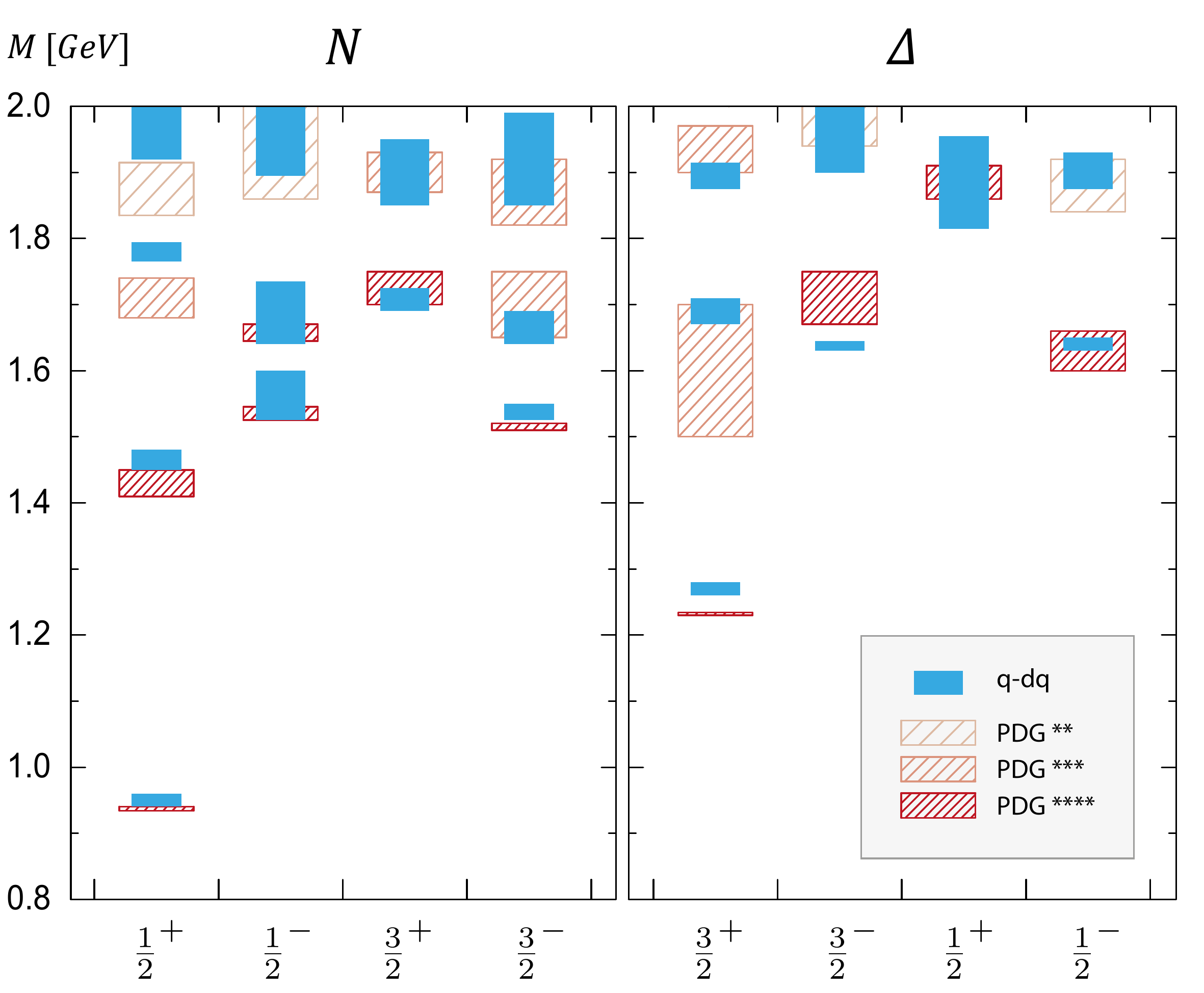}
        \caption{\textit{Left:} Masses of baryons in selected parity partner channels
        determined within a rainbow-ladder three-body and quark-diquark framework \cite{roper:2016}.
        \textit{Right:} Spectrum in the diquark-quark framework with compensated rainbow-ladder deficiencies \cite{roper:2016}.} \label{nucleon_delta}
        \end{center}
\end{figure*}

A different strategy has been followed in the very recent work \cite{roper:2016} within the rainbow-ladder three-body framework. Guided by the
discussion of the merits and drawbacks of the rainbow-ladder approximation in
the meson sector, one may expect the radial excitations in `good'
channels for rainbow-ladder to work reasonably well. In the baryon sector these are the nucleon and to some extent the $\Delta$ channel, which already showed
the correct hyperfine splitting in the ground states. Indeed, a similar feature is seen for the first radial excitations as displayed in Fig.~\ref{nucleon_delta}.
The masses of the excited states are
\begin{align}
m_{N^*}= 1.45(5) \,\mbox{GeV} \hspace{3cm} m_{\Delta^*}= 1.49(6) \,\mbox{GeV}
\end{align}
and therefore close to the experimental masses of the Roper and the $\Delta(1600)$. The systematic errors correspond to the range $1.6 \le \eta \le 2.0$
for the width parameter in the effective interaction~\eqref{couplingMT}.
The wave functions of these states show distinct zero crossings in the relative
quark momenta which support their interpretation as radial excitations. In other channels, however, in particular for the parity partners of the nucleon
and the $\Delta$, the calculated masses of even the ground states are much too low as is evident from Fig.~\ref{nucleon_delta}. Thus the pattern already
observed in the meson sector is repeated in the baryon sector and underlines the interpretation discussed above. In particular, it seems
that a rainbow-ladder based scheme cannot resolve the level ordering issue between the Roper and the negative parity ground state discussed in the beginning of
this review. One should also note that the superficial disagreement between
the model results of \cite{Roberts:2011cf,Segovia:2015hra} and the rainbow-ladder calculation of \cite{roper:2016} is not a matter of principle
but merely the consequence of a different perspective on setting the scale and interpreting the corresponding results.

For the negative parity ground states, a similar effect is seen in the quark-diquark rainbow-ladder framework also shown in the left panel of
Fig.~\ref{nucleon_delta}.
Here, additional insight may be gained when one analyses the diquark content of the states as in Table~\ref{n-structure}.
As already discussed in Sec.~\ref{spec:approx}, for diquarks the argumentation in the meson sector is valid for the opposite parity channels.
Rainbow-ladder works well for pseudoscalar and vector mesons and produces the correct hyperfine splitting in between,
hence the properties of scalar and axialvector diquarks should be similarly reliable. By contrast, pseudoscalar and
vector diquarks are too light because the same happens for scalar and axialvector mesons.
This translates directly to the corresponding three-body states.

In Ref.~\cite{roper:2016} a simple method has been used to compensate for the deficiencies of the rainbow-ladder framework within
the quark-diquark approximation. To this end, the effective interaction (\ref{couplingMT}) in the BSEs for the pseudoscalar and vector
diquarks are modified by a constant factor $c$ smaller than one. This reduces the binding in the problematic channels. The size of this
extra parameter is fixed from the $\rho-a_1$ splitting in the corresponding meson sector; everything else is left unchanged. The resulting
spectrum is shown in the right panel of Fig.~\ref{nucleon_delta}. One finds a dramatic improvement in the problematic channels, with
hardly any changes in the channels that already had good results before. Overall, the spectrum is in very good agreement with experiment
with a one-to-one correspondence of the number of observed states. The level ordering between the Roper and the $N(1535)$ is correct without
any additional adjustments. Considering that the two relevant parameters are fixed in the meson sector (the scale $\Lambda$ via the pion
decay constant and the parameter $c$ by the $\rho-a_1$ splitting), the agreement with experiment is remarkable and highly non-trivial.
The significance of this very recent result is discussed in more detail in Ref.~\cite{roper:2016} and needs to be explored further in future work.

\begin{table}[b]
\footnotesize
\begin{center}
\begin{tabular}{l @{\qquad} cccccc}
\toprule
                           & $N$      & $N(1535)$  & $\Delta(1232)$    & $\Delta(1700)$  	&  $\Delta(1910)$	& $\Delta(1620)$    \\
\midrule
                sc         & $1$     &  $0.45$     &                   &                    &                   &                     \\
                av         & $0.37$  &  $0.56$     & $1$               & $-0.10$            & $0.39$            & $-0.40$               \\
                ps         & $0.02$  &  $1$        &                   &                    &                   &                     \\
                v          & $0.03$  &  $0.27$     & $-0.02$           &  $1$               &  $1$              & $1$                   \\
\bottomrule
\end{tabular}\qquad\quad
\begin{tabular}{l @{\qquad} ccc}
\toprule
                   \%      & $N$      & $N^\ast(1440)$  & $N(1535)$      \\
\midrule
                $s$ wave     & $66$     &  $15$     & $36$                     \\
                $p$ wave     & $33$     &  $61$     & $58$                        \\
                $d$ wave     & $1$      &  $24$     & $6$                           \\
\bottomrule
\end{tabular}
\end{center}
\caption{\textit{Left: } Strength of the leading (scalar, axialvector, pseudoscalar, vector) diquark components for various nucleon resonances~\cite{Eichmann:2016jqx}. \textit{Right: }
         Magnitude of the orbital angular momentum contributions for the nucleon, Roper and $N(1535)$.}\label{n-structure}
\end{table}

As we have seen, the mass of the Roper is well described in both the three-body rainbow-ladder framework and the diquark-quark approximation.
Its wave function suggests an interpretation
in terms of the first radial excitation of the nucleon. It is also interesting to study the internal structure of this state by comparing the relative
size of different partial wave contributions in the Bethe-Salpeter wave functions.
Whereas the nucleon is dominated by $s$-wave contributions with $30 \%$ admixture of $p$ waves~\cite{Eichmann:2011vu},
the Roper is predominantly made of $p$ waves together with a sizeable $d$-wave component~\cite{roper:2016}.
Thus, the inclusion of orbital angular momentum in terms of $p$ waves is mandatory for the Roper. This is also true
for the negative-parity ground state $N(1535)$, which has again a different structure than both the nucleon and the Roper.

\begin{figure*}[t]
        \begin{center}
        \includegraphics[width=0.47\textwidth]{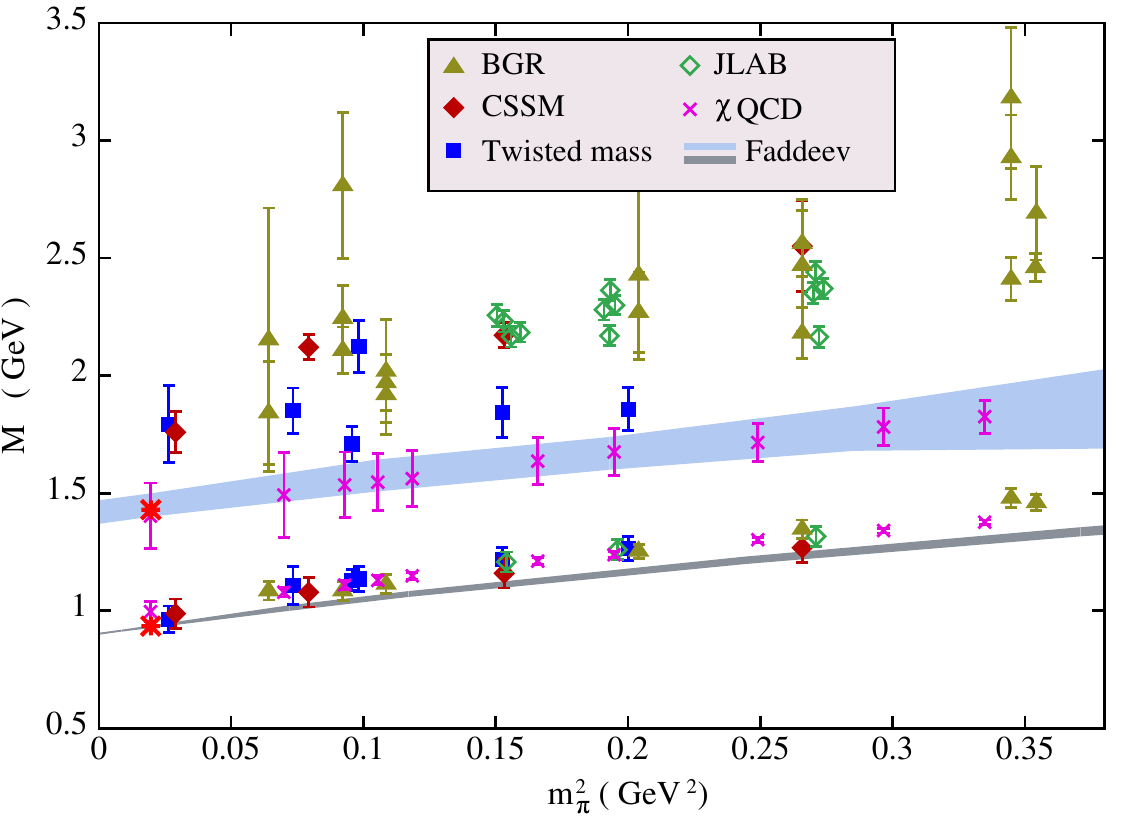}\qquad
        \includegraphics[width=0.47\textwidth]{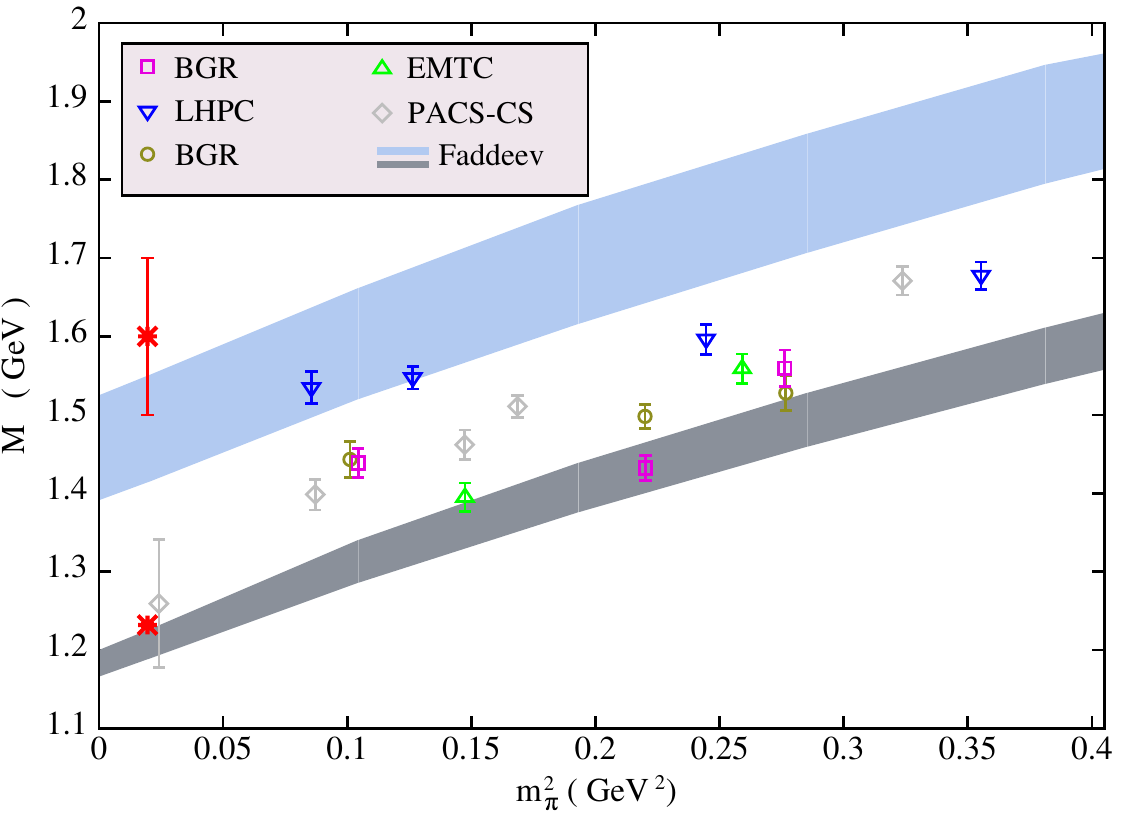}
        \caption{\textit{Left:} Mass evolution of the nucleon and the first radially excited state (Roper) \cite{roper:2016}
                compared to lattice data of \cite{Edwards:2011jj,Mahbub:2010rm,Mahbub:2012ri,Alexandrou:2013fsu,
                Alexandrou:2014mka,Alexandrou:2009hs,Liu:2014jua}.
                \textit{Right:} Mass evolution of the $\Delta$ and its first radial excitation \cite{roper:2016}.
                Only lattice data for the $\Delta$
                ground state are shown; data for the excited $\Delta$ give values too high to be visible in the plot~\cite{Gattringer:2008vj,WalkerLoud:2008bp,Aoki:2008sm,Engel:2013ig}.}\label{roper_delta_evolution}
        \end{center}
\end{figure*}

Finally, let us compare these findings with corresponding ones on the lattice. As already discussed in Sec.~\ref{spec:extracting}, excited states
in general are difficult to extract on the lattice and the Roper makes no difference. This can be seen from the results displayed in the left diagram
of Fig.~\ref{roper_delta_evolution}, where we show the mass evolution of the nucleon and its first radial excitation as a function of the squared pion mass.
We compare the results of different lattice groups with the ones from the three-body Faddeev framework \cite{roper:2016}. The lattice results
seem to be somewhat scattered which reflects the different systematic errors in the respective setups.
In Refs.~\cite{Roberts:2013ipa,Roberts:2013oea} the authors verified that the wave function of the Roper has indeed a single node, thus suggesting
the identification of the observed state with the first radial excitation of the nucleon. Furthermore, it has been  recently argued~\cite{Leinweber:2015kyz}
that the differences of the mass evolution results from the Jefferson Lab HSC~\cite{Edwards:2011jj}, CSSM~\cite{Mahbub:2010rm,Mahbub:2012ri} and Cyprus
groups~\cite{Alexandrou:2013fsu,Alexandrou:2014mka} can be accounted for and brought to consensus with each other. It is, however, not clear from an outsider's
perspective where the results from the $\chi$QCD group~\cite{Liu:2014jua} fit in and therefore it seems fair to state that there may not be
an overall agreement in the lattice community concerning the status of the Roper and the last word is by far not spoken.
In any case, we find that the mass evolution in the three-body Faddeev framework is consistent with the results from $\chi$QCD.
In the $\Delta$ channel, displayed in the right diagram of Fig.~\ref{roper_delta_evolution}, the situation is even less clear. While for the
ground state reasonable agreement may be claimed between the Faddeev results and the lattice evolution, the existing lattice data for the
excited state gives values too high to be visible in our plot. This situation needs to be resolved.

Concerning negative-parity ground states on the lattice, the situation is complicated by the presence of a $\pi N$ scattering state below the mass of the $N(1535)$.
Thus the physical ground state has to be extracted as the first `excitation' on the lattice. Consequently, also here the errors are substantial and
the results by far do not have the quality of the positive-parity ground state calculations. Nevertheless, within error bars there is
agreement between different groups that the $N(1535)$ is seen although the extracted mass is still larger than the experimental one,
see \cite{Edwards:2012fx,Lang:2012db,Mahbub:2013ala,Alexandrou:2013fsu,Alexandrou:2014mka,Leinweber:2015kyz} and references therein.
The structure of this state has been investigated in Ref.~\cite{Braun:2014wpa} with the help of light-cone distribution amplitudes and found to
be very different than that of the nucleon. This ties in with the results displayed in Table~\ref{n-structure}, although the translation of
Bethe-Salpeter wave functions to distribution amplitudes is of course not one-to-one.

\paragraph{Excitations in the strange baryon sector.}
The experimental excitation spectrum in the strange baryon sector shows interesting common features but also differences compared to that of the nucleon
discussed in Sec.~\ref{exp}.
For one, the level ordering between the positive and negative parity states is affected by the appearance of additional flavour-singlet states,
which may also mix with the isosinglet flavour octet $\Lambda$. This problem is connected with the appearance of the $\Lambda(1405)$ in the
negative-parity channel, which is anomalously low when compared with the negative parity nucleon spectrum despite its strange quark content.
A clarification of the underlying mechanism that leads to this difference is very interesting and important and may shed further light on the
flavour dependence of the nonperturbative QCD forces.

With the exception of the contact-interaction calculation of Ref.~\cite{Chen:2012qr},
results on the excitation spectrum in the strange baryon sector are not yet available in the functional framework. On the lattice, there are only a
couple of studies \cite{Menadue:2011pd,Edwards:2012fx,Engel:2012qp}, agreeing with each other in the identification of the $\Lambda(1405)$ with
a dominant flavour singlet component in the operator basis. On the other hand, there is no trace of the radially excited state in the positive parity
sector and therefore nothing can be said at the moment concerning the level ordering. These calculations may be hampered by the restricted operator
basis used so far which employs three-quark operators only, cf. the discussion in Sec.~\ref{spec:extracting}. The structure of the $\Lambda(1405)$
has furthermore been studied in Ref.~\cite{Hall:2014uca} using the interesting idea to measure magnetic form factors in order to extract the quark
content of this state. If the $\Lambda(1405)$ could be described by a $\bar{K}N$-molecule instead of a three-quark state, the strange quark
is trapped inside a spin-0 cluster and cannot contribute to the form factors. Thus by extracting the strange quark contribution to the form factor one
may distinguish between these two possibilities. In fact, the authors of Ref.~\cite{Hall:2014uca} find that close to the physical point this
contribution is highly suppressed, indeed indicating the five quark nature of the $\Lambda(1405)$. If confirmed, the $\Lambda(1405)$ thus may very well
be the lightest pentaquark state found in nature.


\newpage

\section{Form factors}\label{form}

        Since quarks and gluons are confined in hadrons, our best chance to learn about their properties is to perform scattering experiments involving hadrons or leptons.
            Among the simplest observables extracted from such processes, apart from the hadron spectrum, are the form factors of hadrons.
            They encode their momentum-dependent interactions with external currents such as photons, $W$ and $Z$ bosons,
            which are expressed through electromagnetic and axial form factors, but also their coupling to other hadrons as for example the nucleon-pion interaction
            or the electromagnetically induced transitions $N\gamma^\ast\to\Delta$, $N\gamma^\ast\to N^\ast$, etc.
        Consider for example electron-nucleon scattering: due to the smallness of the electromagnetic coupling constant $\alpha_\text{QED}\approx \nicefrac{1}{137}$
        the process is dominated by one-photon exchange, and the corresponding elastic electromagnetic form factors provide intrinsically nonperturbative information on the substructure of the nucleon.
        Similarly, the inelastic reaction $N\gamma^\ast\to X$ allows one to extract nucleon resonances and resolve their electromagnetic properties.

            \begin{figure*}[h!]
            \centerline{%
            \includegraphics[width=0.57\textwidth]{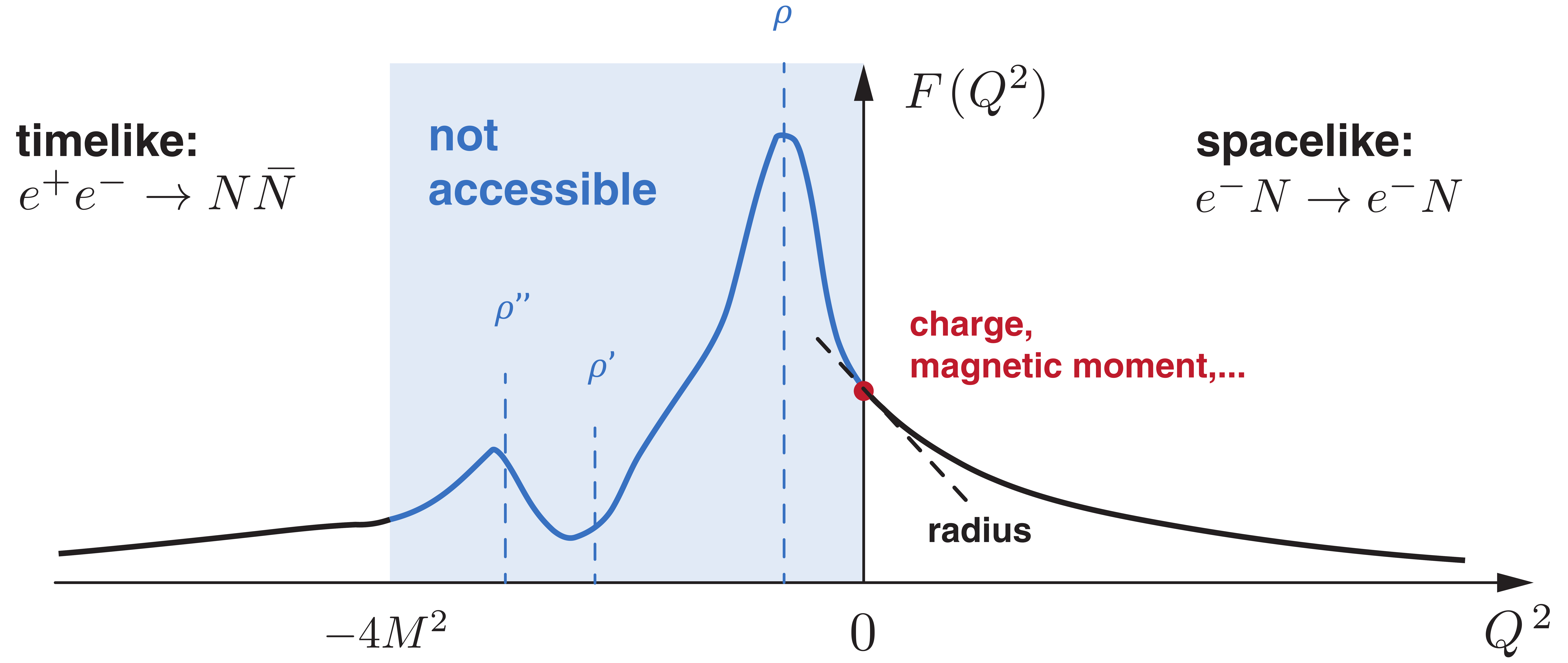}}
            \caption{Sketch of a generic form factor in the spacelike and timelike region.}
            \label{fig:timelike-ffs-1}
            \end{figure*}

             The generic shape of an electromagnetic nucleon or nucleon-to-resonance
             transition form factor is sketched in Fig.~\ref{fig:timelike-ffs-1}.
             Its spacelike behaviour ($Q^2 > 0$) is experimentally extracted from $eN$ scattering or, in the case of a resonance, pion photo- and electroproduction.
             The elastic form factors at $Q^2=0$ encode the charge and magnetic moment but potentially also higher quadrupole and octupole moments, whereas
             the slope at vanishing momentum transfer defines the charge radius which gives a basic measure of the size of the hadron.
             On the other hand, also the timelike properties of form factors are of interest.
             Experimental information on timelike nucleon form factors above threshold ($Q^2 < -4m_N^2$) comes from the reaction $e^+ e^-\to N\conjg{N}$ or its inverse.
             For nucleon resonances, the region below threshold is only indirectly accessible
             via the Dalitz decays $N^\ast\to Ne^+e^-$ that contribute to the dilepton `cocktail' in $NN$ and heavy-ion collisions.
             The characteristic features in this regime are the vector-meson bumps that are produced when the photon fluctuates into $\rho$ and $\omega$ mesons.
             The bump landscape in Fig.~\ref{fig:timelike-ffs-1} is drawn from the pion form factor, experimentally measured via $e^+ e^-\to\pi^+\pi^-$,
             where this property is exposed due to the much smaller threshold ($2m_\pi < m_\rho$).

             From these considerations it is clear that
             form factors encode a wealth of information on the substructure of hadrons which connects different aspects of QCD\@.
             The collection of experimental data on nucleon elastic and transition form factors provides a benchmark test for theoretical approaches,
             and it is constantly growing thanks to existing and future facilities such as Jefferson Lab, ELSA, MAMI, BES-III and PANDA.
             The traditional picture of electromagnetic form factors in terms of Fourier transforms of charge distributions has already come under scrutiny
             because relativity plays a major role.
             In addition, experiments have presented us with several new puzzles: from the form factor ratio of the proton to the question of the proton radius
             and, if we count in the form factors of leptons too, the muon g-2 puzzle.
             The main task for theorists is then to understand and explain the interplay of the various features that enter into such form factors,
             from model-independent aspects and perturbative constraints to nonperturbative properties which are encoded in their low-$Q^2$ and timelike structure, through the level of quarks and gluons in QCD\@.

            In the following we will first discuss model-independent aspects in the theoretical description of form factors.
            In Sec.~\ref{sec:ff-currents} we discuss current matrix elements in general and in Sec.~\ref{sec:ff-vertices} the
            structure of the vector, axialvector and pseudoscalar vertices which couple to the quarks in a baryon.
            In Sec.~\ref{sec:ff-methods} we give an overview of nonperturbative methods to evaluate form factors
            with a particular focus on lattice QCD and the Dyson-Schwinger
            framework, and we discuss applications and results in the remainder of this chapter.

\subsection{Current matrix elements}\label{sec:ff-currents}

       \paragraph{General properties.}
        The electromagnetic, axial and pseudoscalar form factors of a baryon
        are defined as the Lorentz-invariant, $Q^2-$dependent coefficients in a tensor decomposition of the respective current matrix elements
        \begin{equation}
            \mc{J}^{[\mu]}_{\lambda\lambda'}(p_f,p_i) = \langle \lambda  | \,j^{[\mu]}(0)\,| \lambda'  \rangle\,.
        \end{equation}
        Here, $|\lambda'\rangle$ is the Fock state of the incoming baryon with momentum $p_i$, $|\lambda\rangle$ is the outgoing baryon with momentum $p_f$,
        and $Q=p_f-p_i$ is the four-momentum transfer.
        If the incoming and outgoing states are different, $\mc{J}^{[\mu]}$ describes the transition current matrix element between two different baryons.
        We generalize the notation of~\eqref{qqbar-currents} and
        denote the renormalized quark-antiquark currents by a generic label ${[\mu]}$ that distinguishes vector, axialvector and pseudoscalar currents and also absorbs their flavour indices:
        \begin{equation}\label{current-Gamma-0}
            j^{[\mu]}(z) = \conjg{\psi}(z)\,\Gamma^{[\mu]}_0\,\psi(z) \,, \qquad
            \Gamma^{[\mu]}_0 \in \left\{ \, Z_2  \,i\gamma^\mu\,\mathsf{t}_a, \;  Z_2 \,\gamma_5 \gamma^\mu\,\mathsf{t}_a , \; Z_4 \,i\gamma_5\,\mathsf{t}_a \, \right\}  \quad \Rightarrow \quad
            j^{[\mu]} \in \left\{\, j^\mu_a, \; j^\mu_{5,a}, \; j_{5,a} \,\right\}.
        \end{equation}
        The Dirac-flavour structures $\Gamma^{[\mu]}_0$ contain the $SU(N_f)$ flavour matrices $\mathsf{t}_a$ as given below~\eqref{qqbar-currents}
        and the quark renormalization constants $Z_2$ and $Z_4=Z_2 \,Z_m$ have been defined in~\eqref{renormalization-constants}.
        For example, the electromagnetic current that induces electromagnetic elastic or transition form factors has the form
        \begin{equation}\label{current-em}
             j^\mu_\text{em}(0) = Z_2\,\conjg{\psi}(0) \,i\gamma^\mu \mathsf{Q}\,\psi(0) = iZ_2 \left( q_u  \,\bar{u} \,\gamma^\mu u + q_d \,\bar{d}\,\gamma^\mu d\right),
        \end{equation}
        where $\mathsf{Q} = \text{diag}\,(q_u,q_d) = \tfrac{1}{2}(q_u+q_d)\,\mathsf{t}_0 + (q_u-q_d)\,\mathsf{t}_3$ is the quark charge matrix in the two-flavour case.
        The individual components in~\eqref{current-em} define the flavour contributions to the form factors and the contributions from $\mathsf{t}_0$ and $\mathsf{t}_3$ their
        isoscalar and isovector components, respectively.

        In models based on relativistic quantum mechanics, current matrix elements of the form $\langle \lambda  | \,j^{[\mu]}(0)\,| \lambda'  \rangle$ are typically realized
        through simple convolutions of wave functions. In fact, this point of view has become so commonly adopted that it is often taken for granted as a generic feature.
        One should nevertheless appreciate that matrix elements in quantum field theory are much more complicated objects.
        Light-cone quantization bears the perhaps closest similarity to quark models in terms of overlaps of (light-cone) wave functions, although
        one requires infinitely many of them to construct matrix elements~\cite{Brodsky:1997de}.
        As we will see below, the manifestly covariant formulation with Bethe-Salpeter wave functions provides another intuitive understanding of matrix elements
        which is conceptually similar but also differs in decisive aspects.
        In any case, the `wave functions' in a quantum field theory are
        gauge-dependent, and so are its Green functions, whereas hadronic matrix elements are gauge-invariant objects.
        What is then the general strategy to determine gauge-invariant current matrix elements in QCD\@?

\begin{figure*}[t]
                    \begin{center}
                    \includegraphics[width=0.7\textwidth]{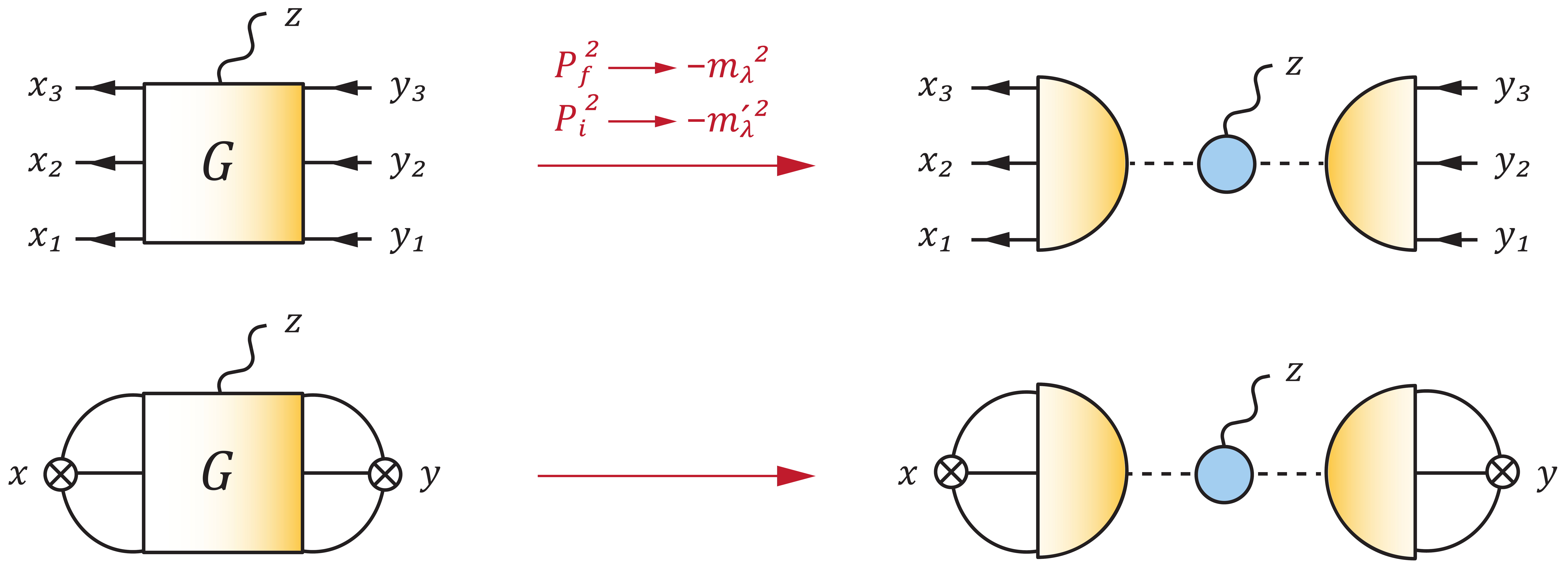}
                    \caption{\textit{Top:} Green function for the coupling of a current to three quarks and its pole behaviour.
                             The half-circles are the baryons' Bethe-Salpeter wave functions, the dashed lines are Feynman propagators,
                             and the residue defines the current matrix element.
                             \textit{Bottom:} Three-point correlator that is evaluated in lattice calculations of form factors. }\label{fig:current-general}
                    \end{center}
            \end{figure*}

        In principle the starting point for extracting baryon form factors is the same as in Sec.~\ref{spec:extracting}:
        baryons appear as poles in higher $n$-point functions through their spectral representation.
        The simplest object where a current matrix element can be extracted from is the analogue of~\eqref{baryon-G6}, which is illustrated in Fig.~\ref{fig:current-general}:
        \begin{equation}\label{baryon-G7}
             \mathbf G^{[\mu]}_{\alpha\beta\gamma,\alpha'\beta'\gamma'}(x_1,x_2,x_3;y_1,y_2,y_3;z) :=
                \langle 0 | \mathsf{T}\,\psi_\alpha(x_1)\,\psi_\beta(x_2)\,\psi_\gamma(x_3)\,\conjg{\psi}_{\alpha'}(y_1)\,\conjg\psi_{\beta'}(y_2)\,\conjg\psi_{\gamma'}(y_3)\,j^{[\mu]}(z)\, | 0 \rangle\,.
        \end{equation}
        It is the seven-point function
        made of three incoming and three outgoing quarks at positions $x_i$ and $y_i$, together with a quark-antiquark current at the point $z$.
        Inserting QCD's completeness relation yields the spectral decomposition in the baryon channels, which takes the following form in momentum space:
        \begin{equation}\label{poles-in-Gmu}
            \mathbf G^{[\mu]}_{\alpha\beta\gamma,\alpha'\beta'\gamma'}(k_f,q_f,p_f;k_i,q_i,p_i) \simeq \sum_{\lambda\lambda'}
                \frac{ \mathbf\Psi^\lambda_{\alpha\beta\gamma}(k_f,q_f,p_f)}{p_f^2+m_\lambda^2}\,
                \mc{J}^{[\mu]}_{\lambda\lambda'}(p_f,p_i)\,
                \frac{ \conjg{\mathbf\Psi}^{\lambda'}_{\alpha'\beta'\gamma'}(k_i,q_i,p_i)}{p_i^2+m_{\lambda'}^2}\,.
        \end{equation}
        The sum over $\lambda$, $\lambda'$ is again formal because it involves integrations over relative momenta; in practice we are mainly interested in the ground-state contribution.
        The residue is given by the baryon's current matrix element
$\mc{J}^{[\mu]}$ and its Bethe-Salpeter wave function
$\mathbf\Psi^\lambda_{\alpha\beta\gamma}(k,q,p)$ defined in~\eqref{eqn:bs-wf-baryon},
        where we denoted the total momenta by $p$ and the relative momenta by $k$ and $q$.

       \paragraph{Current correlators and lattice QCD.}
        Once again, the strategy to compute form factors in lattice QCD is to construct gauge-invariant current correlators by
        contracting the quark fields with suitable projection operators $\Gamma_{\alpha\beta\gamma\sigma}$.
        The resulting interpolating baryon fields are the same as in~\eqref{baryon-current-1}, and
        instead of the full Green function~\eqref{baryon-G7} one then has to deal with the three-point current correlator
        \begin{equation}\label{3-current-correlator}
              \mathbf G^{[\mu]}_{\sigma\sigma'}(x,y,z) = \langle 0 | \mathsf{T}\,J_\sigma(x)\,\bar J_{\sigma'}(y)\,j^{[\mu]}(z)\, | 0 \rangle\,.
        \end{equation}
        It depends on the interpolating baryon fields at the source $y$ and sink $x$ which are coupled to a current at the space-time point $z$.
        In momentum space its spectral decomposition becomes (see Fig.~\ref{fig:current-general})
        \begin{equation}
            \mathbf G^{[\mu]}_{\sigma\sigma'}(p_f,p_i) \simeq
           \sum_{\lambda\lambda'}
                \frac{ r_\lambda\,u_{\sigma}^{(\lambda)}(p_f)}{p_f^2+m_\lambda^2}\,
                \mc{J}^{[\mu]}_{\lambda\lambda'}(p_f,p_i)\,
                \frac{ r_{\lambda'}\,\conjg{u}_{\sigma'}^{(\lambda')}(p_i)}{p_i^2+m_{\lambda'}^2}\,,
        \end{equation}
        from where the current matrix element can be extracted.
        As before, $r_\lambda$ is the integrated Bethe-Salpeter wave function of~\eqref{rlambda-def} which determines the overlap with the baryon.

        In practice, form factors in lattice QCD are extracted from the large Euclidean time behaviour of the
        three-point correlator~\eqref{3-current-correlator}.
        Due to translation invariance it only depends on two coordinates; for example, setting $z=0$ and taking two three-dimensional Fourier transforms
        yields
        \begin{equation}\label{3-current-correlator-2}
            \mathbf G^{[\mu]}_{\sigma\sigma'}(\vect{p}_f,\tau_f,\vect{p}_i,\tau_i) \simeq \sum_{\lambda\lambda'} \frac{e^{-E_f |\tau_f|}}{2E_f}\,\frac{e^{-E_i |\tau_i|}}{2E_i}\,r_\lambda\,r_{\lambda'}\,
            u_\sigma^{(\lambda)}(p_f)\,\mc J_{\lambda\lambda'}^{[\mu]}(p_f,p_i)\,\conjg u_{\sigma'}^{(\lambda')}(p_i) \,.
        \end{equation}
        For large Euclidean time separations $\tau_f-\tau_i$ this expression is dominated by the ground state contribution.
        The time dependence approximately cancels when taking appropriate ratios of three- and two-point correlation functions,
        and in the most basic version the current matrix element can be extracted by fitting to the resulting plateaus (see also the discussion in Sec.~\ref{sec:ff-methods} below).

       \paragraph{Microscopic decomposition.}
       As we have seen, extracting current matrix elements from the large Euclidean time decay
       effectively amounts to singling out the pole residue of the corresponding
       current correlator $\mathbf G^{[\mu]}$.
       Given sufficiently sophisticated tools, its determination from the path integral on the lattice
       is conceptually straightforward as it does not require knowledge of the microscopic interactions apart from what is contained in the QCD action -- in other words,
       the complete dynamics enters through the path integral itself.
       The strategy to calculate form factors in functional approaches is complementary in the sense that
       one would first have to generate the original $n$-point function~\eqref{baryon-G7} dynamically from the quark level
       by solving a set of functional equations and isolate the pole behaviour therein.
       This would be a rather challenging task because, after all, these equations relate the $n$-point functions and not the current correlators derived from them.
       Fortunately, such a course is also not necessary because it is possible to derive an expression for the current matrix element $\mc{J}^{[\mu]}$ directly.

       The basic observation is that $\mathbf G^{[\mu]}$
       is obtained from the six-point function $\mathbf G$ by insertion of an external current $j^{[\mu]}(z)$.
       In the path-integral language this amounts to a functional derivative, which entails that the current couples linearly to all diagrams that appear in $\mathbf G$.
       In that way the operation $\mathbf G \rightarrow \mathbf G^{[\mu]}$ carries the properties of a derivative, i.e., it is linear and satisfies the Leibniz rule,
       which is referred to as `gauging of equations'~\cite{Kvinikhidze:1997wp,Kvinikhidze:1997wn,Kvinikhidze:1998xn,Kvinikhidze:1999xp}.
       Hence we can formally write
             \begin{equation}
                  \mathbf G^{[\mu]}   =  -\mathbf G\left(\mathbf G^{-1}\right)^{[\mu]} \mathbf G \; \stackrel{p_f^2=-m_f^2, \,\,p_i^2=-m_i^2}{\longlonglonglongrightarrow}  \;
                           -\frac{\mathbf\Psi_f \conjg{\mathbf\Psi}_f}{p_f^2+m_f^2}\left(\mathbf G^{-1}\right)^{[\mu]}\frac{\mathbf\Psi_i \conjg{\mathbf\Psi}_i}{p_i^2+m_i^2}\,,
             \end{equation}
             where we have employed a compact notation: we omitted momentum arguments and integrals; $p_i$ and $p_f$ are the baryon momenta and
             $\mathbf\Psi_i = \mathbf\Psi(k_i,q_i,p_i)$, $\mathbf\Psi_f = \mathbf\Psi(k_f,q_f,p_f)$ the respective baryon wave functions with different
       momentum dependencies.
             By comparison with~\eqref{poles-in-Gmu} one obtains the current matrix element as
             the gauged \textit{inverse} Green function between the onshell hadron wave functions:
             \begin{equation}\label{emcurrent-gauging}
                \mc{J}^{[\mu]} = -\conjg{\mathbf\Psi}_f \left(\mathbf G^{-1}\right)^{[\mu]} \mathbf\Psi_i \,.
             \end{equation}

            \begin{figure*}[t]
            \centerline{%
            \includegraphics[width=0.9\textwidth]{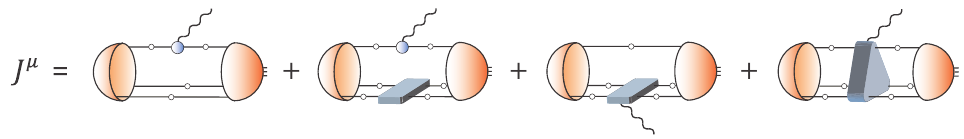}}
            \caption{Elastic or transition current matrix element of a baryon.}
            \label{fig:current-faddeev}
            \end{figure*}

        The relation can be worked out explicitly by applying~\eqref{dyson-eq}, which relates the Green function with the kernel:
        $\mathbf{G = G_0 + G_0\,K\,G}$ or, equivalently, $\mathbf G^{-1} = \mathbf{G_0}^{-1} - \mathbf K$.
        Therefore
        \begin{equation}\label{G-1mu}
             \left(\mathbf G^{-1}\right)^{[\mu]} = \left(\mathbf{G_0}^{-1}\right)^{[\mu]} - \mathbf K^{[\mu]} \,,
        \end{equation}
        where $\left(\mathbf{G_0}^{-1}\right)^{[\mu]}$ is obtained by gauging the product of three inverse quark propagators:
        \begin{equation}
           \left(\mathbf{G_0}^{-1}\right)^{[\mu]}  = \left( S^{-1} \otimes S^{-1} \otimes S^{-1} \right)^{[\mu]} =  \Gamma^{[\mu]} \otimes S^{-1} \otimes S^{-1} + \text{perm.}
        \end{equation}
        The quark-antiquark vertex $\Gamma^{[\mu]}$ will be discussed in detail in Sec.~\ref{sec:ff-vertices}. It
        is obtained by inserting a current $j^{[\mu]}(z)$ into the quark propagator,
        \begin{equation}\label{S-Smu}
            S_{\alpha\beta}(x,y) = \langle 0 |\, \mathsf{T} \,\psi_\alpha(x)\,\conjg{\psi}_\beta(y) \,|0\rangle \quad \rightarrow \quad
           S^{[\mu]}_{\alpha\beta}(x,y,z) = \langle 0 | \mathsf{T}\,\psi_\alpha(x)\,\conjg{\psi}_\beta(y)\,j^{[\mu]}(z)\, | 0 \rangle\,,
        \end{equation}
        and removing two dressed propagators so that in momentum space $S^{[\mu]} = -S\,\Gamma^{[\mu]} S \Rightarrow \Gamma^{[\mu]} = (S^{-1})^{[\mu]}$.
        Hence we arrive at the current matrix element that is visualized in Fig.~\ref{fig:current-faddeev}.
        $\mc{J}^{[\mu]}$ is the sum of impulse-approximation diagrams, where the current couples to the
        quarks only, plus terms where it couples to the kernel of the Bethe-Salpeter equation, namely
        \begin{equation}
            \mathbf{K} = \left( S^{-1} \otimes K_{(2)} + \text{perm.}\right)  + K_\text{(3)}  \quad  \Rightarrow \quad
            \mathbf K^{[\mu]} = \left( \Gamma^{[\mu]} \otimes K_{(2)}  + S^{-1} \otimes K_{(2)}^{[\mu]} + \text{perm.}\right) + K_\text{(3)}^{[\mu]}\,.
        \end{equation}
        $K_\text{(2)}$ and $K_\text{(3)}$ are the irreducible two- and three-body kernels.
        The Bethe-Salpeter wave functions are the bound-state amplitudes with dressed quark propagators attached, so the inverse propagators that appear in $\left(\mathbf{G_0}^{-1}\right)^{[\mu]}$ and $\mathbf K^{[\mu]}$
        cancel with the propagators in the spectator legs.

        The resulting formula is complementary to the discussion above but completely equivalent.
        Instead of extracting the pole residue of $\mathbf{G}^{[\mu]}$,
        we have derived the microscopic decomposition of the current matrix element directly in terms of baryon Bethe-Salpeter amplitudes, quark propagators and two- and three-quark kernels.
        Fig.~\ref{fig:current-faddeev} also provides an intuitive understanding of form factors. The incoming baryon
        splits into its valence quarks which emit and reabsorb gluons in all possible ways, obtain a boost from the current (photons, $W$ and $Z$ bosons, etc.)
        and finally recombine into the outgoing baryon. This is conceptually very similar to the quark model point of view
        and thereby establishes a link between model approaches and the exact expression of the current matrix element in quantum field theory.

        One should keep in mind, however, that the expression is only consistent in combination with the baryon's Faddeev equation in Fig.~\ref{fig:faddeev},
        which must be solved beforehand
        together with the remaining ingredients.
        Only then the symmetries of QCD are rigorously preserved: electromagnetic currents will be conserved, axial form factors will satisfy the Goldberger-Treiman relation, etc.
        For example, if we neglect three-body interactions by setting $K_{(3)}=0$ and work in a rainbow-ladder truncation where $K_{(2)}^{[\mu]}=0$,
        the two rightmost diagrams in Fig.~\ref{fig:current-faddeev} disappear. However,
        there is still a remaining graph that contains the two-body kernel as a spectator.
        Hence the impulse approximation alone is not sufficient for baryons;
        we cannot set all kernels simultaneously to zero because in that way
        we would strictly speaking not even obtain a bound state.\footnote{By comparison, the analogous current matrix element for mesons has only two topologies: the impulse approximation and a diagram with $K_{(2)}^{[\mu]}$.
             In that case a rainbow-ladder truncation does indeed induce the impulse approximation. }

\begin{figure}
 \centering
   \includegraphics[width=0.9\textwidth]{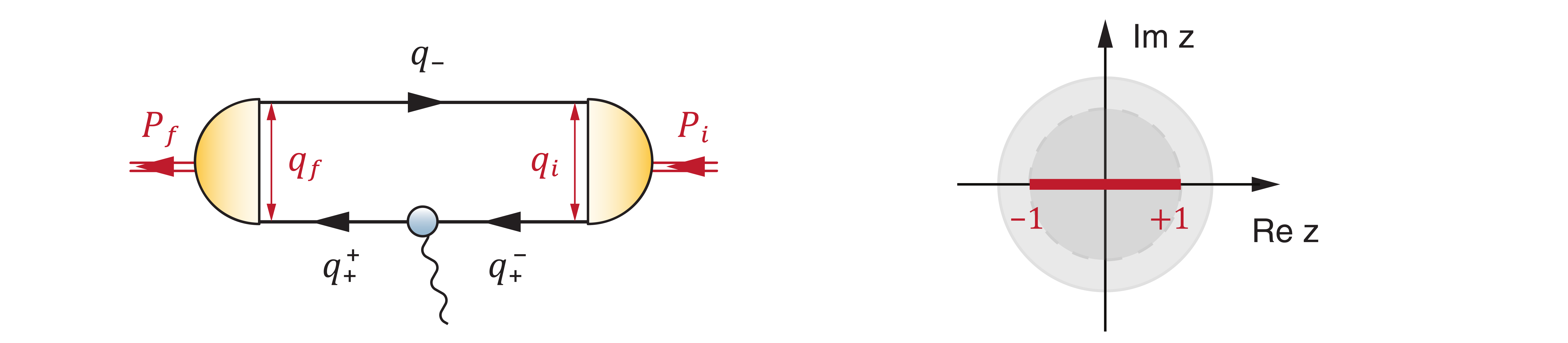}
 \caption{\textit{Left:} Scalar form factor integral of~\eqref{scalar-ff-model}. \textit{Right:} Domain in the angular variable $z=\hat{q}\cdot\hat{P}$ needed in the form factor integral
                          as opposed to that of the rest-frame solution ($-1<z<1$).   }
 \label{fig:ffs-bse-1}
 \end{figure}

       \paragraph{An explicit example.}
       In concluding this section we discuss some properties of Euclidean current matrix elements such as those in Fig.~\ref{fig:current-faddeev}.
       We refrain from giving the explicit three-body formulas since they can be found in the literature~\cite{Eichmann:2011vu} and do not provide
       any insight over the diagrammatic expressions.
       Let us instead revisit the simple scalar example of Fig.~\ref{fig:scalar-bse-1} in Sec.~\ref{spec:green} describing a bound state of mass $M$ with
       two constituents with equal masses $m$.
       Consider a scalar two-body amplitude $\mathbf\Gamma(q,P) = \mathbf\Gamma(q^2,z,-M^2)$ with $z=\hat{q}\cdot\hat{P}$, and assume that we have found it by solving its Bethe-Salpeter equation.
       A typical impulse-approximation form factor integral is illustrated in Fig.~\ref{fig:ffs-bse-1}:
        \begin{equation}\label{scalar-ff-model}
            \mc{J}^{\mu}(P,Q) = \int \!\! \frac{d^4q}{(2\pi)^4}\,\conjg{\mathbf\Gamma}(q_f,P_f) \,D(q_+^+)\,\Gamma^\mu(q_+,Q)\,D(q_+^-)\,\mathbf\Gamma(q_i,P_i)\,D(q_-)\,,
        \end{equation}
        where $\Gamma^\mu(q_+,Q)$ is for example the dressed vector vertex that depends on the relative momentum $q_+$ and the total (photon) momentum $Q$, $D(k)$ are the scalar propagators,
        and we have routed the momenta symmetrically so that
        \begin{equation}
            P_{f,i} = P \pm \frac{Q}{2}\,, \qquad
            q_{f,i} = q \pm \frac{Q}{4}\,, \qquad
            q_\pm = q \pm \frac{P}{2}\,, \qquad
            q_+^\pm = q_+ \pm \frac{Q}{2}\,.
        \end{equation}
        If the constituents were fermions then $\conjg{\mathbf\Gamma}(q,P) = \mc{C}\,\mathbf\Gamma(-q,-P)^T \mc{C}^T$ would be the charge-conjugate amplitude, where $\mc{C}=\gamma_4 \gamma_2$ is the charge-conjugation matrix;
        for our scalar case it is just $\conjg{\mathbf\Gamma}(q,P)=\mathbf\Gamma(-q,-P)$.

        The first question concerns the `Euclidean domain' of the diagram which is inferred from the poles in the propagators.
        Observe that because $P_{f,i}$ are the onshell momenta one has
        \begin{equation}
           P_{f,i}^2 = P^2 + \frac{Q^2}{4} \pm P\cdot Q = -M^2 \quad \Rightarrow \quad P\cdot Q = 0\,, \quad P^2 = -M^2\,(1+\tau)\,,
        \end{equation}
        where $\tau=Q^2/(4M^2)$, and therefore only $Q^2$ remains as an independent variable.
        Hence, for spacelike photon virtualities $Q^2>0$ the momenta $Q$ and $q$ are real (the loop momenta are always real) whereas the average momentum $P$ is imaginary.
        $P$ enters in the propagators so once again we need them in the complex plane.
        Applying the pole analysis around~\eqref{scalar-poles} to the present case yields for $M<2m$:
       \begin{equation}\label{ff-limits}
           0 < 1 + \frac{Q^2}{4M^2} < \frac{4m^2}{M^2}\,,
       \end{equation}
       and therefore a practical limit for both timelike \textit{and} large spacelike $Q^2$.
       The same restrictions apply for meson form factors in QCD if $M$ is identified with the meson mass and $m=m_P$ with the quark `pole mass'
       discussed below~\eqref{theory:bsetensorstructures},
       and analogous considerations hold for baryons (cf.~App.~B.3 of Ref.~\cite{Eichmann:2009zx}). The quark propagator singularities
       pose restrictions on the accessible $Q^2$ interval; beyond those limits residue calculus becomes mandatory (see, e.g., \cite{Oettel:2000ig})
which would be analogous to a calculation in Minkowski space.

        The second point concerns `boosting' the wave functions, which is a rather delicate issue in quark models as discussed in Sec.~\ref{sec:ff-methods} below.
        In a manifestly covariant approach, Green functions and Bethe-Salpeter amplitudes are covariant objects
        whose irreducible representation matrices under Lorentz transformations cancel each other to leave an overall Lorentz transformation for the current matrix element --
        or in other words, covariant loop integrals transform covariantly.
        As a consequence, one only needs to evaluate them for the new momenta as in~\eqref{scalar-ff-model}: $\conjg{\mathbf\Gamma}(q_f,P_f)$ and $\mathbf\Gamma(q_i,P_i)$ instead of $\mathbf\Gamma(q,P)$.
        Still, there are associated issues: $\mathbf\Gamma(q,P)$ was calculated in the rest frame of the total momentum $P$, where $q^2 > 0$ and $z$ are real with $|z|<1$.
        In the form factor integral $q_{f,i}^2 >0$ is still real (for $Q^2>0$), but because the average momentum $P$ is imaginary the angular variables $z_{f,i}$ become complex: $|z_{f,i}|<\sqrt{1+\tau}$, $\tau = Q^2/(4M^2)$,
        which is illustrated in Fig.~\ref{fig:ffs-bse-1}.
        Since the angular dependencies are usually small (as demonstrated in Fig.~\ref{fig:scalar-bse-1}), analytic continuations for instance via Chebyshev expansions are well justified as long as $Q^2$ is not too large.
        In principle the problem can be rigorously solved by calculating the amplitudes directly in the moving frame, so that instead of one complex variable $z_f$ or $z_i$ one solves
        for two real variables in the domain $[-1,1]$.
        However, this is a numerically expensive procedure especially for baryons and so far it has  only been implemented for the pion~\cite{Maris:2005tt}.

\subsection{Vector, axialvector and pseudoscalar vertices} \label{sec:ff-vertices}   %

        The crucial ingredient of the current matrix element in Fig.~\ref{fig:current-faddeev}
        is the dressed quark-antiquark vertex $\Gamma^{[\mu]}$ that appears whenever the external current is coupled to a quark propagator.
        It shows up directly in the first two diagrams but
        indirectly also in the last two diagrams when the kernels are resolved diagrammatically.
        As a consequence, the electromagnetic, axial or pseudoscalar structure of the various form factors is entirely carried by the respective vertices,
        whereas the remainder of these diagrams is universal --
        it is also what appears in the definition of generalized parton distributions or the `handbag' topologies in Compton scattering, cf. Sec.~\ref{compton}.
        Electromagnetic current matrix elements depend on the quark-photon vertex, which is a vector vertex, whereas axial and pseudoscalar form factors test the respective
        axialvector and pseudoscalar vertices.
        As we will see below, the structure properties of these vertices have several model-independent consequences for the form factors, including
        timelike meson poles and the implications of symmetry relations induced by vector current conservation and the PCAC relation.

       \paragraph{General properties.}
       With the help of the Dirac-flavour matrices $\Gamma^{[\mu]}_0$ in~\eqref{current-Gamma-0}, we can define the corresponding vector, axialvector and pseudoscalar quark-antiquark vertices
       $\Gamma^{[\mu]} \in \{ \, \Gamma^\mu, \; \Gamma^\mu_5, \; \Gamma_5 \,\}$
       by inserting the respective current $j^{[\mu]}(z)$ into the quark propagator, cf.~Eq.~\eqref{S-Smu}.
       In analogy to the distinction between Bethe-Salpeter amplitudes and wave functions,
       $\Gamma^{[\mu]}$  is obtained from $S^{[\mu]}$ by removing dressed quark propagators in momentum space,
        \begin{equation}\label{vertex-def-1}
            -S^{[\mu]}(k,Q) =  S(k_+)\,\Gamma^{[\mu]}(k,Q)\,S(k_-)\,,
        \end{equation}
        where $Q$ is the incoming total momentum, $k$ is the average momentum of the quarks, and $k_\pm = k \pm Q/2$.

            \begin{figure*}[t]
                    \begin{center}
                    \includegraphics[scale=0.12]{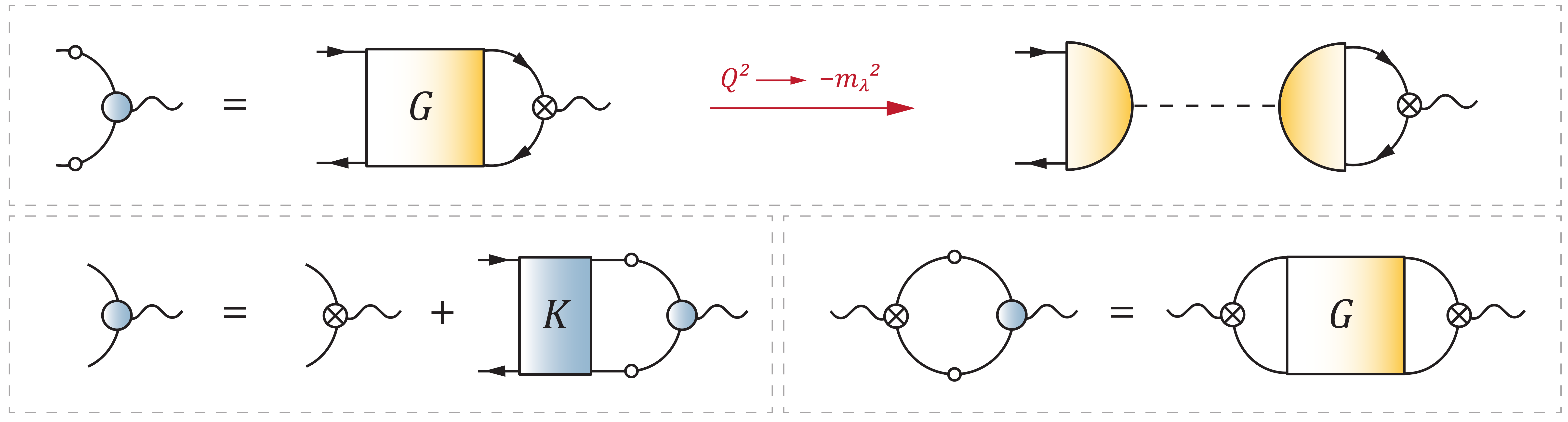}
                    \caption{\textit{Top:} Definition of the quark-antiquark vertex as the contracted four-point function, together with the meson poles it contains.
                             The half-circles are the mesons' Bethe-Salpeter wave functions, the dashed line is a Feynman propagator,
                             and the residue defines the decay constant.
                             \textit{Bottom left:} Inhomogeneous Bethe-Salpeter equation for the vertex.
                             \textit{Bottom right:} Vector-vector current correlator which defines the hadronic vacuum polarisation. }\label{fig:vertex-def}
                    \end{center}
            \end{figure*}

        The definition~\eqref{S-Smu} also makes clear that $S^{[\mu]}$ originates from the quark-antiquark four-point function
        where one $q\bar{q}$ pair has been converted into a current by contraction with $\Gamma^{[\mu]}_0$.
        This is illustrated in Fig.~\ref{fig:vertex-def} and we write it in compact notation as
       \begin{equation}\label{vertex-def-2}
          -S^{[\mu]} = \mathbf{G}_0\,\Gamma^{[\mu]} = \mathbf{G}\,\Gamma^{[\mu]}_0\,.
       \end{equation}
       $\mathbf{G}$ and $\mathbf{G}_0$ denote the full and disconnected four-point functions, respectively, and we dropped the momentum arguments and integrations for brevity.
       As a consequence, a two-point current correlator follows from another contraction of $S^{[\mu]}$ with $\Gamma_0^{[\nu]}$ together with a momentum integration.
       On the other hand, the Green function satisfies $\mathbf{G} = \mathbf{G_0} + \mathbf{G_0\,K\,G}$ and thus one obtains
       an inhomogeneous Bethe-Salpeter equation for the vertex by multiplying~\eqref{vertex-def-2} with $\mathbf{G_0^{-1}}$ from the left, which is also shown in Fig.~\ref{fig:vertex-def}:
       \begin{equation}\label{inhom-vertex-bse}
          \Gamma^{[\mu]} = \Gamma^{[\mu]}_0 + \mathbf{K\,G_0}\,\Gamma^{[\mu]} \,.
       \end{equation}
       It allows one to determine the vertex selfconsistently and in consistency with other Bethe-Salpeter equations.

       \paragraph{Timelike meson poles.}
       The four-point Green function $\mathbf G$ contains all meson poles. Therefore, the vertices $\Gamma^{[\mu]}$ will inherit those poles for which
       the overlap of $\Gamma_0^{[\mu]}$ with the respective Bethe-Salpeter wave function is nonzero.
       For example, the isovector-vector vertex has $\rho$-meson poles in its transverse part:
       \begin{equation}
          \Gamma^\mu(k,Q) \stackrel{Q^2\to-m_\rho^2}{\longlonglongrightarrow} \frac{if_\rho\,m_\rho}{Q^2+m_\rho^2}\,T^{\mu\nu}_Q\,\mathbf\Gamma^\nu_{\rho}(k,Q)\,,
       \end{equation}
       where $\mathbf\Gamma^\nu_\rho(k,Q)$ is the $\rho-$meson Bethe-Salpeter amplitude and $T^{\mu\nu}_Q = \delta^{\mu\nu} - Q^\mu Q^\nu/Q^2$ the transverse projector.
       As indicated in Fig.~\ref{fig:vertex-def}, the residue is determined from the overlap of the $\rho-$meson Bethe-Salpeter wave function with $\Gamma^{[\mu]}_0 = Z_2\,i\gamma^\mu$.
       Since this is the transition matrix element of the isovector-vector current between the vacuum and the $\rho-$meson Fock state, it defines the $\rho-$meson decay constant:
       \begin{equation}
          \text{Tr} \int \!\!\frac{d^4k}{(2\pi)^4}\,Z_2\,i\gamma^\mu\, S(k_+)\,\mathbf\Gamma^\nu_\rho(k,Q)\,S(k_-)  = if_\rho\,m_\rho\,T^{\mu\nu}_Q\,.
       \end{equation}
       Similarly, the axialvector and pseudoscalar vertices contain pseudoscalar poles, such as the pion pole:
       \begin{equation}
          \Gamma^\mu_5(k,Q) \stackrel{Q^2\to-m_\pi^2}{\longlonglongrightarrow} Q^\mu \frac{2if_\pi}{Q^2+m_\pi^2}\,\mathbf\Gamma_\pi(k,Q)\,, \qquad
          \Gamma_5(k,Q) \stackrel{Q^2\to-m_\pi^2}{\longlonglongrightarrow} \frac{2i r_\pi}{Q^2+m_\pi^2}\,\mathbf\Gamma_\pi(k,Q)\,,
       \end{equation}
       where $\mathbf\Gamma_\pi(k,Q)$ is the onshell pion amplitude.
       One can formally separate the vertices into pseudoscalar pole contributions and remainders which are finite at these poles
       (`formally' because the respective Bethe-Salpeter  amplitudes are only meaningful objects on their mass shells $Q^2=-m_\lambda^2$):
       \begin{equation}\label{vertices-pole-contributions}
          \Gamma^\mu_5 = Q^\mu \sum_\lambda \frac{2if_\lambda}{Q^2+m_\lambda^2}\,\mathbf\Gamma_\lambda + \widetilde\Gamma^\mu_5\,, \qquad
          \Gamma_5 = \sum_\lambda \frac{2i r_\lambda}{Q^2+m_\lambda^2}\,\mathbf\Gamma_\lambda + \widetilde\Gamma_5\,.
       \end{equation}
       Pseudoscalar poles only appear in the longitudinal part of the axialvector vertex.
       The remainder $\widetilde\Gamma^\mu_5$ is regular for $Q^\mu \rightarrow 0$ and it is the sum of a non-transverse and a transverse part, where
       the latter also contains axialvector poles.
       Similarly, the pseudoscalar vertex can be split into pole contributions and non-resonant terms.

       From inspection of the current-matrix element in Fig.~\ref{fig:current-faddeev} it is clear that
       the timelike pole structure of these vertices, which only depends on the external variable $Q^2$, will be inherited by the form factors as well.
       The quark-photon vertex that enters electromagnetic current matrix elements has vector-meson poles, and thus any electromagnetic form factor will inherit them --
       which is the microscopic origin of `vector-meson dominance'.
       Similarly, the transverse parts of axial currents contain axialvector poles whereas their longitudinal parts carry pseudoscalar poles,
       and pseudoscalar poles will also appear in any pseudoscalar form factor.
       Therefore, the timelike resonance structure sketched in Fig.~\ref{fig:timelike-ffs-1} originates from the underlying vertices
       (and ultimately from the four-point function that appears in their definition) along with the photon-induced
       cut structure: the photon fluctuates into $\rho_0$ and $\omega$ mesons, which decay into pions, etc.

       \paragraph{Ward-Takahashi identities.}
       QCD's global symmetries provide further model-independent constraints on the structure of the quark-antiquark vertices apart from their singularity structure.
       The vector current conservation and PCAC relations from~\eqref{vcc-pcac-0} translate into Ward-Takahashi identities (WTIs) for the Green functions at the quantum level.
       For the vector, axialvector and pseudoscalar vertices they take the form
       \begin{align}
          Q^\mu \,\Gamma^\mu(k,Q) &= S^{-1}(k_+) - S^{-1}(k_-)\,, \label{vwti} \\
          Q^\mu \,\Gamma_5^\mu(k,Q) +2m\,\Gamma_5(k,Q) &= S^{-1}(k_+)\,i\gamma_5 + i\gamma_5\,S^{-1}(k_-)\,.  \label{axwti}
       \end{align}
       The first relation is the vector WTI which can be equally derived from electromagnetic gauge invariance.
       The second is the axial WTI and it only holds for flavour non-singlet vertices, because in the singlet channel there would be an
       additional term from the axial anomaly (see e.g.~\cite{Bhagwat:2007ha}).
       Both of them have numerous consequences for hadron physics.
       The vector WTI states that the longitudinal part of the vector vertex is completely determined by the quark propagator.
       Its solution is the \textit{Ball-Chiu vertex}~\cite{Ball:1980ay,Ball:1980ax}, together with a further transverse part which is not constrained by gauge invariance:
       \begin{equation}\label{qpv-bc+t}
           \Gamma^\mu(k,Q) = i\gamma^\mu\,\Sigma_A + 2k^\mu (i\slashed{k}\,\Delta_A + \Delta_B) + \sum_{j=1}^8 if_j(k^2, z^2, Q^2)\,\tau_j^\mu(k,Q)\,.
       \end{equation}
       The functions $\Sigma_A$, $\Delta_A$ and $\Delta_B$ are averages and difference quotients of the dressing functions $A(k^2)$ and $B(k^2)=M(k^2)\,A(k^2)$
       of the quark propagator $S^{-1}(k) = i\kslash \,A(k^2) + B(k^2)$, cf.~Eq.~\eqref{quark-propagator}:
       \begin{equation}\label{SigmaA-DeltaA}
           \Sigma_A = \frac{A(k_+^2)+A(k_-^2)}{2}\,, \qquad
           \Delta_A = \frac{A(k_+^2)-A(k_-^2)}{k_+^2-k_-^2}\,, \qquad
           \Delta_B = \frac{B(k_+^2)-B(k_-^2)}{k_+^2-k_-^2}\,.
       \end{equation}
       The remaining transverse part
       depends on eight dressing functions $f_j(k^2,z^2,Q^2)$, with $z=\hat{k}\cdot\hat{Q}$, and supplements the vector-meson poles in the timelike region.
       It satisfies further transverse WTIs which, however, do not provide closed constraints in practice~\cite{Qin:2013mta}.
       The transverse part vanishes linearly for $Q^\mu \to 0$ due to transversality and analyticity;
       a tensor basis that automatically implements these features can be found in App.~\ref{app:hvp}.
       As a consequence,
       the Ball-Chiu vertex alone is what ensures current conservation and therefore charge conservation at $Q^2=0$.
       It also encodes the perturbative limit of the vertex for $Q^2 \to\infty$, where $\Gamma^\mu(k,Q) \to Z_2\,i\gamma^\mu$.

            \begin{figure*}[t]
                    \begin{center}
                    \includegraphics[width=0.88\textwidth]{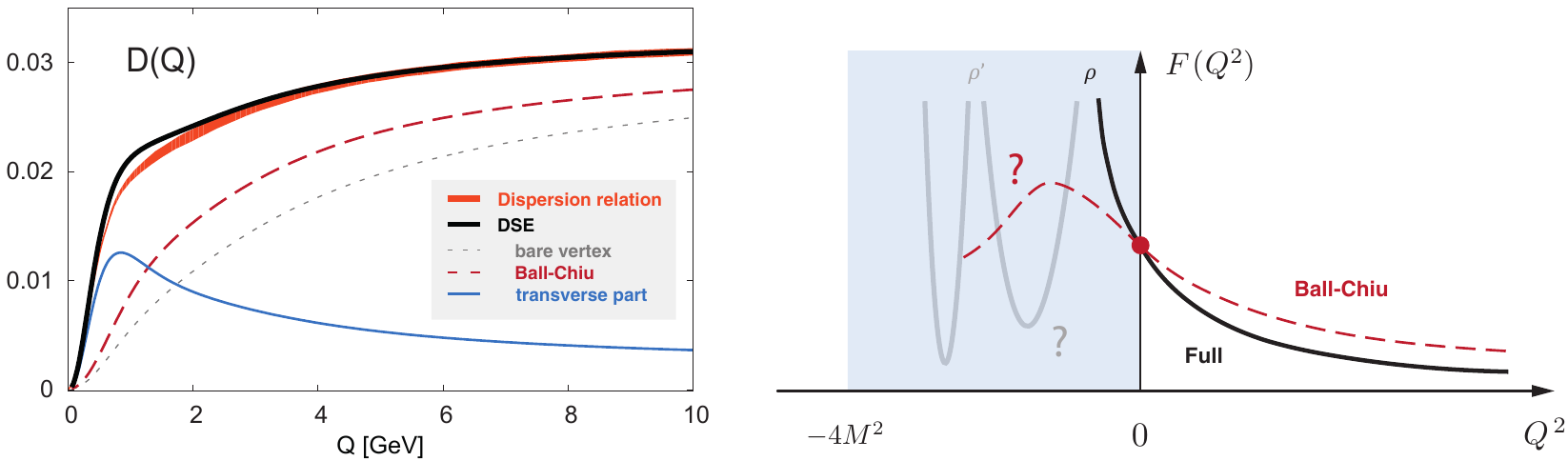}
                    \caption{\textit{Left:} Adler function obtained from dispersion relations~\cite{Eidelman:1998vc} compared to a Dyson-Schwinger calculation in rainbow-ladder truncation~\cite{Goecke:2011pe}.
                    The Ball-Chiu and transverse contributions sum up to the full result; the bare vertex is only included for illustration.
                    \textit{Right:} Sketch of the analogous decomposition for an electromagnetic form factor.}\label{fig:hvp}
                    \end{center}
            \end{figure*}

       \paragraph{Hadronic vacuum polarisation.}
       Let us illustrate these features with an explicit example, namely the hadronic vacuum polarisation which
       constitutes the simplest system testing the structure of the quark-photon vertex.
       It is defined as the Fourier transform of the hadronic part of the vector-vector current correlator, cf.~Fig.~\ref{fig:vertex-def}:
       \begin{equation}\label{hvp}
           \Pi^{\mu\nu}(Q) = \int d^4x\,e^{iQ\cdot x}\,\langle 0 |\,\mathsf{T} j^\mu(x) j^\nu(0)\,|0\rangle\,.
       \end{equation}
       From the discussion above (cf.~Fig.~\ref{fig:vertex-def}) this is just the loop diagram with one dressed quark-photon vertex, one tree-level vertex and two dressed quark propagators:
       \begin{equation}\label{hvp-0}
          \Pi^{\mu\nu}(Q) =   \text{Tr}\int \!\!\frac{d^4k}{(2\pi)^4}\, Z_2 \,i \gamma^\mu \,S(k_+)\,\Gamma^\nu(k,Q)\,S(k_-)  =  \Pi(Q^2)\,Q^2\,T^{\mu\nu}_{Q} + \widetilde\Pi(Q^2)\,\delta^{\mu\nu} .
       \end{equation}
       The second equality states its most general Poincar\'e-covariant form in momentum space.
       Gauge invariance enforces $\widetilde \Pi(Q^2)=0$; however, with a cutoff regularization this is not the case and $\widetilde \Pi(Q^2)$ diverges quadratically (see also the discussion around~\eqref{vac-pol}).
       The physical, transverse contribution $\Pi(Q^2)$ only has a logarithmic divergence.
       Taking the trace gives
       \begin{equation}\label{hvp-1}
           \Pi(Q^2) = \frac{1}{(2\pi)^3}\int dk^2\,k^2\int dz\sqrt{1-z^2}\,\sigma_v(k_+^2)\,\sigma_v(k_-^2)\,\Big( K_\text{BC} + \sum_{i=1}^8 f_j(k^2,z^2,Q^2)\,K_j\Big),
       \end{equation}
       where the kernels $K_\text{BC}$ and $K_j$ originate from the Ball-Chiu and the transverse part of the vertex, respectively.
       Their explicit expressions are collected in App.~\ref{app:hvp}.  They are scalar functions of $k^2$, $z^2 = (\hat{k}\cdot\hat{Q})^2$ and $Q^2$ and completely determined
       by the quark propagator dressing functions $A(k^2)$ and $M(k^2)$,
       and the same is true for the vector dressing function $\sigma_v(k^2)$ defined in~\eqref{quark-propagator}.

       Eq.~\eqref{hvp-1} is a model-independent, exact result for the vacuum polarisation.
       It depends on 10 functions which must be known beforehand: the quark dressing functions $A(k^2)$ and $M(k^2)$
       and the eight transverse form factors $f_i(k^2,z^2,Q^2)$ of the offshell quark-photon vertex.\footnote{Upon
       neglecting the transverse part of the vertex and inserting a tree-level propagator with $A(p^2)=Z_2$ and $M(p^2)=m_q$, the Ball-Chiu vertex turns into a bare vertex and
       one recovers the textbook result for the vacuum polarisation in perturbation theory.}
        In Fig.~\ref{fig:hvp} we show the Dyson-Schwinger result for the Adler function $D(Q^2) = -Q^2\,d\Pi(Q^2)/dQ^2$
        obtained by calculating the quark propagator and quark-photon vertex consistently in a rainbow-ladder truncation~\cite{Goecke:2011pe}.
        Apart from small deviations in the mid-momentum region it agrees very well with the result obtained from dispersion relations~\cite{Eidelman:1998vc}.
        Notice, however, how the Ball-Chiu vertex and the transverse part containing the vector-meson poles
        both conspire to give the final result. This is a quite non-trivial feature which highlights the necessity to take into account all parts of the
        dressed quark-photon vertex. Furthermore, the results agree although the vector mesons obtained in rainbow-ladder are stable bound states
        and their poles do not have widths. The resulting timelike structure therefore differs from that of full QCD, where
        the hadronic vacuum polarisation is an analytic function in the entire complex plane except for a branch cut starting at $Q^2 = -4m_\pi^2$ and extending to minus infinity
        (with poles in the second Riemann sheet),
        which is also what goes into the dispersion relations via the cross section $e^+e^- \to$ hadrons.

        In any case, the separation into Ball-Chiu plus transverse part itself is model-independent and also holds for electromagnetic form factors, as illustrated in the right panel of Fig.~\ref{fig:hvp}.
        The Ball-Chiu part recovers the electric charge, so that charge normalization follows automatically and does not have to be imposed by hand,
        whereas the transverse part produces the timelike vector-meson structure and in principle also the various physical cuts.
        An example is the pion's electromagnetic form factor, where the rainbow-ladder Dyson-Schwinger result~\cite{Maris:1999bh} reproduces
        the spacelike experimental data very well although both contributions are individually large.
        The transverse part vanishes at $Q^2=0$ but produces roughly half of the squared pion charge radius,
        and similarly to Fig.~\ref{fig:hvp} it drops only slowly with $Q^2$.
        The same principles apply for the form factor results to be discussed in the following sections.

       \paragraph{Goldstone theorem.}
       Let us return to the axial Ward-Takahashi identity~\eqref{axwti}, which constrains a particular combination of the axialvector and pseudoscalar vertex.
       With its help we will take a quick detour to prove the Goldstone theorem in QCD.
       Recall the relation $f_\lambda \, m_\lambda^2 = 2m_q \,r_\lambda$
       from~\eqref{fm=2mr}, which is a simple consequence of the PCAC relation and holds for all pseudoscalar mesons.
       So far it tells us nothing about spontaneous chiral symmetry breaking: it only states that in the chiral limit ($m_q=0$) either the mass or
       the decay constant of a pseudoscalar meson vanishes. Hence, if we can show that the pion decay constant does \textit{not} vanish in the chiral limit (as a
       consequence of spontaneous chiral symmetry breaking), we must have a massless pion.
       To prove this, we insert the relation together with~\eqref{vertices-pole-contributions} into the axial WTI, which yields
       \begin{equation}
          Q^\mu \,\Gamma_5^\mu +2m_q\,\Gamma_5 = \sum_\lambda 2if_\lambda\,\mathbf\Gamma_\lambda + Q^\mu \,\widetilde\Gamma^\mu_5 + 2m_q\,\widetilde\Gamma_5 = S^{-1}(k_+)\,i\gamma_5 + i\gamma_5\,S^{-1}(k_-)\,.
       \end{equation}
       Observe that all hadronic poles contained in the vertices have disappeared, which is consistent because the right-hand side of the axial WTI does not exhibit any such poles.
       The pseudoscalar poles have canceled with the numerator and the axialvector poles drop out because they are transverse.
       In the limit $Q^\mu \rightarrow 0$ and $m_q\rightarrow 0$ this becomes the chiral-limit relation
       \begin{equation}\label{B/fpi-0}
           \sum_\lambda f_\lambda\,\mathbf\Gamma_\lambda(k,0) = B(k^2) \gamma_5\,.
       \end{equation}
       The sum goes over all pseudoscalar $0^{-+}$ mesons with identical flavour quantum numbers, i.e., ground states and radial excitations.
       In the chiral limit, $B(k^2)$ is only nonzero if chiral symmetry is spontaneously broken.
       If chiral symmetry were realized and $B(k^2)=0$, all combinations $f_\lambda\,\mathbf\Gamma_\lambda(k,0)$ would have to vanish as well;
       if it is spontaneously broken there is at least one mode with $f_\lambda \neq 0$. From~\eqref{fm=2mr} we must have $m_\lambda\rightarrow 0$ in that case, i.e.\ a massless Goldstone boson.

          In turn, the decay constants $f_\lambda$ vanish for the remaining excited states with $m_\lambda \neq 0$, so we can remove the sum in the equation above to arrive at the celebrated relation~\cite{Maris:1997hd}
       \begin{equation}\label{B/fpi}
           f_\pi\,\mathbf\Gamma_\pi(k,0) = B(k^2) \gamma_5\,.
       \end{equation}
       It states that the chiral-limit pion amplitude is entirely determined from the quark propagator.
       The remaining dressing functions do not vanish but they decouple from the physics because their tensor structures vanish in the chiral limit.
       Finally, taking the trace of~\eqref{B/fpi} with $S(k)\,\gamma_5\,S(k)$ and integrating over the momentum $k$ yields
        $ f_\pi \,r_\pi=  -\langle\overline\psi \psi\rangle/N_f$ in the chiral limit, and substituting this back
       into~\eqref{fm=2mr} gives us the Gell-Mann-Oakes-Renner relation:
       \begin{equation}
          f_\pi^2 \,m_\pi^2 = -2m_q\, \langle\overline\psi \psi\rangle/N_f\,.
       \end{equation}

\subsection{Methodological overview} \label{sec:ff-methods}

       In the following we give a brief overview of some of the theoretical tools used in the calculation of baryon form factors.
       Naturally, our own bias is towards the Dyson-Schwinger/Bethe-Salpeter approach and therefore we will mainly concentrate on pointing out its
       connections with other methods and the inherent similarities and differences.
       We apologize to those colleagues whose work is not mentioned here;
       for more comprehensive discussions we refer to the existing reviews focusing on nucleon spacelike electromagnetic form factors~\cite{Gao:2003ag,HydeWright:2004gh,Arrington:2006zm,Perdrisat:2006hj,Arrington:2011kb,Pacetti:2015iqa,Punjabi:2015bba},
       axial form factors~\cite{Bernard:2001rs,Schindler:2006jq}, transition form factors~\cite{Pascalutsa:2006up,Aznauryan:2011qj,Tiator:2011pw,Aznauryan:2012ba}, and timelike form factors~\cite{Denig:2012by,Pacetti:2015iqa}.

       \paragraph{Lattice QCD.} Nucleon form factor calculations in lattice QCD have made substantial progress in recent years.
       Whereas the early quenched calculations are summarized in the review articles~\cite{Arrington:2006zm,Hagler:2009ni}, nowadays
       essentially all calculations are performed using dynamical fermions.
       Studies of electromagnetic nucleon form factors are available from various collaborations for $N_f=2$~\cite{Alexandrou:2006ru,Lin:2008uz,Alexandrou:2011db,Collins:2011mk,Capitani:2015sba},
       $N_f=2+1$~\cite{Yamazaki:2009zq,Lin:2008mr,Lin:2010fv,Bratt:2010jn,Syritsyn:2009mx,Green:2014xba,Shanahan:2014uka,Shanahan:2014cga}
       and $N_f=2+1+1$ flavours~\cite{Alexandrou:2013joa,Bhattacharya:2013ehc},
       including calculations nearly at the physical pion mass~\cite{Green:2014xba} which agree reasonably well  with the experimental data.
       However, they are so far restricted to moderately low $Q^2$ and still carry sizeable statistical errors;
       for example, the necessary precision for an unambiguous determination of the proton's charge radius is not yet within reach.
       Among other nucleon properties of interest are the axial coupling constant $g_A$,
       which has been traditionally hard to reproduce on the lattice, as well as moments of structure functions and generalized
       parton distributions~\cite{Green:2012ud,Green:2014xba,Abdel-Rehim:2015owa,Gupta:2015tpa}.

	   Concerning form factors, most efforts up to now have concentrated on the ground state
       octet baryons in the positive parity sector.
       With the exception of the $\Delta$~\cite{Alexandrou:2008bn,Alexandrou:2010uk,Alexandrou:2013opa} and $\Omega$~\cite{Alexandrou:2010jv}, unquenched
       calculations of decuplet baryon form factors have not yet been performed to our knowledge. First results for Roper as well as negative-parity nucleon
       and $\Lambda(1405)$ form factors~\cite{Lin:2011da,Menadue:2013xqa,Owen:2014txa,Hall:2014uca} await
       confirmation from other groups.

       In principle, the systematic difficulties associated with the calculation of form factors on the lattice are the
       same as for the hadron spectrum discussed in Sec.~\ref{spec:extracting}: a noise-to-signal ratio that exponentially grows with time,
       finite-volume effects, contaminations from excited states and so on.
       For example, the three-point correlator in~\eqref{3-current-correlator-2} singles out the ground state matrix element
       for asymptotically large Euclidean time separations, but in practice
       one has to deal with the admixture from excited states whose magnitude is encoded in the overlaps~$r_\lambda$.
       Several methods have been developed to address the problem such as the summation method or
       employing two-state fits; see~\cite{Capitani:2015sba,Green:2014xba,Bhattacharya:2013ehc} for details.
       For form factors involving unstable particles such as the $\Delta$ or the Roper resonance, disentangling the physical from the scattering states becomes an additional obstacle.
       The $\Delta$ form factor studies available so far were obtained by fitting to plateaus,
       which is justified as long as the pion masses are sufficiently large and the $\Delta$ is a stable bound state, i.e., for $m_\pi > m_\Delta-m_N \sim 300$~MeV.
       For lighter pion masses the $\Delta$ acquires a width, and in a finite volume the $N\pi$ continuum becomes a discrete set of scattering states
       which in principle requires a generalization of the L\"uscher method to form factors~\cite{Luscher:1990ux,Agadjanov:2014kha}.

       Another issue in form factor calculations are disconnected quark-loop diagrams.
       The current can couple directly to one of the valence quarks but also to sea-quark loops which
       connect points all over the lattice (`all-to-all propagators'). The computational costs to determine them are extremely large
       and, consequently, disconnected diagrams have not been considered in the baryon sector until
       recently, see e.g.~\cite{Bali:2011ks,Abdel-Rehim:2013wlz,Green:2015wqa}. It turns out that they are important for many observables
       related to nucleon structure, giving contributions of more than ten percent to the total values. Others, like the
       electromagnetic form factors are almost insensitive to disconnected contributions~\cite{Abdel-Rehim:2013wlz} and
       their smallness has also been studied~\cite{Shanahan:2014tja} and confirmed~\cite{Green:2015wqa} in calculations
       of the strangeness content of nucleon form factors.

       \paragraph{Timelike properties.}
       An important aspect of form factors is their timelike structure at negative $Q^2$ which is governed by particle production off the photon; see~\cite{Denig:2012by,Pacetti:2015iqa} for reviews.
       Form factors are analytic functions in the physical sheet apart from a branch cut starting at the two-pion threshold $Q^2 < -4m_\pi^2$ and extending to minus infinity.
       Knowledge of their imaginary parts along the cut, i.e., their spectral functions, thus allows one to employ dispersion relations to determine them everywhere in the complex $Q^2$ plane.
       Unfortunately, experiments in the timelike region cannot measure the imaginary parts of individual form factors but only relative phases,
       and the `unphysical' window $-4m_\pi^2 < Q^2 < 0$ as well as the large timelike $Q^2$ domain are not directly accessible in experiments.
       Several successful parametrizations are based on vector-meson dominance models where form factors are expressed by a sum of vector-meson poles~\cite{Iachello:2004aq,Bijker:2004yu,Lomon:2012pn}.
       More sophisticated approaches aim to reconstruct the spectral function via the optical theorem from summing over intermediate states ($\pi\pi$, $K\conjg{K}$, etc.);
       the resulting parametrizations are fitted to the data and have been used to predict, e.g.,
       the proton charge radius~\cite{Lorenz:2012tm} -- which was found to agree with the muonic hydrogen result, cf.~Sec.~\ref{sec:cs-overview}.

       Timelike form factors have traditionally not received as much attention as their spacelike counterparts from the theoretical point of view, which is
       perhaps not surprising considering the experimental difficulties mentioned above.
       Timelike properties are also difficult to extract in lattice QCD~\cite{Meyer:2011um,Feng:2014gba} and
       they are hard~to reproduce in microscopic calculations
       because an infinite summation of gluons is required to dynamically~generate the characteristic meson poles as described in Sec.~\ref{sec:ff-vertices}.
       Dispersion theory makes it clear that the timelike and spacelike properties are intrinsically related,
       and in this respect it is curious that many quark models capable of providing a good spacelike description
       do not have such mechanisms implemented at all (with exceptions discussed further below).
       In any case, the experimental situation may change in the future with new precision data on timelike nucleon form factors expected from Novosibirsk, BESIII/Beijing and PANDA/FAIR~\cite{Denig:2012by},
       and timelike reactions are also promising tools to extract the properties of nucleon resonances~\cite{Agakishiev:2014wqa,Salabura:2014wla,Briscoe:2015qia}.

\newpage

\begin{figure*}[t]
            \centerline{%
            \includegraphics[width=0.95\textwidth]{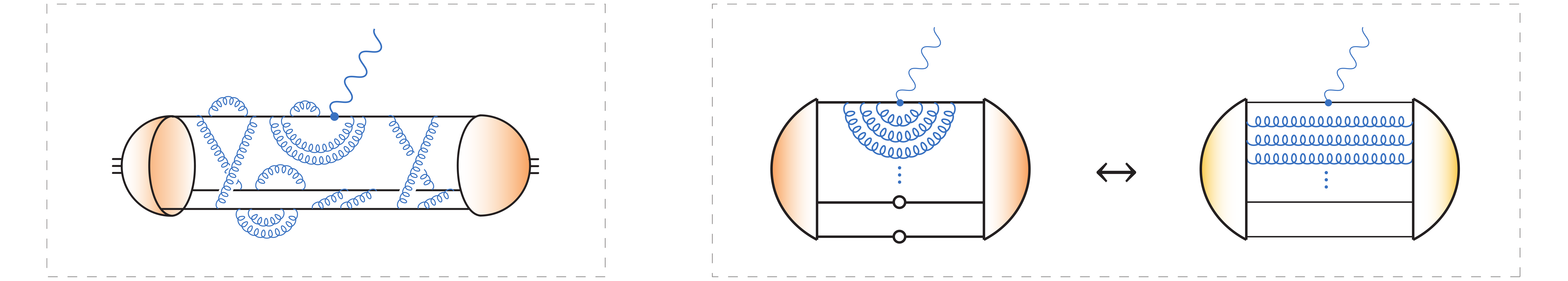}}
            \caption{\textit{Left:} Representative rainbow-ladder-like contribution to a current matrix element.
                     \textit{Right:} Timelike meson poles in the Bethe-Salpeter approach emerge from the dressed vertex through an infinite summation of gluons encoded in the vertex equation~\eqref{inhom-vertex-bse},
                    whereas in light-cone quantized QCD they would arise from summing over the light-cone wave functions of infinitely many Fock states.}
            \label{fig:ffs-rl}
            \end{figure*}

       \paragraph{Three-quark Bethe-Salpeter approach.}
       In contrast to lattice QCD,
       the goal in Dyson-Schwinger approaches is to reconstruct the current matrix element from the underlying Green functions in QCD.
       The relevant equations are given by Figs.~\ref{fig:faddeev} and~\ref{fig:current-faddeev} which can be solved if the ingredients are determined beforehand.
       In practice, truncations must be made: irreducible three-quark interactions have not  been implemented yet, and all form factor studies so far have employed a rainbow-ladder
       truncation where the $qq$ interaction is reduced to an effective gluon exchange.
       In that case the two rightmost diagrams in Fig.~\ref{fig:current-faddeev} disappear,
       and after settling on the form of the interaction the current matrix element is determined by solving the quark propagator, the baryon's Bethe-Salpeter amplitude and the quark-current vertex.
       The quark-gluon topologies arising from the iterative nature of these equations are exemplified by Fig.~\ref{fig:ffs-rl}.
       Since the $qq$ interaction is entirely fixed from the meson sector, one is equipped with a parameter-free calculation for baryons because no further approximations or assumptions have to be made.

       One can say that rainbow-ladder has performed relatively well for the form factors calculated so far, including
       the electromagnetic form factors of the nucleon~\cite{Eichmann:2011vu} and $\Delta$~\cite{Sanchis-Alepuz:2013iia} as well as the remaining octet and decuplet baryons~\cite{Sanchis-Alepuz:2015fcg},
       the $N\to\Delta\gamma$ transition~\cite{Alkofer:2014bya}, and the nucleon's axial and pseudoscalar form factors~\cite{Eichmann:2011pv} which we will review in the following sections.
       This owes to several features.
       Rainbow-ladder reproduces perturbation theory at large momenta, so the correct large-$Q^2$ behaviour of form factors is implicit.
       It preserves chiral symmetry through the vector and axialvector WTIs~(\ref{vwti}--\ref{axwti}),
       which translate into the consistency relations between the quark self-energy and the Bethe-Salpeter kernel discussed earlier in connection with Fig.~\ref{fig:axwti}.
       As long as symmetry-preserving kernels are employed in their BSEs, the resulting vertices satisfy their WTIs automatically, thus ensuring
       the structure of~\eqref{qpv-bc+t} for the quark-photon vertex which guarantees electromagnetic current and charge conservation.
       Similarly, the axial and pseudoscalar form factors satisfy the Goldberger-Treiman relation as a consequence of the axial WTI for the vertices.
       In addition, the  vertices calculated from their BSEs dynamically acquire meson poles which govern the timelike behaviour of the form factors;
       and as we discussed earlier, their impact on the spacelike domain is sizeable.
       Another noteworthy aspect is the implementation of the full covariant structure of the Bethe-Salpeter amplitudes as detailed in Sec.~\ref{spec:BS}, also including $p$ waves which
       naturally appear in a covariant formulation. They have several important consequences for observables as they
       provide the major fraction of orbital angular momentum. As discussed in connection with Table.~\ref{n-structure}, $p$ waves contribute roughly
       $30\%$ to the nucleon and $60\%$ to the Roper, and they turn out to be important for the
       $N\to\Delta\gamma$ transition as well~\cite{Eichmann:2011aa}.

       The main missing effects in these calculations can be attributed to pion-cloud corrections which are important at low $Q^2$ and for quark masses close to the chiral limit.
       The nucleon's interaction with its pion cloud leads to logarithmically divergent charge radii in the chiral limit and the decay of vector mesons into pions is important for timelike form factors.
       In principle, pion cloud effects would correspond to complicated quark-gluon topologies
       (resummation of gluons in crossed channels, production of decay channels etc.) and thus they are naturally absent in rainbow-ladder calculations.
       One possible avenue is to implement them through effective pion exchanges in the quark DSE and the Bethe-Salpeter kernel,
       which originate from the $q\bar{q}$ four-point function that enters in the quark-gluon vertex DSE~\cite{Fischer:2007ze,Fischer:2008wy,Sanchis-Alepuz:2014wea} as discussed in Sec.~\ref{spec:approx}.
       Their consequences have been investigated for meson and baryon spectra but their implementation in form factors remains a task for the future.
       In comparison to lattice QCD, rainbow-ladder bears some similarities to quenched QCD although the analogy has its limitations:
       quark loops in the gluon propagator are absorbed by the effective interaction whereas those appearing in higher $n$-point function are missing.
       The same can be said for the disconnected diagrams discussed earlier which contribute to isoscalar quantities;
       corresponding efforts in the DSE/BSE community are so far limited to the flavour-singlet meson sector~\cite{Alkofer:2008et,Bhagwat:2007ha}.

       While first steps beyond rainbow-ladder have been made  for the baryon spectrum ~\cite{Sanchis-Alepuz:2014wea,Sanchis-Alepuz:2015tha},
       calculating form factors poses an additional challenge due to the additional diagrams in Fig.~\ref{fig:current-faddeev}
       which increase the conceptual and numerical effort drastically.
       For example, the implementation of diagrammatic expression for the kernel such as in Figs.~\ref{fig:kernel3PIexample} or~\ref{fig:DSEwopions2}  has become feasible for spectrum
       calculations, but to obtain form factors one still needs to determine the coupling of the kernel to the photon and calculate the quark-gluon-photon vertex, etc.
       Such efforts have not yet been undertaken and remain a task for the future.
       To this date, the existing rainbow-ladder studies are therefore still the most sophisticated Dyson-Schwinger form factor calculations available.

            \begin{figure*}[t]
            \centerline{%
            \includegraphics[width=0.97\textwidth]{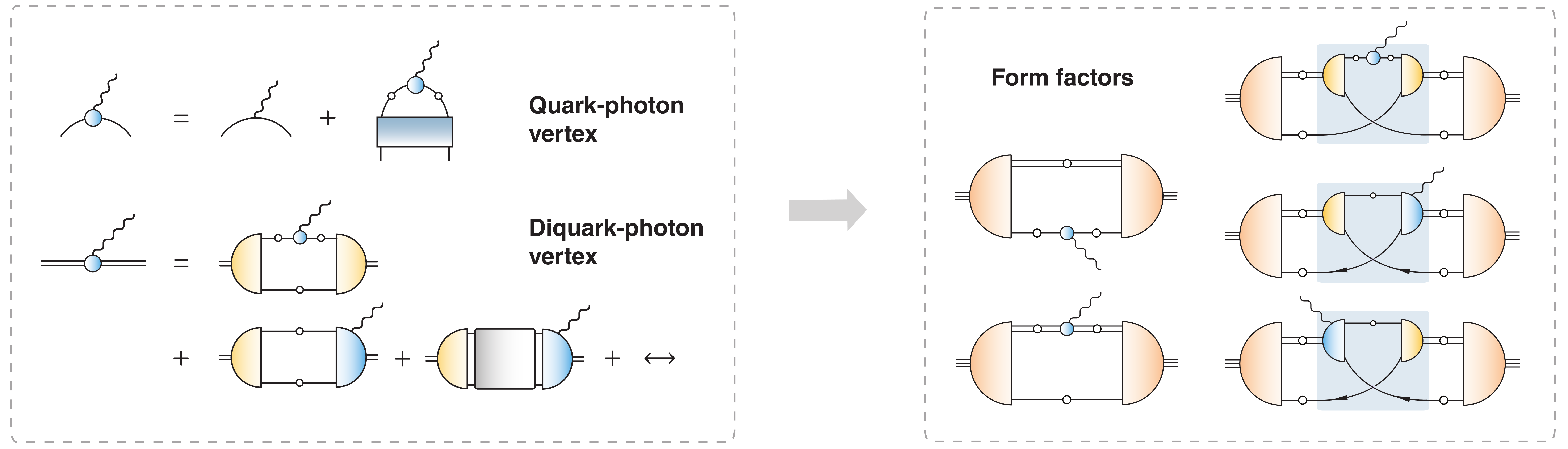}}
            \caption{Current matrix element in the quark-diquark approach (\textit{right panel}).
                     The quark propagator, diquark amplitudes and diquark propagators form the input of the quark-diquark BSE, cf.~Fig.~\ref{fig:quark-diquark}, and
                     the current matrix element needs in addition the quark-photon vertex, the diquark-photon vertices and the seagull amplitudes.
                     The ingredients can be calculated from their own equations in rainbow-ladder truncation, as shown in the left panel.}
            \label{fig:ff-quark-diquark}
            \end{figure*}

       \paragraph{Quark-diquark approaches.}

       As discussed in Sec.~\ref{spec:BS}, the quark-diquark model follows from the three-body equation upon neglecting irreducible three-body interactions
       and assuming that the $qq$ scattering matrix factorizes into diquark correlations.
       The quark-diquark results for nucleon and $\Delta$ masses are similar to their three-body analogues and thus one may ask how well the approximation holds up for form factors.
       The electromagnetic current in the quark-diquark model was constructed
and then applied in Refs.~\cite{Oettel:1999gc,Oettel:2000jj} and is depicted in
       the right panel of Fig.~\ref{fig:ff-quark-diquark}:
       it is the sum of impulse-approximation diagrams, where the photon couples to the quarks and diquarks, and the `gauged' kernel where it couples
       to the exchanged quark and the scalar and axialvector diquark Bethe-Salpeter amplitudes. As before, the sum of all diagrams is electromagnetically gauge invariant.
       While the resulting expressions are simpler to handle in practice, they also depend on quite a number of building blocks
       and therefore the associated systematic model uncertainty is typically larger.

       In Refs.~\cite{Eichmann:2007nn,Eichmann:2008ef,Eichmann:2009zx,Eichmann:2016jqx} an effort was made to calculate these ingredients systematically from the quark level, employing the same rainbow-ladder quark-gluon interaction as above.
       This is visualized in the left panel of Fig.~\ref{fig:ff-quark-diquark}: in addition to the quark propagator and quark-photon vertex,
       also the diquark amplitudes and propagators were calculated, together with resolving
       the scalar, axialvector and scalar-axialvector transition diquark-photon vertices. The seagull amplitudes were
       constructed from their Ward-Takahashi identities involving the full diquark tensor structure and supplemented by a prescription for their transverse part.
       In that way one arrives again at an essentially parameter-free calculation for baryon form factors, now within the quark-diquark approach, where all building blocks react consistently
       to a change of the underlying quark-gluon interaction.
       So far the approach has been applied to the calculation of nucleon and $\Delta$ electromagnetic form factors~\cite{Eichmann:2009zx,Eichmann:2010je,Nicmorus:2010sd},
       the $\gamma N\to\Delta$ transition form factors~\cite{Eichmann:2011aa}
       and the $N\to\Delta\pi$ transition~\cite{Mader:2011zf}.
       It turns out that also here the overall results are similar to the three-body calculations (where available) although
       the details may differ. An example discussed below in Sec.~\ref{form-res-delta} concerns the $\Delta$ electromagnetic form factors, where Fig.~\ref{fig:delta-ffs} shows the
       comparison between the three-quark and quark-diquark results.
       Similar findings apply to the nucleon elastic form factors, so that aside from details
       one can say that the quark-diquark approximation also works reasonably well for form factors.

       An advantage of the Dyson-Schwinger approach is that it offers the opportunity for simplifications.
       Electromagnetic gauge invariance still necessitates the solution of the quark-diquark BSE and the inclusion of all diagrams in the current matrix element of Fig.~\ref{fig:ff-quark-diquark}.
       However,  instead of determining the various ingredients from their own equations, for instance in a rainbow-ladder truncation,
       one may equally employ model parametrizations for the quark and diquark propagators, amplitudes and photon vertices.
       Extending early quark-diquark form factor calculations~\cite{Hellstern:1997pg,Oettel:1999gc,Oettel:2000jj,Bloch:1999rm,Bloch:2003vn,Oettel:2002wf,Alkofer:2004yf},
       this is the avenue taken in the QCD-based model developed in Refs.~\cite{Cloet:2008re}
       where the parameters were adjusted to reproduce the nucleon and $\Delta$ masses (potentially leaving room for pion cloud effects) and to some extent also their electromagnetic form factors.
      This makes it possible to study the effects of the individual diquark components
      and investigate the response to changing ingredients such as the quark mass function~\cite{Cloet:2013gva,Segovia:2015ufa}.
      Here the quarks and diquarks are still momentum-dependent (i.e., non-pointlike) and modelled using realistic parametrizations.
      The photon vertices are constructed from their Ward-Takahashi identities,
      together with transverse parts either based on $\rho$-meson pole ans\"atze or implementing a quark anomalous magnetic moment~\cite{Cloet:2013jya}.
      Several studies are available for the nucleon electromagnetic form factors~\cite{Cloet:2008re} as well as for the $\Delta$ electromagnetic
      and $\gamma N\to\Delta$ transition form factors~\cite{Segovia:2014aza}, among other observables; they are summarized in Refs.~\cite{Bashir:2012fs,Cloet:2013jya,Segovia:2014aza}.
      Recently the approach was applied to calculate the nucleon-to-Roper transition form factors~\cite{Segovia:2015hra} which, using the established inputs for the ingredients,
      are then genuine predictions.

      Further insight can be gained by consulting
      the simplest version of the Dyson-Schwinger approach retaining its basic features, namely the NJL model.
      In that case the quark-diquark equation in Fig.~\ref{fig:quark-diquark} and the current matrix element in Fig.~\ref{fig:ff-quark-diquark} remain the same,
      but since the gluon propagator is now a constant the quark and diquark propagators become constituent propagators,
      the diquark Bethe-Salpeter amplitudes become point-like and the seagull contributions to the form factors vanish.
      In that way one arrives again at a systematic description starting at the level of the quark propagator
      since the model parametrizations discussed above are eliminated in favor of self-consistent Dyson-Schwinger and Bethe-Salpeter solutions, however with reduced momentum dependencies.
      For example, the rainbow-ladder truncation becomes a vector-vector contact interaction and
      in the quark-diquark equation an additional static approximation for the exchange quark propagator is made.
      The NJL model was recently employed to calculate elastic nucleon and hyperon form factors~\cite{Cloet:2014rja,Carrillo-Serrano:2016igi},
      building upon a body of work on other structure properties such as parton distribution functions and nuclear medium effects; see also~\cite{Cloet:2015tha} and references therein.
      Closely related is the contact-interaction model developed in Refs.~\cite{GutierrezGuerrero:2010md,Roberts:2011cf} which requires some modifications to the kernels to make it applicable to excited states;
      it has since been applied to nucleon, $\Delta$ and Roper elastic and transition form factors~\cite{Wilson:2011aa,Segovia:2013rca,Segovia:2013uga}.
      A compilation of results together with a comparison with the quark-diquark model using momentum-dependent ingredients can be found in~\cite{Segovia:2014aza}.

       \paragraph{Quark models.}
       Turning from quantum field theoretical to quantum mechanical approaches, we continue with a brief overview on quark models.
       There is a long history of form factor calculations in quark models, beginning with the early non-relativistic constituent-quark models~\cite{DeRujula:1975qlm,Isgur:1978xj}.
       The non-relativistic approaches turned out to be problematic since form factors acquire relativistic recoil corrections at nonzero $Q^2$ which already affect their charge radii.
       The main question in `relativizing' quark models is then how the wave function obtained from a given Hamiltonian potential changes when going from the rest frame to the moving frame.
       This can depend on the interactions and leads to the distinction between instant, point and light-front forms,
       where certain generators of the Poincar\'e group remain `kinematical' and others become dynamical and interaction-dependent~\cite{Dirac:1949cp}; see~\cite{Punjabi:2015bba} for a pedagogical overview.
       The point and light-front forms are advantageous for form factor calculations because they allow one to boost quark wave functions
       independently of the details of the interaction. In the former the whole Lorentz group is kinematical
       whereas the latter has the maximum number of kinematical generators, including longitudinal boosts and light-front transverse boosts.

       The typical strategy is then to start with a wave function that is either modelled or obtained from a quark model designed to study the baryon spectrum,
       obtain its point or light-front form by appropriate transformations, and calculate the current matrix element from the overlap of two wave functions
       in combination with a quark current. The currents are
       either pointlike or dressed with Dirac and Pauli form factors, potentially based on chiral dynamics or vector-meson dominance.
       In the case of light-front constituent quark models this can be viewed as a Fock-space truncation of QCD in light-cone quantization,
       where hadronic matrix elements do have exact representations in terms of light-cone wave functions~\cite{Brodsky:1997de}.
       In the Drell-Yan-West frame where $q^+=0$, the contribution from pair creation or annihilation is forbidden and form factors
       can be expressed as overlaps of light-cone wave functions with the same number of Fock space constituents~\cite{Drell:1969km,West:1970av,Brodsky:1980zm}, which is also illustrated in~Fig.~\ref{fig:ffs-rl}.

       Nucleon form factors have been calculated in light-front constituent-quark models~\cite{Chung:1991st,Schlumpf:1992vq,Schlumpf:1992pp,Frank:1995pv,Miller:2002qb,Ma:2002ir},
       also including dressed quark form factors which effectively add higher Fock components to the wave functions~\cite{Cardarelli:1995dc,Pace:1999as,Cardarelli:2000tk,deMelo:2008rj}.
       Light-front quark model calculations are also available for nucleon resonances~\cite{Capstick:1994ne,Aznauryan:2012ec,Obukhovsky:2013fpa}.
       Form factor calculations in the point form approach have been performed within the chiral quark model based on Goldstone-boson exchange~\cite{Wagenbrunn:2000es,Glozman:2001zc,Boffi:2001zb,Rohrmoser:2011tw},
       the hyperspherical constituent-quark model~\cite{DeSanctis:2005kt} and a recent quark-diquark model~\cite{DeSanctis:2011zz}.
       The Goldstone-boson exchange model achieved remarkably good results for a range of baryon properties
       also beyond nucleon form factors~\cite{Choi:2009pw,Choi:2010ty,Melde:2008dg,Plessas:2015mpa} using pointlike constituent quarks only.
       The comparative calculation of Ref.~\cite{JuliaDiaz:2003gq}
       in all three forms of dynamics demonstrated that this is not necessarily exclusive to the point form but rather depends on the spatial extent of the wave functions.
       A simple algebraic ansatz of the wave function allowed for a good combined description of nucleon electromagnetic, axial and transition form factors, however with very different parameters:
       instant and front form demand a spatially extended wave function whereas in point-form kinematics it can be narrow.

       A formally covariant calculation was performed using the Bethe-Salpeter equation with instantaneous forces and an instanton-induced interaction~\cite{Merten:2002nz}.
       In that case all model parameters were fixed in the calculation of the baryon mass spectrum, and the resulting nucleon electroweak and transition form factors
       are in qualitative agreement with experiment.
       Another covariant approach is the covariant spectator model of Ref.~\cite{Gross:2006fg},
       which has been actively explored for a range of observables including various elastic and transition form factors of
       octet and decuplet baryons~\cite{Ramalho:2008ra,Ramalho:2008dp,Ramalho:2010js,Ramalho:2010xj,Ramalho:2011ae,Ramalho:2013uza}; a recent overview can be found in Ref.~\cite{Ramalho:2016ada}.
       Here the baryon wave functions are not determined dynamically but instead modelled based on internal symmetries,
       and the current matrix elements are calculated in the impulse approximation.
       Despite this, the approach is perhaps the most similar one with respect to the Dyson-Schwinger form factor calculations mentioned above, and in turn it
       has been applied to a large number of observables including also timelike transition form factors~\cite{Ramalho:2015qna}.

       A possible strategy to improve quark models is to include pion cloud effects.
       The interaction of baryons with their light meson cloud is a natural consequence of QCD's spontaneous chiral symmetry breaking.
       Pions are light and abundantly produced and therefore they modify the structure of baryons at large distances.
       Since the early days of the cloudy bag model~\cite{Theberge:1980ye,Thomas:1981vc}
       it has become a rather popular view that baryons are composed of a short-range three-quark `core' that is augmented with long-range meson cloud effects.
       Several models have been constructed in this spirit to implement the coupling of quarks to the pion and applied to nucleon form factors,
       from the cloudy bag model~\cite{Lu:1996rj,Lu:1997sd,Miller:2002ig}, the chiral quark models of Refs.~\cite{Oset:1984tva,Jena:1992qx,Faessler:2005gd,Faessler:2006ky},
       the chiral quark soliton model~\cite{Christov:1995hr,Christov:1995vm,Holzwarth:1996xq,Weigel:1995jc}, and the light-front quark-diquark model of Ref.~\cite{Cloet:2012cy}.
       More generally, this ties into the question to what extent baryons can be described as three-quark states and what percentage owes to molecular components which
       may even generate new resonances dynamically.

       In summary one can say that many models are able to reproduce the qualitative and in several cases also the quantitative behaviour of form factors.
       It is then the question whether one prefers a pointwise description of experimental data or genuine insight in the underlying dynamics, which requires at least some connection to QCD.
       Dyson-Schwinger approaches are defined in quantum field theory,
       but the inherent possibility of making approximations and the formal similarity of the diagrams in Figs.~\ref{fig:current-faddeev} or~\ref{fig:ff-quark-diquark}
       to quark models allows one to identify common features.
       For example, one could retain the impulse approximation only, although this would break electromagnetic gauge invariance (and gauge invariance is indeed not automatic in quark models).
       One could simplify the Bethe-Salpeter amplitudes to their leading components and
       set the quarks onshell as in the NJL model; in that case the quark-photon vertex becomes the sum of a pointlike vertex together with a transverse $\rho-$meson part obtained from its bubble sum.
       With pointlike quarks only one would also lose the vector-meson poles, but as exemplified by the hadronic vacuum polarisation in Fig.~\ref{fig:hvp}
       the vector-meson tails typically extend to large $Q^2$. Even if spacelike form factors could be reproduced in such a setup it would still fail spectacularly in the timelike region --
       but as we argued above, the spacelike and timelike properties cannot truly be separated from each other.
       From this perspective one should perhaps begin to measure the quark model landscape by more rigorous standards,
       although one has to acknowledge that the amount of insight they have provided so far is indeed remarkable.

\subsection{Nucleon electromagnetic form factors}\label{form-res-octet}

       In the following subsections we discuss form factors of the ground-state nucleon, $\Delta$ and other octet and decuplet baryons on a case-by-case basis.
       We report on results obtained in the Dyson-Schwinger/Bethe-Salpeter framework and contrast
       them with experimental data and lattice results where appropriate.
       We start with the electromagnetic and axial form factors
       for the nucleon, continue with the $\Delta$ baryon, discuss selected examples for nucleon transition form factors such as the $\gamma N\to\Delta$ transition and $\Delta\to N\pi$ decay, and finally
       conclude with elastic hyperon form factors.

       \paragraph{Electromagnetic current.}
       The electromagnetic form factors of hadrons parametrize the matrix elements of the electromagnetic current in~\eqref{current-em}.
            A spin-$\nicefrac{1}{2}$ particle has two electromagnetic form factors, the Dirac and Pauli form factors $F_1(Q^2)$ and $F_2(Q^2)$,
            which are encoded in the electromagnetic current matrix element
            \begin{equation}\label{ffs-nucleon-current-1}
                \mc{J}^\mu(p,Q) = i \conjg{u}(p_f) \left( F_1(Q^2)\, \gamma^\mu + \frac{iF_2(Q^2)}{4m}\,[\gamma^\mu,\slashed{Q}]\right) u(p_i)\,.
            \end{equation}
            Here (and in the following subsections) $p_i$ and $p_f$ denote the incoming and outgoing momenta, $Q=p_f-p_i$ is the four-momentum transfer, $p=(p_f+p_i)/2$ is the average momentum, and $m$ is the nucleon mass.
            Since both nucleons are onshell only $Q^2$ remains an independent variable: $p_f^2 = p_i^2 = -m^2$ and therefore $p\cdot Q = 0$ and $p^2 = -m^2 \,(1+\tau)$, with $\tau=Q^2/(4m^2)$.
            The nucleon spinors are solutions of the Dirac equation and thus eigenspinors of the positive-energy projector:\footnote{In practice
            it is often more convenient to replace the spinors by the positive-energy projectors and work with a matrix-valued current $\mc{J}^\mu(p,Q) = i\Lambda_+(p_f) \left( \dots \right) \Lambda_+(p_i)$.
            The Sachs form factors are then extracted from the traces
            $G_E(Q^2) = \text{Tr}\,\big\{ \mc{J} \cdot \hat{p}  \big\}/(2i\sqrt{1+\tau})$ and
            $G_M(Q^2) = \text{Tr} \,\big\{  \mc{J}^\mu \gamma^\mu_\perp  \big\} \,i/(4\tau)$, where $\gamma^\mu_\perp$ is transverse to the average momentum $p$.
            In the Breit frame one has for example $\hat{Q}^\mu=e_3^\mu$, $\hat{p}^\mu = e_4^\mu$ and therefore $\mc{J} \cdot \hat{p} = \mc{J}^4$ and $ \mc{J}^\mu \gamma^\mu_\perp = \vect{\mc{J}}\cdot\vect{\gamma}$.\label{footnote-sachs-ffs}}
            \begin{equation}
               \Lambda_+(p)\,u(p) = u(p)\,, \qquad \conjg{u}(p)\,\Lambda_+(p) = \conjg{u}(p)\,, \qquad \Lambda_+(p) = \tfrac{1}{2} (\mathds{1} + \hat{\slashed{p}})\,,
            \end{equation}
            where $\hat{p}^\mu$ denotes a normalized four-momentum, i.e.\ $\hat{p}^\mu = p^\mu/\sqrt{p^2} = p^\mu/(im)$
            for an onshell momentum.

             The nucleon's Dirac and Pauli form factors are related to the electric and magnetic Sachs form factors $G_E(Q^2)$ and $G_M(Q^2)$ via
              \begin{equation} \label{ff:sachs}
                      G_E = F_1 - \tau F_2\,, \qquad
                      G_M = F_1 + F_2\,, \qquad \tau=\frac{Q^2}{4m^2}\,.
              \end{equation}
             The Sachs form factors are experimentally more convenient because they diagonalize the Rosenbluth cross section, cf.~Eq.~\eqref{rosenbluth-cs}:
        \begin{equation}\label{rosenbluth-1}
           \frac{d\sigma}{d\Omega} = \left(\frac{d\sigma}{d\Omega}\right)_\text{Mott}\, \frac{\varepsilon\,G_E^2 + \tau\,G_M^2}{\varepsilon\,(1+\tau)}\,, \qquad
           \varepsilon = \left( 1 + 2\,(1+\tau)\,\tan^2\frac{\theta}{2}\right)^{-1},
        \end{equation}
        where $\theta$ is the laboratory scattering angle.
        The Mott cross section describes lepton scattering off a pointlike scalar particle and the deviation from it
        accounts for the nucleon's spin $\nicefrac{1}{2}$
        together with its composite nature: for a pointlike proton, $G_E=G_M=1$.

       \paragraph{Form factor phenomenology.}
           The Dirac and Pauli form factors at vanishing photon momentum transfer $Q^2=0$ encode the proton and neutron charges and their anomalous magnetic moments:
           \begin{equation*}
               F_1^p(0) = 1\,, \qquad
               F_1^n(0) = 0\,, \qquad
               F_2^p(0) = \kappa_p = 1.79\,, \qquad
               F_2^n(0) = \kappa_n = -1.91\,.
           \end{equation*}
           Their slopes at $Q^2=0$ are related to the Dirac and Pauli `charge radii':
           \begin{equation}\label{charge-radii}
               F_1(Q^2) = F_1(0) - \frac{r_1^2}{6}\,Q^2 + \dots\,, \qquad
               F_2(Q^2) = F_2(0) \left[ 1 - \frac{r_2^2}{6}\,Q^2 + \dots \right],
           \end{equation}
           which correspond to the standard definition of a charge radius through the slope of the form factor at vanishing momentum transfer:
           \begin{equation}\label{charge-radii-2}
              r_1^2 = -6\,\frac{dF_1(Q^2)}{dQ^2}\bigg|_{Q^2=0}\,, \qquad
              r_2^2 = -\frac{6}{F_2(0)}\,\frac{dF_2(Q^2)}{dQ^2}\bigg|_{Q^2=0}\,.
           \end{equation}

       \begin{figure*}[t]
                    \begin{center}

                    \includegraphics[width=0.75\textwidth]{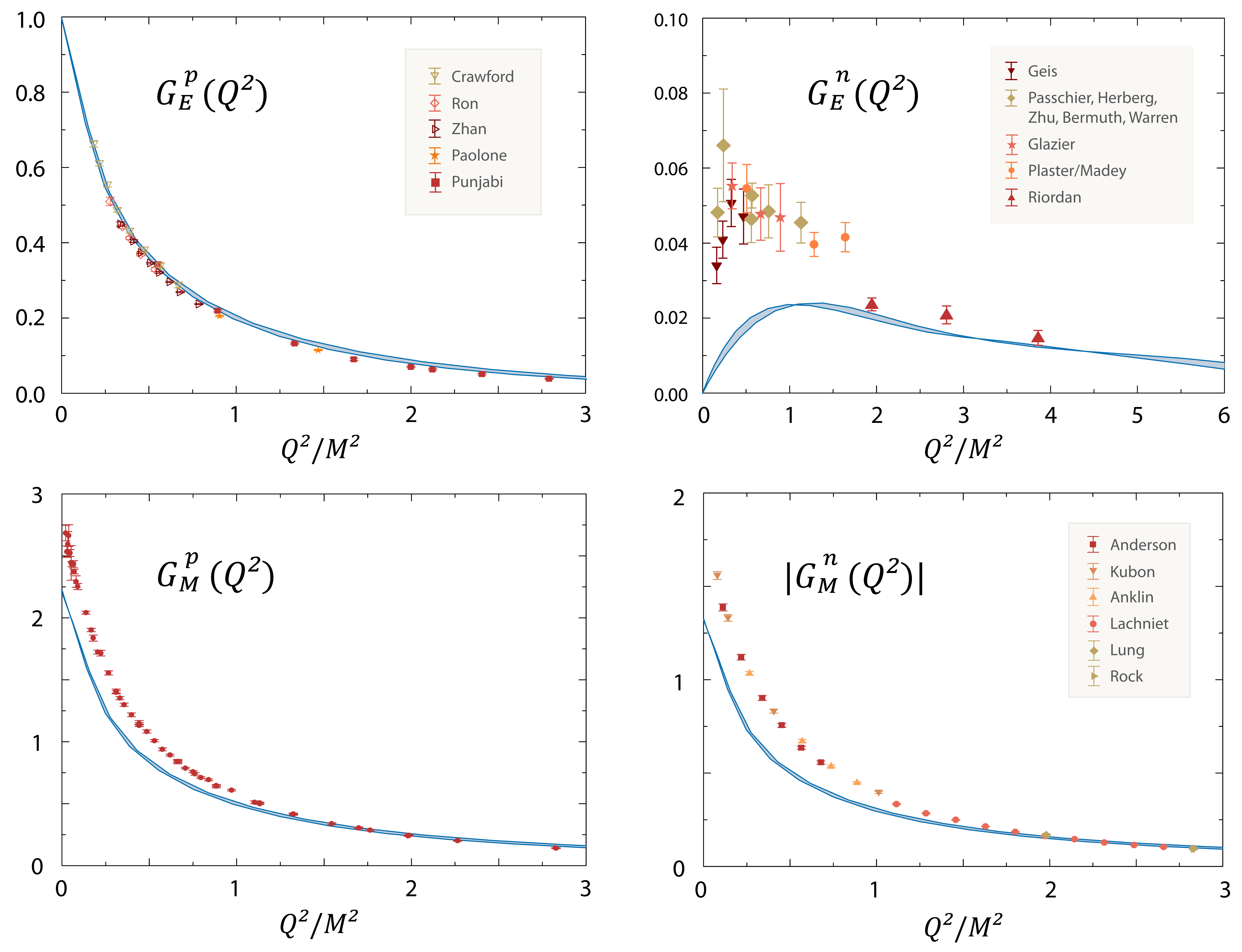}   %
                    \caption{Three-body Dyson-Schwinger results for the nucleon's electromagnetic form factors as a function of the photon momentum transfer $Q^2$~\cite{Eichmann:2011vu}.
                    The references for the experimental data can be found therein.} \label{fig:sachs-ffs}

                    \end{center}
        \end{figure*}

       The Sachs form factors $G_E$ and $G_M$ play a special role because in the Breit frame the temporal component of the vector current matrix element $\mc{J}^\mu(p,Q)$ is proportional to $G_E$
       and its spatial component to $G_M$, see footnote~\eqref{footnote-sachs-ffs}; hence the name `electric' and `magnetic' form factors.
           They contain the charges and magnetic moments
           \begin{equation*}
               G_M^p(0) = \mu_p = 1+\kappa_p = 2.79\,, \qquad
               G_M^n(0) = \mu_n = \kappa_n = -1.91\,,
           \end{equation*}
           and the electric and magnetic Sachs radii of proton and neutron ($r_E^{p,n}$ and $r_M^{p,n}$) are defined accordingly.
           Empirically it turns out that the magnetic form factors are reasonably well described by a dipole form over a wide range of $Q^2$,
           which also agrees with the (naive) scaling predictions of perturbative QCD~\cite{Brodsky:1974vy}.
           For this reason the form factors are frequently divided by the dipole to magnify their deviations from it:
           \begin{equation}\label{ff-dipole}
               G_D(Q^2) =    \frac{1}{( 1 +Q^2/\Lambda^2)^2}\,, \qquad \Lambda = 0.84\,\text{GeV}\,.
           \end{equation}
           The charge radius of a pure dipole form factor is given by
           $r_i = \hbar c  \,\sqrt{12}/\Lambda \sim 0.8\,\text{fm}$,
           with $\hbar c = 0.197$ GeV fm, which already provides a crude estimate for the magnetic charge radii of proton and neutron.
           The electric form factors, on the other hand, have not yet been measured up to quite as high values of $Q^2$ and also behave differently.
           $G_E^n$ vanishes at the origin because the neutron carries no charge,
           whereas $G_E^p$ deviates from the dipole behaviour and even points towards a zero crossing at larger $Q^2$ (to which we will return below).
           The electric charge radius $r_E^p$ of the proton is also subject to the ongoing proton radius puzzle which is discussed in more detail in Sec.~\ref{sec:cs-overview}.
           The experimental data for the electromagnetic Sachs form factors and their dipole ratios are shown in Figs.~\ref{fig:sachs-ffs} and~\ref{fig:nucleon-emffs-ratios}.

       Form factors are Lorentz invariant and uniquely specify the electromagnetic structure of a hadron,
       but their interpretation can depend on the reference frame.
       Their traditional non-relativistic interpretation is that of
       three-dimensional Fourier transforms
       of charge distributions in the Breit frame, where the energy transfer from the photon vanishes and thus $Q^2=-\vect{q}^2$.
       Assume that we can write a generic form factor $F(Q^2)$ as
       \begin{equation}
          F(Q^2) = \int d^3x\,\rho(r)\,e^{i\vect{q}\cdot\vect{x}}
          = 4\pi\int dr\,r^2 \rho(r)\,\frac{\sin |\vect{q}|  r}{|\vect{q}|  r}\,,
       \end{equation}
       where $r=|\vect{x}|$ and $\rho(r)$ is a spherically symmetric charge distribution. In that way a pointlike charge
       produces a constant form factor, an exponential charge distribution corresponds to a dipole form factor,
       and a homogeneous sphere (similar to the typical charge densities of heavy nuclei) produces an oscillating form factor.
       The coefficients in a Taylor expansion at small $|\vect{q}|$ define the charge and the mean-square charge radius:
       \begin{equation}
          F(Q^2) = 4\pi\int dr\,r^2 \rho(r) \left[ 1 - \frac{1}{6}\,|\vect{q}|^2  r^2 + \dots\right]
          = 1 - \frac{1}{6}\,|\vect{q}|^2 \langle r^2\rangle + \dots\,,
       \end{equation}
       which motivates the definitions~(\ref{charge-radii}--\ref{charge-radii-2}).
       Plugging in the measured Sachs form factors then gives a positive charge density for the proton and a neutron density that is
       positive at its core with a negative long-range tail, which has led to the picture of the neutron  sometimes being a proton surrounded by a pion cloud~\cite{Kelly:2002if}.

       Unfortunately these relations only hold non-relativistically because relativistic boost corrections enter with $Q^2/m^2$ and obscure the interpretation
       in terms of charge densities. Such effects are negligible as long as the binding energies are tiny, which is still the case for nuclei, but the typical masses
       of hadrons are small enough to induce corrections at all values of $Q^2$ and even affect their charge radii~\cite{Miller:2009sg}.
       In this sense, the Sachs form factors are not directly related to charge distributions and the definitions~\eqref{charge-radii-2} only reflect a generic measure for the size of a hadron.
       On the other hand, an unambiguous interpretation can be found in the infinite momentum frame: there, the Dirac form factor $F_1(Q^2)$ is the
       two-dimensional Fourier transform of the transverse charge distribution, and the resulting transverse charge densities
       are positive for the proton and negative for the neutron at the center~\cite{Burkardt:2000za,Miller:2007uy,Miller:2008jc}.

       \paragraph{Experimental status.}
       The proton's electromagnetic form factors are experimentally measured in elastic $e^- p$ scattering, whereas those of the neutron are extracted from
       electron-deuteron or electron-helium scattering due to the lack of a free neutron target in nature.
       The current world data for $G_E^p$, $G_E^n$, $G_M^p$ and $G_M^n$ extend up to photon virtualities $Q^2$ of about $8.5$ GeV$^2$, $3.5$ GeV$^2$, $30$~GeV$^2$ and $10$~GeV$^2$, respectively; see~\cite{Puckett:2011xg} for a compilation.
       Traditionally they have been extracted under the assumption of one-photon exchange
       via the Rosenbluth cross section in~\eqref{rosenbluth-1}.
       It came as quite a surprise when the polarisation transfer measurements at Jefferson Lab revealed a ratio $G_E^p/G_M^p$
       that falls off and points towards a zero crossing  at larger $Q^2$~\cite{Jones:1999rz,Gayou:2001qd,Punjabi:2005wq,Puckett:2010ac},
       in stark contrast to the Rosenbluth measurements which had predicted a constant ratio.
       The question of the zero crossing for $G_E^p$ is not yet settled because the experimental errors are still too large, but
       the situation is expected to change in the near future with the Jefferson Lab 12 GeV program where measurements in Halls A, B and C
       will extend the data range for $G_E^p$, $G_E^n$, and $G_M^n$ up to the $10$~--~$14$~GeV$^2$ region~\cite{Dudek:2012vr}.
       It is established by now that two-photon exchange corrections can explain the discrepancy in principle~\cite{Guichon:2003qm,Arrington:2007ux},
       although the effect is still not completely understood because it cannot be calculated model-independently
       and the existing model calculations rely on different underlying mechanisms.
       More surprises have also been found at low $Q^2$, where the present data for $G_E^p$ are incompatible with the muonic hydrogen result for the proton's charge radius.
       We will discuss these points in more detail in Sec.~\ref{compton}, but what they clearly tell us
       is that our knowledge of the nucleon's electromagnetic structure is still not very well understood.

       \paragraph{Dyson-Schwinger calculations.}
       With precise experimental data available, nucleon form factors provide a stringent test for theory calculations whose current status has been discussed in Sec.~\ref{sec:ff-methods}.
       In Fig.~\ref{fig:sachs-ffs} we show the results obtained from the three-body Dyson-Schwinger calculation in Ref.~\cite{Eichmann:2011vu}, whose
       only input is the effective quark-gluon interaction of~\eqref{couplingMT} and no further approximations are made.
       That is, all ingredients of the current matrix element in Fig.~\ref{fig:current-faddeev}
       are calculated self-consistently from their rainbow-ladder Dyson-Schwinger and Bethe-Salpeter equations,
       and the bands show the sensitivity to the model dependence of the interaction.

       The plots show that the behaviour at larger $Q^2$ is reasonably well reproduced,
       whereas the results at low $Q^2$ exhibit discrepancies. The electric proton form factor agrees with the data
       but the bump in the electric neutron form factor is absent, and both magnetic form factors of proton and neutron
       are underestimated. As demonstrated in Ref.~\cite{Eichmann:2011vu}, this can be attributed to the absence of chiral
       meson-cloud effects. The leading-order chiral corrections estimated from heavy-baryon chiral effective field theory
       are compatible with the discrepancies for the magnetic moments in Fig.~\ref{fig:sachs-ffs}, which suggests an interpretation of rainbow-ladder
       as the `quark core' of the nucleon that is stripped of its pion cloud. Qualitatively similar features have been found in
       quark-diquark approaches, using either model inputs~\cite{Cloet:2008re} or self-consistently calculated ingredients~\cite{Eichmann:2009zx}.

\begin{figure}[t]
                    \begin{center}

                    \includegraphics[width=1\textwidth]{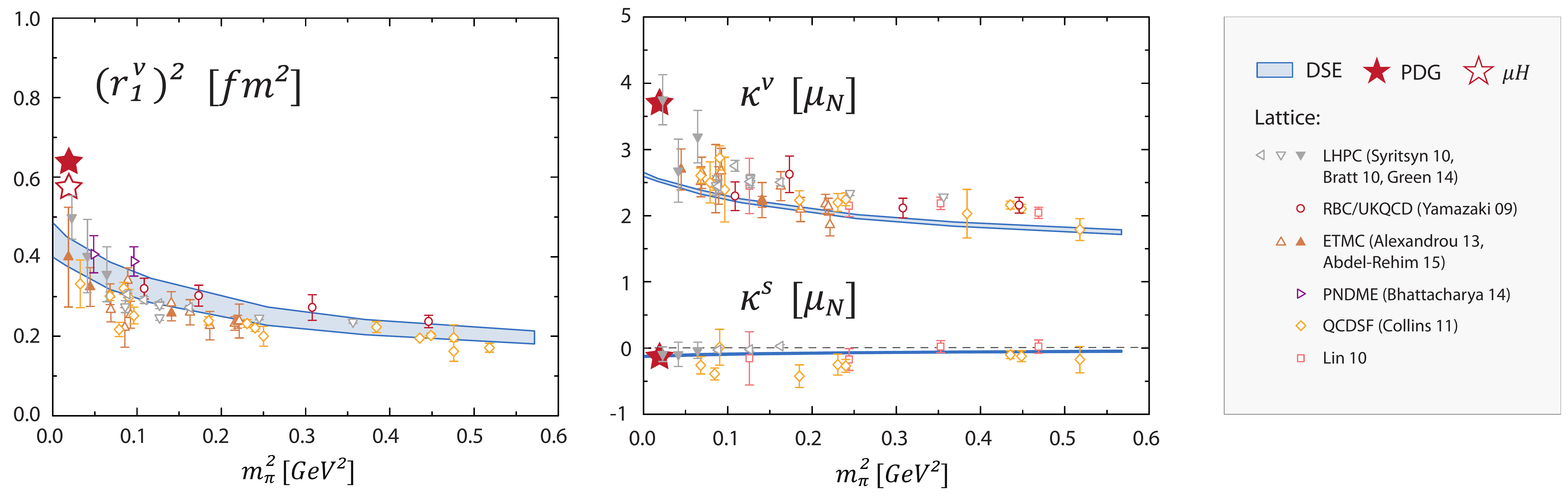}
                    \caption{Quark-mass dependence of nucleon static electromagnetic properties compared to lattice
                    results~\cite{Syritsyn:2009mx,Bratt:2010jn,Green:2014xba,Yamazaki:2009zq,Alexandrou:2013joa,Abdel-Rehim:2015jna,Bhattacharya:2013ehc,Collins:2011mk,Lin:2010fv}.
                             \textit{Left panel:} squared isovector Dirac radius $(r_1^v)^2$.
                             \textit{Right panel:} isovector and isoscalar anomalous magnetic moments $\kappa^v$ and $\kappa^s$ in units of nuclear magnetons.
                             Stars denote the experimental values, either from the PDG~\cite{Agashe:2014kda} or using the muonic hydrogen value for the electric proton radius~\cite{Antognini:1900ns}.
                             Figures adapted from Ref.~\cite{Eichmann:2011vu}. }\label{fig:nucleon-emffs-static}

                    \end{center}
        \end{figure}

\begin{figure}[t]
                    \begin{center}

                    \includegraphics[width=0.8\textwidth]{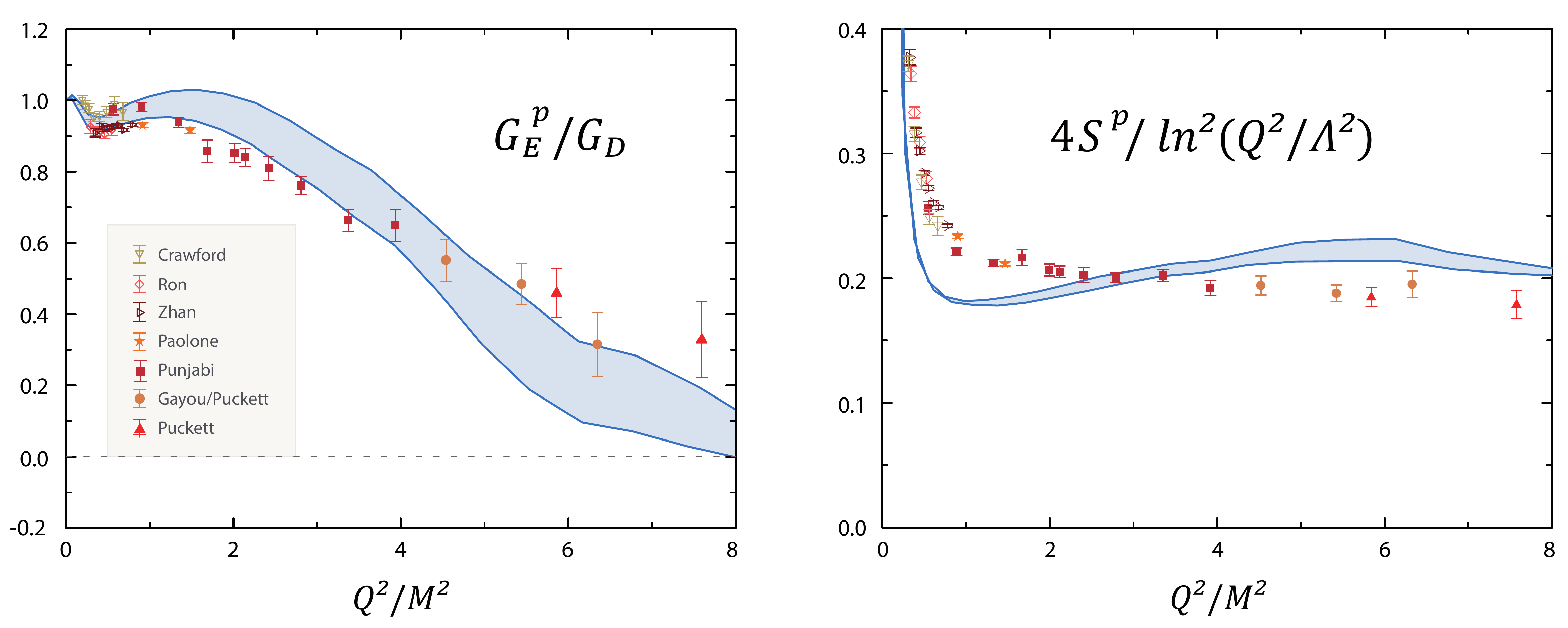}
                    \caption{\textit{Left:} Electric proton form factor normalized by the dipole and compared to experimental data~\cite{Eichmann:2011vu}.
                    \textit{Right:} Pauli over Dirac ratio for the proton and its precocious scaling behaviour.}\label{fig:nucleon-emffs-ratios}

                    \end{center}
        \end{figure}

       The absence of chiral corrections is especially striking in Fig.~\ref{fig:nucleon-emffs-static}, which shows examples for nucleon charge radii and magnetic moments as a function
       of $m_\pi^2$.
       The plots cover the domain from the chiral limit to the strange-quark mass and show
       isovector and isoscalar quantities as linear combinations of proton and neutron.
       Isovector form factors are directly comparable to lattice QCD since they are free of disconnected contributions.
       Meson-cloud effects would increase the charge radii close to the chiral limit where they eventually diverge because of the massless pion,
       and they contribute to the magnetic moments of baryons.
       The absence of pion cloud effects is clearly visible in the isovector Dirac radius which agrees with lattice data at larger pion masses but remains finite in the chiral limit.
       On the other hand, leading-order pion loops in chiral perturbation theory come with an opposite sign
       for the proton and neutron anomalous magnetic moments~\cite{Young:2004tb,Wang:2007iw}, so they are enhanced in the isovector combination $\kappa_v=\kappa_p-\kappa_n$ and cancel out in the isoscalar quantity $\kappa_s=\kappa_p+\kappa_n$.
       The experimental and theoretical values $\kappa_s=-0.12$ are in excellent agreement, which suggests that the dominant corrections to rainbow-ladder form factors
       are indeed of pionic nature.

       Another interesting topic is the origin of the nucleon's magnetic moments.
       Quark models often implement phenomenological Dirac and Pauli form factors for the vertex dressing of the quark, where the latter encodes the quark anomalous magnetic moment (AMM).
       As discussed in Sec.~\ref{sec:ff-vertices} and App.~\ref{app:hvp},
       in general such a dressing is contained in the offshell quark-photon vertex with its Ball-Chiu part and eight transverse components.
       It turns out that the quark AMM in the rainbow-ladder solution for the quark-photon vertex (Fig.~\ref{fig:vertex-def}) is negligible,
       which has the interesting consequence that the nucleon's magnetic moments extracted from its current matrix element are essentially reproduced by the Ball-Chiu vertex alone.
       Therefore, the feature sketched in Fig.~\ref{fig:hvp} of the transverse part not contributing at $Q^2=0$
       effectively also applies to the nucleon \textit{magnetic} form factors even though they are not subject to charge conservation.

       In Ref.~\cite{Chang:2010hb} it was argued that a large chromomagnetic AMM  in the quark-gluon vertex beyond rainbow-ladder
       would also translate into a large electromagnetic AMM for the quark-photon vertex.
       Such AMMs have since been introduced in nucleon form factor model calculations where they were found to improve agreement with the data at moderately low $Q^2$~\cite{Cloet:2013jya}.
       One should note that in the quark-diquark approximation also the diquarks carry AMMs,
       but since the diquarks in the nucleon are always offshell their effective contributions to $\kappa_p$ and $\kappa_n$ are small~\cite{Cloet:2008re}.
       The effect is explicitly demonstrated in Fig.~6.5 of Ref.~\cite{Eichmann:2009zx}, where the diquark-photon vertices were resolved
       from the quark level according to Fig.~\ref{fig:ff-quark-diquark}.
       Recently Dyson-Schwinger solutions for the quark-gluon vertex have become available which do exhibit a sizeable chromomagnetic AMM in its transverse part (see Fig.~9 in Ref.~\cite{Williams:2014iea}),
       but self-consistent calculations for the quark-photon vertex beyond rainbow-ladder, and form factors in general, have not yet been performed to our knowledge.

       \paragraph{Large $Q^2-$behaviour.}
       The transition from nonperturbative to perturbative QCD is implicit in the large-$Q^2$ behaviour of form factors.
       At sufficiently high momentum transfer, asymptotic scaling sets in and elastic form factors follow simple counting rules
       based on the minimum number of gluon exchanges that are required to distribute the momentum transfer equally among the quarks.
       In the nucleon, at least two gluon exchanges are required and thus perturbative QCD predicts the scaling behaviour of the Dirac and Pauli form factors as~\cite{Brodsky:1974vy}
              \begin{equation}
                  F_1 \rightarrow 1/Q^4\,, \qquad F_2 \rightarrow 1/Q^6\,, \qquad S := \tau  F_2/F_1 \rightarrow c,
              \end{equation}
       where $\tau=Q^2/(4m^2)$ and $c$ is an undetermined constant.
       Correspondingly, the Sachs form factors would scale as $G_E, G_M \rightarrow 1/Q^4$ which implies
       that the ratio $G_E/G_M$ should become constant, which is indeed what had been traditionally observed
       using the Rosenbluth separation method.
       However, these predictions have come under scrutiny with the polarisation-transfer measurements at Jefferson Lab,
       where the ratio $G_E^p/G_M^p$ shows a roughly linear decrease with $Q^2$ and points toward a zero crossing at some larger value of $Q^2$.
       Because the magnetic form factor is still reasonably well described by a dipole, the falloff and possible zero are entirely due to $G_E^p$ as shown in Fig.~\ref{fig:nucleon-emffs-ratios}.

       The discrepancy between the perturbative prediction and the experimental data has been attributed to
       helicity non-conservation and the presence of
       non-zero quark orbital angular-momentum in the nucleon wave function~\cite{Miller:2002qb,Ralston:2003mt,Holl:2005zi}.
       Indeed, with an updated perturbative prediction for $F_2(Q^2)$ that accounts for wave-function components with orbital angular momentum~\cite{Belitsky:2002kj} one finds
       $S \rightarrow c\,\ln^2{ \left(Q^2/\Lambda^2\right)}$,
       and the onset of perturbative scaling in the proton's Pauli to Dirac ratio already appears to happen at moderately low $Q^2$ of a few GeV$^2$, cf.~Fig.~\ref{fig:nucleon-emffs-ratios}.
       Here $\Lambda$ is a soft scale parameter that is related to the size of the nucleon.
       Such a relation can accommodate a zero crossing for both $G_E^p/G_M^p$ and $G_E^n/G_M^n$
       through~\eqref{ff:sachs}:
       \begin{equation}
           \frac{G_E}{G_M} = \frac{1-S}{1+S/\tau} \rightarrow  1-c\,\ln^2{ \left(Q^2/\Lambda^2\right)} \,.
       \end{equation}
       Such a zero has been found in many model calculations
       and it is also a typical feature of Dyson-Schwinger approaches. Fig.~\ref{fig:nucleon-emffs-ratios} shows the rainbow-ladder DSE result for the
       proton electric form factor normalized by the dipole. The calculation stops at $Q^2/m^2 \sim 8$ due to the quark singularities
       in the integrands (see the discussion at the end of Sec.~\ref{sec:ff-currents}), but the curve indeed points towards a zero at large $Q^2$.
       Also the neutron ratio $G_E^n/G_D$ exhibits a turnover at $Q^2/m^2 \approx 5$~--~$6$, indicating a possible zero crossing as well.
       Similarly, the quark-diquark model calculation of Refs.~\cite{Cloet:2008re,Cloet:2013gva}
       predicts zeros for both $G_E^p$ and $G_E^n$ at $Q^2 \sim 8$~--~$10$~GeV$^2$, where the exact location depends on the momentum dependence of the microscopic ingredients.

       Additional insight can be gained by separating the proton and neutron form factors into their flavour contributions.
       They are defined from the matrix element of the electromagnetic current~\eqref{current-em}, where one can neglect the small strange-quark admixture to good approximation.
       Exploiting isospin symmetry, the up/down-quark contributions in the proton equal the down/up-quark contributions in the neutron
        and thus they can be directly extracted from experiment:
       \begin{equation}
       \begin{array}{rl}
        F^p &= q_u\,F^u + q_d\,F^d\,,  \\
        F^n &= q_d\,F^u + q_u\,F^d
       \end{array}\qquad \Rightarrow \qquad
       \begin{array}{rl}
       F^u &= 2F^p + F^n\,, \\
       F^d &= F^p + 2F^n\,.
       \end{array}
       \end{equation}
       Consider for the moment the case where $F_1^u = 2F_1^d$ and $F_2^u = -F_2^d$. This entails
       a vanishing Dirac form factor $F_1^n = 0$ for the neutron and Pauli form factors for proton and neutron
       with same magnitude but opposite signs: $F_2^p = -F_2^n$, and hence also $\kappa_p = -\kappa_n$. Experimentally, $F_1^n(Q^2)$ is uniformly negative which
       can only happen if the down quark contribution to the Dirac form factor is suppressed relative to the up quark: $2F_1^d/F_1^u < 1$.
       This suppression was explicitly confirmed in Ref.~\cite{Cates:2011pz} for both $F_1$ and $F_2$ using the latest data for the electric neutron form factor.
       The faster falloff of the $d$-quark form factors implies that the $u$ quarks have a significantly tighter distribution than the $d$ quarks in impact-parameter space~\cite{Miller:2008jc}.
       A quark-diquark picture provides an intuitive explanation for this: in addition to the doubly represented $u$ quark in the proton (and $d$ quark in the neutron),
       the photon predominantly couples to $u$ quarks in the proton because the direct coupling to the $d$ quark only comes in combination with an axialvector diquark correlation~\cite{Roberts:2007jh,Cloet:2008re}.
       However, similar results have been obtained in Ref.~\cite{Eichmann:2011vu} as well as in the symmetric quark model of Ref.~\cite{Rohrmoser:2011tw}.

\subsection{Nucleon axial and pseudoscalar form factors}\label{form-res-octet-axial}

\begin{figure*}[t]
                    \begin{center}

                    \includegraphics[width=1\textwidth]{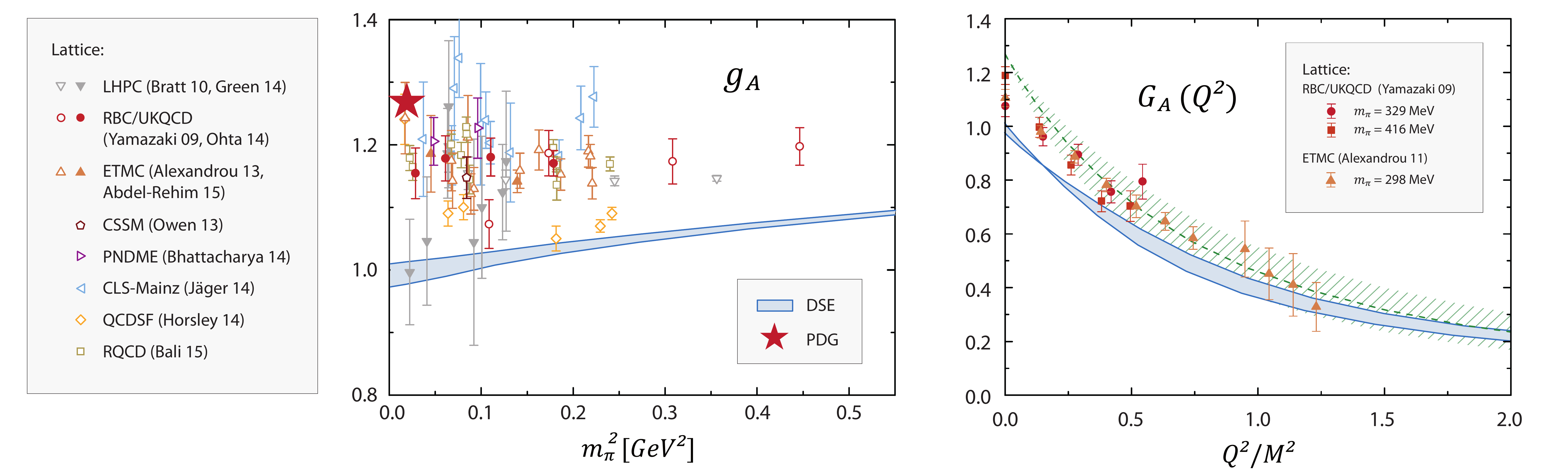}
                    \caption{\textit{Left:} Dyson-Schwinger result for the nucleon's axial charge $g_A$ as a function of the pion mass~\cite{Eichmann:2011pv} compared to recent lattice
                    data~\cite{Bratt:2010jn,Green:2012ud,Yamazaki:2009zq,Ohta:2013qda,Alexandrou:2013joa,Abdel-Rehim:2015owa,Owen:2012ts,Bhattacharya:2013ehc,Jager:2013kha,Horsley:2013ayv,Bali:2014nma}.
                             The star denotes the experimental values.
                    \textit{Right:} Dyson-Schwinger result for $G_A(Q^2)$~\cite{Eichmann:2011pv} compared to a selection of lattice results at different pion masses~\cite{Bratt:2010jn,Alexandrou:2007xj,Alexandrou:2010hf}.
                    The dashed line is the experimental dipole form~\eqref{GA-Dipole} with an axial mass $m_A=1.15 \pm 0.15$ GeV.}\label{fig:GA}

                    \end{center}
        \end{figure*}

              The nucleon's axial and pseudoscalar form factors are of fundamental significance for the properties
              of the nucleon that are probed in weak interaction processes; see~\cite{Bernard:2001rs,Gorringe:2002xx,Schindler:2006jq} for reviews.
              They are experimentally hard to extract and therefore considerably less
              well known than their electromagnetic counterparts. Precisely measured is only the low-momentum
              limit of the axial form factor, the nucleon's isovector axial charge $g_A=1.2723(23)$, which is a key quantity in nuclear physics and
              determined from neutron $\beta$ decay~\cite{Agashe:2014kda}.
          Its determination in lattice QCD has been traditionally difficult:
          the axial charge may be quite sensitive to finite-volume effects and excited state contaminations, making
          the extrapolation to the physical point difficult.
          Although simulations close to the physical pion mass are now available most of them still underestimate the experimental value, see Fig.~\ref{fig:GA} and
          Ref.~\cite{Bhattacharya:2013ehc} for a compilation of results,
          although agreement may be within reach~\cite{Horsley:2013ayv,Bali:2014nma,Abdel-Rehim:2015owa}.
          It is interesting to note that also in chiral perturbation theory one finds strong cancellations at leading and next-to-leading order in the chiral expansion for $g_A$~\cite{Bernard:2001rs,Bernard:2006te}.

             \paragraph{Axial and pseudoscalar current-matrix elements.}
             In the isospin-symmetric limit the nucleon's axialvector current is characterized by two form factors:
             the axial form factor $G_A(Q^2)$ whose value at zero momentum transfer is the nucleon's axial coupling constant $g_A$,
             and the induced pseudoscalar form factor $G_P(Q^2)$. They constitute the nucleon matrix elements of the isovector axialvector current in~\eqref{current-Gamma-0} via
             \begin{equation}\label{ax-current}
                 \mc{J}^\mu_{5}(p,Q) =  \conjg u(p_f)\,\gamma_5 \left( G_A(Q^2)\, \gamma^\mu + G_P(Q^2)\,\frac{i  Q^\mu}{2m}  \right) u(p_i)\,,
             \end{equation}
             where we suppressed the flavour matrices and employed the same kinematics as in Sec.~\ref{form-res-octet}.
             Defining a longitudinal form factor $G_L(Q^2)$ via $G_A = G_L +\tau G_P$~\cite{Eichmann:2011pv}, where $\tau = Q^2/(4m^2)$,
             the current-matrix element can be written as
             \begin{equation}
                 \mc{J}^\mu_5(p,Q) = \conjg u(p_f)  \,\gamma_5 \left( G_A(Q^2)\, \gamma_T^\mu  + G_L(Q^2)\, \frac{Q^\mu  \slashed{Q}}{Q^2} \right) u(p_i)\,,
             \end{equation}
             which can be verified from the identity $\conjg u(p_f) \, \gamma_5\,(\slashed{Q}+2im)\,u(p_i) =  0$.
             Therefore, $G_A$ and $G_L$ correspond to the purely transverse and longitudinal contributions with respect to the momentum transfer,
             and the properties of the underlying vertices in~\eqref{vertices-pole-contributions} entail that axialvector poles can only appear in $G_A$
             and pseudoscalar poles only in $G_L$, whereas $G_P$ contains both and thus it is dominated by the pion pole.

             Similarly, the pseudoscalar current is parametrized by the pseudoscalar form factor $G_5(Q^2)$:
             \begin{equation}\label{ps-current}
                 \mc{J}_5(P,Q) = G_5(Q^2)\,\conjg u(p_f)  \,i\gamma_5\, u(p_i)\,.
             \end{equation}
             On the pion's mass shell, the residue of the pseudoscalar form factor $G_5$ is the pion-nucleon coupling constant,
             which can be made explicit by defining a form factor $G_{\pi NN}(Q^2)$ as
             \begin{equation}\label{GpiNN-def}
                  G_5(Q^2) = \frac{m_\pi^2}{Q^2+m_\pi^2}\,\frac{f_\pi}{m_q}\,G_{\pi NN}(Q^2)\,,
             \end{equation}
             with $G_{\pi NN}(-m_\pi^2) = g_{\pi NN}$.
             It is straightforward to work out the PCAC relation in~\eqref{vcc-pcac-0} at the level of the current matrix elements
             using translation invariance of the quark fields. It connects the longitudinal with the pseudoscalar form factor
             and leads to the Goldberger-Treiman relation, which is valid for all current-quark masses:
             \begin{equation}\label{gtr}
                 Q^\mu \mc{J}^\mu_5  + 2 m_q\,\mc{J}_5= 0 \quad \Rightarrow \quad
                 G_L = \frac{m_q}{m}\,G_5 = \frac{m_\pi^2}{Q^2+m_\pi^2}\,\frac{f_\pi}{m}\,G_{\pi NN} \quad \Rightarrow \quad
                 G_A(0) =  \frac{f_\pi}{m}\,G_{\pi NN}(0)\,.
             \end{equation}
             Thus, the axial and pseudoscalar currents~\eqref{ax-current}--\eqref{ps-current}
             are described by only two independent form factors, $G_A(Q^2)$ and $G_{\pi NN}(Q^2)$, which at $Q^2=0$ are additionally related through the Goldberger-Treiman relation.

          \paragraph{Axial and pseudoscalar form factors.}
          Experimental data for the momentum dependence of $G_A(Q^2)$ come from quasielastic neutrino scattering off nucleons or nuclei
          and charged pion electroproduction~\cite{Bernard:2001rs}.
          Similarly to the electromagnetic form factors, the data can be parametrized by a dipole ansatz which determines the `axial mass' $m_A$
          and the axial radius $r_A$ from the slope of the form factor at $Q^2=0$:
          \begin{equation}\label{GA-Dipole}
              G_A(Q^2) = \frac{g_A}{\left( 1 + Q^2/m_A^2 \right)^2}\,, \qquad
              G_A(Q^2) = g_A \left( 1 - \frac{r_A^2}{6}\,Q^2 + \dots\right) \; \Rightarrow \; r_A^2 = -6\,\frac{G_A'(0)}{g_A} = \frac{12}{m_A^2}\,.
          \end{equation}
          The experimental value for $m_A$ is not well constrained due to model-dependent extractions of $G_A(Q^2)$ from the respective cross sections.
          Pion electroproduction and older neutrino scattering experiments suggest values around $m_A \sim 1$~GeV~\cite{Bernard:2001rs,Lyubushkin:2008pe}
          whereas more recent data from MiniBooNE and K2K favor higher central values up to $m_A \sim 1.3$~GeV~\cite{Gran:2006jn,AguilarArevalo:2010zc}.
          The origin of this discrepancy is unclear and could be a consequence of nuclear medium effects~\cite{Nieves:2011yp}
          or a deviation from the dipole form~\cite{Bhattacharya:2011ah}.

          The $Q^2-$evolution of $G_A$ is depicted in the right panel of Fig.~\ref{fig:GA}, where the Dyson-Schwinger result from the rainbow-ladder three-body calculation~\cite{Eichmann:2011pv}
          is compared to lattice data and the phenomenological dipole parametrization of~\eqref{GA-Dipole}.
          At low $Q^2$, the Dyson-Schwinger result underestimates the lattice data and the dipole curve whereas they converge above $Q^2 \sim 1$~--~$2$~GeV$^2$.
          This is similar to what happens for magnetic form factors and therefore it would also point towards missing chiral cloud effects in the low-momentum region.
          The corresponding axial charge is shown in the left panel as a function of the squared pion mass.
          At the physical $u/d$-quark mass it underestimates the experimental value by $20-25\%$ but slowly increases with the pion mass.
          In principle an interpretation in terms of missing chiral corrections is compatible with finite-volume effects on the lattice:
          if the lattices are too small, they might not be able to hold the full extent of the pion cloud~\cite{Ohta:2011nv}.

          What is interesting to note is that the observed spread in the lattice results for $G_A(Q^2)$ for different values of the pion mass~\cite{Yamazaki:2009zq,Alexandrou:2010hf}
          disappears when plotting them over $Q^2/m^2$, so that they are in fact well compatible with the dipole parametrization.
          This can be easily understood from the timelike structure: form factors are dimensionless functions of dimensionless variables,
          and $G_A(Q^2)$ is dominated by the axialvector meson $a_1(1260)$ with quantum numbers $J^{PC}=1^{++}$
          whose mass is roughly proportional to that of the nucleon when changing the current-quark mass. Thus the expected bump position at $Q^2/m^2 = -m_{a_1}^2/m^2$
          changes only weakly  with the quark mass and so does the spacelike shape of the form factor.

          The pseudoscalar form factor $G_{\pi NN}(Q^2)$ is not directly observable except at the onshell point $Q^2=-m_\pi^2$
          where it becomes the pion-nucleon coupling constant $g_{\pi NN}$, and at $Q^2=0$ where it is related to $g_A$.
          The form factor is often parametrized by extending the Goldberger-Treiman relation~\eqref{gtr} to non-zero $Q^2$ using the dipole parametrization for $G_A(Q^2)$.
          However, due to the different pole structures -- axialvector poles in $G_A$, pseudoscalar poles in $G_{\pi NN}$ starting with the first radial excitation $\pi(1300)$ of the pion --
          there is no reason for this to hold at spacelike momentum transfer.
          The Dyson-Schwinger result for the pseudoscalar form factor in Ref.~\cite{Eichmann:2011pv} indeed deviates from the Goldberger-Treiman relation at $Q^2>0$
          whereas it is compatible with the lattice data of Ref.~\cite{Alexandrou:2007xj}.

          The induced pseudoscalar form factor $G_P(Q^2)$, on the other hand, is dominated by the pion pole at $Q^2=-m_\pi^2$,
          and therefore pion pole dominance describes its spacelike properties reasonably well:
          from the above relations between the form factors one derives $G_P(Q^2) = 4m^2\,G_A(Q^2)/(Q^2+m_\pi^2) + \mathcal{O}(m_\pi^2)$.
          Since $G_P$ is the linear combination of the axial and longitudinal form factors $G_A$ and $G_L$,
          it must include all further pseudoscalar ($0^{-+}$) and axialvector ($1^{++}$) pole structures as well.
          Experimental data for $G_P(Q^2)$ are similarly sparse. Its $Q^2$ dependence can be extracted from pion electroproduction, whereas
          ordinary muon capture on the proton ($\mu^-+ p \rightarrow \nu_\mu + n$) determines
          the induced pseudoscalar coupling constant $g_p := m_\mu/(2m)\,G_P(Q^2=0.88 \,m_\mu^2)$.
          The value reported by the MuCap Collaboration, $g_p = 7.3 \pm 1.1$~\cite{Andreev:2007wg}, is consistent with
          chiral perturbation theory but smaller than the previous world average~\cite{Bernard:2001rs,Gorringe:2002xx}.

\subsection{Delta electromagnetic form factors}\label{form-res-delta}

       The lowest-lying nucleon resonance is the $\Delta$ baryon with mass
$m_\Delta = 1.232$ GeV, width $\Gamma_\Delta=114$~--~$120$~MeV and
$I(J^P)=\tfrac{3}{2}(\tfrac{3}{2}^+)$.
       It contributes a prominent and well-separated peak to the inelastic $N e^-$ cross section, to $N\pi$ scattering, electroproduction and other processes, and
       it plays an important role in Compton scattering where it
gives a large contribution to the nucleon's magnetic polarisability.
Unfortunately the $\Delta$ is very short-lived which implies that measuring its elastic electromagnetic properties is extremely
challenging. Therefore, most of our experimental knowledge on its electromagnetic structure comes
from the $\gamma N\to\Delta$ transition (discussed in
Sec.~\ref{form-res-trans}),
which in itself contributes only a tiny fraction to the total decay
width which is dominated by the $\Delta\to N\pi$ decay.
Due to the strong $\Delta\to N\pi$ coupling, pionic effects are
expected to contribute significantly to the properties of the $\Delta$ baryon.

Experimental information on the $\Delta$ electromagnetic form factors is sparse.
To date, only the magnetic moments of $\Delta^{+}$ and $\Delta^{++}$ are known, although with large uncertainties:
\begin{equation}\label{delta-mm}
    \mu_{\Delta^{++}}=3.7~\text{--}~7.5 \,\mu_N\,, \qquad
    \mu_{\Delta^+}=2.7^{+1.0}_{-1.3}\,\textrm(stat.)\pm 1.5\,\textrm(syst.)\pm 3\,\textrm(theo.)\,\mu_N\,.
\end{equation}
The value for the $\Delta^{++}$ is extracted from radiative pion-nucleon scattering ($\pi^+p\longrightarrow\pi^+p\gamma$)~\cite{Nefkens:1977eb,Bosshard:1991zp,LopezCastro:2000cv,Agashe:2014kda},
and that for the $\Delta^+$ from radiative photoproduction of neutral pions ($\gamma p\longrightarrow \pi^0 p\gamma'$)~\cite{Kotulla:2002cg}.
Information on the $\Delta^0$ and $\Delta^-$ magnetic moments and
all other electromagnetic properties of the $\Delta$ isomultiplet is totally
missing. In particular, there are no experimental data for the $Q^2$ evolution
of its electromagnetic properties.

In principle such knowledge would allow one to study the deformation of baryons: whereas for the nucleon (being a
spin-$\nicefrac{1}{2}$ particle) such information is not directly accessible
via its elastic form factors, for spin-$\nicefrac{3}{2}$ baryons deviations
from sphericity are signaled by non-vanishing values of the electric quadrupole
and magnetic octupole form factors.
Understanding the structure of the $\Delta$ baryon and its deformation, together with the study of the $\gamma N \to
\Delta$ quadrupole transitions, indirectly also allows one to investigate the
deformation of the nucleon, see e.g.~\cite{Pascalutsa:2006up} for a review.

       \paragraph{Electromagnetic current.}
             The $\Delta$ elastic electromagnetic current (or, in general, for
any spin-$\nicefrac{3}{2}$ baryon with positive parity) can be written as
             \begin{equation}\label{eq:current2}
                 \mc{J}^{\mu}(p,Q) = i \conjg u^\alpha(p_f) \left[
                          \left( F_1^\star\,\gamma^\mu +\frac{iF_2^\star}{4m_\Delta}\, [\gamma^\mu,\slashed{Q}]\right)\delta^{\alpha\beta}
                           -\left( F_3^\star\,\gamma^\mu +\frac{iF_4^\star}{4m_\Delta}\,[\gamma^\mu,\slashed{Q}] \right) \frac{Q^\alpha Q^\beta}{4m_\Delta^2}
                           \right]u^\beta(p_i)\,,
             \end{equation}
             see for example App.~B.2 of Ref.~\cite{Nicmorus:2010sd} for a derivation. The kinematics are the same as in the previous subsections except for the replacement $m_N\to m_\Delta$.
        The Rarita-Schwinger vector-spinors are eigenspinors of the
        Rarita-Schwinger projector $\mathds{P}^{\alpha\beta}(p)$ onto positive energy and spin $J=3/2$:
                    \begin{equation}\label{RSprojector}
                        \conjg{u}^\alpha(p) = \conjg{u}^\rho(p)\,\mathds{P}^{\rho \alpha}(p)\,, \quad
                        u^\beta(p) = \mathds{P}^{\beta \sigma}(p)\,u^\sigma(p)\,, \quad
                        \mathds{P}^{\alpha\beta}(p) = \Lambda_+(p)\left( T^{\alpha\beta}_p - \tfrac{1}{3} \gamma_\perp^\alpha\,\gamma_\perp^\beta\right)
                     \end{equation}
        with the positive-energy projector $\Lambda_+(p)$,
        the transverse projector $T^{\alpha\beta}_p = \delta^{\alpha\beta} - \hat{p}^\alpha \hat{p}^\beta$ and
        $\gamma_\perp^\alpha = T^{\alpha\beta}_p \gamma^\beta$.

             The electromagnetic current is expressed in terms of four form factors $F_i^\star(Q^2)$, whose linear combinations constitute the
             Coulomb monopole $G_{E0}$, magnetic dipole $G_{M1}$, electric quadrupole $G_{E2}$, and magnetic octupole $G_{M3}$ form factors~\cite{Nozawa:1990gt,Pascalutsa:2006up}:
             \begin{equation}
             \begin{array}{rl}
                G_{E0} &= \left(1+\frac{2}{3} \tau\right) \Delta_{12} - \frac{1}{3} \tau (1+\tau)\,\Delta_{34}\,, \\
                G_{M1} &= \left(1+\frac{4}{5} \tau\right) F_{12} - \frac{2}{5} \tau (1+\tau)\,F_{34}\,,
             \end{array}\qquad
             \begin{array}{rl}
                G_{E2} &= \Delta_{12} - \frac{1}{2} (1+\tau)\,\Delta_{34}\,, \\
                G_{M3} &= F_{12} - \frac{1}{2} (1+\tau)\,F_{34}\,,
             \end{array}
             \end{equation}
             with $\tau=Q^2/(4m_\Delta^2)$, $F_{ij} = F_i^\star + F_j^\star$ and
$\Delta_{ij} = F_i^\star - \tau F_j^\star$.
In a non-relativistic picture,
the electric monopole and magnetic dipole form
factors can be seen as the analogues of $G_E$ and $G_M$ for the nucleon,
describing the momentum-space distribution of the nucleon’s charge and
magnetisation. The electric quadrupole and magnetic octupole are associated
with the deformation (non-sphericity) of those distributions. Their static
(dimensionless) values
             \begin{equation}
                 G_{E0}(0) = e_\Delta\,, \qquad
                 G_{M1}(0) = \mu_\Delta \,, \qquad
                 G_{E2}(0) = \mathcal{Q}\,, \qquad
                 G_{M3}(0) = \mathcal{O}
             \end{equation}
             define the charge $e_\Delta \in \{  2,1,0,-1 \}$ of the $\Delta$ baryon, its magnetic dipole moment $\mu_\Delta$,
             the electric quadrupole moment $\mathcal{Q}$, and the magnetic octupole moment $\mathcal{O}$.
             Expressed in terms of the dimensionless quantity $G_{M1}(0)$, the numbers in~\eqref{delta-mm} correspond to
             $\mu_\Delta = 7.4\pm 2.5$ for the $\Delta^{++}$  and $\mu_\Delta=3.6^{+1.3}_{-1.7} \pm 2 \pm 4$ for the $\Delta^+$.

        \begin{figure*}[t]
        \begin{center}
\includegraphics[width=0.95\textwidth]{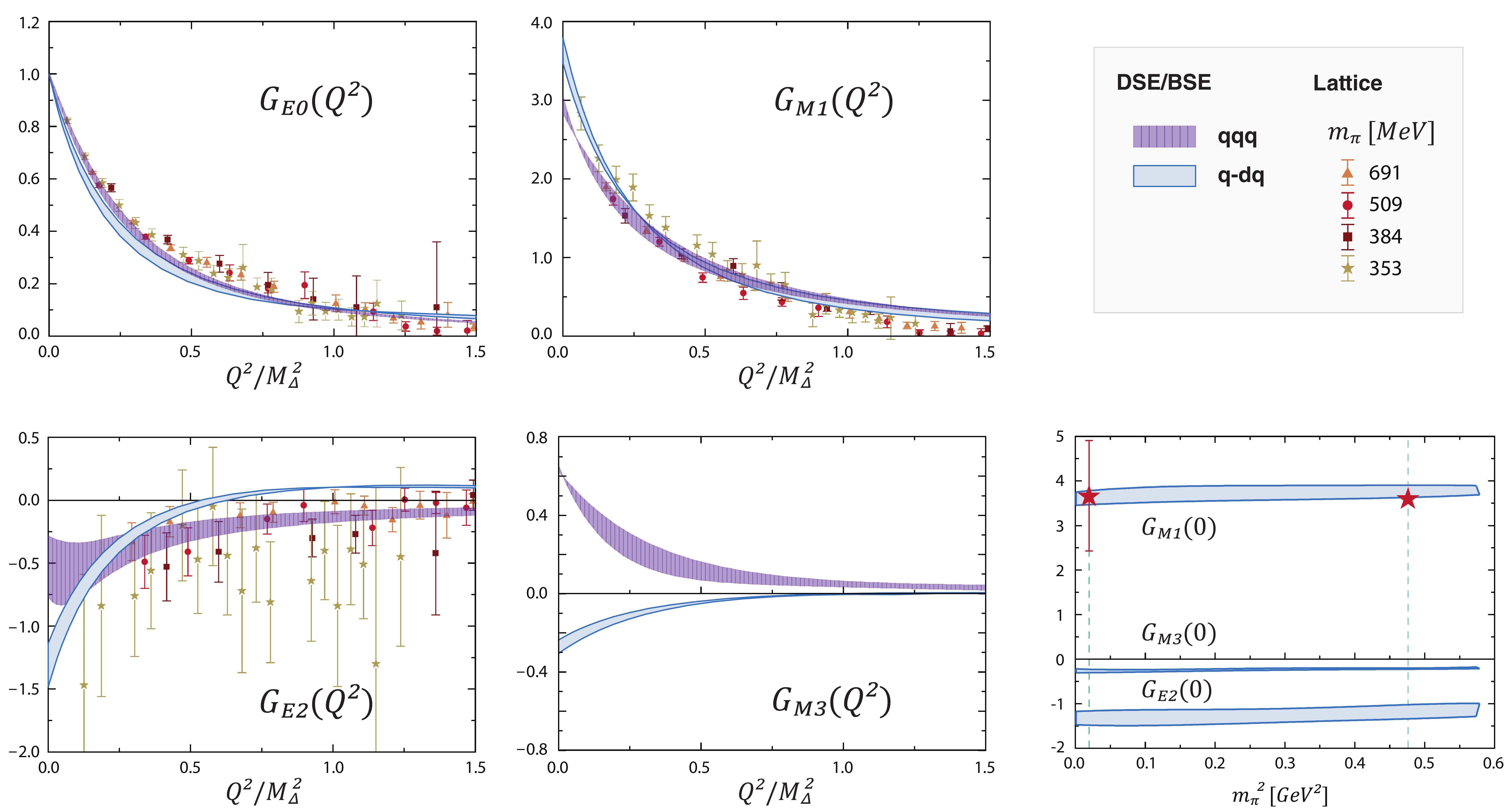}
                    \caption{Electric monopole, magnetic dipole, electric quadrupole
and magnetic octupole form factors for the $\Delta^+$ baryon. We compare lattice data~\cite{Alexandrou:2009hs}
with the Dyson-Schwinger three-quark~\cite{Sanchis-Alepuz:2013iia} and quark-diquark calculations~\cite{Nicmorus:2010sd}; the coloured
bands show the sensitivity to varying the $\eta$ parameter in \eqref{couplingMT}.
The plot on the bottom right shows the current-mass evolution of the static quantities from the quark-diquark calculation~\cite{Nicmorus:2010sd},
where stars denote the experimental magnetic moments of the $\Delta^+$ and $\Omega^-$ baryon.} \label{fig:delta-ffs}
                    \end{center}
        \end{figure*}

       \paragraph{Form factors.}


Given the lack of experimental information, model
calculations are usually compared to lattice QCD simulations. The most recent
lattice calculations of $G_{E0}$, $G_{M1}$ and $G_{E2}$
using dynamical quarks were presented in
~\cite{Alexandrou:2008bn,Alexandrou:2009hs,Alexandrou:2010jv} for
different values of the pion mass, the lowest one being $\sim 350$~MeV, whereas $G_{M3}$
has only been calculated using the quenched approximation~\cite{Alexandrou:2007we}.
The data provided by these calculations still show
large uncertainties which are visible in Fig.~\ref{fig:delta-ffs}.
When taking them as a benchmark for the situation at physical pion masses one must keep in mind that
the pion mass is still in the region where $m_\pi > m_\Delta-m_N$, so the $\Delta$ is a stable bound state below $N\pi$ threshold.
Note that under the assumption of exact isospin
symmetry and due to the symmetry properties of the $\Delta$ flavour
amplitudes, the form factors of all other members of the $\Delta$
isomultiplet can be obtained by multiplying those of the $\Delta^+$
with the appropriate baryon charge, which implies in
particular that all form factors of the $\Delta^0$ vanish identically.

In Fig.~\ref{fig:delta-ffs} we compare the lattice data for the $\Delta^+$ electromagnetic form factors of Ref.~\cite{Alexandrou:2009hs}
with results from two rainbow-ladder Dyson-Schwinger calculations.
One of them solves the three-body equation without further approximations~\cite{Sanchis-Alepuz:2013iia}\footnote{We present here an update of the
results in~\cite{Sanchis-Alepuz:2013iia} where the numerical accuracy has been increased.} and the other uses the quark-diquark approximation~\cite{Nicmorus:2010sd};
in both cases the input is the effective quark-gluon interaction of~\eqref{couplingMT}.
The comparison is interesting because it can shed light on the reliability of the quark-diquark approximation.
A first inspection of Fig.~\ref{fig:delta-ffs} reveals a fair agreement
for $G_{E0}$ and $G_{M1}$ in the whole $Q^2$ range,
whereas $G_{E2}$ and $G_{M3}$ show interesting differences.

       The deformation of the $\Delta$ is encoded in its electric quadrupole and magnetic octupole form factors.
       Non-relativistically, a negative sign for the electric quadrupole moment $G_{E2}(0)$
       indicates an oblate charge distribution for the $\Delta$ in the Breit frame, with
       a different interpretation arising in the infinite-momentum frame~\cite{Alexandrou:2009hs}.
       From the measurement of the $\gamma N\to\Delta$ transition one can infer
       the value $G_{E2}(0)=-1.87(8)$ in the large-$N_C$ limit~\cite{Buchmann:2002mm,Alexandrou:2009hs};
       comparable values are predicted by a range of constituent-quark models~\cite{Ramalho:2009vc}
       and they are in the ballpark of the lattice results.
Similar numbers arise from the Dyson-Schwinger calculations in Fig.~\ref{fig:delta-ffs}.
However, in contrast to the three-body result the quark-diquark calculation $G_{E2}(Q^2)$ develops a zero crossing,
which is an unexpected feature but not clearly excluded from the lattice data.

Also the  magnetic octupole form factor $G_{M3}(Q^2)$ appears to be a very sensitive quantity.
Its lattice signal is weak and plagued by large error bars but compatible with zero.
At next-to-leading order in a chiral expansion, the magnetic octupole
moment also vanishes~\cite{Arndt:2003we}.
Both Dyson-Schwinger calculations yield non-zero values but they differ in their signs,
as can be seen in Fig.~\ref{fig:delta-ffs}: the
quark-diquark result is negative whereas the three-body result is positive.
A similar observation applies to the quark-diquark model calculations of Ref.~\cite{Segovia:2014aza},
where depending on the pointlike or non-pointlike nature of the ingredients the result is either negative or positive but in both cases larger in magnitude.
In the covariant spectator quark model of Refs.~\cite{Ramalho:2009vc,Ramalho:2010xj} the magnetic octupole form factor
is strongly dependent on the $d$-wave content of the wave function;
two different parametrizations for the $\Delta$ wave function lead to different signs
although in both cases the result is negative at zero photon momentum.
We note again that in the Dyson-Schwinger
calculations the partial-wave content of the nucleon and $\Delta$ amplitudes in terms of $s$, $p$, $d$ and $f$ waves is not
restricted in any way but determined dynamically when solving the
bound-state equations. This is certainly desirable, as the details of the form
factors which could depend on the precise internal structure of the state
act as a probe of the quark-gluon interaction.

Another observation concerns the similarity of the Dyson-Schwinger and lattice results below $Q^2 \sim 1$ GeV$^2$,
which is somewhat in contrast to the previously discussed nucleon form factors
where a certain discrepancy is seen which can be attributed to
the absence of pion cloud effects in rainbow-ladder calculations.
Although such a disagreement is also expected to occur for the $\Delta$
baryon, the lattice calculations were performed for unphysical
pion masses above $350$~MeV. Therefore, non-analyticities arising from the $\Delta\to N\pi$ decay
and the associated `pion-cloud effects' are absent in both approaches.
On the lattice, the static electromagnetic properties of the $\Delta$ remain more or
less constant for pion masses above $\sim 400$~MeV~\cite{Boinepalli:2009sq}.
From the bottom right panel in Fig.~\ref{fig:delta-ffs} one can see the same is true for the Dyson-Schwinger results:
the static values for the magnetic dipole, electric quadrupole and magnetic octupole moments
are practically independent of the current-quark mass.

        The evolution with the quark mass (or equivalently $m_\pi^2$) has another interesting consequence.
        For two quark flavours with equal masses the $\Delta^-$ becomes identical to the $\Omega^-$ baryon
        when evaluated at the strange-quark mass because the $\Omega^-$ is a pure $sss$ state.
        Experimentally, $\mu_{\Omega^-}=-2.02(5)\, \mu_N$ which implies $|G_{M1}(0)| = 3.60(9)$.
        It is encouraging that the quark-diquark result for $G_{M1}(0)$ agrees reasonably well with that value,
        because pionic effects should have diminished in the vicinity of the strange-quark mass above which the baryon is increasingly dominated by its `quark core'.
        The same can be said for the $Q^2-$dependence of the form factors:
        when plotted over $Q^2/m_\Delta^2$, the overall shape of Fig.\,\ref{fig:delta-ffs} persists throughout the current-quark mass range
        and the same behaviour is visible in the lattice results.
        Thus, in the absence of chiral effects the $\Omega^-$ form factors are not materially different from those of the $\Delta$ once their inherent mass dependence is scaled out.

\subsection{Nucleon transition form factors}\label{form-res-trans}

            Coming back to our discussion in Sec.~\ref{sec:electroproduction}, let us now discuss results for the electromagnetically induced transition form factors from the nucleon to its resonances.
            Nucleon transition form factors are of particular recent interest because they can be accurately extracted from meson electroproduction experiments
            and provide additional information on the electromagnetic structure of resonances in addition to their masses and widths.
            In recent years the data collected with CLAS at Jefferson Lab over a wide $Q^2$ range have enlarged the data pool by a substantial amount.
            In addition to the $\Delta$ resonance, data for several other states have become available as well, including the Roper and the negative-parity
            resonances $N(1535)$ and $N(1520)$; see~\cite{Aznauryan:2011qj,Tiator:2011pw} for reviews.

 \paragraph{$ \gamma N\to \Delta(1232)$ transition.}

             The $ \gamma N\to \Delta(1232)$ transition current is determined by the three Jones-Scadron form factors $G_M^\star(Q^2)$, $G_E^\star(Q^2)$ and $G_C^\star(Q^2)$
             which are experimentally extracted from the multipole amplitudes in pion electroproduction.
            There are various equivalent ways to define the onshell $\nicefrac{1}{2}^+ \to \nicefrac{3}{2}^+$ transition current;
            the most common one is the Jones-Scadron current itself~\cite{Jones:1972ky}:
        \begin{equation} \label{onshell-3/2-current-0}
           \mc{J}^{\mu}(p,p') = \frac{\sqrt{6}\,\delta_+}{4m^4 \,\lambda_+ \lambda_-}\,\conjg{u}^\alpha(p)\,i\gamma_5\left(
              m^2 \lambda_-\,(G_M^\ast-G_E^\ast)\,\varepsilon^{\alpha\mu}_{pQ}  -G_E^\ast\,\,\varepsilon^{\alpha\beta}_{pQ}\,\varepsilon^{\beta\mu}_{pQ} + \frac{G_C^\ast}{2}\,\,Q^\alpha  Q^\beta \,t^{\beta\mu}_{pQ}\right)u(p') \,.
        \end{equation}
        The spinors are the same as before; however,
        for transition form factors we slightly change our notation compared to the previous sections: $p$ is now the momentum of the resonance, $p'$ is the nucleon momentum
        and $Q=p-p'$ the momentum transfer. The current depends on the quantities
             \begin{equation}
                   t_{ab}^{\mu\nu} = a\cdot b\,\delta^{\mu\nu} - b^\mu a^\nu\,,  \quad
                   \varepsilon^{\mu\nu}_{ab} = \gamma_5\,\varepsilon^{\mu\nu\alpha\beta}a^\alpha b^\beta, \quad
                   \delta_\pm = \frac{m_\Delta \pm m}{2m} \,, \quad
                   \lambda_\pm = \delta_\pm^2 + \tau\,, \quad
                   \tau=\frac{Q^2}{4m^2}\,.
             \end{equation}

        The amplitudes that appear in effective field theory expansions and dynamical reaction models also depend on offshell currents and
        in that case additional requirements apply. Apart from electromagnetic gauge invariance, the offshell transition current must also be invariant under spin-$\nicefrac{3}{2}$ gauge transformations
        and point transformations~\cite{Pascalutsa:1999zz,Pascalutsa:2002pi} and satisfy the analyticity conditions that those imply. The Jones-Scadron current does not meet these criteria but the following form~\cite{Pascalutsa:2007wz,inprep-er} does:
        \begin{equation} \label{onshell-3/2-current}
           \mc{J}^{\mu}(p,p') = \frac{\sqrt{6}\,\delta_+}{4m^2}\,\conjg{u}^\alpha(p)\,i\gamma_5\left(
               g_M\,\varepsilon^{\alpha\mu}_{pQ} -  g_E\,t^{\alpha\mu}_{pQ}  - \frac{ g_C}{m_\Delta}\,i t^{\alpha\beta}_{p\gamma}\,t^{\beta\mu}_{QQ}\right)u(p') \,.
        \end{equation}
        An equivalent form has been used in Ref.~\cite{Kondratyuk:2001qu}.
        The form factors $g_M$, $g_E$ and $g_C$ are free of kinematic constraints\footnote{We
        trivially modified the expression in~\cite{Pascalutsa:2007wz} by replacing $\{g_M, g_E, g_C\} /\lambda_+ \to \{g_M, g_E, g_C\}$, so that
        the form factors do not have kinematic zeros at $\lambda_+=0$.}
        and the onshell relations between the two sets of form factors are (with $\delta=\delta_+\delta_-$)
        \begin{equation}\label{delta-ff-relations}
            G_M^\ast-G_E^\ast = \lambda_+  g_M\,, \qquad
            G_E^\ast =  \frac{\tau-\delta}{2}\,g_E +  \tau \,g_C\,,  \qquad
            G_C^\ast =  \left(\tau-\delta\right)g_C -  \frac{m_\Delta^2}{2m^2}\,g_E\,.
        \end{equation}

            \begin{figure*}[t!]
            \centerline{%
            \includegraphics[width=1\textwidth]{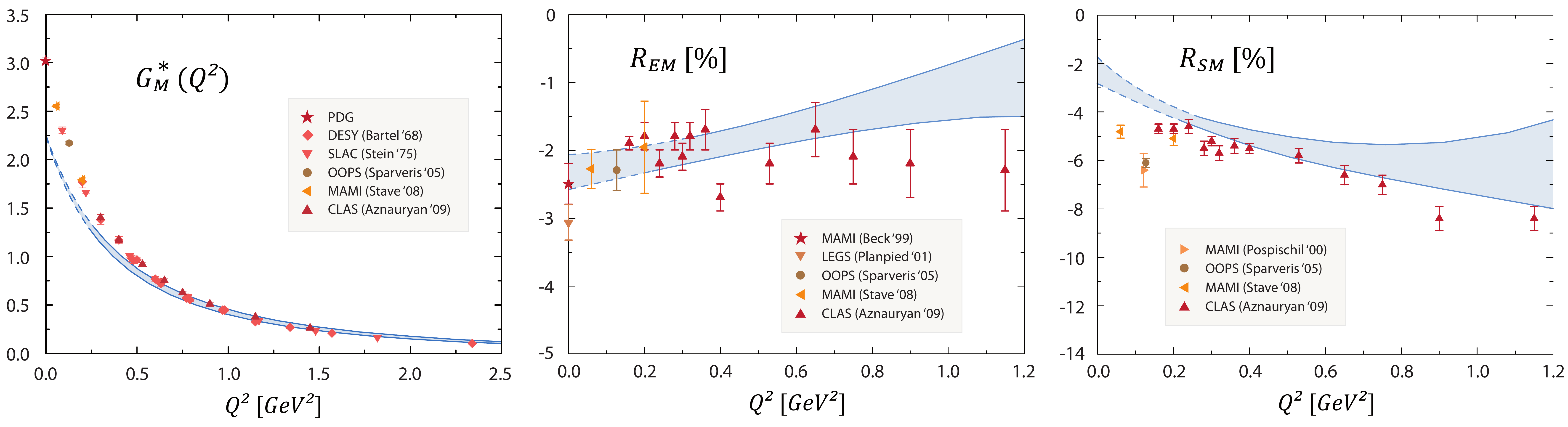}}
            \caption{$\gamma N\to\Delta$ transition form factors from the Dyson-Schwinger approach compared to experimental data~\cite{Eichmann:2011aa}.}
            \label{fig:ndg}
            \end{figure*}

             The $\gamma N\to \Delta(1232)$ transition is dominated by a magnetic dipole transition which,
             in a quark-model picture, amounts to a spinflip of a quark and is encoded in the form factor $G_M^\star(Q^2)$.
             Its static experimental value is $G_M^\star(0)=3.02(3)$;
             experimental data exist in the range up to $Q^2\sim 8$~GeV$^2$ (see~\cite{Pascalutsa:2006up,Aznauryan:2011qj} for detailed reviews).
             The remaining electric and Coulomb quadrupole contributions are much smaller and measure the deformation in the transition.
             They are expressed by the form factors $G_E^\star(Q^2)$ and $G_C^\star(Q^2)$ which are
             related to the magnetic dipole form factor through the form factor ratios
             \begin{equation}\label{ratios}
                 R_{EM} = -\frac{G_E^\star}{G_M^\star}\,, \qquad
                 R_{SM} = -\frac{m^2}{m_\Delta^2} \sqrt{\lambda_+\lambda_-}\,\frac{G_C^\star}{G_M^\star}\,.
             \end{equation}
             Note that the Jones-Scadron form factors are kinematically related at $\lambda_+=0$: $G_E^\ast=G_M^\ast$ and therefore $R_{EM}=-1$,
             and the ratio $R_{SM}$ has kinematic zeros at $\lambda_\pm=0$ (which corresponds to $Q^2 \approx -4.7$ GeV$^2$ and  $Q^2 \approx -0.09$ GeV$^2$) and becomes imaginary in between.

             In Fig.~\ref{fig:ndg} we show results from the rainbow-ladder Dyson-Schwinger approach in the quark-diquark approximation~\cite{Eichmann:2011aa}.
             Dyson-Schwinger based results are also available from the contact-interaction and momentum-dependent quark-diquark models~\cite{Segovia:2014aza}
             and first three-body rainbow-ladder results have been reported in~\cite{Sanchis-Alepuz:2013iia}.
             The magnetic dipole transition form factor $G_M^\star(Q^2)$ follows the
             characteristics of the previously discussed magnetic and axial form factors: it agrees with experimental data at larger $Q^2$ and underestimates them by $\sim 25\%$
             at $Q^2 =0$. This is consistent with quark-model results and the expected behaviour of the pion cloud from coupled-channel analyses.
           What is particularly interesting are the ratios $R_{EM}$ and $R_{SM}$ because
           the former encodes the orbital angular momentum in the transition.
           Perturbative QCD predicts $R_{EM} \rightarrow +100\%$ at $Q^2\rightarrow \infty$ which is, however, not seen in the available data:
           $R_{EM}$ remains small and negative up to $Q^2 \sim 7$~GeV$^2$.
       In non-relativistic quark models, non-zero values for these ratios would require $d-$wave components in the nucleon and $\Delta$ wave functions
       or the inclusion of pion-cloud effects.
       Indeed, the analysis of pion electroproduction data via dynamical reaction models suggests that these ratios are almost
       entirely dominated by meson-baryon interactions~\cite{JuliaDiaz:2006xt}.

           By contrast, Fig.~\ref{fig:ndg} shows that $R_{EM}$ and $R_{SM}$ are quite well reproduced even in the absence of a pion cloud.
       In the case of $R_{EM}$, this behaviour originates from the relativistic $p$ waves in the nucleon and $\Delta$ bound-state amplitudes which are a consequence of
       Poincar\'e covariance as discussed in Sec.~\ref{spec:BS}. They carry the dominant fraction of orbital angular momentum but are absent in quark models.
           If $p$ waves are switched off entirely, the ratio starts at $\sim-1\%$, crosses zero and increases rapidly
           with a trend towards the asymptotic prediction $R_{EM} \to 1$ for $Q^2 \to\infty$;
            its value at $Q^2=1.2$ GeV$^2$ is already $+9\%$~\cite{Eichmann:2011aa}.
       The contribution from $d$ waves, on the other hand, is almost negligible.
           Hence, $R_{EM}$ is presumably \textit{not} a pion-cloud effect; instead, the results entail that the form factors $G_E^\ast$
           and $G_C^\ast$ compare to their experimental values in a similar fashion as $G_M^\ast$ and chiral effects would contribute a similar fraction to all of them at low $Q^2$.
           Above $Q^2 \sim 2$ GeV$^2$ the model uncertainties stemming from the effective interaction~\eqref{couplingMT}
           and the seagull amplitudes in Fig.~\ref{fig:ff-quark-diquark} become too large to make reliable predictions.
           Nevertheless these results make clear that $p$ waves do have observable consequences even if they are subleading,
           which underlines the necessity of implementing complete tensor bases for the Poincar\'e-covariant nucleon and $\Delta$ Bethe-Salpeter amplitudes.

 \paragraph{$ \gamma N\to$ Roper  transition.}
  Another interesting case is the $ \gamma N\to$ Roper  transition discussed already in Sec.~\ref{sec:exp-nucleon-res},
  as it has the potential to clarify the underlying composition of the Roper resonance. A recent quark-diquark model calculation~\cite{Segovia:2015hra}
  produces indeed results that are qualitatively similar to the experimental data shown in Fig.~\ref{fig:helicity-amps},
  including the observed zero crossing and a corresponding zero crossing in the Pauli-like form factor $F_2(Q^2)$.
  Together with the results from the three-body Bethe-Salpeter equation displayed in Fig.~\ref{nucleon_delta} this provides
  further evidence that the Roper is indeed the first radial excitation of the nucleon, although meson-cloud effects may play
  a significant role. Work on nucleon to resonance transition form factors in the Dyson-Schwinger approach has only begun
  and we may expect to see more results in the future.

            \begin{figure*}[t!]
            \centerline{%
            \includegraphics[width=0.94\textwidth]{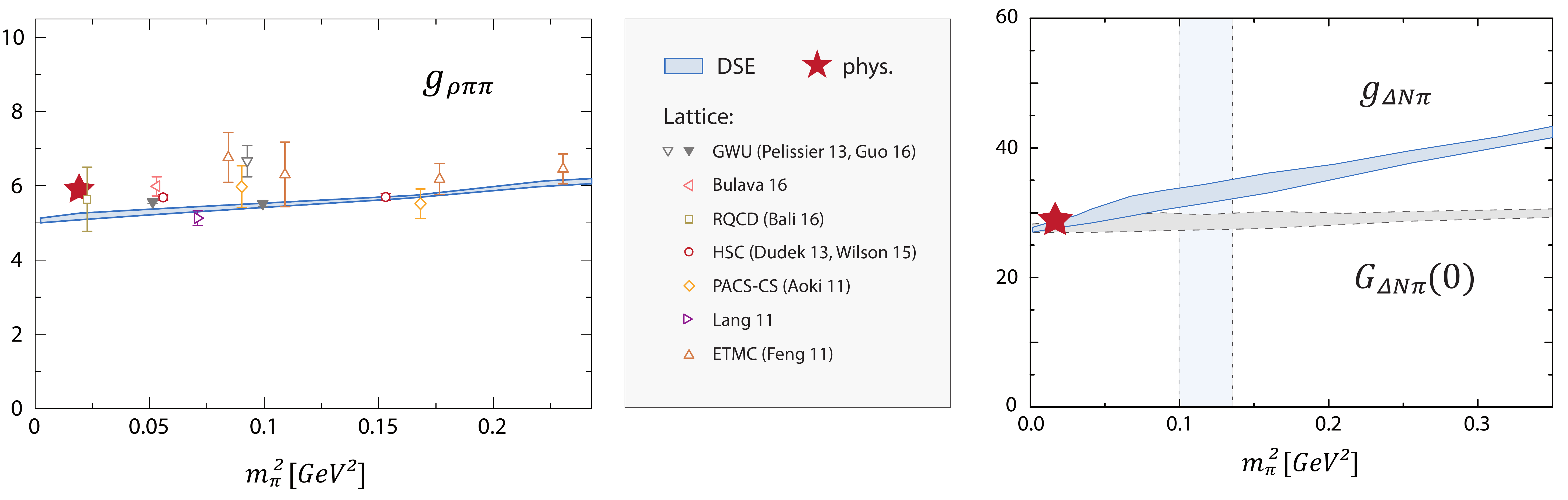}}
            \caption{\textit{Left:} Current-mass evolution of the $\rho\to\pi\pi$ coupling constant extracted from the rainbow-ladder electromagnetic current matrix element analogous to Fig.~\ref{fig:ffs-bse-1}
                     and compared to lattice results~\cite{Pelissier:2012pi,Guo:2016zos,Bulava:2016mks,Bali:2015gji,Dudek:2012xn,Wilson:2015dqa,Aoki:2011yj,Lang:2011mn,Feng:2010es}.
                     \textit{Right:} Analogous $\Delta\to N\pi$ coupling obtained from a quark-diquark calculation in rainbow-ladder. Figures adapted from Ref.~\cite{Mader:2011zf}.}
            \label{fig:delta-n-pi}
            \end{figure*}

 \paragraph{$ \Delta \to N\pi$ decay.}
 Let us finally return to the comment we made in Sec.~\ref{spec:results}. Most hadrons are resonances and they decay: the $\rho$ meson decays into two pions, the $\Delta$ and Roper mainly decay into $N\pi$, etc.
 A complete description of hadrons should incorporate these properties and dynamically generate widths for such resonances.
 However, none of the truncations that are currently employed in functional methods are yet capable of doing so and instead of resonances they produce `bound states' without widths.
 On the other hand, it is important to note that this does \textit{not} mean that one cannot \textit{calculate} the $\rho\to\pi\pi$, $\Delta\to N\pi$ decays in rainbow-ladder.
 The $\rho\to\pi\pi$ decay is proportional to the residue of the pion's electromagnetic form factor at the timelike $\rho$-meson pole ($Q^2=-m_\rho^2$).
 The $\Delta\to N\pi$ decay is the residue of the pseudoscalar $N\to\Delta$ transition form factor at the pion pole ($Q^2=-m_\pi^2$)
 constructed from the same diagrams as in Fig.~\ref{fig:current-faddeev}, and the same is true for other resonances.
 As it turns out, the existing calculations yield quite reasonable values for these decays.
 Ultimately, the decay mechanisms would have to be backfed into the Bethe-Salpeter equations and this is what would shift their T-matrix poles into the complex plane and thereby generate the desired widths.
 This is perhaps also the point where the hadronic coupled-channel and quark-level philosophies most clearly differ:
 while in the former the resonant nature is a fundamental component of a state whose underlying quark-gluon properties are inaccessible,
 in the latter acquiring a width is often viewed as a `correction' that comes on top of dynamically generating a hadron as a $q\bar{q}$, $qqq$ or multiquark system in the first place.

       The decay width of the $\Delta(1232)$ is almost exclusively produced by the $\Delta\to N\pi$ decay,
       whereas the only other decay channel, the electromagnetic $\Delta\rightarrow N\gamma$ transition, has a branching fraction of less than $1\%$.
       The experimental decay width is $\Gamma_\Delta= 117(3)$~MeV from where the coupling strength $g_{\text{\tiny{$\Delta N$}}\pi}=29.2(4)$
       can be inferred. Theoretically it can be extracted from the residue of the $Q^2$-dependent pseudoscalar transition form factor $G_{\text{\tiny{$\Delta N$}}\pi}(Q^2)$
       at the pion pole $Q^2=-m_\pi^2$.
       In Ref.~\cite{Mader:2011zf} that form factor was calculated in the rainbow-ladder quark-diquark approach using the decomposition in Fig.~\ref{fig:ff-quark-diquark},
       however without the (pseudoscalar) seagull terms because in contrast to the electromagnetic case there is no Ward-Takahashi identity to constrain their structure.
       The resulting transition form factor was found to be compatible with lattice calculations~\cite{Alexandrou:2010uk} and the resulting value $g_{\text{\tiny{$\Delta N$}}\pi}=28.1$
       for the coupling constant turned out to be remarkably close to the experimental number, cf.~Fig.~\ref{fig:delta-n-pi}. Similarly to the analogous $\rho\to\pi\pi$ decay also shown in the figure,
       $G_{\Delta N\pi}(Q^2=0)$ is almost independent of the current-quark mass. The moderate rise in $g_{\Delta N\pi}$ is simply due to the rise of the form factor
       toward the pion pole which moves away from the origin with increasing pion masses.
       By contrast, the phase space for the decay closes at $m_\pi = m_\Delta-m_N$ (shown by the vertical band) and thus the actual decay width of the $\Delta$
       vanishes rapidly with the current-quark mass; hence it is mainly governed by the kinematic phase space factor multiplying the coupling constant $g_{\Delta N\pi}$.

\subsection{Hyperon form factors}\label{form-res-octet-decuplet}

The study of electromagnetic properties of baryons has been historically
focused, both theoretically and experimentally, on the nucleon, the $\Delta$ and
their transitions, whereas the strange members of the octet and the decuplet
(the hyperons) have received much less attention. Knowledge of their structure, however, could shed
light on the role played by strangeness in the determination of baryon
properties or, more generally, the quark mass and flavour dependence of QCD
interactions. In particular, as we discuss below, the comparison of model
predictions for form factors with experimental or lattice
results (where available) allows one to evaluate the effect of the meson cloud and, for example,
the interplay of kaon and pion clouds.

From the experimental point of view, the information on the electromagnetic
properties of hyperons is scarce -- which is natural as they are more difficult
to produce and decay very fast. To date, experimental data exist for the
magnetic moments of all octet members except $\Sigma^0$ and for the electric radius of the $\Sigma^-$ (and,
of course, the nucleon). For the strange members of the decuplet, however, nothing
is known.
First experimental results for the hyperon form factors at large timelike
 momentum ($|Q^2|\sim 14$ GeV$^2$) have been reported by the CLEO
collaboration~\cite{Dobbs:2014ifa,Seth:2014dxa,Seth:2014taa}. As part of the $N^*$ program of the CLAS12 upgrade at Jefferson Lab, it
is planned to measure $KY$  electroproduction ($Y$ being a hyperon) and to
extract from there some hyperon transition form factors at non-vanishing
$Q^2$~\cite{LeeJLAB}.

The calculation of electromagnetic properties of hyperons began in earnest once
experimental investigations were planned. A number of
quark models provided predictions for their static properties, such as
magnetic moments and electric radii, see e.g.~\cite{Wagner:1998fi,Wagner:2000ii}
for a collection of results and references. The validity of such models for
non-vanishing momentum transfer is questionable and more sophisticated quark
models have been developed. For example, a reduction of the Bethe-Salpeter
equation by using instantaneous interactions was used in
\cite{VanCauteren:2005sm,VanCauteren:2004fe} to study electromagnetic form
factors of octet hyperon resonances. A reduction to a quark-diquark system has
been used in~\cite{Carrillo-Serrano:2016igi} within an NJL model for the
interactions and in the covariant spectator formalism in~\cite{Ramalho:2011pp}.
In the latter the emphasis was placed on studying the role of pion cloud effects in
form factors. Static properties have also been studied using chiral
perturbation theory~\cite{Kubis:2000aa,Geng:2008mf,Geng:2009ys} and at low
momentum transfer in~\cite{Kubis:1999xb}.
The different role played by pion, kaon and eta
clouds as a tool for the extrapolation of lattice calculations to the physical
regime has been studied in full, quenched and partially quenched chiral
perturbation theory in~\cite{Leinweber:2002qb,Boinepalli:2006xd}.

A calculation of the elastic
electromagnetic form factors for the octet and decuplet form factors using DSEs
has been presented in~\cite{Sanchis-Alepuz:2015fcg}.
From lattice QCD, very precise data for the electromagnetic form factors of
octet baryons up to $Q^2=1.3$~GeV$^2$ have been recently published~\cite{Shanahan:2014uka,Shanahan:2014cga}. This calculation is performed using
$N_f=2+1$ dynamical flavours and omitting  disconnected diagrams. The
lattice results were obtained at unphysical pion masses and
extrapolated to the physical point using partially quenched chiral perturbation
theory. We show these results for the $\Sigma$ and $\Xi$ isomultiplets and
compare with the corresponding DSE results from~\cite{Sanchis-Alepuz:2015fcg}
as well as with the experimental values for the magnetic moments~\cite{Agashe:2014kda}.
In interpreting the differences between these results it
proves useful to make use of the works cited above
\cite{Leinweber:2002qb,Boinepalli:2006xd} on the role played (such as sign of
the effect, its
magnitude, etc.) by kaon, eta and pion clouds.

        \begin{figure*}
        \begin{center}
\includegraphics[width=0.35\textwidth]{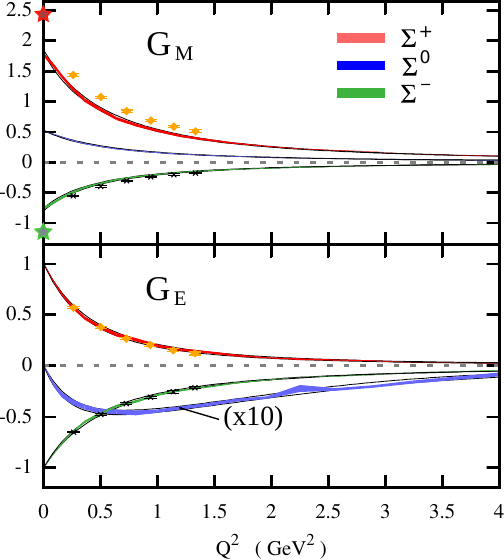}   %
\includegraphics[width=0.35\textwidth]{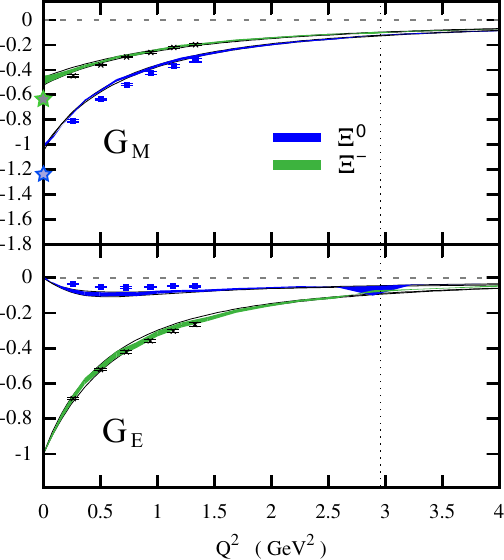}   %
                    \caption{Magnetic (\textit{upper panel}) and electric (\textit{lower panel})
form factors for the baryon
octet's $\Sigma$ triplet (\textit{left}) and $\Xi$ doublet (\textit{right}).
The magnetic form factor is given in units of the nuclear magneton $\mu_N$.
Coloured bands represent the result of the numerical DSE calculation
\cite{Sanchis-Alepuz:2015fcg} for $\eta=[1.6,2.0]$ in~\eqref{couplingMT}.
Vertical dotted lines indicate the region beyond which the singularities of
the quark propagator are probed.
Stars indicate experimental values. Lattice
data for non-vanishing $Q^2$ are from~\cite{Shanahan:2014uka,Shanahan:2014cga}.}
\label{fig:octet-ffs}
                    \end{center}
        \end{figure*}

For the charged $\Sigma$ baryons and for the $\Xi$ doublet the DSE results show an
overall good agreement with the corresponding lattice data. This is particularly true for the
electric form factors, which are protected in the infrared by charge conservation. It
is in the magnetic sector where one can more easily see the effect of the meson
cloud. While $G_M$ from the $\Sigma^-$ and $\Xi^-$ Dyson-Schwinger results
agrees well with the lattice data at finite $Q^2$, there is a significant deviation for the
$\Sigma^+$ and the $\Xi^0$. Also, at $Q^2=0$ the deviation of the DSE result for the  magnetic
moment with respect to the experimental
value is similar in relative size for the $\Sigma^+$ and the $\Sigma^-$, whereas
it is much smaller for the $\Xi^+$ than for the $\Xi^0$.
This pattern may be understood from
the different influence and the interplay of pion and kaon cloud effects on the
various states~\cite{Leinweber:2002qb,Boinepalli:2006xd}. For the $\Sigma^\pm$ states,
pion cloud effects play an important role for both states at small $Q^2$, leading to the
observed discrepancies between the DSE magnetic moments and the experimental values. For
larger values of $Q^2$ also kaon cloud effects are important, which are strong only in the
$\Sigma^+$-channel, where one sees a larger discrepancy of the DSE results
with lattice QCD\@. For the $\Xi$ states, pion cloud effects are small in all cases and kaon cloud
effects are much larger in the $\Xi^0$ than in the $\Xi^-$. As a consequence we observe only small
deviations of the magnetic moment of the $\Xi^-$ from the DSE calculation with both
experiment and lattice QCD\@.

\begin{figure*}[t!]
        \begin{center}
\includegraphics[width=0.70\textwidth]{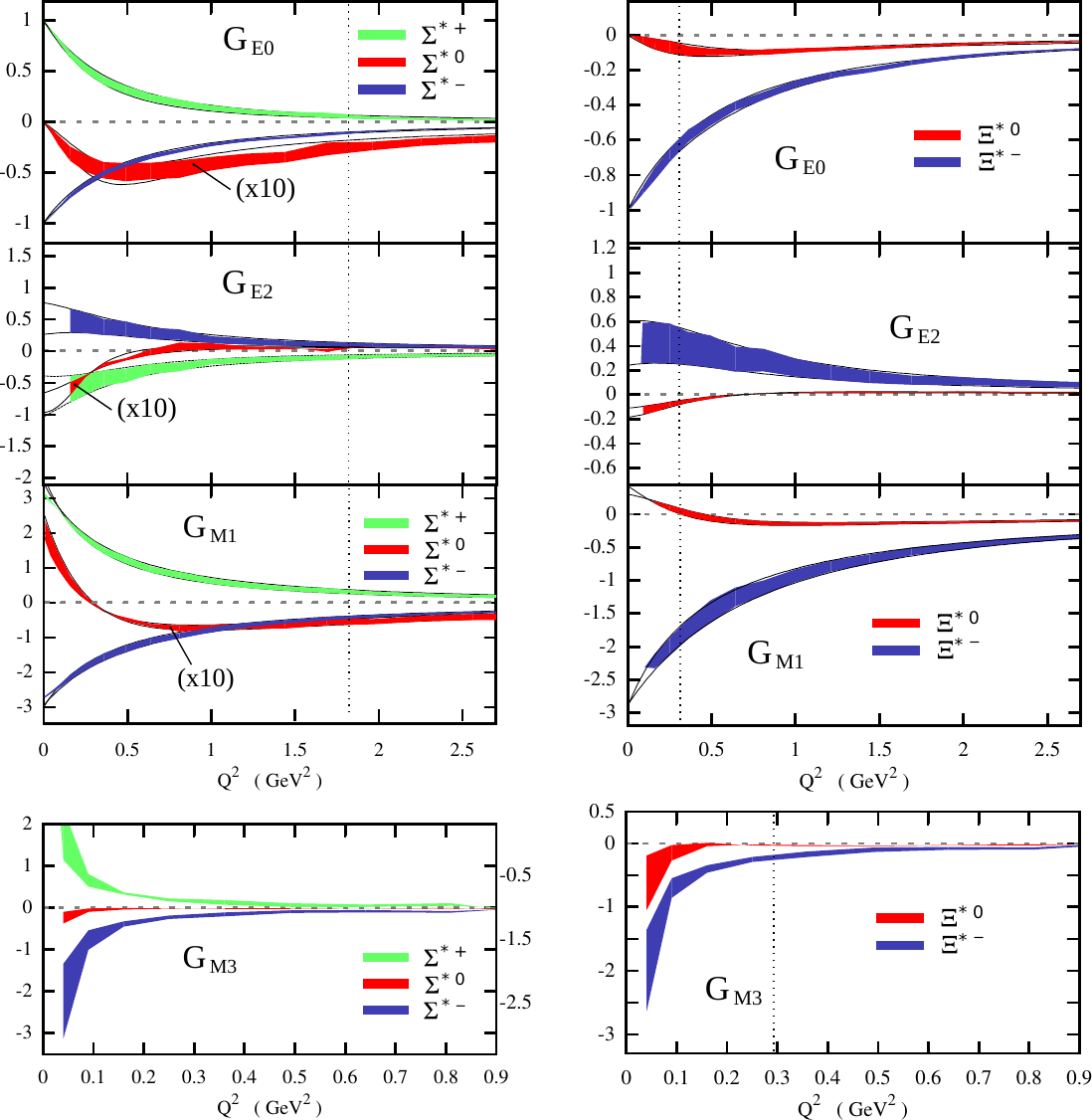}
                    \caption{Electromagnetic
form factors for hyperons in the baryon decuplet.
Coloured bands represent the result of the numerical DSE calculation
\cite{Sanchis-Alepuz:2015fcg} for $\eta=[1.6,2.0]$ in~\eqref{couplingMT}.
Vertical dotted lines indicate the region beyond which the singularities of
the quark propagator are probed.}
\label{fig:decuple-ffs}
                    \end{center}
        \end{figure*}

It is interesting to bring back a technical issue discussed earlier. As
explained in Sec.~\ref{sec:ff-currents} in the context of Eq.~\eqref{ff-limits}, form
factor calculations in the DSE approach are limited  to a maximum value
of $Q^2$ due to the presence of complex conjugate
  poles in the quark propagators. The specific values of these limits
depend on the masses of the baryons of interest. We show those limits in
Figs.~\ref{fig:octet-ffs} and~\ref{fig:decuple-ffs} as dotted
vertical lines. Although, in principle, calculations beyond those limits are
rigorously forbidden, it turns out that in most cases
  results are still smooth beyond these regions. On a phenomenological level,
and with due caution, extrapolations of DSE results for high-$Q^2$ values seem
thus possible.

Coming back to the discussion of results, it is clear that a pressing issue in
DSE calculations of baryon form factors is the inclusion of the
effects of the pion and kaon clouds. However, judging from the results
presented in Fig.~\ref{fig:octet-ffs} as well as from the corresponding results
for the nucleon and $\Delta$ presented earlier, it seems fair to say that even
a simplistic truncation of the DSE system, namely the rainbow-ladder
truncation, already provides an excellent description of (ground-state) baryon
structure. With this idea in mind, we tackle now the interpretation of the
decuplet form factor results from~\cite{Sanchis-Alepuz:2015fcg}, shown in
Fig.~\ref{fig:decuple-ffs}.

As already mentioned above, for decuplet hyperons there is no information from
experiment or from any realistic lattice QCD calculation to rely upon. The
only lattice results available to date~\cite{Boinepalli:2009sq} concern static
properties only and are calculated in the quenched approximation. With the
information gathered so far, one can thus interpret the results in
Fig.~\ref{fig:decuple-ffs} as qualitative predictions, especially for moderate
$Q^2$.

As a first remark, it is interesting to note that the form
factors of neutral states are, in contrast with those of the $\Delta^0$,
 not vanishing due to the presence of strange quarks. Moreover, the electric
monopole form
factors of both $\Sigma^{*0}$ and $\Xi^{*0}$ are of the same order as the
analogous $G_E$ of their octet counterparts.

As for the $\Delta$, the electric quadrupole form factor provides a measure of
the deviation of the
charge distribution from sphericity.
Keeping in mind that the extraction of $G_{E2}$ and $G_{M3}$ for
very low $Q^2$ is numerically difficult (as explained in
\cite{Sanchis-Alepuz:2015fcg}),
the DSE calculation
predicts a prolate  shape (positive $G_{E_2}$) for the
charge distribution in
$\Sigma^{*~+}$ and an oblate one (negative sign) for the $\Sigma^{*-}$. In the case
of
the $\Sigma^{*0}$ one obtains a zero crossing at $Q^2\sim 0.7$~GeV${}^2$,
the charge distribution changing from an oblate to a prolate shape
as
the electromagnetic probe increases in energy. It will be interesting to check
whether this feature survives a more sophisticated calculation in an extension
of the present framework including meson cloud effects.
An analogous result is seen for the $\Xi^{*~0}$ and $\Xi^{*~-}$ hyperons.
The quadrupole form factor $G_{E_2}$ also features a zero crossing for the
$\Xi^{*~0}$,
but this time in the region in which the singularities of the quark propagator
are
probed and the DSE predictions are less reliable.

As for the the magnetic dipole form factor $G_{M_1}$,
the most interesting feature is again a zero crossing for both neutral
members $\Sigma^{*0}$ and $\Xi^{*0}$. Finally, with respect to the magnetic
octupole form factors, extracting accurate results is again impossible
with the current accuracy. Only the sign of the corresponding form factor
seems to be stable. As with
$G_{E_2}$, a non-vanishing $G_{M_3}$ indicates a deformation of the magnetic
distribution, with a positive (negative) sign signalling a prolate (oblate)
shape.


\newpage

\section{Compton scattering}\label{compton}
       Nucleon resonances have been traditionally extracted from $N\pi$ scattering, and only in recent years meson photo- and electroproduction experiments have
       become the main tools for gathering information on the baryon excitation spectrum.
       Understanding the structure and dynamics of scattering amplitudes is clearly of great importance:
       dynamical reaction models and amplitude analyses based on general principles such as unitarity, analyticity and crossing symmetry are necessary to organize the experimental data
       and disentangle the various partial-wave contributions in order to extract resonance properties from $N\pi$, photo- and electroproduction amplitudes.
       On the other hand, scattering amplitudes contain an abundance of information in addition to hadronic poles and thus their study
       also serves a purpose beyond the extraction of resonance physics.

       An especially interesting example is nucleon Compton scattering (CS). It encodes a broad and fascinating range of physical applications
       spanning from nucleon (generalized) polarisabilities to structure functions, $p\bar{p}$ annihilation, two-photon corrections to form factors
       and the proton radius puzzle. The CS amplitude also provides access to structure properties that cannot be accessed in deep inelastic scattering,
       for example generalized parton distributions (GPDs)
       that encode the transverse spatial and spin structure and allow one to establish a three-dimensional tomography of the nucleon.
       Our experimental knowledge of the CS amplitude is mainly restricted to a few kinematic limits including the
       (generalized) polarisabilities in real and virtual CS, the forward limit,
       and deeply virtual CS (DVCS) from where generalized parton distributions are extracted.
       In addition, the crossed process $p\conjg{p}\to\gamma\gamma$ will be measured by PANDA.
       We will give a brief overview of some of these topics in Sec.~\ref{sec:cs-overview}.

\begin{figure}[t]
                    \begin{center}
                    \includegraphics[width=1\textwidth]{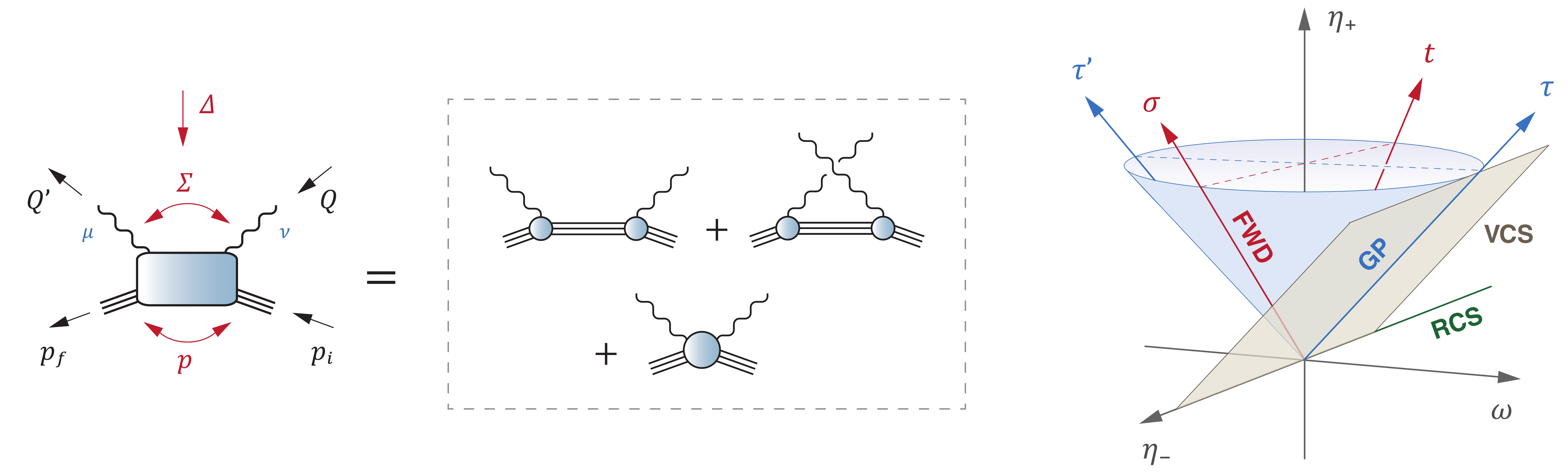}
                    \caption{\textit{Left:} Kinematics of the nucleon Compton scattering amplitude together with its decomposition into Born terms and a structure part.
                             \textit{Right:} Compton scattering phase space in the variables $\eta_+$, $\eta_-$ and $\omega$.
                             The interior of the cone is the spacelike region that is integrated over to obtain the two-photon exchange contribution to nucleon-lepton scattering.
                             Real Compton scattering (RCS) lives on the $\eta_-$ axis
                             and virtual Compton scattering (VCS) on the plane $\tau'=0$. The boundary of the cone contains the forward (FWD) limit at $t=0$ as well as the
                             VCS limit where the generalized polarisabilities (GPs) are defined.}\label{fig:phasespace}
                    \end{center}
            \end{figure}

        Before doing so, let us fix the kinematics and discuss some basic properties.
        The nucleon CS amplitude depends on three independent momenta (see Fig.~\ref{fig:phasespace}):
        the average nucleon momentum $p=(p_f+p_i)/2$, the average photon momentum $\Sigma = (Q+Q')/2$, and the momentum transfer $\Delta=Q-Q'$.
        The nucleon is onshell ($p_f^2=p_i^2=-m^2$) and therefore the process is described by four Lorentz-invariant kinematic variables which we define as
        \begin{align}
           \eta_+ = \frac{Q^2  +  {Q'}^2}{2m^2}\,, \quad
           \eta_- = \frac{Q\cdot Q'}{m^2}\,, \quad
           \omega = \frac{Q^2-{Q'}^2}{2m^2}\,, \quad
           \lambda = -\frac{p\cdot\Sigma}{m^2}\,,
       \end{align}
        where $m$ is the nucleon mass. This is completely analogous to pion electroproduction\footnote{Once again,
        Lorentz-invariant scalar products differ from their Minkowski counterparts only by minus signs and therefore
        these variables are the same in Minkowski space if one defines them as
        $\eta_+ = -(q^2 + {q'}^2)/(2m^2)$, $\eta_- = -q\cdot q'/m^2$ and so on.} discussed in Sec.~\ref{sec:electroproduction}, except that ${Q'}^2$ is arbitrary and not fixed to be ${Q'}^2=-m_\pi^2$.
        Occasionally we will make use of the alternative variables
        \begin{align}
            t = \frac{\Delta^2}{4m^2} = \frac{\eta_+-\eta_-}{2}\,, \quad
            \sigma = \frac{\Sigma^2}{m^2} = \frac{\eta_++\eta_-}{2}\,,\quad
            \tau = \frac{Q^2}{4m^2} = \frac{\eta_++\omega}{4}\,, \quad
            \tau' = \frac{{Q'}^2}{4m^2}=\frac{\eta_+-\omega}{4}
        \end{align}
        and refer to the Mandelstam variables
        \begin{align}  \renewcommand{\arraystretch}{1.0}
           \left\{ \begin{array}{c} s \\ u \end{array}\right\} = -(p \pm \Sigma)^2 = m^2 \,( 1- \eta_- \pm 2\lambda).
       \end{align}
        The kinematic phase space in the variables $\{\eta_+, \eta_-, \omega\}$ is illustrated in the right of Fig.~\ref{fig:phasespace}.
        The spacelike region that is integrated over in nucleon-lepton scattering (see Fig.~\ref{fig:nucleon-lepton-scattering} below) forms a cone around the $\eta_+$ axis.
        In that case $t$ remains an external variable, so the relevant phase space is the intersection of the plane $t=const.$ with the interior of the cone.
        The apex of the cone is where the static polarisabilities are defined, with momentum-dependent extensions to
        real CS ($\eta_+=\omega=0$),
        virtual CS ($\tau'=0 \Rightarrow \eta_+=\omega$) including the generalized polarisabilities at $\eta_-=0$ and $\lambda=0$,
        the doubly-virtual forward limit ($t=0$, $\tau=\tau'$ $\Rightarrow$ $\eta_+=\eta_-$, $\omega=0$),
        the timelike process $p\bar{p}\to\gamma\gamma$ for $t<-1$, and so on.

       Fig.~\ref{fig:phasespace} shows the decomposition of the CS amplitude into an `elastic' part, namely the Born terms which depend on the nucleon electromagnetic form factors,
       and an inelastic one-particle-irreducible (1PI) part that carries the structure and dynamics.
       Such a separation is formally always possible but in general not gauge invariant.
       This point is important because defining polarisabilities as the coefficients of the structure part
       is only sensible if the structure part satisfies electromagnetic gauge invariance alone; otherwise the polarisabilities have to be extracted
       from the complete CS amplitude.
       The intermediate nucleons in the Born terms are offshell and thus it is the half-offshell nucleon-photon vertex that appears in these diagrams,
       which has a richer structure than the onshell current.
       Individual gauge invariance of both Born and structure parts is only guaranteed by implementing an onshell Dirac current with $Q^2$-dependent Pauli and Dirac form factors
       (see~\cite{Scherer:1996ux} for a discussion),
       and in the following such a choice is implicitly understood.
       We will return to this issue in Sec.~\ref{sec:cs-hadronic-vs-quark} below.

\subsection{Overview of two-photon physics}\label{sec:cs-overview}

        \paragraph{Polarisabilities.}
       There has been much recent interest in a precision determination of the nucleon's polarisabilities, which is reflected in a number of reviews on the
       topic~\cite{Drechsel:2002ar,HydeWright:2004gh,Drechsel:2007sq,Downie:2011mm,Griesshammer:2012we,Schumacher:2013hu,Hagelstein:2015egb}.
       The electric polarisability $\alpha$ and magnetic polarisability $\beta$ tell us how the nucleon responds to an external electromagnetic field,
       with current PDG values $\alpha=11.2(4)\times 10^{-4}$ fm$^3$ and $\beta=2.5(4)\times 10^{-4}$ fm$^3$ for the proton~\cite{Agashe:2014kda}.
       The polarisabilities are proportional to the volume and their smallness indicates that the proton is a rigid object due to the strong binding of its constituents.
       Whereas $\alpha+\beta$ is constrained by a sum rule (see~\eqref{sum-rules} below), the small value for $\beta$ is commonly believed to be due to a cancellation
       between the paramagnetic nucleon `quark core' and the interaction with its diamagnetic pion cloud.

       While lattice calculations for polarisabilities are underway, see
       e.g.~\cite{Engelhardt:2007ub,Detmold:2009dx,Detmold:2010ts,Primer:2013pva,Hall:2013dva,Chang:2015qxa} and Refs.\ therein,
       the main theoretical tools to address CS are `hadronic' descriptions such as chiral perturbation theory, which provides
       a systematic expansion of the CS amplitude at low energies and low momenta, and dispersion relations which establish a direct link to experimental data.
       The main ideas behind these approaches are best understood by considering the forward CS amplitude as an example,
       which is related to the inelastic photoabsorption cross section via unitarity.
       In the following we will sketch the basic principles and refer to the aforementioned reviews for detailed discussions and a comprehensive list of references.

       In the forward limit the two photons are still virtual but the momentum transfer vanishes, $\Delta^\mu=0$, and thus $Q=Q'$ and $p_i=p_f$.
       In that case only two independent variables remain, $\eta_+ = \eta_- = \eta = Q^2/m^2$ and $\lambda = -p\cdot Q/m^2$ (whereas $\omega=0$), or equivalently the Mandelstam variables $s$ and $u$.
       The CS amplitude reduces to four independent tensor structures and it is given by 
       \begin{align}\label{cs-amp-fwd}
          \mc{M}^{\mu\nu}(p,Q) = \frac{1}{m}\,\conjg{u}(p) \,\bigg( \frac{c_1}{m^4}\,t^{\mu\alpha}_{Qp}\,t^{\alpha\nu}_{pQ} + \frac{c_2}{m^2}\,t^{\mu\nu}_{QQ}
          + \frac{c_3}{m}\,i \varepsilon^{\mu\nu}_{Q\gamma} + \frac{c_4}{m^2}\,\lambda\,\big[ t^{\mu\alpha}_{Q\gamma}, t^{\alpha\nu}_{\gamma Q}\big] \bigg)\, u(p)\,.
      \end{align}
       Here $u(p)$, $\conjg{u}(p)$ are nucleon spinors and we defined
       \begin{equation}\label{t-and-epsilon}
           t^{\mu\nu}_{AB}=A\cdot B\,\delta^{\mu\nu} - B^\mu A^\nu\,, \qquad
           \varepsilon^{\mu\nu}_{AB} = \gamma_5\,\varepsilon^{\mu\nu\alpha\beta} A^\alpha B^\beta\,,
       \end{equation}
       where $A$ and $B$ can denote four-vectors as well as $\gamma-$matrices.
       The Compton form factors (CFFs) $c_i(\eta,\lambda)$ are Lorentz-invariant dimensionless functions which are even in $\lambda$ due to crossing symmetry.\footnote{Their
       relation to the standard forward amplitudes, as defined for example in Ref.~\cite{Hagelstein:2015egb}, is given by
       $\{ T_1,\, T_2,\, S_1,\, S_2\} = 4\pi \alpha_\mathrm{QED}/m \times \{\lambda^2 c_1 + \eta\, c_2, \, \eta\, c_1,\, c_3,\, -2\lambda\,c_4\}$.}
       The four tensor basis elements have the lowest possible powers in the photon momenta without introducing kinematic singularities, and thus the only singularities contained in the CFFs
       are physical poles and cuts.

        \begin{figure}
        \centering
          \includegraphics[width=0.85\textwidth]{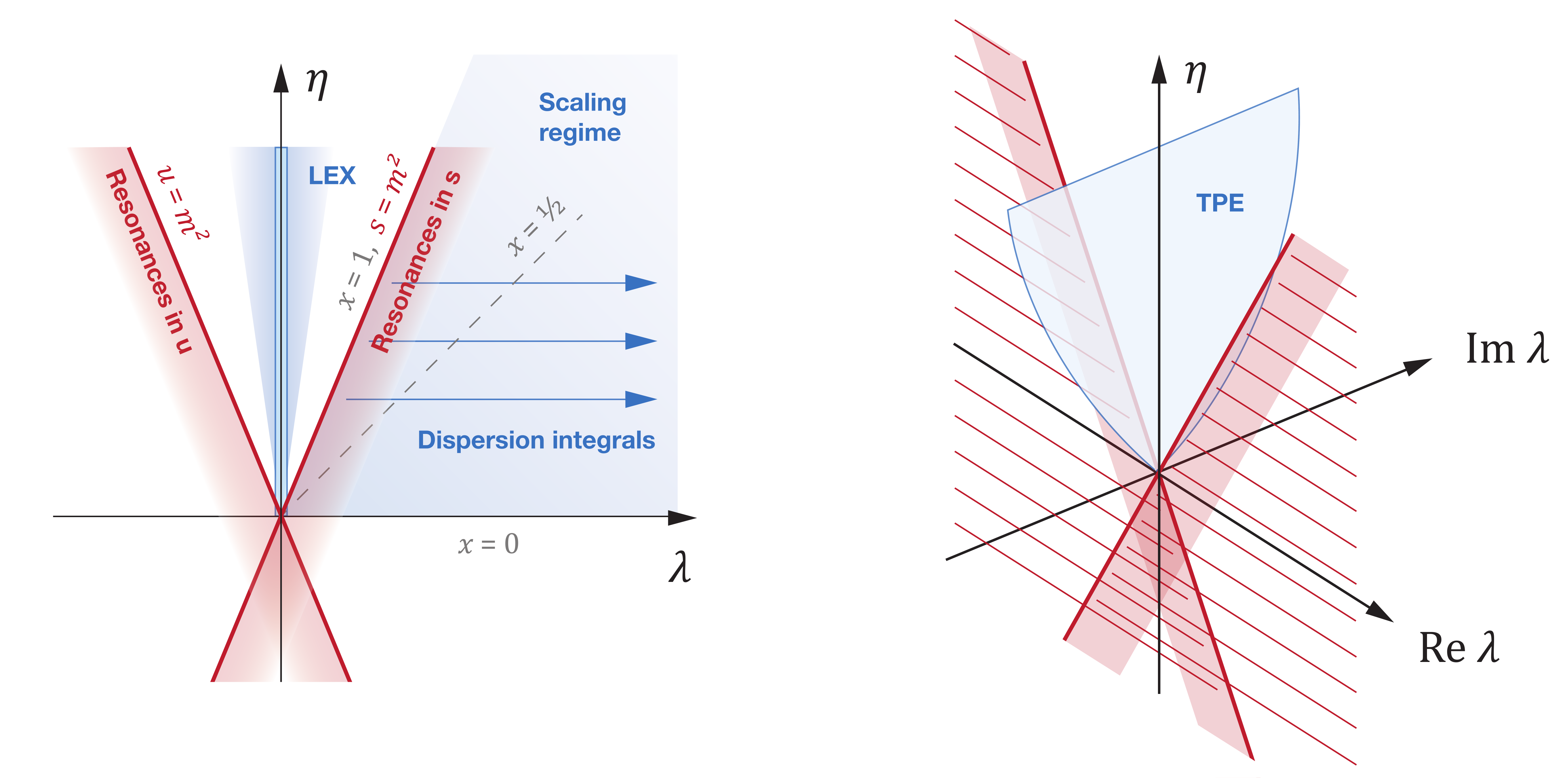}
        \caption{Phase space of the forward CS amplitude in the variables $\eta = Q^2/m^2$ and $\lambda = -p\cdot Q/m^2$. The $s-$ and $u-$channel nucleon Born poles are shown by the thick (red) lines together with
                 the nucleon resonance regions. The domain $x \in [0,1]$ in the upper right quadrant is the physical region in nucleon photoabsorption $N\gamma^\ast\to X$.
                 The diagram on the right illustrates the right- and left-hand cuts stemming from particle production as well as the integration region for
                 two-photon exchange (TPE) corrections to the nucleon-lepton scattering amplitude.}
        \label{fig:fwd-cs-amplitude}
        \end{figure}

       The forward phase space in the variables $\eta$ and $\lambda$ is sketched in Fig.~\ref{fig:fwd-cs-amplitude}.
       Nucleon resonances appear at fixed $s$ and $u$, starting with the Born poles at $s=m^2$ and $u=m^2$ (or $\lambda = \pm \eta/2$).
       The resonance regions are indicated by the shaded (red) areas in the plot, where
       at larger $s$ and $u$ the resonances are eventually washed out.
       In addition, at fixed $\eta$ one has branch cuts from multiparticle production:
       the right-hand cut starts at the first threshold $s=(m+m_\pi)^2$ and extends to infinity and the left-hand cut begins at $u=(m+m_\pi)^2$.
       The cut structure is visualised in the figure on the right.
       The cuts overlap in the timelike region $\eta < 0$, where in principle one has additional vertical cuts due to particle production off the photons.
       In any case, for spacelike photon momenta the CFFs are analytic functions in the physical sheet, apart from the Born poles and branch cuts
       which are confined to the real $\lambda$ axis. The right figure also shows the `Euclidean' domain $|\text{Im}\,\lambda| < \sqrt{\eta}$ along the imaginary axis
       where the CFFs are purely real. This is the analogue of the spacelike cone in Fig.~\ref{fig:phasespace} and it
       contributes to the two-photon exchange (TPE) integral in the nucleon-lepton cross section.

       The forward CS amplitude is of special interest because
       the optical theorem, which follows from unitarity, relates its imaginary part to the total photoabsorption cross section $\gamma^\ast N\to X$
       and thus to the nucleon's structure functions $f_{1,2}(x,Q^2)$ and $g_{1,2}(x,Q^2)$.
       In the physical region the Bjorken variable $x=\eta/(2\lambda)$ goes from $x \in [0,1]$ which corresponds to $\lambda \in [\eta/2, \infty)$.
       Since at fixed $\eta$ the imaginary parts of the CFFs along the cuts are known  from the cross section data,
       one can exploit Cauchy's formula to determine them everywhere in the complex $\lambda$ plane.
       This leads to dispersion relations of the form
       \begin{equation}\label{dispersion-relations}
          c_i(\eta,\lambda) = \frac{1}{\pi} \int_{\lambda_\text{el}^2}^\infty d{\lambda'}^2 \,\,\frac{\text{Im}\,\,c_i(\eta,\lambda')}{{\lambda'}^2 - \lambda^2-i\epsilon}\,,
       \end{equation}
       where the dispersion integrals extend from the elastic threshold $\lambda_\text{el} = \eta/2$ to infinity.
       Depending on whether the integrals converge or not for $|\lambda'|\to\infty$, it may be necessary to use subtracted dispersion relations for the quantities $c_i(\eta,\lambda) - c_i(\eta,0)$.
       Hence, apart from potential subtraction functions $c_i(\eta,0)$
       the forward CS amplitude is completely determined from the experimental data.

       On the other hand, for small values of $\lambda$ the CFFs can be expanded in powers of $\lambda^2$.
       This is technically also how the polarisabilities are defined, namely as the coefficients in a low-energy expansion (LEX). For instance,
       the $Q^2-$dependent generalized electric and magnetic polarisabilities appear in the expansion of the scalar CFFs $c_1$ and $c_2$:
       \begin{equation}\label{lex}
          c_1(\eta,\lambda) = c_1^\text{Born}(\eta,\lambda) + \frac{m^3}{\alpha_\mathrm{QED}}\, (\alpha+\beta) + \mc{O}(\lambda^2)\,,  \qquad
          c_2(\eta,\lambda) = c_2^\text{Born}(\eta,\lambda) + \frac{m^3}{\alpha_\mathrm{QED}}\, \beta + \mc{O}(\lambda^2)\,.
       \end{equation}
       The Born terms are singular in the limits $\lambda = \pm \eta/2$ and depend on the electromagnetic Dirac and Pauli form factors which contain the nucleon's magnetic moments and charge radii.
       Upon expanding also the right-hand sides of the dispersion integrals and comparing the coefficients one obtains sum rules that relate these quantities with moments of the structure functions,
       for example:
       \begin{equation}\label{sum-rules}
           \alpha+\beta = \frac{\alpha_\mathrm{QED}}{m^3}\int_{\lambda_0}^\infty d\lambda^2\,\frac{f_1(x,Q^2)}{\lambda^4}\,, \qquad
           \frac{\kappa^2}{2} = \lim_{\eta\to 0}\,\int_{\lambda_0}^\infty d\lambda^2\,\frac{\eta \,g_2(x,Q^2) -\lambda^2 \,g_1(x,Q^2)}{\lambda^5}\,,
       \end{equation}
       where $\lambda_0 = \eta/2 + m_\pi/m + m_\pi^2/(2m^2)$ is the pion production threshold.
       The first relation is the Baldin sum rule~\cite{Baldin1960310} for the polarisability sum $\alpha+\beta$ and
       the second is the Gerasimov-Drell-Hearn (GDH) sum rule~\cite{Gerasimov:1965et,Drell:1966jv} for the nucleon's anomalous magnetic moment $\kappa$.
       In principle both relations hold for arbitrary $\eta$, which defines the generalized polarisabilities $\alpha(\eta)$ and $\beta(\eta)$ and the generalized GDH integral.
       For extensive discussions of the various sum rules derived from CS we refer to the reviews~\cite{Schumacher:2013hu,Hagelstein:2015egb}.

       On the other hand, at low momenta and low energies (which translates into low powers of $\eta$ and~$\lambda$) the CFFs can be calculated dynamically from chiral effective field theory.
       In the forward limit this is particularly relevant for determining the subtraction function $c_2(\eta,0)$ that appears in the dispersion relations
       and is related to the magnetic polarisability; see our comments on the proton radius puzzle below.
       The role of chiral effective field theory and its application to Compton scattering is discussed in the recent reviews~\cite{Griesshammer:2012we,Hagelstein:2015egb}.

       We have discussed the forward limit but the situation in real and virtual CS is similar.
       In real CS there are six CFFs which depend on two independent variables ($\eta_-$ and $\lambda$), whereas
       in virtual CS one has twelve CFFs and three Lorentz invariants ($\eta_+$, $\eta_-$ and $\lambda$).
       In the off-forward case $t\neq 0$ the link to the photoabsorption cross section and the nucleon structure functions is lost.
       However, one can still set up dispersion relations analogous to~\eqref{dispersion-relations}, where unitarity relates the imaginary part of the CS amplitude
       to the sum over all intermediate hadronic ($N\pi$, $N\pi\pi$, etc.) states.
       In addition to the $s$- and $u$-channel poles and cuts induced by $N\gamma^\ast$, one must also take into account
       timelike production off single photons as well as two-photon singularities in the $t$ channel, which complicate the analyses considerably.
       Real and virtual CS are conceptually similar to the pion photo- and electroproduction processes discussed in Sec.~\ref{sec:electroproduction}, including
       the makeup of the phase space in Fig.~\ref{fig:electroprod-phasespace} which would be identical for a vanishing pion mass.
       Deeply virtual CS is especially important in this respect as it allows one to probe the partonic nature of the Compton scattering process
       and extract generalized parton distributions; see~\cite{Belitsky:2005qn,Guidal:2013rya,Mueller:2014hsa,Kumericki:2016ehc} for reviews.

        \paragraph{Two-photon corrections to form factors.}

\begin{figure}
        \centering
          \includegraphics[width=0.89\textwidth]{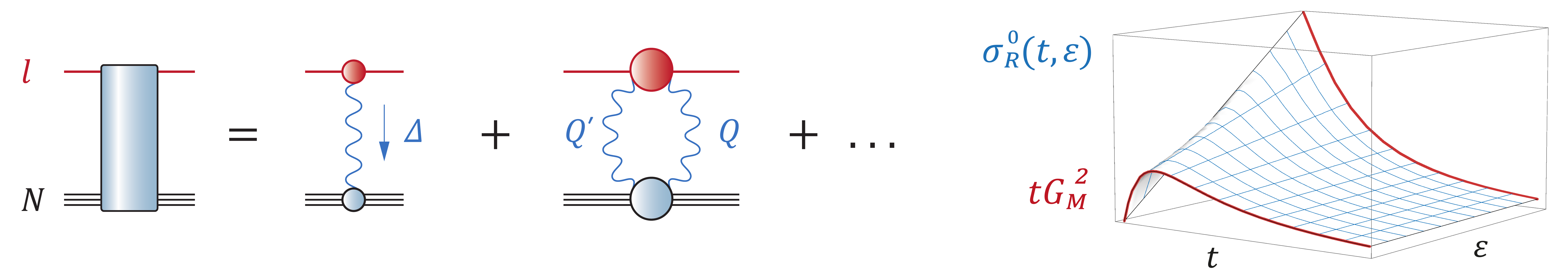}
        \caption{\textit{Left:} Nucleon-lepton scattering amplitude and its contributions in terms of one- and two-photon exchanges.
                 Small circles are the nucleon and lepton electromagnetic currents and large circles denote their Compton amplitudes.
                 \textit{Right:} Sketch of the reduced Rosenbluth cross section in the variables $t>0$ and $0<\varepsilon<1$. At fixed $t$ the $\varepsilon-$dependence is linear.}
        \label{fig:nucleon-lepton-scattering}
        \end{figure}

        An important application of CS concerns two-photon exchange (TPE) contributions to nucleon form factors. 
        The issue resurfaced after measurements of the proton's form factor ratio $\mu_p\,G_E/G_M$ and its deviation from unity; see~\cite{Carlson:2007sp,Arrington:2011dn} for reviews.
        The main problem is visualised in Fig.~\ref{fig:nucleon-lepton-scattering}.
        The electromagnetic form factors of the nucleon are experimentally extracted
        from $Ne^-$ scattering, traditionally under the assumption of dominant one-photon exchange (Born approximation) since each additional photon contributes a factor $\alpha_\mathrm{QED} \approx \nicefrac{1}{137}$.
        If we stick to the kinematics in Fig.~\ref{fig:phasespace} and denote the average nucleon and lepton momenta by $p$ and $k$, respectively, then
        the scattering amplitude depends on two Lorentz invariants: the momentum transfer $t=\Delta^2/(4m^2)$
        and the crossing variable $\nu = -p\cdot k/m^2$. For convenience one defines the variable
        \begin{equation}
           \varepsilon = \frac{\nu^2 - t\,(1+t)}{\nu^2+t\,(1+t)} = \left( 1 + 2\,(1+t)\,\tan^2\frac{\theta}{2}\right)^{-1}\; \in \; [0,1]\,,
        \end{equation}
        where $\theta$ is the laboratory scattering angle.
        The Born approximation then yields the Rosenbluth cross section:
        \begin{equation}\label{rosenbluth-cs}
           \frac{d\sigma}{d\Omega} = \left(\frac{d\sigma}{d\Omega}\right)_\text{Mott}\, \frac{\sigma_R^0(t,\varepsilon)}{\varepsilon\,(1+t)}\,, \qquad \sigma^0_R(t,\varepsilon) = \varepsilon\,G_E^2(t) + t\,G_M^2(t)\,.
        \end{equation}
        The Mott cross section describes lepton scattering off a pointlike scalar particle, and the deviation from it
        accounts for the spin-$\nicefrac{1}{2}$ nature of the nucleon
        together with its electric and magnetic form factors.
        The reduced cross section $\sigma^0_R$ is sketched in Fig.~\ref{fig:nucleon-lepton-scattering}. At fixed $t$,
        its dependence on $\varepsilon$ is linear and thus allows one to extract the magnetic form factor from the intercept at $\varepsilon = 0$
        and the electric form factor from the slope in $\varepsilon$. This is referred to as the Rosenbluth method~\cite{Rosenbluth:1950yq}, and
        the extracted form factor ratio satisfies $\mu_p\,G_E/G_M \approx 1$ in agreement with perturbative scaling arguments.

        The drawback of the Rosenbluth method is that the contribution from $G_E$ becomes small at larger $t$,
        which makes $G_E$ sensitive to (even small) $\varepsilon-$dependent corrections to the cross section.
        In this respect, new experiments with polarised beams and/or polarised targets performed at Jefferson Lab
        allowed to extract the ratio $G_E/G_M$ directly.
        However, the resulting ratio disagrees with the scaling predictions and shows
        a falloff with $t$, even pointing towards a zero crossing at larger $t$~\cite{Jones:1999rz,Gayou:2001qd,Punjabi:2005wq,Puckett:2010ac}.

        Can TPE effects resolve the discrepancy?
        From an analysis of the general structure of the $Ne^-$ scattering amplitude it was concluded in Ref.~\cite{Guichon:2003qm}
        that TPE corrections can indeed generate large corrections to the ratio obtained with the Rosenbluth separation,
        whereas they would have a minimal impact on the ratio extracted from the polarisation transfer measurements.
        Explicit TPE calculations were subsequently performed using hadronic approaches including the elastic Born terms~\cite{Blunden:2003sp,Blunden:2005ew}
        and inelastic resonance contributions of the CS amplitude~\cite{Kondratyuk:2005kk,Kondratyuk:2007hc}, as well as employing dispersion relations~\cite{Borisyuk:2008es,Gorchtein:2006mq,Tomalak:2014dja}
        and perturbative analyses at the quark level based on handbag and perturbative QCD mechanisms~\cite{Chen:2004tw,Afanasev:2005mp,Carlson:2007sp,Borisyuk:2008db,Kivel:2009eg}.

        Because the polarisation experiments measure the ratio and not $G_M$ itself, they only determine the $\varepsilon$ slope at fixed $t$  in Fig.~\ref{fig:nucleon-lepton-scattering},
        which turns out to be flatter than the one obtained from the Rosenbluth separation.
        TPE effects on the cross section are most pronounced at small $\varepsilon$, which
        implies that the true Born contribution~$\sigma_R^0$ inferred from the polarisation measurements should be larger than the full cross section:
        $\sigma_R = \sigma^0_R\,(1+\delta)$, with a negative TPE contribution $\delta(t,\varepsilon)$ for small $\varepsilon$.
        It turns out that the sign of $\delta$ is clearly sensitive to structure effects:
        whereas the Born terms for pointlike nucleons (and thus also pointlike quarks) produce a positive sign,
        the hadronic calculations yield negative values for $\delta$ at larger $t$ which are in the right ballpark to explain the discrepancy.
        In summary, most of these studies can explain parts of the difference in terms of TPE effects, although
        the precise interplay between hadronic and partonic effects remains an open question.
        To this end, the ratio of the elastic $e^+p/e^-p$ cross sections has been measured by several experiments to
        provide a definitive test of the magnitude of TPE effects~\cite{Rachek:2014fam,Adikaram:2014ykv,Rimal:2016toz,Milner:2013daa}.

        \paragraph{Proton radius puzzle.}
        A closely related issue that also depends on the magnitude of TPE corrections is the proton radius puzzle. We will only briefly sketch the problem here and refer
        to~\cite{Jentschura:2010ej,Antognini:2013jkc,Pohl:2013yb,Karshenboim:2015kk,Carlson:2015jba,Hagelstein:2015egb} for detailed reviews.
        The basic outline is the same as in Fig.~\ref{fig:nucleon-lepton-scattering}:
        both the scattering of electrons and muons on nucleons as well as their energy levels in electronic and muonic atoms
        are sensitive to the structure of the proton.
        This entails in particular a dependence on the proton charge radius $R_E$, as defined in analogy to~\eqref{charge-radii-2} from the slope of the proton's electric form factor.
        The values for $R_E$ extracted from electron measurements (electron-proton scattering and hydrogen spectroscopy) essentially
        agree with each other and are reflected in the CODATA value $R_E=0.8775(51)$ fm~\cite{Mohr:2012tt}.

        Since the muon is about 200 times heavier than the electron, experiments with muons are more sensitive to proton structure effects.
        Phenomenologically speaking, a muon orbits closer to the proton than an electron and its energy levels are more strongly affected by the proton size.
        The CREMA collaboration has performed precision measurements of the $2P$\,--\,$2S$ transitions
        in muonic hydrogen~\cite{Pohl:2010zza,Antognini:1900ns}. The experimental result for the Lamb shift is $E_\text{exp} = 202.3706(23)$ meV, which has to be
        compared to the theoretical value~\cite{Antognini:2013jkc}
        \begin{equation}
           E_\text{th} = 206.0336(15) - 5.2275(10)\,(R_E/\text{fm})^2 + E_\text{TPE}\,.
        \end{equation}
        The first term contains QED effects including vacuum polarisation and self-energy corrections which are independent of the proton radius.
        The third term comprises two-photon effects and its currently accepted value is $E_\text{TPE} = 0.0332(20)$~meV~\cite{Birse:2012eb},
        with similar numbers obtained by various authors (see Table~6.1 in Ref.~\cite{Hagelstein:2015egb} for a compilation).
        Therefore, fitting theory to experiment allows one to extract the proton charge radius $R_E=0.84087(39)$.
        Its value is an order of magnitude more precise than the one obtained with electron measurements,
        but it is also by $7\sigma$ smaller than the CODATA value -- hence the proton radius puzzle.

        Possible origins of the discrepancy include:
        (1) a problem with the electron measurements in the sense that the uncertainty in the proton radius extraction may be larger than expected;
        (2) missing hadronic effects; or (3) new physics beyond the Standard Model through so far undiscovered new interactions that violate lepton universality.
        The first point may seem obsolete in light of the available precise low-$Q^2$ electron scattering data which support the upper value for the proton radius~\cite{Bernauer:2010wm}.
        It has been argued, however, that the data may be equally compatible with the smaller radius~\cite{Lorenz:2014vha,Griffioen:2015hta,Higinbotham:2015rja} although the issue is disputed~\cite{Bernauer:2016ziz}.
        In this respect, forthcoming data at extremely low $Q^2$ from the PRAD experiment at Jefferson Lab~\cite{PRad} may help to resolve the problem.
        In any case, even if the smaller radius turned out to be the correct one this does still not explain the larger radius inferred from atomic spectroscopy.
        The second and third explanations, on the other hand, highlight the similarity of the proton radius puzzle with the muon g-2 problem~\cite{Jegerlehner:2009ry}.
        In both cases it is the muon measurement that deviates from the established or expected result.
        This could either point towards a common origin beyond the Standard Model, or also to missing QCD effects since they are greatly magnified in the muonic system
        compared to the electronic environment.
        Whereas in the g-2 case the QCD contributions
        are encoded in the hadronic vacuum polarisation and the light-by-light scattering amplitude,
        the proton radius puzzle is sensitive to TPE effects whose origin is the nucleon CS amplitude.

        Since it is essentially the forward limit of the CS amplitude that is relevant for the proton radius problem,
        its two contributions, elastic Born terms and inelastic structure part, are both constrained by experimental data.
        The Born terms are determined by the nucleon electromagnetic form factors and the inelastic part is related to the nucleon structure functions via dispersion relations, as discussed above.
        The bigger contribution to $E_\text{TPE}$ comes from the Born terms; it is encoded in the 3rd Zemach moment~\cite{Friar:1978wv} together with its recoil corrections.
        The $Q^2$-dependent subtraction function that appears in the inelastic part can be determined from the magnetic polarisability in combination with
        low-energy expansions, chiral perturbation theory and perturbative QCD constraints~\cite{Pachucki:1999zza,Carlson:2011zd,Hill:2011wy,Birse:2012eb,Alarcon:2013cba}.
        However, the resulting value for $E_\text{TPE}$ is too small to resolve the problem --
        to obtain a proton radius consistent with the CODATA value, the TPE contributions would have to be larger by an order of magnitude (requiring $E_\text{TPE} \sim 0.360$~meV).
        Therefore, at present
        it seems unlikely that TPE effects can resolve the proton radius puzzle unless one is willing to account for `haywire' hadronic effects~\cite{Miller:2012ne,Carlson:2015jba}.
        In any case, given that our overall knowledge of the CS amplitude is still rather sparse
        more detailed investigations are certainly desirable.

\subsection{Hadronic vs.\ quark-level description}\label{sec:cs-hadronic-vs-quark}

\begin{figure}
 \centering
   \includegraphics[width=1\textwidth]{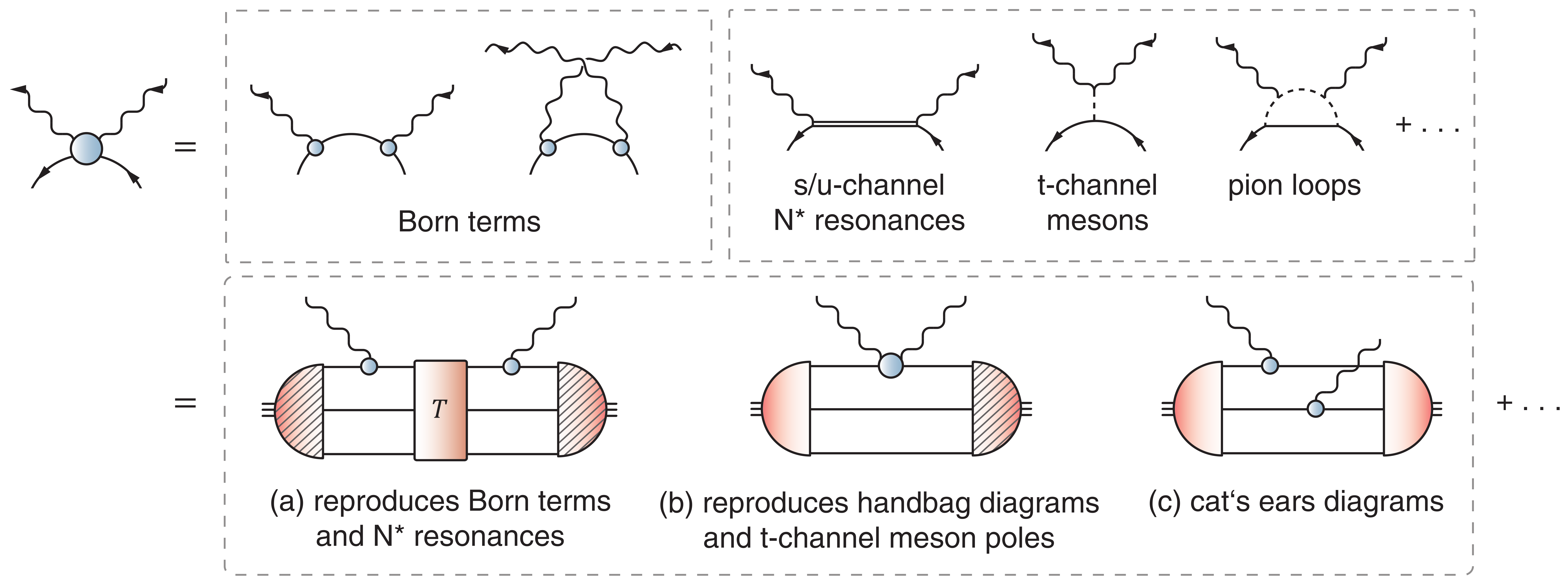}
 \caption{Hadronic vs. quark-level decomposition of the nucleon Compton scattering amplitude. The first row depicts the hadronic contributions as the sum of Born terms and a 1PI structure part.
          The latter encodes the polarisabilities and contains $s/u-$channel nucleon resonances, $t-$channel meson exchanges and pion loops.
          The second row shows the microscopic decomposition (in rainbow-ladder) featuring Bethe-Salpeter amplitudes, quark propagators, quark-photon and quark Compton vertices, and the three-quark scattering matrix~\cite{Eichmann:2012mp}.}
 \label{fig:cs-decomposition}
 \end{figure}

       A question that is common to all systems discussed so far is the interplay of hadronic and quark-level effects.
       At the hadronic level the CS amplitude is given by the sum of Born terms, which are determined by the nucleon form factors,
       and the 1PI structure part that carries the dynamics and encodes the polarisabilities, see Fig.~\ref{fig:cs-decomposition}.
       The latter contains $s/u-$channel nucleon resonances beyond the nucleon Born terms (including the $\Delta$, Roper, etc.),
       $t-$channel meson exchanges (pion, scalar, axialvector, $\dots$), and pion loops,
       with well-established low-energy expansions in chiral effective field theory.
       This is usually viewed as the `correct' description at low energies.
       On the other hand, handbag dominance in DVCS is well established and a key ingredient to factorization theorems,
       and the handbag picture is interpreted as the `correct' approach at large photon virtualities.
       Is it then possible to connect these two facets by a common, underlying description at the level of quarks and gluons
       that is valid in \textit{all} kinematic regions and reproduces all established features, from hadronic poles to the handbag picture?

 \paragraph{Microscopic expression for the scattering amplitude.}
       In analogy to the form factor diagrams in Fig.~\ref{fig:current-faddeev} one can derive a closed nonperturbative expression for the CS~amplitude
       and other scattering amplitudes at the quark level~\cite{Eichmann:2011ec,Eichmann:2012mp}.
       The onshell scattering amplitude $\mc{M}^{\mu\nu}$ is the residue of the quark six-point function that is coupled to \textit{two} external currents with $q\bar{q}$ quantum numbers:
        \begin{equation}
            \mathbf G^{\mu\nu} \; \stackrel{P_f^2=-m_f^2, \,P_i^2=-m_i^2}{\longlonglonglongrightarrow}  \; \frac{\mathbf\Psi_f\,\mc{M}^{\mu\nu} \,\conjg{\mathbf\Psi}_i}{(P_f^2+m_f^2)(P_i^2+m_i^2)}\,,
        \end{equation}
        where $\mathbf\Psi_f$ and $\mathbf\Psi_i$ again denote the Bethe-Salpeter wave functions,
        and $m_i=m_f$ if the incoming and outgoing baryons are the same.
        Following similar steps as in Sec.~\ref{sec:ff-currents} one arrives at the following expression for the scattering amplitude:
        \begin{equation}\label{scattering-amplitude-1}
            \mc{M}^{\mu\nu} = \conjg{\mathbf\Psi}_f \left[ \left(\mathbf{G}^{-1}\right)^{\{\mu} \mathbf{G} \left(\mathbf{G}^{-1}\right)^{\nu\}} - \left(\mathbf{G}^{-1}\right)^{\mu\nu} \right] \mathbf\Psi_i \,.
        \end{equation}
        The curly brackets denote a symmetrization of the indices and
        the quantities $\left(\mathbf G^{-1}\right)^{\mu}$ and $\left(\mathbf G^{-1}\right)^{\mu\nu}$ read (in a slightly simplified notation):
        \begin{equation}\label{cs-decomp-kernels}
        \begin{split}
           \left(\mathbf G^{-1}\right)^{\mu} = \left(\mathbf{G_0}^{-1}\right)^{\mu} - \mathbf K^{\mu} &= \Big[\, \Gamma^{\mu} \otimes S^{-1} \otimes S^{-1}
            - \Gamma^{\mu} \otimes K_\text{(2)}  -  S^{-1} \otimes K_\text{(2)}^{\mu}  + \text{perm.} \,\Big] - K_\text{(3)}^{\mu} \\
           \left(\mathbf G^{-1}\right)^{\mu\nu} = \left(\mathbf{G_0}^{-1}\right)^{\mu\nu} - \mathbf K^{\mu\nu} &=
           \Big[\, \Gamma^{\mu\nu} \otimes S^{-1} \otimes S^{-1}  -\Gamma^{\mu\nu} \otimes K_\text{(2)}  + \Gamma^{\{\mu} \otimes \Gamma^{\nu\}} \otimes S^{-1} \\
           &  \qquad  - \Gamma^{\{\mu} \otimes K_\text{(2)}^{\nu\}} - S^{-1} \otimes K_\text{(2)}^{\mu\nu} + \text{perm.}\,\Big] - K_\text{(3)}^{\mu\nu}\,.
        \end{split}
        \end{equation}
        Depending on the types of hadrons and currents involved, the resulting scattering amplitudes describe a variety
        of different reactions such as Compton scattering, pion electroproduction, $N\pi$ scattering, or crossed-channel
        processes such as $p\bar{p}$ annihilation into two photons or meson production.
        The approach can be applied to mesons as well to derive the expressions for pion Compton scattering, $\pi\pi$ scattering (from the residue of the correlator
        of four pseudoscalar currents) or the hadronic light-by-light amplitude (as the correlator of four vector currents).
        This is worked out in detail in Refs.~\cite{Eichmann:2011ec,Goecke:2012qm}.
        For example, for a scattering amplitude with external pion legs one has to take the residue of the pseudoscalar vertex at the pion pole, which has the same effect
        as replacing the vertex $\Gamma^\mu$ by the onshell pion amplitude.

        The physics content of~\eqref{scattering-amplitude-1} is best assessed diagrammatically.
        The second row of Fig.~\ref{fig:cs-decomposition} illustrates the generic topologies (apart from permutations and symmetrizations)
        that survive in a rainbow-ladder truncation where the gauged kernels $K_\text{(2,3)}^{\mu}$ and $K_\text{(2,3)}^{\mu\nu}$ do not contribute.
        The diagrammatics represents a convenient reformulation of~(\ref{scattering-amplitude-1}) along the following lines.
        The first term for $\mc{M}^{\mu\nu}$  in \eqref{scattering-amplitude-1} contains the quark six-point function $\mathbf{G}$ which contributes
        all possible baryon poles. At any such pole location, $\mathbf{G}$ takes the form
             \begin{equation}
                 \mathbf G \; \stackrel{P^2=-m^2}{\longlongrightarrow}  \; \frac{\mathbf\Psi\,\conjg{\mathbf\Psi}}{P^2+m^2}\,,
             \end{equation}
        and by comparison with~\eqref{emcurrent-gauging} one confirms that the resulting contributions to $\mc{M}^{\mu\nu}$ are indeed the
        products of two currents with an intermediate propagator. Hence, this term reproduces the {\it nucleon} Born terms in the CS amplitude together
        with all intermediate $s$- and $u$-channel resonances which contribute to diagram (a) of Fig.~\ref{fig:cs-decomposition}.
        On the other hand, $\mathbf{G}$ also contains disconnected diagrams which produce {\it quark} Born terms.
        In Fig.~\ref{fig:cs-decomposition} those
        were combined with the 1PI quark two-photon (Compton) vertex $\Gamma^{\mu\nu}$, which appears
        in the second term of~\eqref{scattering-amplitude-1}, into the full Compton vertex shown in diagram (b). Thus this quantity
        has the same decomposition as in Fig.~\ref{fig:phasespace}, i.e., it is given by the sum of quark Born terms and a 1PI part containing
        all further structure~\cite{Eichmann:2012mp}. The quark Born terms are important in that they provide the handbag topologies which
        reproduce the perturbative handbag diagrams in DVCS. Furthermore, note that the hatched amplitudes in both diagrams abbreviate the Bethe-Salpeter wave
        functions of the baryons multiplied with $S^{-1}\otimes S^{-1} - K_{(2)}$, i.e., the combination that appears in~\eqref{cs-decomp-kernels}
        denoting the interactions of the spectator quarks.
        The remaining cat's-ears topologies in diagram (c) originate from contributions of the form $\Gamma^{\{\mu} \otimes \Gamma^{\nu\}} \otimes S^{-1} $
        together with further disconnected parts that originate from $\mathbf{G}$ in~\eqref{scattering-amplitude-1}. Here the photons couple to
        different quark lines without a T-matrix insertion.

        We mentioned in the discussion around~\eqref{vertex-def-2} that the quark-photon vertex is the quark-antiquark Green function that is contracted with the tree-level tensor $\gamma^\mu$
        (modulo propagator legs).
        Similarly, one can show that the quark Compton vertex is the quark-antiquark Green function contracted with the quark Born terms plus a further piece that vanishes in rainbow-ladder~\cite{Eichmann:2012mp}.
        As a consequence, it contains all possible intermediate mesons in the $t$ channel which reproduce the meson exchanges in the first row of Fig.~\ref{fig:cs-decomposition},
        but for example also quark-disconnected diagrams in the form of multi-gluon exchange.
        This property also allows one to derive an inhomogeneous Bethe-Salpeter equation for the vertex from where it can be determined dynamically, including its Born and 1PI parts,
        in consistency with the dynamical equations for other vertices and wave functions.

        In summary, one finds that the microscopic decomposition reproduces all onshell hadron pole contributions, whereas
        ambiguities stemming from intermediate offshell hadrons never arise because hadronic degrees of freedom do not appear explicitly.
        On the other hand, neither the handbag nor the cat's-ears contributions from diagram (c) have a direct analogue in the hadronic expansion where they are rather absorbed into counterterms.
        In any case, the sum of all graphs in the box of Fig.~\ref{fig:cs-decomposition}
        satisfies electromagnetic gauge invariance  so that the resulting CS amplitude is purely transverse; it is $s/u-$channel crossing symmetric;
        it reproduces all known hadronic poles; and it contains the handbag contributions which are perturbatively (and presumably also nonperturbatively) important.
        In this respect it is perhaps not surprising that the aforementioned hadronic and partonic TPE calculations for form factors each reproduce roughly \textit{half} of the required effect,
        because they correspond to different quark-gluon topologies: the `hadronic' aspects are encoded in diagram (a) whereas the `partonic' contributions essentially come from diagram (b).

        A more fundamental question concerns the identification of pion loops and hadronic coupled-channel effects, because
        this is where the hadronic and quark-gluon descriptions begin to overlap. In the microscopic decomposition
        such offshell hadronic loops never appear; at best they can be viewed as an effective way of dealing with more complicated quark-gluon topologies.
        For example, the gauged kernel diagrams in~\eqref{cs-decomp-kernels}, which drop out in rainbow-ladder,
        contain topologies with four- and six-point Green functions that can conspire to reproduce the $N\pi$ loops shown in the top row of Fig.~\ref{fig:cs-decomposition},
        but pion effects are also inherent in diagram (a) which contains the nucleon resonances.
        While the hadronic and quark-level approaches should be ultimately equivalent, one still has to keep in mind that
        hadronic equations are derived from non-renormalizable effective Lagrangians which depend upon input from experiment or the underlying theory which is QCD.

 \paragraph{Gauge invariance.}
 Among the theoretical cornerstones of Compton scattering is electromagnetic gauge invariance, which
 requires that the CS amplitude is transverse with respect to the photon momenta: ${Q'}^\mu \mc{M}^{\mu\nu} = 0$ and $Q^\nu \mc{M}^{\mu\nu}= 0$.
 In combination with analyticity, it entails that the 1PI part in Fig.~\ref{fig:phasespace} must be at least quadratic in the photon momenta.
 Transversality and analyticity allow one to establish low-energy expansions for the non-Born parts, and they form the
 origin of the low-energy theorem in CS which states that in the limit where the static polarisabilities are defined,
 the cross section (which is then completely determined by the Born terms) reproduces the Thomson limit of a structureless Dirac particle.

       A problem in practice is the following:  what if the CS amplitude is decomposed into several parts as in Fig.~\ref{fig:phasespace}, where only the sum is gauge invariant but not the individual terms alone?
       For example, a sensible definition of polarisabilities requires both the Born and 1PI parts to be individually gauge invariant,
       but this is not guaranteed from the definition.
       Similarly, what happens if the microscopic calculation is approximated by a subset of diagrams, so that gauge invariance is broken by an \emph{incomplete} calculation?
 Since transversality is connected with analyticity, a simple transverse projection does not suffice because
 an approximation that breaks electromagnetic gauge invariance can induce kinematic singularities that render its results meaningless.

 The problem can be illustrated with a textbook example, namely the photon vacuum polarisation whose general form is
 $\Pi^{\mu\nu}(Q) = a\,\delta^{\mu\nu} + b\,Q^\mu Q^\nu$. The coefficients $a$ and $b$ are functions of $Q^2$ and
 must be analytic at $Q^2=0$; poles would correspond to intermediate massless particles but since the vacuum polarisation is 1PI
 intermediate propagators are excluded by definition. Gauge invariance entails transversality, $Q^\mu  \Pi^{\mu\nu}=0$, which fixes $a=-b\,Q^2$.
 The vacuum polarisation can then be written as the sum of a transverse part and a `gauge part' (which is \textit{not} longitudinal):
 \begin{equation}\label{vac-pol}
    \Pi^{\mu\nu}(Q) = \Pi(Q^2)\,t_{QQ}^{\mu\nu} + \widetilde{\Pi}(Q^2)\,\delta^{\mu\nu}\,,
 \end{equation}
 with $t^{\mu\nu}_{AB}$ defined in~\eqref{t-and-epsilon}.
 The transverse dressing function $\Pi(Q^2)$ is free of kinematic singularities and zeros.
 The gauge part $\delta^{\mu\nu}$ is the tensor that we eliminated in the first place, so $\widetilde{\Pi}(Q^2)$ must vanish due to gauge invariance.
 This is what happens in dimensional regularization, whereas a cutoff breaks gauge invariance and induces a quadratic divergence (only) in the gauge part.
 If we did not know about the decomposition~\eqref{vac-pol} and performed a transverse projection,
\begin{equation}
 \left(\delta^{\mu \alpha} - \frac{Q^\mu Q^\alpha}{Q^2}\right) \Pi^{\alpha\nu}(Q) = \left(\Pi(Q^2) + \frac{\widetilde{\Pi}(Q^2)}{Q^2} \right) t_{QQ}^{\mu\nu}\,,
\end{equation}
 we would pick up a $1/Q^2$ pole from the gauge part which invalidates the extraction at zero momentum.
 The transverse/gauge separation is also necessary if gauge invariance is broken by more than a cutoff, for instance by an incomplete calculation:
 ultimately the sum of all gauge parts must vanish, but the partial result for $\Pi(Q^2)$
 is still free of kinematic problems and -- ideally -- not strongly affected by gauge artifacts.

 This example also provides the template for the CS amplitude, where a complete decomposition into transverse and gauge parts free of kinematic singularities is necessary as well.
 Extending the forward expression in~\eqref{cs-amp-fwd} to the general case, the CS amplitude can be written as
 \begin{equation}\label{cs-basis}
   \mathcal{M}^{\mu\nu}(p,Q',Q) = \frac{1}{m}\,\conjg{u}(p_f)\, \bigg[ \, \underbrace{\bigg( \frac{c_1}{m^4}\,t^{\mu\alpha}_{Q'p}\,t^{\alpha\nu}_{pQ} + \frac{c_2}{m^2}\,t^{\mu\nu}_{Q'Q} + \dots\bigg)}_{\text{transverse part, 18 tensors}}
                                                                   \;\; +  \underbrace{\bigg(\widetilde c_1 \,\delta^{\mu\nu} + \dots\bigg)}_{\text{\scriptsize gauge part,} \atop \text{\scriptsize 14 tensors}} \,\bigg] \,u(p_i)\,,
 \end{equation}
 which is the analogue of~\eqref{ep-amp} and~\eqref{ep-basis} for pion electroproduction.
 The transverse part depends on 18 tensors which have been derived in~\cite{Tarrach:1975tu}, see also Ref.~\cite{Drechsel:2002ar}.
 In analogy to the forward case, one can insert factors of $\omega$, $\lambda$ and $m$ where necessary so that all CFFs $c_i(\eta_+, \eta_-, \omega, \lambda)$ are
 dimensionless and symmetric under photon crossing and charge conjugation (and thus quadratic in $\omega$ and $\lambda$).
 The gauge part is worked out in Ref.~\cite{inprep-er} and vanishes for the full CS amplitude.
 However, even if one breaks gauge invariance by retaining only a subset of diagrams, the transverse CFFs
 still yield a well-defined prediction as we will see shortly.

  Returning to the separation of Born and 1PI terms, electromagnetic gauge invariance generally entails that
       the CS amplitude can be decomposed into~\cite{Eichmann:2012mp}
       \begin{equation}
           \mc{M}^{\mu\nu} = \mc{M}^{\mu\nu}_\text{Born} + \mc{M}^{\mu\nu}_\text{1PI} = \mc{M}^{\mu\nu}_\text{Born} + \mc{M}^{\mu\nu}_\text{WTI} + \mc{M}^{\mu\nu}_{\perp}\,.
       \end{equation}
       Here the structure part is further split into $\mc{M}^{\mu\nu}_\text{WTI}$, which satisfies a Ward-Takahashi identity and is thus related to the offshell nucleon propagator
       and nucleon-photon vertex, and a fully transverse contribution $\mc{M}^{\mu\nu}_{\perp}$.
       Because $\mc{M}^{\mu\nu}$ and $\mc{M}^{\mu\nu}_{\perp}$ are both gauge invariant,
       also the sum $\mathcal{M}^{\mu\nu}_\text{Born} + \mathcal{M}^{\mu\nu}_\text{WTI}$ must be gauge invariant -- but not the Born terms alone.
       In any case, the expansion~\eqref{cs-basis} holds for each individual contribution and allows one to isolate the gauge part.
       A simple example is tree-level Compton scattering on a scalar particle:
       the Born terms alone are not gauge invariant, but upon projecting onto the basis~\eqref{cs-basis} the gauge part is $-2 \delta^{\mu\nu}$, which is cancelled by the seagull term
       $\mathcal{M}^{\mu\nu}_\text{WTI} = 2 \delta^{\mu\nu}$ stemming from the Lagrangian of scalar QED~\cite{Bakker:2015cba}.

 \begin{figure}
 \centering
   \includegraphics[width=0.85\textwidth]{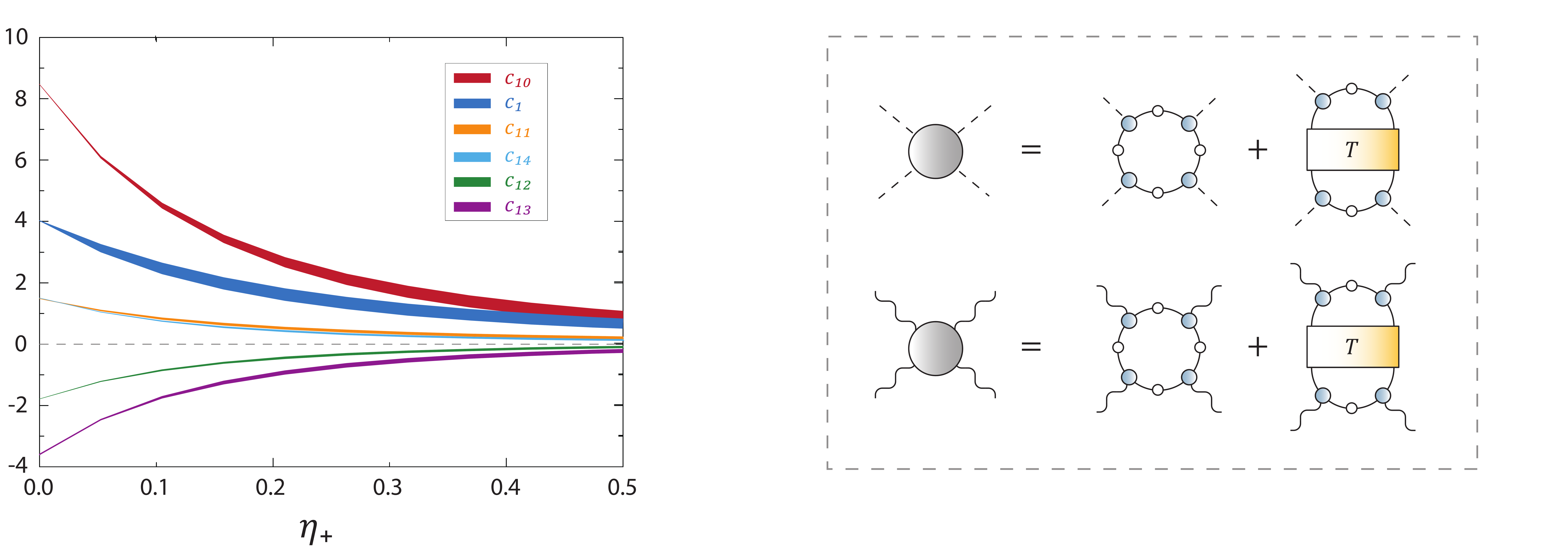}
 \caption{\textit{Left:} Dominant Compton form factors corresponding to the residues of the nucleon Born terms after removing the common pole factor~\cite{Eichmann:2016tbi}.
          The bands contain the full kinematic dependence on all four variables inside the spacelike cone in Fig.~\ref{fig:phasespace}.
          \textit{Right:} Microscopic expressions for $\pi\pi$ and hadronic light-by-light scattering in rainbow-ladder (permutations are not shown). }
 \label{fig:born-terms}
 \end{figure}

       Considering nucleon Compton scattering,
       a gauge-invariant Born term can be ensured by implementing a nucleon-photon vertex with $Q^2$-dependent Pauli and Dirac form factors only,
       as given in~\eqref{ffs-nucleon-current-1} without the positive-energy projectors.
       In that case $\mathcal{M}^{\mu\nu}_\text{WTI}=0$ and thus the Born and structure parts are individually transverse.
       The resulting CFFs stemming from the Born terms can be worked out analytically and they share common pole factors $1/(\eta_-^2-4\lambda^2)$.
       Their residues for the leading CFFs are plotted in Fig.~\ref{fig:born-terms}. As required, the gauge part is exactly zero.
       The implementation of offshell form factors destroys this property:
       $\mathcal{M}^{\mu\nu}_\text{Born}$ then carries a gauge part which cancels with that of $\mathcal{M}^{\mu\nu}_\text{WTI}$,
       but it turns out that within a reasonable range of model parametrizations the transverse CFFs remain almost unchanged~\cite{Eichmann:2016tbi}.

       The bottom line of the discussion is the following: dealing with gauge-dependent expressions is not a problem as long as one uses a basis decomposition
       that can separate gauge artifacts from the physical content. As a surplus, the transverse amplitudes are simple
       functions whose only singularities are physical poles and cuts. This is visible in Fig.~\ref{fig:born-terms}: all CFFs are well-behaved and approach constant values for $\eta_+\to 0$.
       Note that the bands contain the \textit{full} kinematic dependence
       on all four variables $\eta_+$, $\eta_-$, $\omega$ and $\lambda$ inside the spacelike cone in Fig.~\ref{fig:phasespace}, but effectively they only depend  on $\eta_+$.
       The residues of the nucleon Born terms therefore scale with $\eta_+$, which reflects the symmetric makeup of the phase space \textit{and}
       the choice of a `good' tensor basis whose amplitudes are free of kinematic artifacts.

\subsection{Applications}\label{compton-results}

  In practice, the expression for the scattering amplitude in~\eqref{scattering-amplitude-1} is only truly useful if it is amenable to calculations.
  There are several obstacles involved, including the appearance of higher $n-$point functions (such as the quark six-point function $\mathbf G$) which are difficult to calculate,
  or limited kinematic access to the relevant physical regions.
  Similarly to Fig.~\ref{fig:fwd-cs-amplitude}, the spacelike (or `Euclidean') region is directly accessible whereas an extension to the physical (or `Minkowski')
  domain, where experimental data are available, requires more effort: solving Bethe-Salpeter equations for complex relative momenta, implementing residues associated with quark singularities, etc.
  These are not fundamental obstructions but they will require substantial investments in future developments.
  Nevertheless, Fig.~\ref{fig:phasespace} shows that several applications are directly accessible: the Euclidean cone is identical to the integration domain for two-photon exchange and it contains
  the limits of forward CS and the generalized polarisabilities in virtual CS\@.
  Similar conclusions apply to other processes such as pion electroproduction (whose domain for massless pions is identical to the virtual CS plane), $\pi\pi$ scattering etc.
  In the following we will give a brief overview of existing applications based on the expression in~\eqref{scattering-amplitude-1}.

 \paragraph{$\pi\pi$ scattering.}
 Among the technically simpler applications are the correlators of four $q\bar{q}$ currents,
 from where for instance the $\pi\pi$ scattering amplitude can be determined.
 In that case the decomposition in Fig.~\ref{fig:cs-decomposition} becomes a sum of symmetrized dressed quark loops (`impulse approximation')
 together with diagrams that contain the $q\bar{q}$ scattering matrix, cf.~Fig.~\ref{fig:born-terms}. Once again the quark four-point function contains all possible intermediate meson poles
 and thus they are automatically inherited by the $\pi\pi$ scattering amplitude.

\begin{figure}
 \centering
   \includegraphics[width=0.99\textwidth]{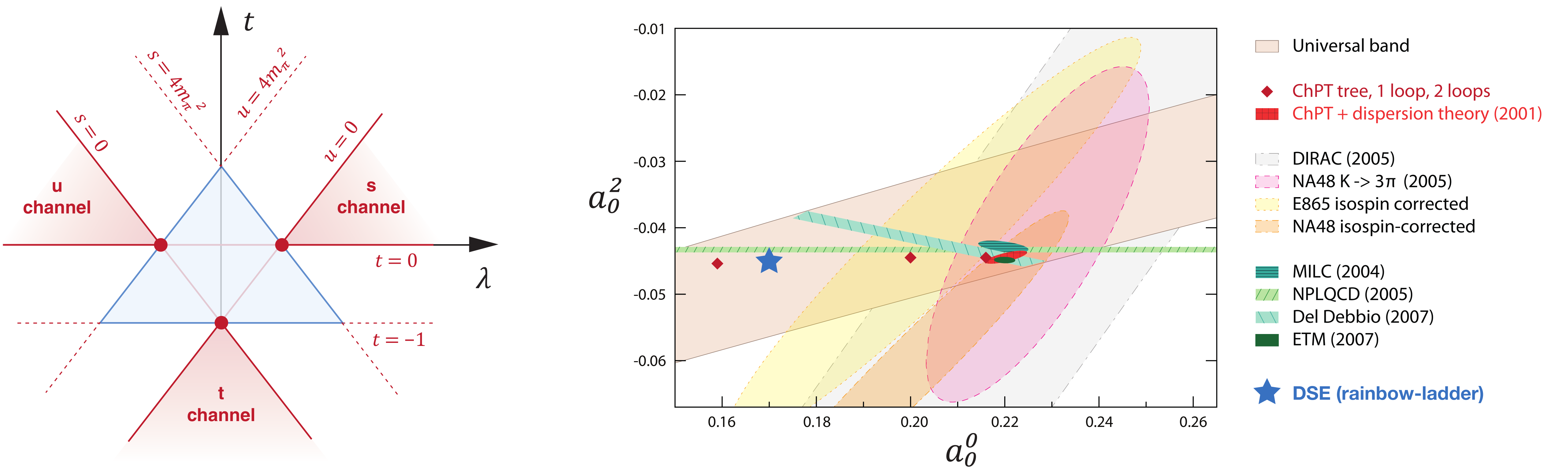}
 \caption{\textit{Left panel:} Phase space in $\pi\pi$ scattering; the scattering lengths are extracted at the threshold $s=4m_\pi^2$, $u=t=0$.
          \textit{Right:} $\pi\pi$ scattering lengths as determined from chiral perturbation theory, experiments, lattice QCD and Dyson-Schwinger equations. Figure adapted from Ref.~\cite{Colangelo:2007df}.
          }
 \label{fig:pipi-scattering}
 \end{figure}

 The phase space for $\pi\pi$ scattering is the Mandelstam plane in Fig.~\ref{fig:pipi-scattering}.
 If we employ the same kinematics as in Compton scattering it is spanned by the variables $t$ and $\lambda$.
 The physical $s$, $t$ and $u$-channel regions above the respective $\pi\pi$ thresholds are illustrated together with the `Euclidean' region
 that is free from singularities.
 In Refs.~\cite{Bicudo:2001jq,Cotanch:2002vj} the rainbow-ladder expression for the $\pi\pi$ scattering amplitude in the forward limit $t=0$ was calculated and
 the scattering lengths at threshold ($s=4m_\pi^2$) were extracted.
 (A similar calculation has been recently performed in the covariant spectator approach~\cite{Biernat:2014xaa}.)
 Fig.~\ref{fig:pipi-scattering} shows the theoretical and experimental status for the $\pi\pi$ scattering lengths $a_0^0$ and $a_0^2$ in the two isospin channels.
 The inclusion of pion loops moves the chiral perturbation theory (ChPT) values towards the right where they eventually converge with lattice results,
 which implies that $a_0^0$ is sensitive to loop corrections but $a_0^2$ is not.
 The Dyson-Schwinger results essentially agree with the tree-level ChPT prediction,
 including the Adler zero in the chiral limit,
 whereas at larger energies they reproduce the meson poles in the scattering amplitude.
 Once again this underlines the observation that rainbow-ladder is mainly missing pion-cloud effects,
 which can be as large as $25 \%$ in some observables in consistency with the discrepancies in various form factors at low $Q^2$.
 Clearly, rainbow-ladder is not yet a \textit{precision} tool, but $25\%$ accuracy would still represent a far advance
 towards determining the QCD corrections to the proton radius puzzle or the muon g-2 problem,
 where the current discrepancies between theory and experiment span multiple standard deviations.

 \paragraph{Hadronic light-by-light amplitude.}
 Conceptually closely related is the hadronic light-by-light (HLbL) amplitude, i.e., the correlator of four vector currents, which
 has seen much recent interest due to its relevance for the muon g-2 problem.
 The anomalous magnetic moment of the muon is encoded in the muon-photon vertex, whose
 leading contribution is Schwinger's result for the vertex correction: $a_\mu = \alpha_\mathrm{QED}/(2\pi) + \mathcal{O}(\alpha_\mathrm{QED}^2)$~\cite{Schwinger:1948iu}.
 Its experimental value is extremely precisely measured but differs from the Standard Model prediction by $3\sigma$.
 The main uncertainties enter through QCD via the hadronic vacuum polarisation and HLbL amplitude, with a typical theoretical
 estimate $a_\mu^\text{HVP} = 685.1(4.3) \times 10^{-10}$ and $a_\mu^\text{HLbL} = 11.6(3.9)\times 10^{-10}$~\cite{Jegerlehner:2009ry}.
 Although they are much smaller than the QED corrections they dominate the theoretical uncertainty by far.
 The HLbL estimate comes from model calculations, see e.g. \cite{Benayoun:2014tra} for an overview.
 There are ongoing efforts in lattice QCD~\cite{Blum:2015gfa} and dispersive approaches~\cite{Pauk:2014rfa,Colangelo:2014pva} to provide model-independent
 information on the magnitude of the HLbL contribution.

 Although the microscopic decomposition of the LbL amplitude is identical to that of $\pi\pi$ scattering (see Fig.~\ref{fig:pipi-scattering}), the problem is more complicated in practice.
 The photon four-point function depends on six independent variables instead of two, and instead of one scalar function
 one has to deal with 41 transverse tensor structures (and 136 non-transverse tensors in general). Electromagnetic gauge invariance additionally complicates the problem
 since one needs a Bose-symmetric transverse/gauge separation analogous to~\eqref{cs-basis}.
 These issues were addressed in Ref.~\cite{Eichmann:2015nra}, where
 the LbL phase space was arranged into multiplets of the permutation group $S_4$,
 and a Bose-symmetric transverse tensor basis free of kinematic singularities and with minimal powers in the photon momenta was constructed.
 In addition, the dressed quark loop in Fig.~\ref{fig:pipi-scattering} was calculated in~\cite{Goecke:2010if,Goecke:2012qm}.
 Although no transverse/gauge separation was made therein,
 a transverse projection with different gauge parameters suggested that gauge artifacts were negligible.
 Interestingly, the value for the quark loop turns out to be quite large, $a_\mu^\text{QL} = 10.7(2) \times 10^{-10}$, which
 would already reduce the discrepancy between theory and experiment to about $2\sigma$.
 It remains to be seen whether a full calculation including the T-matrix diagrams (which automatically reproduce the meson exchanges)
 corroborates that result or induces further notable changes.

 \vspace{-2mm}

 \begin{figure}
 \centering
   \includegraphics[width=0.65\textwidth]{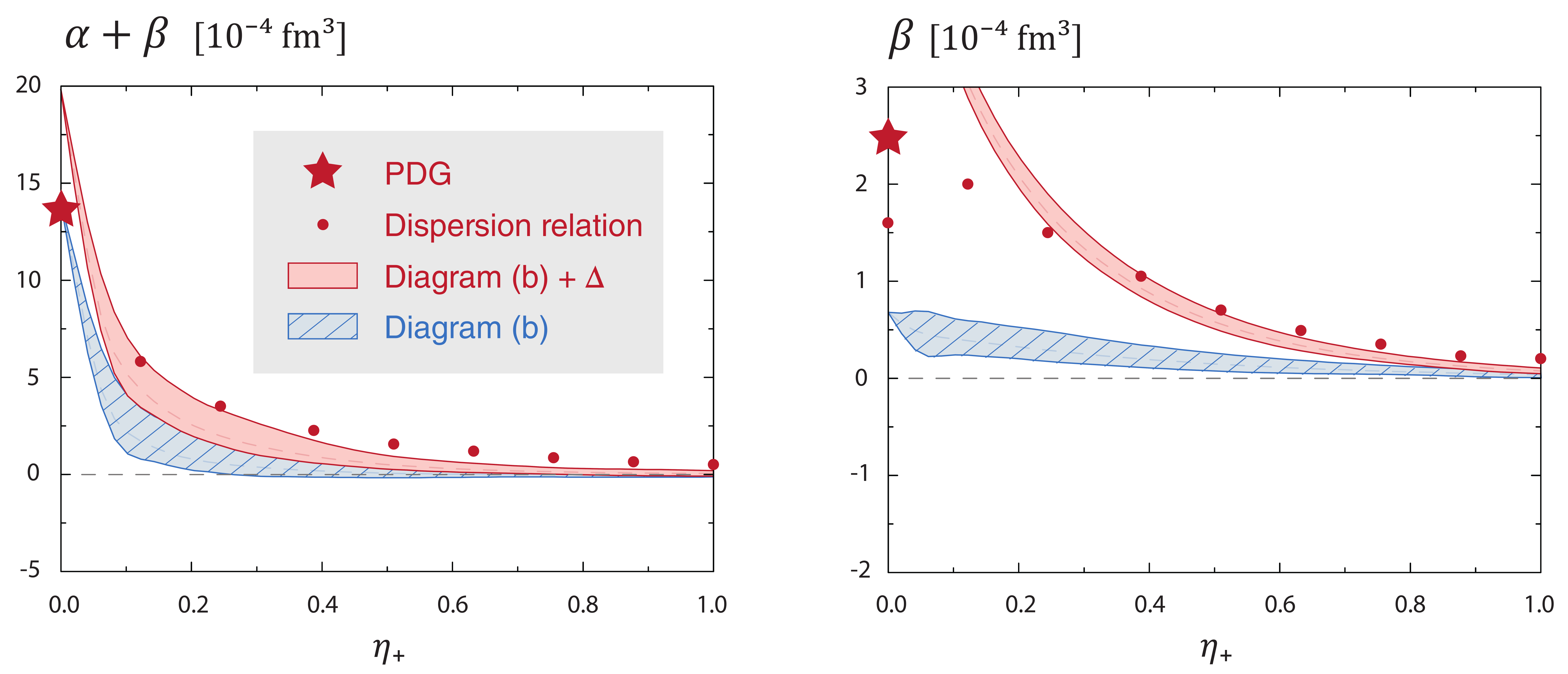}
 \caption{Proton polarisabilities as functions of $\eta_+$~\cite{Eichmann:2016tbi}. The bands were obtained from diagram (b) of Fig.~\ref{fig:cs-decomposition} with and without the $\Delta$ contribution.
          The dots were extracted from Ref.~\cite{Downie:2011mm} and the stars are the experimental values~\cite{Agashe:2014kda}.}
 \label{fig:polarizabilities}
 \end{figure}

 \paragraph{Compton scattering.}
 Coming back to nucleon Compton scattering, what has been done so far is to calculate diagram (b) in the decomposition of Fig.~\ref{fig:cs-decomposition}.
 This involves a self-consistent calculation of the quark Compton vertex which depends on 6 Lorentz invariants and 128 tensor structures (72 of which are transverse)~\cite{Eichmann:2012mp}.
 The resulting CS amplitude reproduces both the Born terms as well as the $t-$channel meson poles at the hadronic level.
 The dominant meson contribution is pion exchange, so in principle one can extract
 the $\pi^0\to\gamma\gamma$ transition form factor from the CS amplitude.
 This was explicitly checked in Ref.~\cite{Eichmann:2012mp}, where the extracted form factor was compared with the direct rainbow-ladder result which has been first calculated in~\cite{Maris:2002mz}.

 First results for polarisabilities were reported in Ref.~\cite{Eichmann:2016tbi}.
 The scalar polarisabilities $\alpha$ and $\beta$ are related to the CFFs $c_1$ and $c_2$ (without Born terms) in the limit where all kinematic variables are zero, cf.~Eq.~\eqref{lex}:
 $\{\alpha+\beta,\beta\} = \{c_1, c_2\} \times \alpha_\mathrm{QED}/m^3$. Note that the nucleon Born terms would come from diagram (a) so they are automatically excluded.
 The hatched bands in Fig.~\ref{fig:polarizabilities}  are the outcome of diagram (b) inside the spacelike cone;
 the total result includes $\Delta$ exchange as the dominant approximation to diagram (a).
 The dispersion relation results for the generalized polarisabilities from Refs.~\cite{Downie:2011mm,Drechsel:2002ar} are shown for comparison.
 It turns out that the sum $\alpha+\beta$, which in the forward limit is constrained by the Baldin sum rule, is dominated by the handbag contributions whereas
 the magnetic polarisability $\beta$ is mainly produced by the $\Delta$ pole.
 The discrepancy for $\beta$ at low $\eta_+$ is presumably due to missing pion loops -- $\beta$ is subject to cancellations between the quark core
 (which then mainly comes from $\Delta$ exchange) and pion cloud effects.

 In the future it will be further interesting to investigate spin polarisabilities and gather knowledge on the spacelike momentum dependence of the CS amplitude,
 which will improve our understanding of two-photon corrections to form factors as well as the proton radius puzzle.
 Finally, the same framework can be adapted to other processes such as pion electroproduction,
 which has contributed much to our knowledge of nucleon resonances and transition form factors.
 For their clean extraction one needs to know the non-resonant `QCD background' beyond hadronic exchanges, which
 is information that a microscopic approach can provide.


\newpage

\section{Outlook}\label{sec:summary}
Baryon spectroscopy, baryon structure and baryon dynamics remain very active
and lively fields. Dedicated experiments at many facilities such as Jefferson Lab,
BESIII, ELSA, J-PARC, LHCb, MAMI and the future PANDA/FAIR experiment
are contributing and will continue to contribute to the rich tapestry that
makes up the baryonic world. New surprises seem to await us around each corner,
ensuring that the field will continue to engage scientists for decades to come.

Challenges remain on the experimental side, in particular the completion of
single and double polarisation experiments, the systematic combination of
hadronic and photoproduction data and the refinement of coupled-channel analyses.
On the theory side, the past years have seen continuous advances in connecting
the underlying quark and gluon world with that of baryon phenomenology. However,
there remains much to be done, with many important questions that have not been
settled so far.
For the lattice community, one goal is to place the extraction of data for the
excited states on the same firm ground as that of the ground states.
This includes in particular the need to perform analogous simulations that
employ physical pion masses as well as further refinements to the methods of
analysis.

For the functional methods community, the challenges are to capitalize on
continuous improvements to the truncation/approximation schemes involved. While
the aim is not to enter the realm of high precision physics on the sub-percent
level, much can be achieved in the way of understanding the underlying physical
mechanisms that generate the observed phenomena. A prime example of this is dynamical
chiral symmetry breaking and the associated dynamical generation of a momentum
dependent quark mass, but also the corresponding effect in the quark-gluon interaction
and higher Green functions. Much progress has been made in this respect.
The recently achieved access to excited states is furthermore crucial for systematic studies
of the details of the underlying QCD interactions generating the rich spectrum of baryons.

An important issue where functional methods can contribute is to pin down the role of diquark correlations
in the baryon structure. This needs to be studied in one of the advanced approximation
schemes discussed in this review, which includes not only quark-quark interaction
terms but also three-body kernels derived from QCD\@. In all modern treatments, the
pointlike diquarks of the early models become extended objects with internal structure.
We have argued that observables like ground state masses and the first radial excitation
in the nucleon and $\Delta$ channel are indeed dominated by the presence of quark-quark
correlations inside the baryons. Whether this is also true
for other excited states remains to be clarified. Moreover, form factors such as
$G_{M3}$ of the $\Delta$ baryon discussed in this review may serve to distinguish the two
pictures beyond mere spectral considerations. Thus it will be very interesting as to what
other methods like lattice QCD -- or even future experiments -- can say about this quantity.

Other still open questions are the nature of the Roper, the $\Lambda(1405)$ and the associated
level orderings between different spin, parity and flavour channels which need to be better understood.
Recent results indicate that the level ordering between the Roper and the $N(1535)$ is reproduced
in the functional approach \cite{roper:2016}.
The precise identification of the origin of flavour dependent forces large enough to generate
the $\Sigma$-$\Lambda$ splitting, however, is still missing. The implementation of terms representing meson
cloud effects can and should be improved and carried over from spectroscopy to form factors.

Furthermore, it will be very interesting whether already existing states and potential new ones
can be cleanly identified as exotic in the sense that they cannot be described in the
simple three-body quark picture.  Distinctive
patterns of ground and excited state masses, salient features in their decay patterns or even
structural properties like form factors are necessary for a clean identification.

Ultimately, deciphering the underlying structure of QCD requires a machinery
that relates the properties of quarks and gluons
with the vast pool of hadron structure observables that contemporary scattering experiments provide us with.
Applications of functional methods to elastic and transition form factors are underway,
and the current efforts to extend them to  hadronic scattering amplitudes
represent first steps towards understanding some of the present puzzles
in hadron physics, from two-photon effects and the proton radius to the anomalous magnetic moment of the muon.
A microscopic description of scattering amplitudes may also be useful
in the experimental extraction of resonance properties, and it can provide a nonperturbative link to the
spin and transverse momentum structure of the quarks and gluons inside baryons.
In this respect we believe that exciting times are ahead of us.


%
%
\section*{Acknowledgements}

We thank Raul Briceno, Kai-Thomas Brinkmann, Stan Brodsky, Ian Cloet, Jozef Dudek, Andreas Krassnigg, Christian Lang, Axel Maas, Volker Metag,
Daniel Mohler, Viktor Mokeev, Ulrich Mosel, Diana Nicmorus, Martin Oettel, Joannis Papavassiliou, Sasa Prelovsek,
Hugo Reinhardt, Craig D. Roberts, Enrique Ruiz-Arriola, Adam Szczepaniak, Ulrike Thoma and Ross Young for useful discussions and valuable insights on the subject.
We are grateful to Kai-Thomas Brinkmann, Axel Maas, Daniel Mohler and Martin Oettel
for a critical reading of (parts of) the manuscript.

This work was partially supported by the DFG collaborative research centre TR 16,
an Erwin Schr\"odinger fellowship J3392-N20 from the Austrian Science Fund (FWF),
by the Helmholtz International Center for FAIR within the LOEWE program of the
State of Hesse and the BMBF project 05H15RGKBA.

\appendix
\renewcommand*{\thesection}{\Alph{section}}


\newpage

\section{Conventions and formulas}\label{app:conventions}

        \paragraph{Euclidean conventions.}
        As noted throughout this review we work with Euclidean conventions.
        In the context of Bethe-Salpeter equations in Sec.~\ref{spec:BS} we argued that they are convenient as they simplify matters in practical calculations.
        More generally,
        they arise from the imaginary-time boundary conditions in the action $S$ of a quantum field theory that
        enters in the generating functional and the corresponding Green functions as in Eq.~\eqref{correlators}.
        In the operator formalism these are necessary when taking expectation values of operators and projecting them
        onto the interacting vacuum, whereas in the path integral formalism they are needed to arrive at a non-negative probability measure.
        In any case, when starting in Minkowski space the spacetime volume becomes
         \begin{equation}
             \lim_{T\rightarrow \infty(1-i\epsilon)} i \!\int\limits_{-T}^T  \! dx_0 \int \! d^3x =
             \lim_{\tau\rightarrow \infty} \int\limits_{-\tau}^\tau  \! dx_4 \int \! d^3x =
             \int \! d^4x_E\,, 
         \end{equation}
         which motivates to introduce a Euclidean vector $x_E^\mu = ( \vect{x}, ix_0 )$ with an imaginary four-component. 
         For a general Euclidean four-vector $a^\mu_E$, this means that its square
         changes sign compared to the Minkowski version: $a^2_E = -a^2$. Therefore, a vector is spacelike if $a^2>0$ and timelike if $a^2<0$.
         Because the metric is now positive,  the distinction between upper and lower indices disappears.
         To preserve the meaning of the slash $\slashed{a} = a^0 \gamma^0 - \vect{a}\cdot\vect{\gamma}$
         we should also redefine the $\gamma-$matrices, so that
         the resulting replacement rules for vectors $a^\mu$, tensors $T^{\mu\nu}$, and $\gamma-$matrices can be summarized as
         \begin{equation}\label{rules}\renewcommand{\arraystretch}{1.0}
             a^\mu_E = \left(\begin{array}{c}\vect{a}\\ia_0\end{array}\right), \qquad
             \gamma^\mu_E = \left(\begin{array}{c}-i\vect{\gamma}\\\gamma_0\end{array}\right), \qquad
             T^{\mu\nu}_E = \left(\begin{array}{cc} T^{ij} & iT^{i0} \\ iT^{0i} & -T^{00}\end{array}\right),  \qquad
             \gamma^5_E = \gamma^5\,,
         \end{equation}
         where `$E$' stands for Euclidean and no subscript refers to the Minkowski quantity.
         As a consequence,
         \begin{equation}\label{consequences}
             a_E \cdot b_E = \sum_{k=1}^4 a_E^k \, b_E^k = -a\cdot b, \qquad
             \slashed{a}_E = a_E \cdot \gamma_E = i\slashed{a}, \qquad
             \{ \gamma^\mu_E, \gamma^\nu_E \} = 2\delta^{\mu\nu}
         \end{equation}
         and we see that the Lorentz-invariant scalar product of any two four-vectors differs by a minus sign from its Minkowski counterpart.


         Our sign convention for the Euclidean $\gamma-$matrices changes all signs in the Clifford algebra relation~\eqref{consequences} to be positive,
         and since this implies $(\gamma_E^i)^2=1$ for $i=1 \dots 4$ we can choose them to be hermitian: $\gamma_E^\mu = \left( \gamma^\mu_E\right)^\dag$.
         For example, in the standard representation they read
            \begin{equation}\renewcommand{\arraystretch}{1.0}
                \gamma^k_E  =  \left( \begin{array}{cc} 0 & -i \tau_k \\ i \tau_k & 0 \end{array} \right) , \quad
                \gamma^4_E  =  \left( \begin{array}{c@{\quad}c} \mathds{1} & 0 \\ 0 & \!\!-\mathds{1} \end{array} \right) , \quad
                \gamma^5  =  \left( \begin{array}{c@{\quad}c} 0 & \mathds{1} \\ \mathds{1} & 0 \end{array} \right)\,,
            \end{equation}
            where the $\tau_k$ are the usual Pauli matrices.
         Also the generators of the Clifford algebra are then hermitian:
         \begin{equation}
             \sigma^{\mu\nu} = \frac{i}{2}\,[\gamma^\mu,\gamma^\nu] \quad \Rightarrow \quad
             \sigma_E^{\mu\nu} = -\frac{i}{2}\,[\gamma^\mu_E, \gamma^\nu_E]\,, \qquad
             (\sigma_E^{\mu\nu})^\dag = \sigma_E^{\mu\nu}\,.
         \end{equation}
         Despite appearances, this does not alter the Lorentz transformation properties and the definition of the conjugate spinor as $\conjg{\psi} = \psi^\dag \gamma^4$
         (which was necessary to make a bilinear $\conjg{\psi} \psi$ Lorentz-invariant) remains intact.
         Denoting the representation matrix $\psi'(x') = D(\Lambda)\,\psi(x)$ of the Lorentz transformation by
         \begin{equation}
             D(\Lambda) = \exp\left[ -\frac{i}{4}\,\omega_{\mu\nu}\,\sigma^{\mu\nu}\right] = \exp \left[ -\frac{i}{4}\,\omega_E^{\mu\nu}\,\sigma^{\mu\nu}_E \right],
         \end{equation}
         then irrespective of $\gamma^4\,(\sigma^{\mu\nu}_E)^\dag \,\gamma^4 \neq \sigma^{\mu\nu}_E$ the relation
         $\gamma^4 \,D(\Lambda)^\dag\,\gamma^4 = D(\Lambda)^{-1}$ still holds, because the infinitesimal Lorentz transformation $\omega^{\mu\nu}_E$ which is related to its Minkowski counterpart via~\eqref{rules} is now complex.
         Hence
         \begin{equation}
            \conjg{\psi}'(x') = \psi^\dag(x)\,D(\Lambda)^\dag\,\gamma^4 = \psi^\dag (x)\,\gamma^4\,D(\Lambda)^{-1} = \conjg{\psi}\,D(\Lambda)^{-1}\,,
         \end{equation}
         and therefore $\conjg{\psi}(x)\,\psi(x)$ is Lorentz-invariant, $\conjg{\psi}(x)\,\gamma^\mu_E\,\psi(x)$ transforms like a Lorentz vector, etc.

         The replacement rules furthermore imply $\partial \cdot a = (\partial\cdot a)_E$, \,$\slashed{\partial} = i\slashed{\partial}_E$ and  $\Box = -\Box_E$,
         so that for example in the case of a fermionic action we arrive at
         \begin{equation}\label{fermionic-action}
         e^{iS} = \exp\left[i\! \int d^4x\,\conjg{\psi}\,(i\slashed{\partial}-m) \,\psi\right] = \exp\left[-\!\int d^4x_E\,\conjg{\psi}\,(\slashed{\partial}_E + m)\,\psi\right] =: e^{-S_E}\,.
         \end{equation}
         In practical applications, this means that when starting from a Euclidean action
         all calculations are performed in Euclidean space until  at the very end one arrives at Lorentz-invariant quantities.
         Those are for example the momentum-dependent dressing functions of Green functions and Bethe-Salpeter amplitudes or observables
         such as masses, form factors or the coefficients of scattering amplitudes.
         Lorentz invariant scalar products only differ by minus signs in Minkowski and Euclidean conventions so they are trivially connected.
         Note, however, that although the transcription rules~\eqref{rules} can always be applied to transform Minkowski formulas into Euclidean ones and vice versa,
         this does not imply that the integrations are trivially possible because usually one  must take care of singularities in the (complex) integration domain as discussed in Sec.~\ref{spec:BS}.
         Expressed in terms of the `phase space' of the resulting Lorentz invariants, e.g., the squared total momentum $P^2$ in a Bethe-Salpeter equation such as in Fig.~\ref{fig:scalar-bse-2},
         this leads to domains where Euclidean calculations are straightforward (`where the Wick rotation is possible')
         and others where they require residue calculus.




        \paragraph{Formulas.}
            Dropping now the index `$E$', we collect some useful formulas. As already stated above, we have
            \begin{equation}
                \sigma^{\mu\nu} = -\frac{i}{2} \left[ \gamma^\mu, \gamma^\nu \right] \qquad \text{and} \qquad
                \gamma^5 = -\gamma^1 \gamma^2 \gamma^3 \gamma^4 = -\frac{1}{24} \,\varepsilon^{\mu\nu\rho\sigma} \gamma^\mu \gamma^\nu \gamma^\rho \gamma^\sigma \quad \text{with} \quad  \varepsilon^{1234} = 1\,.
            \end{equation}
            It is convenient to define the fully antisymmetric combinations of Dirac matrices via the commutators
            \begin{align}
            [A,B] &= AB-BA \,, \\
            [A,B,C] &= [A,B]\,C + [B,C]\,A + [C,A]\,B\,, \label{three-commutator} \\
            [A,B,C,D] &= [A,B,C]\,D + [B,C,D]\,A + [C,D,A]\,B + [D,A,B]\,C\,.
            \end{align}
            Inserting $\gamma-$matrices, this yields the antisymmetric combinations mentioned in Eq.~\eqref{eqn:diracmatrices}:
            \begin{align}
            [\gamma^\mu,\gamma^\nu] &= \gamma_5\,\varepsilon^{\mu\nu\alpha\beta}\,\gamma^\alpha \gamma^\beta \,, \\
            \tfrac{1}{6}\,[\gamma^\mu,\gamma^\nu,\gamma^\rho] &= \tfrac{1}{2}\,(\gamma^\mu \gamma^\nu \gamma^\rho - \gamma^\rho \gamma^\nu \gamma^\mu)   = -\gamma_5\,\varepsilon^{\mu\nu\rho\sigma}  \gamma^\sigma\,, \\
            \tfrac{1}{24}\,[\gamma^\mu,\gamma^\nu,\gamma^\alpha,\gamma^\beta] &= -\gamma_5\,\varepsilon^{\mu\nu\alpha\beta}\,.
            \end{align}
            The various contractions of $\varepsilon-$tensors are useful for working out the squares of the spin and orbital angular momentum operators in Eq.~\eqref{spin-oam}:
            \begin{equation}
            \begin{split}
               \varepsilon^{\mu\nu\rho\lambda}\,\varepsilon^{\alpha\beta\gamma\lambda} &= \delta^{\mu\alpha}\,( \delta^{\nu\beta}\,\delta^{\rho\gamma} - \delta^{\nu\gamma}\,\delta^{\rho\beta})
                                                                                        + \delta^{\mu\beta}\,( \delta^{\nu\gamma}\,\delta^{\rho\alpha} - \delta^{\nu\alpha}\,\delta^{\rho\gamma})
                                                                                        + \delta^{\mu\gamma}\,(\delta^{\rho\beta}\,\delta^{\nu\alpha} - \delta^{\rho\alpha}\,\delta^{\nu\beta})\,, \\
               \tfrac{1}{2}\,\varepsilon^{\mu\nu\lambda\sigma}\,\varepsilon^{\alpha\beta\lambda\sigma} &= \delta^{\mu\alpha}\,\delta^{\nu\beta} - \delta^{\mu\beta}\,\delta^{\nu\alpha}\,, \\
               \tfrac{1}{6}\,\varepsilon^{\mu\lambda\sigma\tau}\,\varepsilon^{\alpha\lambda\sigma\tau} &= \delta^{\mu\alpha}\,, \\
               \tfrac{1}{24}\,\varepsilon^{\lambda\sigma\tau\omega}\,\varepsilon^{\lambda\sigma\tau\omega} &= 1\,.
            \end{split}
            \end{equation}
            Four-momenta such as in Eq.~\eqref{rest-frame} are most conveniently expressed through hyperspherical coordinates:
            \begin{equation}\label{APP:momentum-coordinates}\renewcommand{\arraystretch}{1.0}
                p^\mu = \sqrt{p^2} \left( \begin{array}{l} \sqrt{1-z^2}\,\sqrt{1-y^2}\,\sin{\phi} \\
                                                           \sqrt{1-z^2}\,\sqrt{1-y^2}\,\cos{\phi} \\
                                                           \sqrt{1-z^2}\;\;y \\
                                                           \;\; z
                                         \end{array}\right) =
                        \sqrt{p^2} \left( \begin{array}{l} \sin{\psi} \,\sin{\theta} \,\sin{\phi} \\
                                                           \sin{\psi} \,\sin{\theta} \,\cos{\phi} \\
                                                           \sin{\psi} \,\cos{\theta} \\
                                                           \cos{\psi}
                                         \end{array}\right) ,
            \end{equation}
            and a four-momentum integration reads:
            \begin{equation} \label{hypersphericalintegral}
                \int \!\!\frac{d^4 p}{(2\pi)^4} \;\;
                       = \;\; \frac{1}{(2\pi)^4}\,\frac{1}{2} \int_0^{\infty} dp^2 \,p^2 \int_{-1}^1 dz\,\sqrt{1-z^2}  \int_{-1}^1 dy \int_0^{2\pi} d\phi \,.
            \end{equation}
            If $p$ describes the onshell momentum of a bound state with mass $m$, then $p^2=-m^2$; and
            the rest frame corresponds to $z=1$ which means
         \begin{equation}\renewcommand{\arraystretch}{1.0}
             p^\mu = \left(\begin{array}{c}\vect{0}\\im\end{array}\right).
         \end{equation}
         Note, however, that in Euclidean space the fourth component of a vector is not preferred in any way over any other component.
         For example, a Lorentz-invariant Bethe-Salpeter equation such as in Eq.~\eqref{scalar-bse} remains invariant if we swap all four-components with all three-components of the involved vectors.
         This has the curious consequence that an onshell particle with energy $E=m$ and momentum $\vect{p}=0$ is physically equivalent to a four-vector with energy $E=0$ and imaginary momentum $\vect{p} = im \vect{e}_3$.


        \paragraph{Onshell spinors.}
        From the fermionic action~\eqref{fermionic-action} we can read off the Dirac equations for the positive- and negative-energy onshell spinors with spin $\nicefrac{1}{2}$:
        \begin{equation}
        \begin{split}
            (i\slashed{p}+m)\,u(\vect{p}) &= 0 = \conjg{u}(\vect{p})\,(i\slashed{p}+m)\,, \\
            (i\slashed{p}-m)\,v(\vect{p}) &= 0 = \conjg{v}(\vect{p})\,(i\slashed{p}-m)\,,
        \end{split}
        \end{equation}
        where the conjugate spinor is again $\conjg{u}(\vect{p}) = u(\vect{p})^\dag \gamma^4$.
        Since the onshell spinors only depend on $\vect{p}$ they are the same as in Minkowski space;
        for example in the standard representation:
        \begin{equation}\renewcommand{\arraystretch}{1.0}
            u_s(\vect{p}) =   \sqrt{\frac{E_p+m}{2m}}\left(\begin{array}{c} \xi_s \\ \frac{\vect{p}\cdot\vect{\sigma}}{E_p+m}\,\xi_s \end{array}\right), \qquad
            \xi_+ = \left(\begin{array}{c} 1 \\ 0 \end{array}\right),  \quad
            \xi_- = \left(\begin{array}{c} 0 \\ 1 \end{array}\right),  \qquad
            E_p = \sqrt{\vect{p}^2 + m^2}\,.
        \end{equation}
        We have normalized them to unity,
        \begin{equation}
           \conjg{u}_s(\vect{p})\,u_{s'}(\vect{p}) =  -\conjg{v}_s(\vect{p})\,v_{s'}(\vect{p}) = \delta_{ss'}\,, \qquad
           \conjg{u}_s(\vect{p})\,v_{s'}(\vect{p}) = \conjg{v}_s(\vect{p})\,u_{s'}(\vect{p}) = 0\,,
        \end{equation}
        and their completeness relations define the positive- and negative-energy projectors:
        \begin{equation}\label{positive-energy-projector}
           \sum_s u_s(\vect{p})\,\conjg{u}_s(\vect{p}) = \frac{-i\slashed{p}+m}{2m} = \Lambda_+(p)\,, \qquad
           \sum_s v_s(\vect{p})\,\conjg{v}_s(\vect{p}) = \frac{-i\slashed{p}-m}{2m} = -\Lambda_-(p)\,,
        \end{equation}
        so that $\Lambda_\pm(p) = (\mathds{1} \pm \hat{\slashed{p}})/2$. Therefore, $\Lambda_+(p)\,u(\vect{p}) = u(\vect{p})$, $\Lambda_-(p)\,u(\vect{p}) = 0$ and so on,
        and the spinors for higher spin follow accordingly.

        \paragraph{Charge conjugation.}
            The charge conjugation matrix is defined as
            \begin{equation}
                \mc{C} = \gamma^4 \gamma^2, \qquad \mc{C}^T = \mc{C}^\dag = \mc{C}^{-1} = -\mc{C} \qquad \Rightarrow \qquad \mc{C}\,\gamma_5^T\,\mc{C}^T = \gamma_5\,, \quad \mc{C}\,\gamma_\mu^T\,\mc{C}^T = -\gamma_\mu\,.
            \end{equation}
        The conjugate Bethe-Salpeter amplitudes and wave functions that appear throughout the main text are in fact the charge-conjugate amplitudes;
        for example for $q\bar{q}$ bound states:
        \begin{equation}
           \conjg{\mathbf\Gamma}(p,P) = \mc{C}\,\mathbf{\Gamma}(-p,-P)^T\,\mc{C}^T\,, \qquad
           \conjg{\mathbf\Gamma}^\mu(p,P) = -\mc{C}\,\mathbf{\Gamma}^\mu(-p,-P)^T\,\mc{C}^T\,.
        \end{equation}
        For eigenstates of $C$ parity with eigenvalue $C=\pm 1$, the amplitudes satisfy
        \begin{equation}
           \conjg{\mathbf\Gamma}(p,P) = C\,\mathbf\Gamma(p,-P)\,, \qquad
           \conjg{\mathbf\Gamma}^\mu(p,P) = -C\,\mathbf\Gamma^\mu(p,-P)\,.
        \end{equation}
        This explains the prefactors $p\cdot P$ in the tensor bases~\eqref{pion-basis} and~\eqref{exotic-basis}: they ensure
        that each basis element is charge-conjugation invariant by itself, which implies $f_i(p^2,p\cdot P,P^2) = f_i(p^2,-p\cdot P,P^2)$ for the dressing functions
        and thus they can only depend on $(p\cdot P)^2$. The same reasoning applies for the factors $\lambda$ in the tensor bases~\eqref{ep-basis} for pion electroproduction
        and~\eqref{cs-amp-fwd} for Compton scattering.


\newpage

\section{Quark-photon vertex and hadronic vacuum polarization}\label{app:hvp}

       \paragraph{Quark-photon vertex.}
       In the following we give the explicit form of the offshell quark-photon vertex in Eq.~\eqref{qpv-bc+t}.
       The vertex $\Gamma^\mu(k,Q)$ depends on the relative momentum $k$ and total momentum $Q$ and the fermion momenta are $k_\pm = k \pm Q/2$.
       Its Ball-Chiu part~\cite{Ball:1980ay}
       \begin{equation}\label{qpv-bc-2}
          \Gamma^\mu_\text{BC}(k,Q) = i\gamma^\mu\,\Sigma_A + 2k^\mu (i\slashed{k}\,\Delta_A + \Delta_B)
       \end{equation}
       follows from a straightforward evaluation of the Ward-Takahashi identity~\eqref{vwti}.
       The transverse part is derived by starting from the most general Poincar\'e-covariant form of the vertex in momentum space (see Eq.~(75) in Ref.~\cite{Eichmann:2012mp}),
       written such that all tensors are symmetric under charge conjugation and thus all dressing functions depend on $k^2$, $Q^2$ and $(k\cdot Q)^2 = k^2\,Q^2\,z^2$.
       One then applies the transversality condition $Q^\mu\,\Gamma^\mu(k,Q)=0$, which leads to four constraints that have to be solved so that no kinematic singularities
       are introduced in the process. The resulting eight transverse tensors are given by~\cite{Kizilersu:1995iz,Skullerud:2002ge,Eichmann:2012mp}
             \begin{equation}\label{qpv-transverse-basis}
             \begin{split}
                 \begin{array}{rl}
                 \tau_1^\mu \!\!\!\!\!\! &= t^{\mu\nu}_{QQ}\,\gamma^\nu\,, \\
                 \tau_2^\mu \!\!\!\!\!\! &= t^{\mu\nu}_{QQ}\,k\!\cdot\! Q\,  \tfrac{i}{2} [\gamma^\nu,\slashed{k}]\,, \\
                 \tau_3^\mu \!\!\!\!\!\! &= \tfrac{i}{2}\,[\gamma^\mu,\slashed{Q}]\,, \\
                 \tau_4^\mu \!\!\!\!\!\! &= \tfrac{1}{6}\,[\gamma^\mu, \slashed{k}, \slashed{Q}]\,,
                 \end{array}\quad
                 \begin{array}{rl}
                 \tau_5^\mu \!\!\!\!\!\! &= t^{\mu\nu}_{QQ}\,ik^\nu\,, \\
                 \tau_6^\mu \!\!\!\!\!\! &= t^{\mu\nu}_{QQ}\,k^\nu \slashed{k}\,, \\
                 \tau_7^\mu \!\!\!\!\!\! &= t^{\mu\nu}_{Qk}\,k\!\cdot\! Q\,\gamma^\nu\,, \\
                 \tau_8^\mu \!\!\!\!\!\! &= t^{\mu\nu}_{Qk}\,\tfrac{i}{2}\,[\gamma^\nu,\slashed{k}]\,,
                 \end{array}
             \end{split}
             \end{equation}
             where $t_{AB}^{\mu\nu} = A\cdot B\,\delta^{\mu\nu} - B^\mu A^\nu$ and the three-commutator
             is defined in Eq.~\eqref{three-commutator}.
             Transversality and analyticity are manifest in this basis. It is also `minimal' in the sense
             that the eight tensors have the lowest possible powers in the photon momentum:
             $\tau_3^\mu$, $\tau_4^\mu$ and $\tau_8^\mu$ are linear in $Q^\mu$ whereas all others depend on higher powers.
             As a consequence, the dressing functions $f_j(k^2, z^2, Q^2)$ are free of kinematic singularities and constraints
             and go to constants as $Q^2\to 0$ or $k^2 \to 0$.
             A simple parametrization for the $f_j$ from the rainbow-ladder solution of the quark-photon vertex can be found in Ref.~\cite{Eichmann:2014qva};
             it turns out that~$f_3$, which encodes the anomalous magnetic moment of the quark, is practically zero therein.

       \paragraph{Dirac and Pauli form factors.}
       We would further like to establish a connection between the offshell fermion-photon vertex $\Gamma^\mu(k,Q)$ and the onshell Dirac and Pauli form factors $F_1(Q^2)$ and $F_2(Q^2)$.
       On the one hand, the nucleon's electromagnetic current~\eqref{ffs-nucleon-current-1} can be derived as the onshell projection of an offshell nucleon-photon vertex with the form given above.
       On the other hand,  in quark models Dirac and Pauli form factors often serve as a means to introduce structure into the photon's coupling to the quark,
       assuming that the quarks can be expressed in terms of onshell spinors. Thus, relating them with the offshell vertex that appears
       in the current-matrix element in Fig.~\ref{fig:current-faddeev} also helps to connect the quark model with QCD.

        The onshell current is obtained by sandwiching the offshell vertex between positive-energy projectors
        and taking the fermion momenta onshell:
        \begin{equation}\label{onshell-nucleon}
           \mc{J}^\mu(k,Q) = \Lambda_+(k_+)\,\Gamma^\mu(k,Q)\,\Lambda_+(k_-)\big|_\text{onshell} \,,
        \end{equation}
        where the limit $k_+^2=k_-^2=-m^2$ entails $k^2=-m^2-Q^2/4$ and $k\cdot Q=0$, and so the only remaining independent variable is $Q^2$.
        In principle the above form follows from a pole condition: in a hadronic loop diagram the offshell nucleon-photon vertex is contracted with nucleon propagators,
        whose residues on the mass shell are proportional to the positive-energy projector $\Lambda_+$ defined in Eq.~\eqref{positive-energy-projector}.
        For the analogous quark-photon vertex this is not legitimate since the quarks do not have a mass shell, but let us nevertheless continue under such an assumption.
        First of all, note that $\mc{J}^\mu(k,Q)$ is automatically transverse because also the Ball-Chiu part~\eqref{qpv-bc-2} becomes transverse in the projection; for example:
        \begin{equation}
            Q^\mu\,\Lambda_+(k_+) \,\gamma^\mu \Lambda_+(k_-) \big|_\text{onshell} = 0\,,
        \end{equation}
        and the same is true for the remaining tensors in Eq.~\eqref{qpv-transverse-basis} because $k\cdot Q=0$ on the mass shell.
        As a result, the 12 offshell tensor structures 
        collapse into two tensors and the current takes the standard Dirac form:
             \begin{equation}\label{current-onshell-standard}
                 \mc{J}^\mu(k,Q) =  i\Lambda_+(k_+)\left[ F_1(Q^2)\,\gamma^\mu + \frac{iF_2(Q^2)}{4m}\,[\gamma^\mu,\slashed{Q}]\right]\Lambda_+(k_-)\,.
             \end{equation}
        The Dirac and Pauli form factors $F_1(Q^2)$, $F_2(Q^2)$
        are related to the offshell dressing functions via~\cite{Eichmann:2012mp}
             \begin{equation}\label{qpv-onshell-dressing-functions}
             \begin{split}
                 F_1(Q^2) &= \Sigma_A + 2m\,(\Delta_B-m \Delta_A) + Q^2 \left[ f_1 - \frac{f_4}{2} - m \left(f_5 + m f_6 - \frac{f_8}{2}\right) \right] , \\
                 \frac{F_2(Q^2)}{2m} &= f_3-m f_4 - (\Delta_B-m \Delta_A) + \frac{Q^2}{2} \left( f_5 + m f_6 - \frac{f_8}{2}\right) ,
             \end{split}
             \end{equation}
        with the dressing functions evaluated at the onshell point.
         On the mass shell, however, the fermion propagator is that of a free particle which implies via Eq.~\eqref{SigmaA-DeltaA}:
        \begin{equation}
            \Sigma_A = A(-m^2)=1, \qquad
            \Delta_A = A'(-m^2) = 0, \qquad
            \Delta_B = B'(-m^2) =0
        \end{equation}
        and therefore the combination $f_3-m f_4$ is what contributes to the anomalous magnetic moment at $Q^2=0$.

       \paragraph{Hadronic vacuum polarization.}
       As explained in Sec.~\ref{sec:ff-vertices}, the hadronic vacuum polarization is the simplest observable that tests the structure of the quark-photon vertex,
       since apart from the vertex it only involves the dressed quark propagator and the tree-level vertex.
       Inserting the expressions~\eqref{qpv-bc-2} and~\eqref{qpv-transverse-basis} into Eq.~\eqref{hvp-0} and taking traces yields
       \begin{equation}\label{hvp-2}
       \begin{split}
           \Pi(Q^2) &= \frac{1}{(2\pi)^3}\int_0^{\Lambda^2} dk^2\,k^2\int_{-1}^1 dz\sqrt{1-z^2}\,\sigma_v(k_+^2)\,\sigma_v(k_-^2)\,\Big( K_\text{BC} + \sum_{j=1}^8 f_j(k^2,z^2,Q^2)\,K_j\Big), \\
           \widetilde\Pi(Q^2) &= \frac{1}{(2\pi)^3}\int_0^{\Lambda^2} dk^2\,k^2\int_{-1}^1 dz\sqrt{1-z^2}\, \sigma_v(k_+^2)\,\sigma_v(k_-^2)\,\widetilde K_\text{BC}\,,
       \end{split}
       \end{equation}
       where the kernels coming from the Ball-Chiu part are given by
       \begin{equation}
       \begin{split}
           K_\text{BC} &= -\frac{\Sigma_A}{2} \, ( 1 + 4k^2 z^2 \Delta_M^2) - \Delta_A\,k^2 z^2\,\mathbf{Y}  -2k^2\,\frac{1-4z^2}{3Q^2}\,(\Sigma_A \mathbf Y + \Delta_A \,\mathbf X_+)\,, \\
          \widetilde K_\text{BC} &= \Sigma_A\,( \mathbf X_+ - 2k^2 z^2\,\mathbf Y) -  k^2 z^2 \Delta_A\left( 2\mathbf X_+ - Q^2 \,\mathbf Y\right)
       \end{split}
       \end{equation}
       and those from the transverse part read
       \begin{equation}
        \begin{array}{rl}
          K_1 &= \mathbf X_- - w \,, \\
          K_2 &= -k^2 \,Q^2 \,z^2\,(\mathbf Z-2w\,\Delta_M), \\
          K_3 &= -\mathbf Z\,, \\
          K_4 &= -w\,,
       \end{array}\qquad
       \begin{array}{rl}
          K_5 &= w\,\Sigma_M\,, \\
          K_6 &= w\,\big(\tfrac{1}{2}\,\mathbf X_- - k^2\big), \\
          K_7 &= k^2  \,z^2\,\mathbf X_-, \\
          K_8 &= w\,\Sigma_M-k^2 \,\mathbf Z\,,
       \end{array}\qquad\quad
          w = \frac{2k^2}{3}\, (1-z^2)\,.
       \end{equation}
       $\Sigma_M$ and $\Delta_M$ are defined from the quark mass function in analogy to~\eqref{SigmaA-DeltaA} and we abbreviated
       \begin{equation}
          \mathbf X_\pm = k^2 \pm \frac{Q^2}{4} + \Sigma_M^2  \pm k^2\,z^2\,Q^2\,\Delta_M^2 \,, \qquad
          \mathbf Y = 1 + 2\,\Sigma_M\,\Delta_M\,, \qquad
          \mathbf Z = \Sigma_M - 2k^2\,z^2\,\Delta_M\,.
       \end{equation}
       The kernels only depend on the quark propagator's dressing functions $A(k^2)$ and $M(k^2)$.
       In principle $\widetilde\Pi(Q^2)$ must vanish due to gauge invariance; however, with a momentum cutoff it is nonzero and produces a quadratic divergence.
       For a tree-level propagator with $A(p^2)=Z_2$ and $M(p^2)=m_q$ one has
       \begin{equation}
          \Sigma_A=Z_2, \qquad \Sigma_M=m_q, \qquad \Delta_A=\Delta_M=0, \qquad \mathbf X_\pm = k^2 \pm \frac{Q^2}{4} + m_q^2, \qquad \mathbf Y = 1,
       \end{equation}
       which yields
       \begin{equation}
          K_\text{BC} = -Z_2\left(\frac{1}{2}+2k^2\,\frac{1-4z^2}{3Q^2}\right), \qquad \widetilde K_\text{BC} = Z_2\left( k^2  + \frac{Q^2}{4} + m_q^2 - 2\,k^2 z^2\right),
       \end{equation}
       and upon neglecting the transverse part of the vertex one recovers the expression for the vacuum polarization in perturbation theory.
       Note that the $1/Q^2$ term in $K_\text{BC}$ is not singular at $Q^2\to 0$ because
       \begin{equation}
          \int_{-1}^1 dz \sqrt{1-z^2}\,(1-4z^2) = 0\,.
       \end{equation}



\bibliography{baryons}

\end{document}